\documentclass[12pt,twoside,openright]{book}
\usepackage[spanish,english]{babel}
\usepackage[latin1]{inputenc}
\usepackage{axodraw4j}
\usepackage{feynarts}
\usepackage{pstricks}
\usepackage{color}
\usepackage[T1]{fontenc}
\usepackage[a4paper]{geometry}
\usepackage{verbatim}
\usepackage{pifont}
\usepackage{varioref}
\usepackage{float}
\usepackage{url}
\usepackage{amssymb}
\usepackage{graphicx}
\usepackage{setspace}
\usepackage{amsmath}

\makeatletter

\newcommand{\noun}[1]{\textsc{#1}}
\newcommand\Var[1]{{\ensuremath{\mathit{#1}}}}
\newcommand{\mathsym}[1]{{}}
\newcommand{\unicode}[1]{{}}
\newcommand{\bc}{\begin{center}}
\newcommand{\ec}{\end{center}}

\newcommand{\be}{\begin{eqnarray}}
\newcommand{\ee}{\end{eqnarray}}
\newcommand{\bi}{\begin{itemize}}
\newcommand{\ei}{\end{itemize}}

\newcommand{\Mtexp}{173.34}

\newcommand{\Mhexp}{125.15}

\newcommand{\MWexp}{80.384}
\newcommand{\MWerr}{0.014}

\newcommand{\Mtdiff}{ \bigg(\frac{m_t}{{\rm GeV}}-\Mtexp\bigg)}
\newcommand{\asdiff}{\, \frac{\alpha_3(m_Z)-0.1184}{0.0007} }
\newcommand{\Mhdiff}{\bigg(\frac{m_h}{{\rm GeV}}-\Mhexp\bigg)}
\newcommand{\MWdiff}{\frac{m_W - \MWexp{\rm GeV}}{\MWerr{\rm GeV}}  }

\usepackage{lettrine}
\usepackage{hyperref}

\AtBeginDocument{

}

\makeatother

\usepackage{babel}

\title{ \vspace*{-1in}
\begin{figure}[htb]
\begin{center}
\includegraphics[scale=0.2]{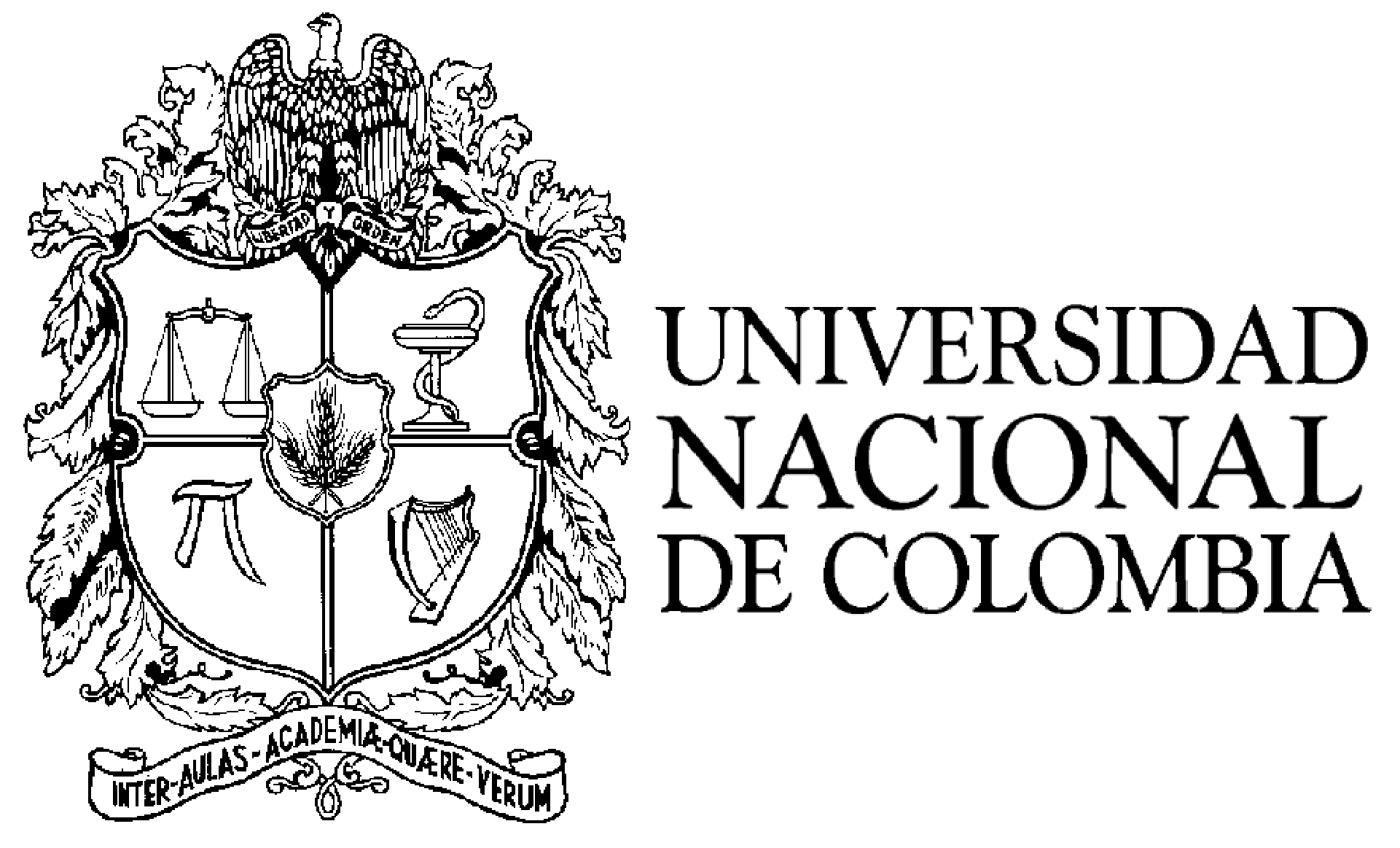}
\end{center}
\end{figure} \textbf{The Higgs Boson at LHC and the Vacuum Stability of the Standard Model}}

\author{\textbf{Edilson Alfonso Reyes Rojas}\\
\\
\\
\\
A thesis submitted to the Universidad Nacional de Colombia\\
for the degree of Magister en Ciencias F\'{i}sicas\\
 in the Faculty of Sciences \\
\\
\\
\\
Supervisor:\\
\textbf{PhD. Angelo Raffaele Fazio} \\
\\
\\
\\
\rule{80mm}{0.1mm}\\
Universidad Nacional de Colombia\\
Facultad de Ciencias, Departamento de F\'{i}sica \\
Grupo de Campos y Part\'{i}culas\\
Bogot\'{a} D.C. Colombia}

\begin{document}

\maketitle

\pagenumbering{Roman}

\section*{Acknowledgements}
\addcontentsline{toc}{chapter}{Acknowledgments} 
\selectlanguage{spanish}
Primero me gustar\'{i}a agradecer a mi director de tesis A. Raffaele Fazio, por brindarme la oportunidad de investigar junto con \'{e}l en el \'{a}rea de la teor\'{i}a de campos cu\'{a}nticos. El me ha instruido durante la etapa mas importante de mi formaci\'{o}n acad\'{e}mica, brind\'{a}ndome su conocimiento en tres cursos de f\'{i}sica te\'{o}rica y su apoyo incondicional no solo cuando he necesitado de su asesor\'{i}a profesional sino tambi\'{e}n personalmente. Agradezco toda la paciencia que ha tenido conmigo en este largo proceso, sobretodo con las dificultades que se nos han presentado. Mas que un director de tesis a sido un gran amigo que me ha brindado su apoyo y siempre confi$\acute{o}$ en mi incluso cuando yo mismo he dejado de hacerlo.

Por otra parte quiero agradecer a varias personas que han contribuido con mi formaci\'{o}n acad\'{e}mica. Al profesor Gino Isidori, quien amablemente me invito a realizar una pasant\'{i}a en la Universidad de Zurich, me ayudo a esclarecer varios conceptos del an\'{a}lisis de la estabilidad del vac\'{i}o y me dio nuevas perspectivas de investigaci\'{o}n para el futuro. Al profesor Stefano Di Vita quien siempre estuvo abierto a discutir con nosotros v\'{i}a skype los detalles del calculo que realizamos en este trabajo. Por ultimo a los profesores S. Martin y A. Djouadi por sus amables discusiones en la escuela de verano de f\'{i}sica de part\'{i}culas realizada en el ICTP. 

Deseo agradecer a las instituciones que me han apoyado financieramente. En primer lugar agradezco a la Universidad Nacional por apoyarme otorg\'{a}ndome tres becas auxiliar docente y 
el grant de la DIB: Convocatoria del Programa Nacional de Proyectos Para el Fortalecimiento de la Investigaci\'{o}n, la Creaci\'{o}n y la Innovaci\'{o}n en Posgrados de la Universidad Nacional de Colombia 2013-2015. En segundo lugar quiero agradecer al ICTP por su total apoyo financiero para participar en la escuela: Summer School of Particle Physics - 2013, esto me permiti\'{o} ademas participar en la escuela de f\'{i}sica subnuclear realizada en Erice: The 51st International School of Subnuclear Physics. Por ultimo agradezco a la universidad de Zurich por su apoyo financiero para asistir a una pasant\'{i}a de 15 d\'{i}as realizada con el profesor Gino Isidori, donde fui invitado a un workshop acerca del estado actual del an\'{a}lisis de la estabilidad del vac\'{i}o en el modelo est\'{a}ndar. 

Agradezco a mi familia, que por ser bastantes personas no puedo mencionar individualmente, pero que tambi\'{e}n han sido un apoyo en varias situaciones de mi vida a pesar de las diferencias que hemos tenido y de habernos distanciado tanto. En particular, agradezco a mi hermano Diego, matem\'{a}tico de la Universidad Nacional, por sus diversas explicaciones en esta \'{a}rea del conocimiento.

Finalmente quiero agradecer y dedicar esta tesis a Susana, la persona mas importante y el amor de mi vida, sin ti no tendr\'{i}a sentido nada de lo que estoy haciendo, todo lo hago con el fin de construir el mejor futuro juntos. 

\newpage
\selectlanguage{english}
\section*{Abstract}\addcontentsline{toc}{chapter}{Abstract} 
The main aim of this work is to study the conditions of absolute vacuum stability within the Standard Model (SM) by the knowledge of the behaviour of the Higgs quartic coupling $\lambda$ up to high energy scales and using the new data on the Higgs mass given by ATLAS and CMS  ($m_H\approx 125 - 126$ GeV) as an input parameter. The Higgs mass value observed by ATLAS and CMS leads to a negative value of the coupling $\lambda$ at energies around the scale $\sim10^{10}$~GeV, making metastable the vacuum of the Standard Model, as it is seen by the renormalization group improved (RGI) effective potential. The stability status of SM crucially depends upon the precise values of the top and Higgs masses, a more precision determination of those masses and related uncertainties can modify drastically our conclusions about the SM stability properties. For this reason, we have computed the vacuum bubbles and the Higgs tadpole diagrams, at two-loop level of accuracy, in a renormalization scheme proposed by A. Sirlin and R. Zucchini, where the input parameters are obtained in terms of physical observables related with muon decay, and where the threshold effects are included. In particular, we focus on the analytic computation of the Higgs tadpoles contributions by two different methods. From one side we have computed the sum of the tadpoles as the first derivative of the two-loop Higgs 1PI effective potential in the Sirlin-Zucchini scheme, on the other side we have checked the obtained result toward the direct diagrammatic two-loop computation, by proposing a way of automatization of our procedure based on the generation of Feynman diagrams, reduction of their integrands and evaluation of scalar integrals and sum of all contributions in a non-redundant way. We used the code TARCER that implements the Tarasov method to reduce two-loop tensorial integrals.


\begin{singlespace}
\index{Table of Contents}{\footnotesize \tableofcontents{}}{\footnotesize \par}
\end{singlespace}

\chapter{\noun{\label{cha:Introduction}\index{Introduction}}Introduction}
\pagenumbering{arabic}

\lettrine{O}{ ne} of the most important advances of the last century has been the development of unified models of the weak, electromagnetic and strong interactions within the framework of a gauge theory. The Standard Model (SM) is a quantum field theory invariant under the direct product gauge group of local symmetry $SU(3)_C\otimes SU(2)_{EW}\otimes U(1)_{Y}$, that describes three of the fundamental interactions of nature. Due to the short range of the weak nuclear forces, the vectorial $SU(2)$ gauge bosons $W^{\pm}$ and $Z^{0}$ have to be massive, the perturbative unitarity of the theory can be only preserved by gauge symmetry. In the electro-weak SM of Weinberg \cite{Weinberg}, Glashow \cite{Glashow} and Salam \cite{Salam} (WGS model) the generation of masses in a unitary theory is performed by the Higgs mechanism of spontaneous symmetry breaking (SSB). The Higgs mechanism \cite{Higgs}, called by Peter Higgs himself, the ABEGHHH'tH (by Anderson, Brout, Englert, Guralnik, Hagen, Higgs, Kibble and 't Hooft), owing to the numerous developments of the original idea of Anderson (1962) in the condensed matter physics, introduces in the Lagrangian of EWSM a complex doublet scalar field that generates masses for the three gauge bosons $W^{\pm}$ and $Z^{0}$ while the photon remains massless. 
 
The scalar sector contains the so called  Higgs boson, a neutral spinless particle whose self-interaction potential is dictated by the spontaneous symmetry breaking mechanism. No predictions are given by the theory about the Higgs mass. The detection of the Higgs boson has been the more popular goal of the particle physics in the last thirty years and one of the strong motivation for the LHC CERN program. A new particle with a mass of $125.66 \pm 0.34$ GeV with the characteristics of the Higgs boson of the SM has been discovered in July 2012 by the CMS \cite{CMS-II} and ATLAS \cite{ATLAS-II} collaborations. The discovery was based primarily
on mass peaks observed in the photon-to-photon ($\gamma\gamma$) fusion  and $ZZ \rightarrow \ell^+\ell^-\ell'^+\ell'^-$ 
decay channels, where one or both of the Z bosons can be off-shell and where $\ell$ and $\ell'$ denote an electron or muon. Other decay channels have been measured in  \cite{ATLAS-III, CMS-III}  and no significative signs of new phenomena have been discovered yet. Indeed, the new resonance seems to be the SM Higgs boson i.e. the couplings of this particle with the other SM particles are compatible with the SM predictions, so far no clear deviation from the SM Higgs properties have been detected at the LHC also concerning its spin and parity, being the Higgs boson a $J^P=0^+$ particle \cite{Atlas4}\cite{Atlas5}\cite{CMS4}\cite{CMS5}. With $m_H$ known, all properties of the SM Higgs boson, such as its production cross section and partial decay widths, can be predicted in particular the self quartic Higgs coupling, supposing to be SM like (meaning that it is currently only indirectly measured). If no new phenomena appear at the TeV scale, we can try to explain fundamental problems, like the vacuum stability or naturalness problem, only in terms of the SM. This theoretical possibility, known as the desert scenario, is currently largely explored and is an interesting source of hints of what kind of new physics can emerge for different scales of energy. 

In this thesis we focus in the vacuum stability problem of the minimal Standard Model. This issue can be faced by studying the scale-dependent properties of the Higgs effective potential in the absence of new physics at sub-Planckian energies. For energies higher than electroweak scale the analysis of vacuum stability is reduced to the study of the renormalization group evolution of the Higgs quartic coupling $\lambda$. The observed Higgs mass $m_H=125.66 \pm 0.34$ GeV  (or the more recent estimation from the combined ATLAS and CMS analysis \cite{comb} $m_H= 125.09\pm 0.24$ GeV) leads to a negative value of the Higgs quartic coupling $\lambda$ at some energy scale below the Planck scale, making the Higgs potential instable or metastable. Since the measured value of $m_H$ is in a window of parameters where the Standard Model can be extrapolated till the Planck scale with no problem of consistency but also that instability could arise, a highly precise analysis for the vacuum stability becomes mandatory. With the inclusion of the three-loop RG equations and two-loop matching conditions, the instability scale occurs at $\Lambda_{I}\approx 10^{11}$ GeV (well below the Planck scale) meaning that at that scale the effective potential starts to be unbounded from below or that a new minimum can appear and there is a non-trivial transition probability to that minimum \cite{Espinosa}. The analysis is done under the assumption that new physics shows up only at very high energy scales, possibly the Planck scale. Moreover, it is assumed that, despite the presence of these new physics interactions, the tunneling rate probability can be calculated with the potential obtained with SM interactions only, the contributions coming from very high scale physics should be suppressed. The experimental value of the Higgs mass prospects, actually, scenarios which are at the border between the absolute stability and the metastability, the measured value of $m_H$ puts the SM in the so-called near-critical position. Because of the present experimental uncertainties on the SM parameters (mostly the top quark mass) we cannot conclusively establish the fate of the EW vacuum, although metastability is now preferred at $99.3\%$ CL. The above statement is the motivation for making a refined study, at the next-to-next to leading order (NNLO) of the vacuum stability problem. 

There are some technical problems when one makes a NNLO stability analysis. Up to two-loop level the threshold corrections for all SM parameters need the evaluation of mass dependent radiative corrections like the contributions due to the Higgs tadpoles $T$, the scalar, fermion and vector self-energies, the vertices $V_{W}$ and boxes $B_{W}$ appearing in the radiative corrections to the muon decay process at zero external momentum, etc. The computation of these topologies are hampered by the renormalization of two-loop tensor Feynman integrals \footnote {In the subsequent chapters we will describe in details how those contributions arise.}. The evaluation of two-loop integrals leads to divergent quantities that have to be absorbed in a suitable renormalization procedure, in order to be able to predict measurable quantities. 
Contrarily to what happens at one-loop, at two-loop level and beyond, the non-local divergences of the Feynman integrals are mixed and the procedures learned in a standard course of quantum field theory to remove the divergences usually don't apply. Moreover due to the complicated structure of the integrals and the large number of diagrams at two-loop level, typically thousands, an explicit computation of the many topologies involved in the stability analysis and its renormalization are not found in the literature, where the results are presented numerically. For this master thesis in order to learn and getting expertise with the techniques of two-loop calculations for the high energy physics at high level of precision we decided to address the problem of the vacuum stability of the SM looking at the simplest radiative corrections provided by the two-loop sum of the Higgs tadpoles. Their simplicity is due to the fact that they are not affected by the infrared divergences. They are simple but not trivial, as we hope to be able to communicate in this work. The calculation that this thesis affords is non-trivial, first of all because it is a two-loop one and therefore it is affected by numerous technicalities at many levels. There is the need of generating the one-particle-irreducible (1PI) tadpoles, then one has to reduce and evaluate the related tensorial integrals representing those diagrams, this requires to become familiar with the huge number of codes available in net. We choose TARCER, that will be described in the following chapters. Anyway a two-loop calculation is insidious, it is not a simple operation of cutting divergences as it could appear at one loop. It is not obvious how to extract and treat nested divergences, it is therefore quite easy to make small mistakes that the poles of the dimensional regularization do not cancel. In this specific case the sum of the Higgs tadpoles is not a physical quantity so in principle the poles do not cancel. Therefore at the stage of planning the activities for this thesis we asked ourself, how could we be sure that our diagrammatical result is correct. 

In the advanced  graduate courses of quantum field theory that the author of this thesis took at Universidad Nacional de Colombia in Bogot\'a the relation between the 1PI effective potential and the zero-momentum 1PI Green functions was studied and we decided to exploit that knowledge. The two loop effective potential computed in the Minimal Modified Subtraction Scheme $\overline{MS}$ is known \cite{Ford-Jones} \cite{Ford-Jack}, however the need of studying the vacuum stability as depending on the threshold corrections made the renormalization scheme $\overline{MS}$ unsuitable and an on-shell scheme was preferred, the so called Sirlin-Zucchini (SZ) scheme. The job has been therefore to make the appropriate shifts to pass from the two-loop effective potential in $\overline{MS}$ to the one in SZ, again straightforward but not trivial. The sum of the two-loop Higgs tadpoles in this scheme has been obtained from the first derivative of the Higgs effective potential with respect to the classical Higgs field. The operation is straightforward but not-trivial, since the effective potential is made of complicated functions and there are dedicated tricks to make the first derivative. Finally our aim is to compute the non-trivial and fundamental contribution of the two-loop Higgs tadpoles sum to study the SM vacuum stability and we perform that calculation by using two methods. One diagrammatical and the other exploiting the functional methods in quantum field theory. We personally believe in fact that the criterium of self consistency of the results in research are the best way to be sure about the new obtained results and to probably guess where and how to go to continue our research.  

The thesis is organized as follows. In Chapter \ref{cha:Effective Potential} we firstly address the consistency of the Standard Model within the perturbative unitarity, then we present the main properties of the scalar sector of the SM. We investigate the consequences to break local gauge invariance of an Abelian theory, and then we develop the Higgs mechanism of Spontaneous Symmetry Breaking (SSB) in the SM. Later, we analyse two different methods to compute the 1PI effective potential and we compute the effective potential of the SM at one-loop level. Finally, we consider the limit of validity of the effective potential and we improve the 1PI effective potential by using the tools of the renormalization group, to extend the analysis of the spontaneous symmetry breaking within perturbation theory also where the amplitudes of the classical fields are large.

In Chapter \ref{cha:The Stability Problem of the SM} we study the scale-dependent properties of the SM. We discuss the problem of SM vacuum stability and its consequences by using the effective potential approach, with the ATLAS and CMS values for the Higgs mass. In particular, we show that the stability analysis can be reduced to the study of the positivity of the coupling $\lambda$ for values of the classical field higher than the electroweak scale. Finally we review the current status of some additional constraints to the Higgs mass related with the high energy behaviour of the running couplings: the triviality constraint and the hierarchy problem.

The evolution of $\lambda$ implies to solve a system of ordinary differential equations and to impose boundary conditions over the input parameters. The boundary conditions are imposed from the relations between the running coupling constants $g_{i}(\mu)$ and the relevant physical observables. In Chapter \ref{cha:Sirlin-Zucchini} we determine the relation between the running parameters $g_{i}(\mu)$ in the modified minimal subtraction ($\overline{MS}$) scheme and the physical parameters related to the on-shell scheme proposed by A. Sirlin and R. Zucchini (SZ scheme), where the principal feature is the use of the Fermi constant of the muon decay as an input parameter. We expose the details of the Sirlin-Zucchini renormalization scheme up to two-loop level. In particular, the problem with the renormalization of the top quark mass is discussed.

In Chapter \ref{cha:TwoLoopCalc} we review the technical details to compute the two-loop tensor Feynman integrals arising from the threshold corrections to the running couplings determined in Chapter \ref{cha:Sirlin-Zucchini}. In particular, we expose the Tarasov method to reduce tensorial integrals to a superposition of some scalar master integrals. Later, the Tarasov method is implemented using the TARCER code of Mathematica, to compute the $\overline{MS}$ effective potential. This computation is needed in the Chapter \ref{cha:TwoLoopTadpoles} to obtain the two-loop tadpoles in the Sirlin-Zucchini scheme. For this reason we present in detail the computation of all sectors of the $\overline{MS}$ effective potential.
  
In Chapter \ref{cha:TwoLoopTadpoles}  we compute the two-loop effective potential in the on-shell renormalization scheme of Sirlin Zucchini, where the threshold corrections of the SM couplings are taken into account, and we derive the two-loop tadpole contribution to the threshold relation between the Higgs quartic coupling $\lambda$ and the Higgs mass $m_{H}$, using two methods: from the first derivative of the Higgs effective potential with respect to the classical field and diagrammatically also automating by the code TARCER. This is the main newest contribution provided by this thesis.

The thesis contains eight appendices. To fix the notation which will be used later on, we present in Appendix \ref{AppSM} a brief introduction to the Standard Model (SM) of the strong and electroweak interactions. First we introduce the SM Lagrangian before the spontaneous symmetry breaking and then we give the kinetic and interaction terms of the Lagrangian at the broken phase. In Appendix \ref{AppBetaF-EP} we present the main renormalization group tools, necessary in the construction of any vacuum stability analysis using the effective potential approach. Appendix \ref{AppFeynArts} is dedicated to expose the principal functions of the FeynArts code to draw Feynman diagrams and generate its respective amplitudes. In Appendix \ref{app-counterterms} we summarize the one-loop corrections to the threshold SM parameters and its numerical values at the EW scale. Appendix \ref{AppTarcer} contains a short summary of the operation of TARCER code, a computer implementation in Mathematica of the Tarasov procedure to reduce tensorial integrals, we expose its operation with a simple example, the calculation of the scalar sector of the Higgs Tadpoles at two-loop level. Appendix \ref{AppDyQ} contain the details of the derivation of the $\alpha$-parametric representation for any m-point two-loop Feynman integral with a tensorial structure, this derivation is fundamental in the construction of the Tarasov method to reduce tensorial integrals exposed in Chapter \ref{cha:TwoLoopCalc}. Finally, the appendices G and H contain the evaluation of the more relevant master integrals used in this work. The two-point one-loop integral $B_{11}^{(d)}$ and its $\varepsilon$-expansion up to first order in $\varepsilon$ is included in Appendix \ref{App1lEpsExpansion}, whereas the evaluation of the zero-point two-loop integral $J_{111}^{(d)}$ is included in Appendix \ref{AppIntegralJ}. Both evaluations are non-trivial, for this reason we decided include here the details of the computation.         

\chapter{\noun{\label{cha:Effective Potential}\index{Effective Potential}}The Higgs Potential in the Standard Model}

\lettrine{I}{n} this chapter we investigate the consequences to break local gauge invariance of a pure abelian theory by introducing a mass term for the gauge bosons by hand. Furthermore, we will study the mechanism of electroweak symmetry breaking, which is known to generate the masses of the gauge vector bosons and fermion masses without spoiling renormalizability. Moreover, we will study the scalar sector of the Standard Model, and in particular we review the method of calculating radiative corrections to the classical potential which can have significant consequences in the stability of the vacuum of the theory. The requirement of vacuum stability leads to severe bounds on Higgs and fermion masses that will be studied in the next chapters. Finally, we will see that in almost all calculations in which the effective potential is needed, the region of field space is so large that the logarithmic terms make the loop expansion to be not reliable and thus using a renormalization group improved potential is essential. 

\section{A Gauge Theory of Weak Interactions}

The Fermi theory of weak interactions is known to correctly describe very low-energy phenomenology, typically up to an energy scale of $\frac{1}{\sqrt{G_F}}\sim 300 \rm{GeV}$, being $G_F$ the universal Fermi coupling constant of weak interactions. However, its extrapolation to higher energies provides inconsistent predictions with the quantum mechanical nature of the microscopic weak interactions phenomena. In fact, in the quantum mechanical description of nuclear weak phenomena by four spin-$\frac{1}{2}$ quantum fields, the unitarity of the S matrix is perturbatively broken at high energy \cite{Horejsi}. For instance, in the weak scattering process $e^{-}+\nu_{e}\rightarrow e ^{-}+\nu_{e}$ the unitarity relations implied by $SS^\dag=1$ are maintained at the lowest perturbative level if
\begin{equation}
E_{c.m.}=\sqrt{s}\leq \sqrt{s_0}\equiv\sqrt{\frac{4\pi}{\sqrt{2}G_{F}}}.
\end{equation} 
The critical value $\sqrt{s_0}$ at which the unitarity condition is saturated is usually called "unitarity bound". For the reaction in question, the unitarity bound amounts to $\sqrt{s_0}\sim 870\, \rm{GeV}$, meaning that for $E_{c.m.}>\sqrt{s_0}$ the tree total level cross section computed in Fermi theory, which for this process amounts to 
\begin{equation}
\sigma=\frac{2}{\pi}G_F^2s, 
\label{cross1}
\end{equation}
ceases to be a good approximation. It is worth to compare an unitarity bound of Fermi theory, typically at $\sqrt{s}\sim G_F^{-\frac{1}{2}}$, with the analogous in spinorial quantum electrodynamics (QED). For instance, the QED scattering process $e^-+e^+\rightarrow \mu^-+\mu^+$ does not break the unitarity at tree level whatever is the high energy regime one is using. Moreover the amplitudes of partial waves in spinor QED grow at most logarithmically with energy due the fact that QED is a renormalizable theory. The Fermi theory of weak interactions is, on the contrary, a non-renormalizable theory, by adopting the power counting criterium, the interaction has a coupling constant with the units of an inverse of a squared mass. Therefore, by simple dimensional arguments one discovers a power growth of tree level scattering amplitudes at high energy and consequently a non-renormalizability at higher orders of the perturbative expansion. 
The problem of the perturbative renormalizability and the perturbative unitarity are, in fact, deeply correlated, therefore one could try to construct a consistent theory of fundamental interactions based on the criterion of renormalizability and related cancellations of infinities. A part the possibility of looking at such cancellation by considering one-loop diagrams it is possible to construct a unitary and renormalizable theory by asking  asymptotic softness or "tree-unitarity". We believe it is important to spend some words about this subject, firstly because it elucidates the strong connection between renormalizability and unitarity and second because it naturally motivates the introduction of the gauge theories. One may require, again by dimensional analysis, that an arbitrary $n$-point tree level amplitude in the high energy limit $E\rightarrow\infty$ behaves (for fixed non-zero scattering angles) at most like 
\begin{equation}
A^{\rm tree}_n=O(E^{4-n}).
\label{tree}
\end{equation}
In a renormalizable theory the high-energy behaviour of the full amplitude $A_n$ has a power law character modified at higher order by an almost logarithmic rise
\begin{equation}
A_{n}|_{E\rightarrow\infty}=O(E^{4-n}\ln^k E)
\label{compl}
\end{equation}
where $k\geq 0$. Now, higher order diagrams are obtained, in a sense, as an iteration of tree diagrams. In fact, because of the optical theorem, the imaginary part of a one loop diagram may be expressed in terms of an appropriated tree level amplitude squared. From tree level and one-loop graphs one may get the imaginary part of a two loop diagram etc. Such an iteration procedure is a consequence of the perturbative unitarity of the $S-$ matrix. Thus, if the tree level amplitude for some two-to-two process behaved for $E\rightarrow\infty$ as $E^\delta$, where $\delta> 0$, then the imaginary part of a one-loop amplitude would behave like $E^{2\delta}$, growing faster than the lowest order approximation in the limit $E\rightarrow\infty$. From the imaginary part of a diagram one may calculate the full amplitude via dispersion relations; in doing this one has to perform appropriate subtractions in order to suppress ultraviolet divergences. Since the power-like growth law is at one-loop worse than that encountered  at tree level, in further iterations the power behaviour of the corresponding imaginary part gets worse, which necessitates introducing more subtractions in dispersion relations. This corresponds to an infinite number of renormalization counterterms. On the other hand, if for a two-to-two process the amplitude behaves for $E\rightarrow\infty$ like $A^{\rm tree}_4\sim O(1)$, the imaginary part of the one-loop amplitude will behave in the same way as the tree-level amplitude and there is a priory no manifest reason to expect that the character of power behaviour would be substantially changed at higher orders. However, it may happen that the high energy asymptotics of the real part of a one-loop amplitude is different from that of the imaginary part; so it may happen a situation in which the condition ($\ref{tree}$) is fulfilled but some loop amplitude grows as a positive power of energy for $E\rightarrow\infty$, as is the case of the Adler-Bell-Jackiw (ABJ) triangle anomaly. The unitarity requirement strongly motivates the existence of gauge symmetry at classical and quantum level.
Based on the previous criterion, it is possible to obtain the renormalizable and unitary WGS (Weinberg-Glashow-Salam) model built on a Yang Mills gauge theory which possesses local invariance under the action of $SU(2)_{L}\times U(1)_{Y}$ group and consequently it is a renormalizable theory. By the previous motivation the gauge invariance is not just an esthetic criterion, it is dictated by the perturbative unitarity requirement of the quantum theory. About the connection between the gauge invariance and the perturbative unitarity at very high energy, more will be said in the next section.  The Standard Model reduces itself to the Fermi theory in the low-energy limit because the four-fermion interaction Lagrangian provides the same results as the exchange of a massive vector boson with a momentum much smaller than its mass. 
The SM describes the electromagnetic and weak interactions between quarks and leptons, combined with the $SU(3)_{C}$ quantum  chromo-dynamics \cite{GellMann} of the strong interactions between quarks and gluons, provides a unified framework to describe those three fundamental forces of nature. A brief description of the kinds of fields (matter fields and gauge fields) and interactions of the Standard Model in the symmetry phase, before introducing the electroweak symmetry breaking mechanism where the gauge fields and the fermion fields been kept massless, will be discussed in the Appendix \ref{AppSM}. In this section we will focus our attention on the broken phase of the Standard Model, and then, on the mechanisms to give mass to the vector and fermion fields.  
  
\subsection{Masses for the Gauge Bosons and Renormalizability}

Consider the muon decay process $\mu\rightarrow e+\nu_\mu+\bar{\nu}_e$, its tree-level amplitude in Fermi's theory amounts 
\begin{eqnarray}
- \dfrac{G_{\mu}}{\sqrt{2}}\bar{\nu}_{\mu}\gamma^{\alpha}(1-\gamma_{5})\mu \bar{e}\gamma_{\alpha}(1-\gamma_{5})\nu_{e}, \label{eq: Fermi_muon}
\end{eqnarray}
where for each particle we have indicated the corresponding wave function, $G_{\mu}$ is the Fermi coupling constant, which is actually deduced from the muon lifetime. By including the first order electromagnetic corrections to (\ref{eq: Fermi_muon}) Sirlin found in \cite{ASirlin0} for the muon time decay 
\begin{eqnarray}
\frac{1}{\tau_\mu} = \frac{G_\mu^2 m_\mu^5}{192 \pi^3} 
\left( 1- \frac{8m_e^2}{m_\mu^2} \right) \left[1 + \frac{\alpha}{\pi}\left(\dfrac{25}{8} - \dfrac{\pi^{2}}{2} \right)  
+  (6.701 \pm 0.002) \left( \frac{\alpha} {\pi} \right)^2 \right],
\end{eqnarray}
where $\alpha$ is the fine structure constant. This leads to the precise value \cite{ParticleData}
\begin{eqnarray}
G_\mu = (1.16637 \pm 0.00001) \cdot 10^{-5}~{\rm GeV}^{-2}. \label{Gmuonvalue}
\end{eqnarray}
In the  Standard Model, the same process is through the exchange of a virtual W boson, giving rise to the amplitude
\begin{eqnarray}
- \dfrac{g^{2}}{2}(1+\Delta r)\left( \bar{\nu}_{\mu}\gamma^{\alpha}\dfrac{(1-\gamma_{5})}{2}\mu\right)\dfrac{1}{q^{2}-m_{W}^{2}} \left(  \bar{e}\gamma_{\alpha}\dfrac{(1-\gamma_{5})}{2}\nu_{e}\right), \label{eq: SM_muon}
\end{eqnarray}
where $q$ is the transferred momentum, $g$ is the $SU(2)$ coupling constant, $m_W$ is the mass of the gauge boson and $\Delta r$ includes the radiative corrections to the muon decay. 
The relation with Fermi theory is seen at low transferred  momentum, $q^{2}\ll m_{W}^{2}$,
\begin{eqnarray}
\dfrac{G_{\mu}}{\sqrt{2}}= \dfrac{g^{2}}{8m_{W}^{2}}(1+\Delta r). \label{muon-g}
\label{G}
\end{eqnarray}
The relation (\ref{G}) of the renormalized constant $g$ and $m_{W}$ with the universal constant $G_{\mu}$ of the weak interactions will play a central role in the Chapter \ref{cha:Sirlin-Zucchini}, where an on-shell renormalization scheme will be used to study the vacuum stability of the SM. The $1 fm$ range weak interactions are mediated by quite heavy vector bosons $m_W=80.385±0.015 \,GeV$ and $m_Z=91.1876±0.0021\, GeV$ \cite{ParticleData}, however an explicit mass term in gauge theory breaks explicitly the gauge invariance \cite{tHooft}. 
Suppose, in fact, to put by hand a mass term in QED for the photon, in a Lagrangian formulation we would have
\begin{eqnarray}
\mathcal{L}(A,\partial A)=-\frac{1}{4}(\partial^{\mu}A^{\nu}-\partial^{\nu}A^{\mu})(\partial_{\mu}A_{\nu}-\partial_{\nu}A_{\mu})+\frac{1}{2}m_{A}^{2}A^{\mu}A_{\mu}, \label{quadraticA}
\end{eqnarray}
the free propagator of the massive field $A$ in the momentum space is
\begin{eqnarray}
D_{\mu\sigma}(k)=\frac{i}{k^{2}-m_{A}^{2}}\left(-g_{\mu\sigma}+\frac{k_{\mu}k_{\sigma}}{m_{A}^{2}}\right) \label{PropA}
\end{eqnarray}
which behaves at high momentum like a constant, rather than vanishing as $k^{-2}$. Therefore the degree of superficial divergence $D$ of a Feynman diagram with a boson line provided by the field $A$ is, in $3+1$ space-time dimensions,
\begin{eqnarray}
D=4-\dfrac{3}{2}F^{E}-2B^{E}+V\left[\dfrac{3}{2}n_{f}+2n_{b}-4 \right], \label{DegreeS} 
\end{eqnarray}
where $V$ is the number of vertices, $F^{E}$ is the number of external fermionic lines, $B^{E}$ the number of external bosonic lines, and $n_{f}$, $n_{b}$ are, respectively, the number of internal fermionic and bosonic lines in a vertex.   
The degree $D$ decreases with increasing number of external lines. Therefore, if the last term in the right hand side of equation (\ref{DegreeS}) is zero or negative, then only a finite number of diagrams will be ultraviolet divergent, and the whole theory can be made finite by renormalizing only these divergent amplitudes, at any order in perturbation theory. The condition for renormalizability then becomes
\begin{eqnarray}
\dfrac{3}{2}n_{f}+2n_{b}\leq 4. \label{ConditionD}
\end{eqnarray} 
In our example of spinor QED with massive photon, there is only one kind of vertex with two fermionic lines, $n_{f}=2$, and one bosonic line, $n_{b}=1$. Therefore the condition (\ref{ConditionD}) breaks down and the theory is non-renormalizable (even if the fine structure constant is still dimensionless). 
The high energy behaviour of the propagator (\ref{PropA}), or better its residue at the physical pole, is also responsible for cross sections increasing with the square of the center of mass energy as (\ref{cross1}) in Fermi theory \cite{LS}, complemented by massive gauge bosons. In a process with massive gauge bosons external lines a violation of unitarity at high energy arises. From the tensor structure of the propagator (\ref{PropA}) we recognize that for the gauge boson momentum $k^\mu$ on shell, the tensor in (\ref{PropA}) is the projector onto the physical polarization states, therefore  we may attribute that bad increasing behaviour to the longitudinal component of the vector field. In fact, the polarization vector of the spin zero component is
\begin{equation}
\epsilon^\mu_L(k)=\left(\frac{|\vec{k}|}{m_A},\frac{k^0}{m_A}\frac{\vec{k}}{|\vec{k}|}\right)
\end{equation}
and in the high energy limit ($|\vec{k}|>>m_A$)
\begin{equation}
\epsilon^\mu_L(|\vec{k}|)=\frac{k^\mu}{m_A}+O\left(\frac{m_A}{k^0}\right).
\end{equation}
To restore the perturbative unitarity the decoupling of the longitudinal component of the gauge boson is needed. Consequently the non-appearence of such degrees of freedom in the physical spectrum, induces gauge invariance. The spin zero polarization corresponds to the freedom of redefining the vector field by a gauge transformation. Moreover in the massless case it is exactly the gauge invariance that allows the decoupling of the state with the longitudinal polarization. To guarantee unitarity a mechanism of gauge bosons mass generation preserving gauge invariance is required. In the next section we will study the Higgs mechanism. 

\subsection{Spontaneous Symmetry Breakdown}

To see how one can introduce a mass term for gauge vector bosons without
spoiling renormalizability, we are going to investigate the spontaneous
symmetry breakdown in the Electro-Weak Standard Model $SU(2)_{L}\times U(1)_{Y}$.
From the famous papers of Higgs~\cite{Higgs} and Englert~\cite{Englert} we have
learned that we must introduce scalar fields in the theory in order
to break spontaneously a gauge symmetry. The simplest way to do this
is by introducing in the Lagrangian of the theory ${\cal L}_{SM}$\footnote{The Lagrangian of SM before introduce the electro-weak symmetry breakdown was obtained in the Appendix A.} the scalar sector $L_S$
\begin{eqnarray}
L_{S}=\left(D_{\mu}\phi\right)\left(D^{\mu}\phi\right)^\dag-U(\phi),
\end{eqnarray}
where the scalar field $\phi$ is a doublet of $SU(2)$ 
\begin{eqnarray*}
\phi=\left(\begin{array}{c}
\phi^{+}\\
\phi^{0}
\end{array}\right) & ; & Y(\phi)=1.
\end{eqnarray*}
$D_{\mu}$ is the covariant derivative defined by
\begin{eqnarray*}
D_{\mu}=\partial_{\mu}-i\frac{g}{2}\tau^{a}W_{\mu}^{a}-i\frac{g'}{2}B_{\mu}.
\end{eqnarray*}
$\tau^a$ are the Pauli's matrices and $Y(\phi)=2(Q-T_{3})$ denotes the quantum number of the weak hypercharge introduced in the Appendix
A. The first component of the doublet, $\phi^{+}$,
is a complex field that has electric charge $Q=1$ and third component
of weak isospin $T_{3}=1/2$, whereas the component $\phi^{0}$ has
the quantum numbers: $Q=0$ and $T_{3}=-1/2$ and it is a real field. 
Moreover, $U(\phi)$ is the so-called scalar classical potential, it
is some function of the field $\phi$, but not of their derivatives%
\footnote{Because of the requirement of translational invariance, we can neglect
the derivative terms. %
}. The gauge invariance and renormalizability constrain the scalar
potential to be of the form
\begin{eqnarray}
U(\phi)=\mu^{2}\left|\phi\right|^{2}+\lambda\left|\phi\right|^{4},
\end{eqnarray}
where $\lambda$ is a positive number, as requested for having an inferiorly bounded spectrum of the Hamiltonian,  instead $\mu^2$ can be either positive
or negative. If $\mu^{2}$ is positive, the potential has a minimum
for $\phi=0$, the symmetry of the Lagrangian is manifest, and $\mu^{2}$
is the mass of the scalar meson $\phi$. If $\mu^{2}$ is negative,
then $\mu^{2}$ can't be interpreted as a mass squared for the field
$\phi$, and the potential has now an infinite number of degenerate
minima, given by all those field configurations for which
\begin{eqnarray}
\left|\phi\right|^{2}=-\frac{\mu^{2}}{2\lambda}\equiv\frac{1}{2}\mathbf{v}^{2}.
\end{eqnarray}
All these minimum configurations are connected by gauge transformations
under the gauge group $SU(2)$, which changes the phase of the field $\phi$ without affecting
its modulus. Hence, which one we choose as the minimum is irrelevant,
but once we have chosen one, the gauge symmetry is broken. The Lagrangian is still gauge invariant, nevertheless the vacuum state, that corresponds to the field configuration minimizing the scalar potential is not gauge invariant. This phenomenon is called \emph{Spontaneous Breakdown of Symmetry} and its implementation is known as Higgs mechanism \cite{Higgs, Englert}. A very deeply review of the Higgs mechanism is given by Coleman in the book "Secret Symmetry" \cite{Colemanbook} and summarized by Sher in \cite{M-Sher}. It is worth to remark that the space time of the quantum field theory in question is formally infinite, therefore, considering the quantum version of the problem, the quantum probability of tunnelling among this set of infinite vacuum is null, so there will be no linear combination of vacuum states that will be formed to be a new vacuum state of the theory. Any perturbation will always be diagonal in the degenerate vacua, related by a gauge transformation.

To investigate physics about the asymmetric minimum, let us define
the new real scalar field $H(x)$, and let us expand the field $\phi$
around one of the infinite minimum configurations
\begin{eqnarray}
\mathbf{v}=\left(\begin{array}{c}
0\\
v
\end{array}\right) & \Rightarrow & \phi(x)=\frac{1}{\sqrt{2}}\left(\begin{array}{c}
\phi^{+}\\
v+H(x)
\end{array}\right).
\end{eqnarray}
We can reparametrize $\phi$ in a convenient way by an $SU(2)$ gauge
transformation and put the scalar field in the form corresponding to the so called unitarity gauge,  always reacheable because to the gauge invariance of the Lagrangian,
\begin{eqnarray*}
\phi(x)=\frac{1}{\sqrt{2}}\left(\begin{array}{c}
0\\
v+H(x)
\end{array}\right).
\end{eqnarray*}
In terms of the new field, the classical potential takes the form
\begin{eqnarray}
U(\phi)=\frac{1}{2}(2\lambda v^{2})H^{2}+\lambda vH^{3}+\frac{1}{4}\lambda H^{4}.
\end{eqnarray}
Then, from the classical potential a new physical particle arises with
quantum field $H(x)$,  quartic self-coupling $\lambda$ and squared mass
\begin{eqnarray}
m_{H}^{2}=2\lambda v^{2}=-2\mu^{2}
\end{eqnarray}
known as the Higgs boson, a neutral particle with
zero spin. Expanding the kinetic term of the Lagrangian $L_{S}$, we have:
\begin{eqnarray*}
(D^{\mu}\phi)^{\dagger}(D_{\mu}\phi)=\frac{1}{2}\partial^{\mu}H\partial_{\mu}H+\left[\frac{1}{4}g^{2}v^{2}W^{\mu+}W_{\mu}^{-}+\frac{1}{8}(g^{2}+g'^{2})v^{2}Z^{\mu}Z_{\mu}\right]\left(1+\frac{H}{v}\right)^{2}.
\end{eqnarray*}
Therefore, the gauge bosons $W$ and $Z$ have acquired masses, whereas
the photon remains massless 
\begin{eqnarray}
m_{W}=\frac{1}{2}vg, & m_{Z}=\frac{1}{2}v\sqrt{\left(g^{2}+g'^{2}\right)}, & m_{A}=0.
\end{eqnarray}
Using the mass of $W$ in (\ref{muon-g})
we can express the value of $v$ in terms of Fermi constant:
\begin{eqnarray}
\frac{1}{v^{2}}=\sqrt{2}G_{\mu}(1-\Delta r).
\end{eqnarray}
At classical level and using the measured valued of the Fermi constant
(\ref{Gmuonvalue}), we get
\begin{eqnarray*}
v=\sqrt{\frac{1}{\sqrt{2}G_{\mu}}}\simeq246.22\: GeV.
\end{eqnarray*}
The described mechanism of mass generation of gauge boson does not affect the gauge invariance of the Lagrangian, the mass terms for $W$ and $Z$ do not spoil the renormalizability of the theory, contrary to what happened when we tried to break the
symmetry explicitly. When one quantized a gauge theory in the functional approach, the gauge fixing term must be added \cite{Peskin,Bardin}
\begin{eqnarray*}
{\cal L}_{GF}=-\frac{1}{2\xi}(\partial^{\mu}Z_{\mu})^{2}
\end{eqnarray*}
bringing to the expression for the gauge $Z$ propagator:
\begin{eqnarray}
D_{\xi}^{\mu\nu}=\frac{i}{k^{2}-m_{Z}^{2}}\left[-g^{\mu\nu}+\frac{(1-\xi)k^{\mu}k^{\nu}}{k^{2}-\xi m_{Z}^{2}}\right].
\end{eqnarray}
This propagator decreases asymptotically as an homogeneous function of degree - 2. For the choice of the gauge parameter $\xi\rightarrow\infty$, we are in the so called unitary gauge and the $Z$ boson propagator takes the form of eq. (\ref{PropA}). Gauge invariance ensures that the theory is still renormalizable even if the propagator falls off more slowly than $1/k^2$. Renormalizability is not manifest, although individual loop diagrams diverge as $\log\xi$ or worse for $\xi\rightarrow\infty$, these divergences must cancel in the sum of all diagrams contributing to a given process due to the gauge invariance of the S-matrix, therefore there is a smooth limit for $\xi\rightarrow\infty$. However when we let $\xi$ to be finite, the propagator has an unphysical singularity at $k^{2}=\xi m_{Z}^{2}$.
This singularity is located at the mass squared of a new unphysical
scalar field $G(x)$ that does not propagate in the unitary gauge \cite{Bardin}. The field $G(x)$ is known in
the literature as the \emph{Goldstone boson}. Its appearance is a
consequence only of the spontaneous breakdown of a continuous symmetry
and its contribution exactly cancels the term with the unphysical
singularity in the $Z$ propagator. So the Goldstone bosons of a spontaneously broken gauge theory do not appear in any gauge.

Different values of $\xi$ refer to different gauges. The most commonly
choices are the Feynman gauge ($\xi=1$), which has a vector propagator
\begin{eqnarray}
D_{F}^{\mu\nu}=-\frac{ig^{\mu\nu}}{k^{2}-m_{V}^{2}}
\end{eqnarray}
with massive Goldstone bosons, and the Landau gauge ($\xi=0$), whose propagator is
\begin{eqnarray}
D_{L}^{\mu\nu}=\frac{i}{k^{2}-m_{V}^{2}}\left[-g^{\mu\nu}+\frac{k^{\mu}k^{\nu}}{k^{2}}\right],
\end{eqnarray}
with massless Goldstone bosons having zero coupling to
physical scalars. In the following the Landau gauge will be used. The reason is because the coupling
of unphysical gauge depedent scalars (Goldstone bosons) to physical scalars is zero
and because the Landau gauge is scale independent, i.e., the renormalized gauge
parameter, unlike the renormalized couplings and masses, does not depend upon the renormalization
scale. Another reason is the vanishing of the interaction between
scalars and ghost fields in that gauge, on the contrary in an $R_{\xi}$ -gauge,
the couplings of these ghost fields to scalars and their squared masses
are proportional to $\xi$. 

\section{The Scalar Sector}

In determining the mass spectrum of the SM and studying how lower bounds to the Higgs mass can arise, it is necessary to find the vacuum expectation values of the scalar fields in the theory by computing the effective potential. To this aim, we will study the scalar sector of the SM, and in particular the phenomenon of spontaneous breaking of the gauge symmetry. In this section we will present several methods of calculation of the one-particle -irreducible effective potential. The formal discussion of the effective potential is most conveniently done in the context of the generating functional formalism, the framework of functional integrals and its renormalization properties are very well known, see for instance \cite{Colemanbook}, \cite{Peskin}, and \cite{Iliopoulos}. 

Consider an interacting  quantum field theory of a single scalar field, $\phi$, the vacuum to vacuum amplitude $\left\langle 0_{out}|0_{in}\right\rangle _{J}$ in presence of a classical source $J(x)$ is
\begin{eqnarray}
Z[J]=\int D\phi exp\left(i\int d^{4}x\left[L(\phi,\partial_{\mu}\phi)+J(x)\phi(x)\right]\right).
\end{eqnarray}
This functional $Z[J]$ can be functionally expanded in powers
of $J(x)$,
\begin{eqnarray}
Z[J]=\sum_{n=0}^{\infty}\frac{1}{n!}\int\prod_{i=1}^{n}\left[d^{4}x_{i}J(x_{i})\right]G(x_{1},\ldots,x_{n}),
\end{eqnarray}
and the coefficients $G(x_{1},\ldots,x_{n})=<0|T(\phi(x_1)\cdots\phi(x_n))|0>$ are the Green's functions
of the theory,
\begin{eqnarray}
G(x_{1},\ldots,x_{n})=\left.\frac{1}{Z[J]}\left(-i\frac{\delta}{\delta J(x_{1})}\right)\cdots\left(-i\frac{\delta}{\delta J(x_{n})}\right)Z[J]\right|_{J=0}.
\end{eqnarray}
In this sense $Z[J]$ is called the generating functional of the
Green's functions of n-points. For a most extended treatment of perturbation theory in the presence of an external source $J$, see for instance the  Schwinger's paper \cite{Schwinger} where many of the concepts of the functional approach were introduced for the first time. \\
A more useful generating functional is provided by the negative of the vacuum energy (times the total evolution time) as a function of external source, $W[J]$, related to $Z[J]$ by $Z[J]=exp(iW[J])$.
The functional
\begin{eqnarray}
W[J]=-ilnZ[J] & = & \left.\sum_{n=0}^{\infty}\frac{1}{n!}\int\prod_{i=1}^{n}\left[d^{4}x_{i}J(x_{i})\right]\frac{\delta^{n}W[J]}{\delta J(x_{1})\ldots\delta J(x_{n})}\right|_{J=0}
\end{eqnarray} 
generates the connected Green's functions. The first functional derivative of
$W[J]$ with respect to $J$ is the vacuum expectation value in the
presence of a non-zero source, so-called classical field
\begin{eqnarray}
\phi_C(J)\equiv\frac{\delta W[J]}{\delta J(x)}=\left\langle \Omega\left|\phi(x)\right|\Omega\right\rangle \left|_{J}\right. .
\label{eq:ClassicalField}
\end{eqnarray}
The full quantum dynamics of the theory is encoded in the effective action, $\Gamma[\phi_{c}]$, the Legendre transform of $W(J)$,
\begin{eqnarray}
\Gamma[\phi_{c}]=W[J]-\int d^{4}xJ(x)\phi_{c}(x),\label{eq:EffectiveAction}
\end{eqnarray}
in fact it satisfies the extremal Euler-Lagrange equation
\begin{equation}
\left.\frac{\delta \Gamma[\phi_C]}{\delta \phi_C}\right|_{\phi_C = < \Omega |\phi(x)|\Omega >}=0.
\label{SSB1}
\end{equation}
In order to compute $\Gamma[\phi_{c}]$, one must invert 
(\ref{eq:ClassicalField}) to obtain $J$ as a function of $\phi_{c}$,
then replace $J$ in equation (\ref{eq:EffectiveAction}). The effective
action has an expansion in powers of the classical fields, whose coefficients
\begin{eqnarray}
\Gamma_{n}(x_{1},\ldots,x_{n})=\frac{\delta^{n}\Gamma[\phi_{c}]}{\delta\phi_{c}(x_{1})\ldots\delta\phi_{c}(x_{n})},
\end{eqnarray}
can be shown to be the n-point one particle irreducible\footnote{An 1PI Feynman diagram is a connected diagram that cannot be disconnected by cutting a single internal line.} (1PI) Green's functions (sometimes called proper vertices), and we therefore refer to $\Gamma[\phi_{c}]$ as the generating functional of connected 1PI Green's functions of the theory \cite{Colemanbook, Peskin, Ryder}.
The full propagator of the theory is obtained as
\begin{equation}
\left.\frac{\delta^2\Gamma[\phi_C]}{\delta\phi_C(x)\delta\phi_C(y)}\right|_{\phi_C=< \Omega |\phi(x)|\Omega >}=iD^{-1}(x,y).
\end{equation}
The functional $\Gamma[\phi_{c}]$ is an appropriate tool to study spontaneous symmetry breaking which will occur if (\ref{SSB1}) has a non-trivial solution.
Instead of expanding in powers of $\phi_{c}$, we can expand $\Gamma[\phi_{c}]$
in powers of momentum, about the point where all external momenta
vanish. To achieve this aim, we use the Fourier transform of the
coefficients $\Gamma_{n}(x_{1},\ldots,x_{n})$:
\begin{eqnarray}
\Gamma_{n}(x_{1},\ldots,x_{n})=\int\frac{dp_{1}}{(2\text{\ensuremath{\pi}})^{4}}\ldots\frac{dp_{n}}{(2\text{\ensuremath{\pi}})^{4}}e^{i(p_{1}x_{1}+\ldots p_{n}x_{n})}(2\pi)^{4}\delta(p_{1}+\ldots p_{n})\tilde{\Gamma}_{n}(p_{1},\ldots,p_{n}),
\end{eqnarray}
and then we expand $\tilde{\Gamma}_{n}$ in powers of momenta around
$p_{i}=0$. In the position space, the effective action becomes 
\begin{eqnarray}
\Gamma[\phi_{c}]=\int d^{4}x\left[-V(\phi_{c})+\frac{1}{2}(\partial_{\mu}\phi_{c})^{2}Z(\phi_{c})+\ldots\right],
\end{eqnarray}
where
\begin{eqnarray}
V(\phi_{c})=-\sum_{n=0}^{\infty}\frac{1}{n!}\tilde{\Gamma}_{n}(0)\phi_{c}^{n}\label{eq:EffPotential}
\end{eqnarray}
is called the effective potential of the theory, because it does not contain derivatives of the classical field. From expansion (\ref{eq:EffPotential}) it is easy to see that the $n$th derivative of $V$ is the sum of all 1PI graphs with $n$
vanishing external momenta. In tree approximation (neglecting all diagrams with closed loops), $V(\phi_{c})$ is just the classical potential. 
Moreover, since we are only interested in the case where the vacuum expectation value is translationally invariant (momentum conservation is not spontaneously broken), we
can simplify the relation (\ref{SSB1}) to 
\begin{eqnarray}
\frac{d}{d\phi_{c}}V(\phi_{c})=0, \label{SSBCondition}
\end{eqnarray}
for some non-zero value of $\phi_{c}$. Spontaneous symmetry breaking will occur 
if $\phi_{c}$ develops a non-zero vacuum expectation value even when the source $J$ is set
equal to zero. As can be read of equation (\ref{SSB1}), it takes place when the classical field that minimized the effective action is different from zero. Therefore, from equation (\ref{SSBCondition}), we conclude that the minimization of $V$ (the location of its minima) is precisely the condition that
determines whether or not spontaneous symmetry breaking occurs \cite{Zichichi}. 

\subsection{The Effective Potential Calculation}

\begin{figure}
\begin{center}
\scalebox{0.4}{
\fcolorbox{white}{white}{
  \begin{picture}(226,130) (239,-111)
    \SetWidth{1.0}
    \SetColor{Black}
    \Line[](240,-46)(336,-46)
    \GOval(400,-46)(64,64)(0){0.882}
  \end{picture}
}}
\end{center}
\caption{Tadpole topology in a shifted theory.}\label{TadpoleShift}
\end{figure}
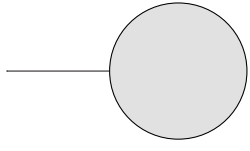

In this section we present two methods to compute the effective potential up to one-loop contribution in the SM. We begin with the loop expansion procedure introduced by Y. Nambu \cite{Nambu} and reviewed by Coleman and Weinberg \cite{Coleman}, and then we will make the explicit sum of the 1PI Green's functions
given by equation (\ref{eq:EffPotential}). The zero-order contribution to the effective potential is just the classical potential $U(\phi)$. The next-order contribution is the sum of all 1PI diagrams with an arbitrary number of external lines and with zero external momenta.
Let us consider the theory with a single real scalar field $\phi$ with the Lagrangian
\begin{eqnarray}
{\cal L}=\frac{1}{2}(\partial_{\mu}\phi)^{2}-\frac{1}{2}m^{2}\phi^{2}-\frac{1}{4}\lambda\phi^{4}.
\end{eqnarray}
The zero-loop potential is just
\begin{eqnarray}
V_{0}=\frac{1}{2}m^{2}\phi_{c}^{2}+\frac{1}{4}\lambda\phi_{c}^{4}.
\end{eqnarray}
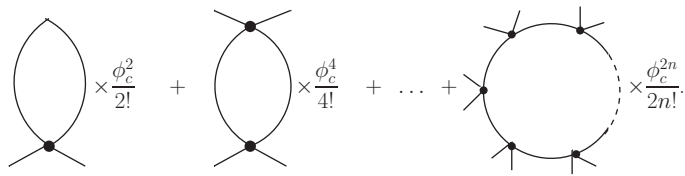
\begin{figure}
\begin{center}
\scalebox{0.5}{
\fcolorbox{white}{white}{
  \begin{picture}(458,128) (116,-203)
    \SetWidth{1.0}
    \SetColor{Black}
    \Arc(521.5,-139.5)(50.621,32.905,327.095)
    \Arc[dash,dashsize=4,clock](522,-140)(50.621,32.905,-39.094)
    \Line(471,-139)(456,-154)
    \Line(468,-136)(456,-127)
    \Line(492,-97)(471,-91)
    \Line(492,-97)(498,-79)
    \Line(492,-178)(477,-190)
    \Line(492,-181)(492,-199)
    \Line(543,-91)(543,-76)
    \Line(543,-94)(561,-88)
    \Line(537,-184)(537,-202)
    \Line(537,-187)(555,-196)
    \Vertex(471,-139){3}
    \Vertex(492,-97){3}
    \Vertex(543,-94){3}
    \Vertex(492,-181){3}
    \Vertex(540,-187){3}
    \Arc(177.592,-132.687)(56.656,122.681,238.512)
    \Arc(119.533,-133.811)(54.6,-59.797,63.377)
    \Line(117,-196)(144,-181)
    \Line(147,-181)(174,-196)
    \Vertex(147,-181){4.243}
    \Arc(278.85,-136)(48.522,-68.034,68.034)
    \Arc(321.844,-135.73)(51.166,119.049,242.224)
    \Vertex(297,-181){4.243}
    \Line(297,-181)(324,-196)
    \Line(270,-196)(297,-181)
    \Line(270,-79)(297,-91)
    \Line(327,-79)(297,-91)
    \Text(180,-153)[lb]{\Large{\Black{$\times \dfrac{\phi_{c}^{2}}{2!}~~~~+~~~~$}}}
    \Text(332,-153)[lb]{\Large{\Black{$\times\dfrac{\phi_{c}^{4}}{4!}~~~+~\dots~+~$}}}
    \Text(580,-153)[lb]{\Large{\Black{$\times \dfrac{\phi_{c}^{2n}}{2n!}$.}}}
    \Vertex(297,-91){4.243}
  \end{picture}
}}
\end{center}
\caption{\label{EP-FM} Sum of one-loop 1PI Green's functions at zero external momenta that contribute to the effective potential $V_{1}$.}
\end{figure}
The one-loop effective potential $V_1$ is given by the sum (\ref{eq:EffPotential}) of all 1PI Green's functions at zero external momenta,
\begin{eqnarray}
\tilde{\Gamma}_{2n}(0)=-i\frac{(2n)!}{2^{n}2n}\left(-4!\frac{i\lambda}{4}\right)^{n}\int\frac{d^{4}k}{(2\pi)^{4}}\left[\frac{i}{k^{2}-m^{2}+i\varepsilon}\right]^{n},
\end{eqnarray}
represented in Fig. \ref{EP-FM}, where Green's functions with odd number of external lines are zero because the Lagrangian hasn't odd powers of the field $\phi$. Therefore 
\begin{eqnarray}
V_{1}=\frac{i}{2}\sum_{n=1}^{\infty}(3\lambda\phi_{c}^{2})^{n}\frac{1}{n}\int\frac{d^{4}k}{(2\pi)^{4}}\frac{i}{(k^{2}-m^{2}+i\varepsilon)^{n}}.
\end{eqnarray}
The first two terms, with $n=1$ and $n=2$, are divergent. For $n=1$ the divergence is quadratic, whereas for $n=2$ the divergence is logarithmic. The terms with $n\geq3$ are finite. Let us first compute the finite part. By the usual method of Wick's rotation in the framework of dimensional regularization (see \cite{Peskin} or \cite{Smirnov}) we have
\begin{eqnarray}
\int\frac{d^{d}l}{(2\pi)^{d}}\frac{1}{(l^{2}-\Delta)^{n}}=\frac{i(-1)^{n}}{(4\pi)^{d/2}}\frac{\Gamma\left(n-\frac{d}{2}\right)}{\Gamma(n)}\left(\frac{1}{\Delta}\right)^{n-\frac{d}{2}},\label{eq:dimreg}
\end{eqnarray}
since for $n=3$ the integral (\ref{eq:dimreg}) converges, we may use $d=4$, and the finite part of the potential is:
\begin{eqnarray}
V_{1}^{fin}=\frac{i}{2}\frac{i}{(4\pi)^{2}}\sum_{n=3}^{\infty}(3\lambda\phi_{c}^{2})^{n}\frac{(-1)^{n}}{n}\frac{\Gamma\left(n-2\right)}{\Gamma(n)}m^{4-2n}.
\end{eqnarray}
If we define he dimensionless parameter $z=\frac{3\lambda\phi_{c}^{2}}{m^{2}}$ and we use the property of the Euler's gamma function
\begin{equation}
\Gamma(n)=(n-1)(n-2)\Gamma(n-2),
\end{equation}
we reach the result
\begin{eqnarray}
V_{1}^{fin}=-\frac{m^{4}}{32\pi{}^{2}}\sum_{n=3}^{\infty}\frac{(-1)^{n}z^{n}}{n(n-1)(n-2)},
\end{eqnarray}
and after partial fractioning 
\begin{eqnarray}
\frac{1}{n(n-1)(n-2)}=\frac{1}{2}\left(\frac{1}{n}-\frac{2}{n-1}+\frac{1}{n-2}\right),
\end{eqnarray}
we obtain
\begin{eqnarray}
V_{1}^{fin}=-\frac{m^{4}}{64\pi{}^{2}}\sum_{n=3}^{\infty}(-1)^{n}z^{n}\left(\frac{1}{n}-\frac{2}{n-1}+\frac{1}{n-2}\right)
\end{eqnarray}
to be rewritten after appropriate shifts as
\begin{eqnarray}
V_{1}^{fin}=-\frac{m^{4}}{64\pi{}^{2}}\left(z-\frac{z^{2}}{2}+\sum_{n=1}^{\infty}\frac{(-1)^{n}z^{n}}{n}+2z^{2}-\sum_{n=1}^{\infty}\frac{(-1)^{n+1}2z^{n+1}}{n}+\sum_{n=1}^{\infty}\frac{(-1)^{n+2}z^{n+2}}{n}\right) \nonumber \\
\nonumber \\
=-\frac{m^{4}}{64\pi{}^{2}}\left(\sum_{n=1}^{\infty}\frac{(-1)^{n}z^{n}}{n}(1+2z+z^{2})+z+\frac{3}{2}z^{2}\right).\:\:\: ~~~~~~~~~~~~ \;\;\, ~~~~~~~
\end{eqnarray}
From the known power series 
\begin{eqnarray*}
ln(1+z)=\sum_{n=1}^{\infty}\frac{(-1)^{n+1}}{n}z^{n} & {\rm if} & \left|z\right|<1,
\end{eqnarray*}
the finite one-loop effective potential amounts to
\begin{eqnarray}
&& V_{1}^{fin}=\frac{m^{4}}{64\pi{}^{2}}\left[(1+z)^{2}ln(1+z)-z-\frac{3}{2}z^{2}\right] \nonumber\\
&&\nonumber \\
&& ~~~~~~~=~\frac{1}{64\pi{}^{2}}\left[(m^{2}+3\lambda\phi_{c}^{2})^{2}ln\left(\frac{m^{2}+3\lambda\phi_{c}^{2}}{m^{2}}\right)-3\lambda\phi_{c}^{2}m^{2}-\frac{3}{2}\left(3\lambda\phi_{c}^{2}\right)^{2}\right].
\end{eqnarray}
Now about the divergent part 
\begin{eqnarray*}
V_{1}^{div}=\frac{i}{2}\left[(3\lambda\phi_{c}^{2})\int\frac{d^{4}k}{(2\pi)^{4}}\frac{1}{k^{2}-m^{2}+i\varepsilon}+\frac{1}{2}(3\lambda\phi_{c}^{2})^{2}\int\frac{d^{4}k}{(2\pi)^{4}}\frac{1}{(k^{2}-m^{2}+i\varepsilon)^{2}}\right],
\end{eqnarray*}
it makes sense in dimensional regularization, by using the formula (\ref{eq:dimreg}) and then making the Laurent expansion around $\varepsilon=0$, if $\varepsilon=(4-d)/2$:
\begin{eqnarray}
&& V_{1}^{div}=\frac{i}{2}\left[(3\lambda\phi_{c}^{2})\frac{m^{2}}{(4\pi)^{2}}i\left(\frac{1}{\varepsilon}+1\right)\left(\frac{4\pi e^{-\gamma_{E}}\mu^{2}}{m^{2}}\right)^{\varepsilon}+\frac{1}{2}(3\lambda\phi_{c}^{2})^{2}\frac{1}{(4\pi)^{2}}i\left(\frac{1}{\varepsilon}\right)\left(\frac{4\pi e^{-\gamma_{E}}\mu^{2}}{m^{2}}\right)^{\varepsilon}\right]\nonumber \\
&& \nonumber \\
&& ~~~~~~=-\frac{3}{2}\lambda\frac{m^{2}}{(4\pi)^{2}}\left(\frac{1}{\varepsilon}+1-log\left(\frac{m^{2}}{\bar{\mu}^{2}}\right)\right)\phi_{c}^{2}-\frac{9}{4}\frac{\lambda^{2}}{(4\pi)^{2}}\left(\frac{1}{\varepsilon}-log\left(\frac{m^{2}}{\bar{\mu}^{2}}\right)\right)\phi_{c}^{4}.
\end{eqnarray}
Where we have defined the new scale of energy $\bar{\mu}^{2}=4\pi e^{-\gamma_{E}}\mu^{2}$ and $\mu^2$ is the introduced mass units to take into account the dimensional changes of the parameters of the theory in dimensional regularization.
In order to cancel the divergences, one can add suitable counter-terms. 
The book of D. Bardin and G. Passarino \cite{Bardin} presents an exhaustive review of the most important mass dependent and independent renormalization schemes. 
We start from the unrenormalised potential, written in terms of bare
quantities $m_{0}^{2}$ and $\lambda_{0}$, and set $m_{0}^{2}\rightarrow m^{2}+\delta m^{2}$
and $\lambda_{0}\rightarrow\lambda+\delta\lambda$. Then, we obtain
\begin{eqnarray*}
V_{bare}=V+\delta V,
\end{eqnarray*}
where
\begin{eqnarray*}
V=\frac{1}{2}m^{2}\phi_{c}^{2}+\frac{1}{4}\lambda\phi_{c}^{4}+V_{1},
\end{eqnarray*}
while, to leading order, 
\begin{eqnarray*}
\delta V=\frac{1}{2}\delta m^{2}\phi_{c}^{2}+\frac{1}{4}\delta\lambda\phi_{c}^{4}.
\end{eqnarray*}
In order to fix the finite part of the counter-terms\footnote{The finite part of the counter-terms are arbitrary. Different choices correspond to different renormalization schemes, and different definitions of the renormalized parameters.}, we define our renormalized parameters $m$ and $\lambda$ by the renormalization prescriptions:
\begin{eqnarray}
\left.\frac{\partial^{2}V}{\partial\phi_{c}^{2}}\right|_{\phi_{c}=0}=m^{2} & ; & \left.\frac{\partial^{4}V}{\partial\phi_{c}^{4}}\right|_{\phi_{c}=0}=6\lambda.
\label{substraction}
\end{eqnarray}
Since the above equations hold for the tree-level potential, and since
the finite part of the one-loop corrections starts with $\phi_{c}^{6}$,
the counter-term $\delta V$ is just the divergent part of the one-loop
potential, $\delta V=V_{1}^{div}$, therefore
\begin{eqnarray}
&&\delta m^{2}=-3\lambda\frac{m^{2}}{(4\pi)^{2}}\left(\frac{1}{\varepsilon}+1-log\left(\frac{m^{2}}{\bar{\mu}^{2}}\right)\right)\\
&&~\delta\lambda ~=-9\lambda^{2}\frac{1}{(4\pi)^{2}}\left(\frac{1}{\varepsilon}-log\left(\frac{m^{2}}{\bar{\mu}^{2}}\right)\right),
\end{eqnarray}
so for the above defined renormalization scheme 
\begin{eqnarray}
V_{1}=V_{1}^{fin}=\frac{1}{64\pi{}^{2}}\left[(m^{2}+3\lambda\phi_{c}^{2})^{2}ln\left(\frac{m^{2}+3\lambda\phi_{c}^{2}}{m^{2}}\right)-3\lambda\phi_{c}^{2}m^{2}-\frac{3}{2}\left(3\lambda\phi_{c}^{2}\right)^{2}\right].
\end{eqnarray}
Another renormalization scheme is the so-called modified minimal subtraction
$\overline{MS}$ renormalization scheme. By definition, the counter-terms
in the $\overline{MS}$ scheme subtracts only the terms proportional
to the pole $(D-4)^{-1}$ plus the terms proportional to $\gamma_{E}-ln(4\pi)$.
Using the new scale $\bar{\mu}$, that we will call $\mu$ in the following, the modified minimal subtraction is equivalent to removing only the pole $1/\varepsilon$.
The $\overline{MS}$ renormalized effective potential up to one-loop order is
\begin{eqnarray*}
V=V_{\rm cl}+V_{1}^{\overline{MS}},
\end{eqnarray*}
where
\begin{eqnarray}
V_{1}^{\overline{MS}}=V_{1}^{fin}-\frac{1}{64\pi^{2}}\left[6\lambda\phi_{c}^{2}m^{2}+6\lambda\phi_{c}^{2}\left(m^{2}+\frac{3}{2}\lambda\phi_{c}^{2}\right)ln\frac{{\mu}^{2}}{m^{2}}\right].
\end{eqnarray}
Using the identity 
\begin{eqnarray*}
6\lambda\phi_{c}^{2}\left(m^{2}+\frac{3}{2}\lambda\phi_{c}^{2}\right)=(m^{2}+3\lambda\phi_{c}^{2})^{2}-m^{4},
\end{eqnarray*}
we obtain 
\begin{eqnarray*}
V_{1}^{\overline{MS}}=\frac{1}{64\pi^{2}}(m^{2}+3\lambda\phi_{c}^{2})^{2}\left[ln\frac{m^{2}+3\lambda\phi_{c}^{2}}{\mu^{2}}\right]-\frac{1}{64\pi^{2}}\left[\frac{3}{2}(6\lambda\phi_{c}^{2}m^{2})+\frac{3}{2}(3\lambda\phi_{c}^{2})^{2}+m^{4}ln\frac{m^{2}}{\mu^{2}}\right].
\end{eqnarray*}
Adding the term $-\frac{1}{64\pi^{2}}\left[\frac{3}{2}m^{4}-\frac{3}{2}m^{4}\right]$
and neglecting the field independent terms $-\frac{3}{2}m^{4}+m^{4}ln\frac{m^{2}}{\mu^{2}}$,
we finally obtain 
\begin{eqnarray}
V_{1}^{\overline{MS}}=\frac{1}{64\pi^{2}}(m^{2}+3\lambda\phi_{c}^{2})^{2}\left[ln\frac{m^{2}+3\lambda\phi_{c}^{2}}{\mu^{2}}-\frac{3}{2}\right].\label{eq:MSbar1lEffPot}
\end{eqnarray}
Note that the one loop effective potential $V_{1}^{\overline{MS}}$ is related with $V_1$ defined in the above renormalization scheme (\ref{substraction}) through the matching condition:
\begin{eqnarray*}
V_{1}=V_{1}^{\overline{MS}}+(\delta V^{\overline{MS}}-\delta V)=V_{1}^{\overline{MS}}-\left.\delta^{(1)}V\right|_{fin},
\end{eqnarray*}
where $\left.\delta^{(1)}V\right|_{fin}$ is the finite part of the
counter-term $\delta V$ after removing the term proportional to $(d-4)^{-1}+\gamma_{E}-ln(4\pi)$,
specifically
\begin{eqnarray}
\delta V^{\overline{MS}}-\delta V=\frac{1}{64\pi^{2}}\left[6\lambda\phi_{c}^{2}m^{2}+6\lambda\phi_{c}^{2}\left(m^{2}+\frac{3}{2}\lambda\phi_{c}^{2}\right)ln\frac{\mu^{2}}{m^{2}}\right]. \label{DiffSchemes}
\end{eqnarray}
Similar matching conditions are used for the renormalization parameters in the two schemes
\begin{eqnarray*}
\lambda(\mu)+\delta\lambda=\lambda^{\overline{MS}}+\delta\lambda^{\overline{MS}} & ; & m^{2}(\mu)+\delta m^{2}=m_{\overline{MS}}^{2}+\delta m_{\overline{MS}}^{2}.
\end{eqnarray*}
It is remarkable that since in all renormalization schemes the terms with logarithms of the classical fields are the same, the difference between schemes are in the terms proportional to powers of $\phi_{c}$ and the constant terms, like in eq. (\ref{DiffSchemes}).
The diagrammatic method described above based on Figure \ref{EP-FM}  is very inefficient if we want to compute the effective potential in more complicated theories, like the Standard Model, where the potential receives contributions also from fermion and vector loops. Fortunately, there is a simple method of calculating higher-order corrections to the effective potential (see Lee and
Sciaccaluga \cite{LeeSci}) which allows one to compute the one-loop
scalar potential in a very simple way. The method is as follows. Consider
that the effective action $\Gamma[\phi_{c}]$ is expanded about some arbitrary shift $\phi_{c}=\omega$. Then, the expression of the effective potential
is now
\begin{eqnarray*}
V'(\phi_{c})=-\sum_{n=0}^{\infty}\frac{1}{n!}\tilde{\Gamma}_{n}(0)(\phi_{c}+\omega)^{n}=-\sum_{n=0}^{\infty}\frac{1}{n!}\tilde{\Gamma'}_{n}(\omega,0)\phi_{c}{}^{n},
\end{eqnarray*}
where $\tilde{\Gamma'}_{n}(\omega,0)$ is the 1PI Green's function
of the shifted theory. Using the binomial theorem we can rewrite the
above relation as:
\begin{eqnarray*}
V'(\phi_{c})=-\sum_{k=0}^{\infty}\frac{1}{k!}\left(\sum_{n=0}^{\infty}\frac{1}{n!}\tilde{\Gamma}_{n}(0)\frac{n!}{(n-k)!}\omega{}^{n-k}\right)\phi_{c}^{k},
\end{eqnarray*}
therefore
\begin{eqnarray}
\tilde{\Gamma'}_{1}(\omega,0)=\sum_{n=0}^{\infty}\frac{1}{n!}\tilde{\Gamma}_{n}(0)n\omega{}^{n-1}.
\end{eqnarray}
The l.h.s. of the above expression is just the tadpole diagram in
the shifted theory to any order, represented by the Figure \ref{TadpoleShift}.
Evaluating this diagram, then integrating with respect to $\omega$, and setting $\omega=\phi_{c}$ gives the effective potential 
\begin{eqnarray}
\int_{0}^{\phi_{c}}d\omega\tilde{\Gamma'}_{1}(\omega,0)=\sum_{n=0}^{\infty}\frac{1}{n!}\tilde{\Gamma}_{n}(0)n\int_{0}^{\phi_{c}}d\omega\omega{}^{n-1} ~~ \nonumber \\
=\sum_{n=0}^{\infty}\frac{1}{n!}\tilde{\Gamma}_{n}(0)n\left.\frac{\omega{}^{n}}{n}\right|_{0}^{\phi_{c}}~~~~~~~~~ \nonumber \\
 =\sum_{n=0}^{\infty}\frac{1}{n!}\tilde{\Gamma}_{n}(0)\phi_{c}^{n}=-V(\phi_{c}).
\end{eqnarray}
Let us see an explicit example. Consider the massive, self-interacting scalar theory. The classical potential of the shifted theory is
\begin{eqnarray}
&& U(\phi)=\frac{1}{2}m^{2}(\phi+\omega)^{2}+\frac{1}{4}\lambda(\phi+\omega)^{4} \nonumber \\
&& ~~~~~~~=~ (m^{2}\omega+\lambda\omega^{3})\phi+\frac{1}{2}(m^{2}+3\lambda\omega^{2})\phi^{2}+\lambda\omega\phi^{3}+\frac{\lambda}{4}\phi^{4}.
\end{eqnarray}
The tree-level contribution to the potential can be computed from the tree-level tadpole in the shifted theory
\begin{eqnarray*}
-m^{2}\omega-\lambda\omega^{3},
\end{eqnarray*}
which, integrated in $\omega$ between $0$ and $\phi_{c}$ gives minus the tree-level potential
\begin{eqnarray*}
V_{0}(\phi)=\frac{1}{2}m^{2}\phi_{c}^{2}+\frac{1}{4}\lambda\phi_{c}^{4}.
\end{eqnarray*}
To compute the one-loop contribution, we need to consider only one diagram,
with one scalar external line and one scalar internal propagator.
In the shifted theory the mass of the $\phi$ field is $m^{2}+3\lambda\omega^{2}$, and
the cubic vertex is $-3\lambda\omega$, therefore
\begin{eqnarray}
\tilde{\Gamma'}_{1}^{(1l)}(\omega,0)=-3\lambda\omega\int\frac{d^{d}k}{(2\pi)^{d}}\frac{i}{k^{2}-m^{2}-3\lambda\omega^{2}} ~~~~~~~~~~~~~~~~~~~~~ \nonumber\\
=\frac{3\lambda\omega}{(4\pi)^{2}}(m^{2}+3\lambda\omega^{2})\left[\frac{1}{\varepsilon}-ln\left(\frac{m^{2}+3\lambda\omega^{2}}{4\pi e^{-\gamma_{E}}\mu^{2}}\right)+1\right].
\end{eqnarray}
Using the $\overline{MS}$ scheme, the renormalized one-loop potential is
\begin{eqnarray*}
V_{1}^{\overline{MS}}(\phi_{c})=\frac{1}{(4\pi)^{2}}\int_{0}^{\phi_{c}}d\omega3\lambda\omega(m^{2}+3\lambda\omega^{2})\left[ln\left(\frac{m^{2}+3\lambda\omega^{2}}{\mu^{2}}\right)-1\right].
\label{effV}
\end{eqnarray*}
The above integral is straightforwardly performed and we obtain the expected result
\begin{eqnarray}
V_{1}(\phi_{c})=\frac{1}{64\pi^{2}}(m^{2}+3\lambda\phi_{c}^{2})^{2}\left(ln\frac{m^{2}+3\lambda\phi_{c}^{2}}{\mu^{2}}-\frac{3}{2}\right).
\end{eqnarray}

\subsection{The One-Loop Effective Potential}

\begin{figure}
\begin{center}
\scalebox{0.5}{
\fcolorbox{white}{white}{
  \begin{picture}(504,231) (146,-64)
    \SetWidth{1.0}
    \SetColor{Black}
    \Line[dash,dashsize=10](190,-39)(190,36)
    \Arc[dash,dashsize=10](190,76)(43.012,144,504)
    \Vertex(190,31){5}
    \Line[dash,dashsize=10](405,-39)(405,36)
    \Arc[arrow,arrowpos=0.5,arrowlength=5,arrowwidth=2,arrowinset=0.2](405,76)(46.098,41,401)
    \Vertex(405,31){5}
    \Line[dash,dashsize=10](600,-39)(600,41)
    \PhotonArc(600,71)(41.231,-166,194){7.5}{13}
    \Vertex(600,36){5}
    \Text(205,-34)[lb]{\Large{\Black{$\phi_{3}$}}}
    \Text(420,-34)[lb]{\Large{\Black{$\phi_{3}$}}}
    \Text(615,-34)[lb]{\Large{\Black{$\phi_{3}$}}}
    \Text(395,146)[lb]{\Large{\Black{$F$}}}
    \Text(610,141)[lb]{\Large{\Black{$V$}}}
    \Text(185,-69)[lb]{\Large{\Black{$(a)$}}}
    \Text(400,-69)[lb]{\Large{\Black{$(b)$}}}
    \Text(595,-69)[lb]{\Large{\Black{$(c)$}}}
    \Text(190,146)[lb]{\Large{\Black{$S$}}}
  \end{picture}
}}
\end{center}
\caption{One-loop contributions to the effective potential in the SM. (a) Scalar contribution. (b) Fermion contribution. (c) Vector contribution.}\label{1LTadpoleSM}
\end{figure}
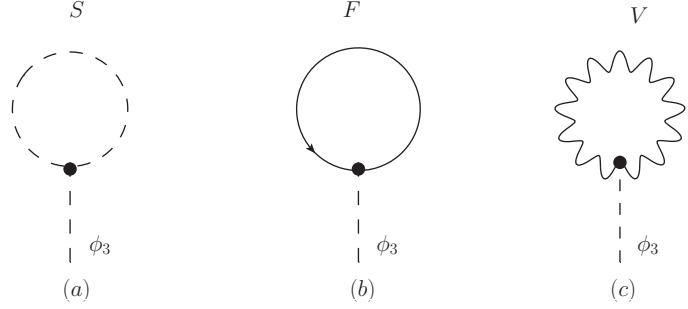

The above procedure can be applied straightforwardly to the non-Abelian gauge theories \cite{Coleman} and to the Standard Model \cite{Mahanthappa}. The scalar field is now a complex doublet given by 
\begin{eqnarray*}
\phi=\left(\begin{array}{c}
G^{\pm}\\
1/\sqrt{2}\left(H+iG_{0}\right)
\end{array}\right)=\frac{1}{\sqrt{2}}\left(\begin{array}{c}
\phi_{1}+i\phi_{2}\\
\phi_{3}+i\phi_{4}
\end{array}\right),
\end{eqnarray*}
and the classical potential in the shifted theory, after the shift $\phi_{j}\rightarrow\phi_{j}+\omega_{j}$, is now
\begin{eqnarray}
V_{0}'=\phi_{j}\omega_{j}(m^{2}+\lambda\omega^{2})+\frac{1}{2}\left[(m^{2}+\lambda\omega^{2})\delta_{ij}+2\lambda\omega_{i}\omega_{j}\right]\phi_{i}\phi_{j}+\lambda\omega_{i}\phi_{i}\phi_{j}\phi_{j}+\frac{1}{4}\lambda(\phi_{i}\phi_{i})^{2}.\label{eq:SMPotShifted}
\end{eqnarray}
In the Standard model, the one-loop effective potential receives contributions from classes of diagrams belonging to the scalar sector, the vector
boson sector, and the fermion sector, 
\begin{eqnarray*}
V_{1}(\phi)=V_{S}(\phi)+V_{V}(\phi)+V_{F}(\phi).
\end{eqnarray*}
These contributions are obtained by the computation of one-loop
tadpole diagrams in the shifted theory, represented by Figure \ref{1LTadpoleSM}.
We make the computation in the Landau gauge. The resulting effective potential is the same
regardless of the values assigned to the shift variables $\omega_{i}$,
therefore we may choose $\omega_{i}=0$ for all $i$ except for $\omega_{3}=\omega$.
In this case, the potential (\ref{eq:SMPotShifted}) of shifted theory describes three
scalar fields, $\phi_{1}$, $\phi_{2}$ and $\phi_{4}$ with mass
$m^{2}+\lambda\omega^{2}$ and one scalar field, $\phi_{3}$, with
mass $m^{2}+3\lambda\omega^{2}$. The trilinear couplings $\phi_{3}\phi_{j}\phi_{j}$
are $\lambda\omega$ for $j\neq3$ and $-3\lambda\omega$ for $j=3$.
Following exactly the same steps that for a single scalar field, but now referring to the four components, we obtain
\begin{eqnarray*}
V_{S}(\phi)=\frac{1}{64\pi^{2}}(m^{2}+3\lambda\phi_{c}^{2})^{2}\left(ln\frac{m^{2}+3\lambda\phi_{c}^{2}}{\mu^{2}}-\frac{3}{2}\right)+\frac{3}{64\pi^{2}}(m^{2}+\lambda\phi_{c}^{2})^{2}\left(ln\frac{m^{2}+\lambda\phi_{c}^{2}}{\mu^{2}}-\frac{3}{2}\right).
\end{eqnarray*}
Let see now the vector boson contribution to the effective potential,
$V_{V}(\phi)$. In this case the only term in the interaction we need is included in  
\begin{eqnarray*}
{\cal L_{D}}=(D^{\mu}\phi)^{\dagger}(D_{\mu}\phi),
\end{eqnarray*}
with 
\begin{eqnarray*}
D_{\mu}\phi=\left(\partial_\mu-\frac{i}{2}gW_{\mu}^{j}\tau^{j}-\frac{i}{2}g'YB_{\mu}\right)\phi.
\end{eqnarray*}
As discussed below, this term contains the mass term for the vector
bosons and the scalar-vector-vector vertices needed to compute the
one-loop tadpole. By gauge invariance $\phi$ can be chosen such that,
making a rotation in the isospin space, the $\phi_{1}$, $\phi_{2}$
and $\phi_{4}$ components vanishes:
\begin{eqnarray*}
\phi=\frac{1}{\sqrt{2}}e^{i\tau\theta(x)}\left(\begin{array}{c}
0\\
\phi_{3}
\end{array}\right).
\end{eqnarray*}
This leads to the scalar field in the shifted theory%
\footnote{We choose again $\omega_{i}\neq0$ only for $i=3$, and we work in
the unitarity gauge, $\theta(x)=0.$%
}
\begin{eqnarray*}
\bar{\phi}=\frac{1}{\sqrt{2}}\left(\begin{array}{c}
0\\
\phi_{3}+\omega_{3}
\end{array}\right).
\end{eqnarray*}
Therefore,
\begin{eqnarray*}
D_{\mu}\bar{\phi}=\frac{1}{\sqrt{2}}\left[\left(\begin{array}{c}
0\\
\partial_{\mu}\phi_{3}
\end{array}\right)-\frac{i}{2}\left(\phi_{3}+\omega_{3}\right)\left(\begin{array}{c}
g\left(W_{\mu}^{1}-iW_{\mu}^{2}\right)\\
-gW_{\mu}^{3}+g'B_{\mu}
\end{array}\right)\right].
\end{eqnarray*}
In terms of the Weinberg's angle $e=gsin\theta_{W}$ and $e=g'cos\theta_{W}$ the result is 
\begin{eqnarray*}
-gW_{\mu}^{3}+g'B_{\mu}=\sqrt{g^{2}+g'^{2}}\left(-cos\theta_{W}W_{\mu}^{3}+sin\theta_{W}B_{\mu}\right)=-\sqrt{g^{2}+g'^{2}}Z_{\mu},
\end{eqnarray*}
and
\begin{eqnarray*}
D_{\mu}\bar{\phi}=\frac{1}{\sqrt{2}}\left[\left(\begin{array}{c}
0\\
\partial_{\mu}\phi_{3}
\end{array}\right)-\frac{i}{\sqrt{2}}\left(1+\frac{\phi_{3}}{\omega_{3}}\right)\left(\begin{array}{c}
g\omega_{3}W_{\mu}^{+}\\
-\sqrt{g^{2}+g'^{2}/2}\omega_{3}Z_{\mu}
\end{array}\right)\right],
\end{eqnarray*}
so that
\begin{eqnarray}
\left(D^{\mu}\bar{\phi}\right)^{\dagger}\left(D_{\mu}\bar{\phi}\right)=\frac{1}{2}\partial^{\mu}\phi_{3}\partial_{\mu}\phi^{3}+\frac{1}{4}\left(1+\frac{\phi_{3}}{\omega_{3}}\right)^{2}\left[g^{2}\omega_{3}^{2}W_{\mu}^{+}W_{\mu}^{-}+\frac{g^{2}+g'^{2}}{2}\omega_{3}^{2}Z_{\mu}Z^{\mu}\right].
\end{eqnarray}
From the second term on the r.h.s. of the above equation we obtain
the mass terms: $g^{2}\omega_{3}^{2}/4$ for $W$ and $(g^{2}+g'^{2}/4)\omega_{3}^{2}$
for $Z$, and the coupling constants: $G_{WW\phi_{3}}=g^{2}\omega_{3}g_{\mu\nu}/2$
and $G_{ZZ\phi_{3}}=(g^{2}+g'^{2})\omega_{3}g_{\mu\nu}/4.$ Using
this information, one notes that the one-loop tadpole receives one
contribution from a loop of a $W$ vector boson and a contribution
from the $Z$ boson. The $W$ contribution in the Landau gauge is:
\begin{eqnarray*}
T_{W}^{(1l)}=\frac{g^{2}\omega_{3}g_{\mu\nu}}{2}\int\frac{d^{4}k}{(2\pi)^{4}}\frac{i}{k^{2}-g^{2}\omega_{3}^{2}/4}\left[-g^{\mu\nu}+\frac{k^{\mu}k^{\nu}}{k^{2}}\right].
\end{eqnarray*}
Using dimensional regularization and the eq. (\ref{eq:dimreg}) we obtain
\begin{eqnarray*}
T_{W}^{(1l)}=\frac{g^{2}\omega_{3}}{2}\frac{\mu^{4-d}}{(4\pi)^{d/2}}\frac{\Gamma(1-d/2)}{\Gamma(1)}\left(\frac{1}{g^{2}\omega_{3}^{2}/4}\right)^{1-d/2}(-d+1).
\end{eqnarray*}
Expanding around $\varepsilon=(4-d)/2$ and using the minimal subtraction, we get 
\begin{eqnarray}
T_{W}^{(1l)}
\longrightarrow
 -3ln\frac{g^{2}\omega_{3}^{2}/4}{\mu^{2}}+1.
\end{eqnarray}
The contribution of $W$ to the effective potential is obtained integrating regarding to $\omega_{3}$
\begin{eqnarray*}
V_{W}(\phi_{c})=\frac{1}{(4\pi)^{2}}\int_{0}^{\phi_{c}}d\omega_{3}\frac{g^{2}\omega_{3}}{2}\left(\frac{g^{2}\omega_{3}^{2}}{4}\right)\left(3ln\frac{g^{2}\omega_{3}^{2}/4}{\mu^{2}}-1\right),
\end{eqnarray*}
which straightforwardly gives the result 
\begin{eqnarray}
V_{W}(\phi_{c})=\frac{6}{64\pi^{2}}\left(\frac{1}{4}g^{2}\phi_{c}^{2}\right)^{2}\left[ln\frac{g^{2}\phi_{c}^{2}/4}{\mu^{2}}-\frac{5}{6}\right].
\end{eqnarray}
The contribution of the vector sector to the effective potential is the sum $V_{V}(\phi_{c})=V_{W}(\phi_{c})+V_{Z}(\phi_{c})$. The contribution of $Z$ can be obtained as follows:
\begin{eqnarray}
V_{Z}(\phi_{c})=\frac{3}{64\pi^{2}}\left(\frac{1}{4}(g^{2}+g'^{2})\phi_{c}^{2}\right)^{2}\left[ln\frac{(g^{2}+g'^{2})\phi_{c}^{2}/4}{\mu^{2}}-\frac{5}{6}\right].
\end{eqnarray}
Finally, we must consider the contribution of fermions and thus the
Yukawa sector. With the same choice for $\omega$ adopted above, the
shifted Lagrangian in the Yukawa sector becomes 
\begin{eqnarray*}
{\cal L}_{F}=-\sum_{f}\frac{h_{f}}{\sqrt{2}}(\phi_{3}+\omega_{3})\overline{\psi}_{f}\psi_{f}.
\end{eqnarray*}
Proceeding as above we find 
\begin{eqnarray}
V_{F}(\phi_{c})=-\frac{12}{64\pi^{2}}\sum_{f}\left(\frac{1}{2}h_{f}^{2}\phi_{c}^{2}\right)^{2}\left[ln\frac{h_{f}^{2}\phi_{c}^{2}/2}{\mu ^{2}}-\frac{3}{2}\right],
\end{eqnarray}
where the color sum has been performed, the global minus is because of the fermion loop. Moreover, the ghost
sector doesn't contribute to the effective potential in the Landau
gauge, because the couplings of the tri-linear vertices involving
FP ghosts are proportional to the gauge parameter $\xi$, that vanishes in this gauge. 

To summarize, the one-loop effective potential of the Standard Model
in the Landau gauge and renormalized in the $\overline{MS}$ subtraction
scheme is the sum of all contributions above computed
\begin{eqnarray}
V^{(1l)}_{\overline{MS}}(\phi_{c})=\frac{1}{64\pi^{2}}\sum_{j}(-1)^{2s_{j}}(2s_{j}+1){\cal M}_{j}^{4}(\phi_{c}^{2})\left[ln\frac{{\cal M}_{j}^{2}(\phi_{c}^{2})}{\mu^{2}}-c_{j}\right], \label{eq:1l-SM-EP}
\end{eqnarray}
where ${\cal M}_{j}^{2}(\phi_{c}^{2})$ denotes the field dependent squared mass of each particle $j$ in the theory
\begin{eqnarray}
{\cal M}_{j}^{2}(\phi_{c}^{2})=\left\{ \begin{array}{c}
m_{H}^{2}(\phi_{c}^{2})=m^{2}+3\lambda\phi_{c}^{2}\\
\\
m_{G}^{2}(\phi_{c}^{2})=m^{2}+\lambda\phi_{c}^{2}~\\
\\
m_{W}^{2}(\phi_{c}^{2})=\frac{1}{4}g^{2}\phi_{c}^{2}~~~~~~\\
\\
~~~m_{Z}^{2}(\phi_{c}^{2})=\frac{1}{4}(g^{2}+g'^{2})\phi_{c}^{2}\\
\\
m_{f}^{2}(\phi_{c}^{2})=\frac{1}{2}h_{f}^{2}\phi_{c}^{2}~~~~~
\end{array}\right.
\end{eqnarray}
The value of ${\cal M}_{j}^{2}(\phi_{c}^{2})$ at $\phi_{c}^{2}=v^{2}$
equals the squared mass of the corresponding particle. Note that the index $j$ is running over all particles in the spectrum of the SM.
Moreover in eq.~(\ref{eq:1l-SM-EP}) $s_{j}$ is the spin of particle $j$, and
$c_{j}$ is a constant depending on renormalization scheme chosen.
For the $\overline{MS}$ scheme, we get
\begin{eqnarray*}
s_{j}=\left\{ \begin{array}{c}
0\rightarrow j=S\\
1\rightarrow j=V\\
\frac{1}{2}\rightarrow j=F
\end{array}\right. & ; & c_{j}=\left\{ \begin{array}{c}
~~~~\frac{3}{2}\rightarrow j=S,\: F\\
\\
\frac{5}{6}\rightarrow j=V
\end{array}\right. . 
\end{eqnarray*}
It is worth to be noted here that the only contribution with a negative
coefficient  is the fermionic one. This negative sign has significant consequences in the analysis of the SM vacuum stability. Depending of the precise value of all coupling constants of the theory, the shape of the effective potential can change drastically for large ranges of the classical field $\phi_{c}$; the potential can be stable, metastable or even unstable \cite{Hung-Sher}. This leads to the very well known instability problem of the effective potential: one can find values of the top Yukawa coupling constant, $h_{t}$, together with a range of values of $\phi_{c}$ which satisfy perturbative conditions, so that $V^{(1l)}(\phi_{c})$ becomes lower than $V^{(1l)}(v)$ for $\phi_{c}>>v$, implying an unstable effective potential. Currently the values of all SM parameters, including the top and Higgs masses, are very well known, allowing us to study the true shape of the effective potential and his implications on the vacuum stability of SM. These ideas will be developed carefully in the first section of the Chapter \ref{cha:The Stability Problem of the SM}. 

\subsection{Renormalization Group Improvement of the Effective Potential}

\begin{figure}
\begin{center}
\scalebox{0.5}{
\fcolorbox{white}{white}{
  \begin{picture}(416,151) (155,-134)
    \SetWidth{1.0}
    \SetColor{Black}
    \Arc[](240,-35)(48.374,7,367)
    \Arc[](336,-35)(48.374,7,367)
    \Vertex(395,-35){1}
    \Vertex(410,-35){1}
    \Vertex(425,-35){1}
    \Arc[](485,-34)(48.374,7,367)
    \Vertex(288,-35){4}
    \Vertex(438,-35){4}
    \Vertex(384,-35){4}
    \Text(234,-125)[lb]{\Large{\Black{$1$}}}
    \Text(336,-125)[lb]{\Large{\Black{$2$}}}
    \Text(400,-120)[lb]{\Large{\Black{$\dots$}}}
    \Text(486,-125)[lb]{\Large{\Black{$n -loops$}}}
  \end{picture}
}}
\end{center}
\caption{\label{nl-VB}An arbitrary n-loops vacuum bubble topology.}
\end{figure}
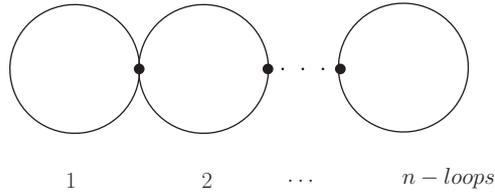

From the definition of the effective potential (Eq. \ref{eq:EffPotential})
we note that if we work in a shifted theory, the first term of the expansion, the zero-order term, is just the effective potential evaluated in the parameter of the shift $V(\omega)$.
According to the equation (\ref{eq:EffPotential}) this term is equal to minus the zero-points 1PI Green function,
\begin{eqnarray*}
V(\omega)=-\tilde{\Gamma}_{0}(0).
\end{eqnarray*}
Therefore, we can compute the n-loop effective potential valued in
the shift parameter $\omega$ just by calculating the n-loop 1PI Feynman
diagrams without external lines, better known as the vacuum bubbles
and represented schematically by fig. \ref{nl-VB}. This alternative
procedure of calculation, known as background field method (BFT) \cite{Background1}\cite{Background2}, 
is specially useful when we want to compute
the n-loops effective potential with $n\geq2$. The main reason is
simply because the number of zero-point diagrams to compute is lower
than the number of diagrams that we need to compute by the tadpole technique of the previous section. From the vacuum bubbles (see fig. \ref{nl-VB}) is very
easy to see that the n-loops potential have terms of order 
\begin{eqnarray}
g_{i}{}^{n+1}\left[ln\left(\frac{\phi_{c}^{2}}{\mu^{2}}\right)\right]^{n},\label{eq:reliability}
\end{eqnarray}
where the $g_{i}$ represent any of the couplings in the SM, $\lambda,\: g^{2},\: g'^{2},\: h_{t}^{2}$, etc. This allows us to see explicitly a common problem when one needs the n-loop effective potential. If we are interested in studying the
potential over a large range for the classical field $\phi_{c}$,
it is necessary to ensure that the potential is still in the perturbative
region of validity. From eq. (\ref{eq:reliability}) we can see that
it is not sufficient that the coupling $g_{i}$ be smaller than one,
instead we need that $g_{i}ln\left(\phi_{c}^{2}/\mu^{2}\right)$ must
be sufficiently small. Owing to the renormalization group (RG) invariance of the theory  it is
always possible to redefine the renormalization scale $\mu$ to make the
logarithmic term as small as possible. However, this would imply perturbativity for just 
a single value of the classical field, if one is interested in the
potential over a large range from $\phi_{1}$ to $\phi_{2}$, then
it is necessary for $g_{i}ln\left(\phi_{1}^{2}/\phi_{2}^{2}\right)$
to be smaller than one. To see this it is sufficient to make the dimensional transmutation of re-expressing $\mu$ or any other mass scale in terms of the extremal of $V$. In almost all calculations of interest the region of field space is so large that $g_{i}ln\left(\phi_{1}^{2}/\phi_{2}^{2}\right)$
is near unity. This situation makes unreliable the loop expansion
and therefore will be essential improve the effective potential. The
appropriate technique to improve it, exploits the renormalization group
invariance of the theory and used the RG methods to build an improved
potential that will be valid just if $g_{i}$ is less than one since will make the resummation of the logarithms \cite{Cabibbo, Flores}. 

In this section we will develop this procedure for the one-loop effective
potential of the SM. We begin with the renormalization group equations for the effective
potential. Defining $e^{t}=\mu^2$ then the invariance of the bare $1PI$ Green functions imply due to (\ref{eq:EffPotential}) an analogous invariance of the effective potential
\begin{eqnarray*}
\frac{dV(\phi_{c})}{dt}=0.
\end{eqnarray*}
Using the chain rule we obtain the Callan-Symanzik equation for the effective potential
\begin{eqnarray}
\left(\frac{\partial}{\partial t}+\beta_{g_{i}}\frac{\partial}{\partial g_{i}}+m^{2}\gamma_{m}\frac{\partial}{\partial m^{2}}+ \phi_{c} \gamma\frac{\partial}{\partial\phi_{c}}\right)V(\phi_{c})=0,\label{eq:RGE-EP}
\end{eqnarray}
with
\begin{eqnarray}
\beta_{g_{i}}=\frac{dg_{i}}{dt}~~,~~m^{2}\gamma_{m}=\frac{dm^{2}}{dt}~~,~~\gamma=\dfrac{1}{\phi_{c}}\frac{d\phi_{c}}{dt}.\label{eq:MC-EP}
\end{eqnarray}
The beta functions of the coupling constants $\beta_{g_{i}}$, the
anomalous dimension $\gamma$ and the function $\gamma_{m}$ are computable
in perturbation theory as a power series expansion in the couplings
$g_{i}$ and they don't include logarithmic terms. Their non-perturbative values
are unknown, therefore the renormalization group
equation (\ref{eq:RGE-EP}) can only be perturbatively solved. However, if the renormalized coupling constants are small, the effective potential, seen as the perturbative solution of the RG equation, can be determined to any level
of accuracy and will not require that $g_{i}ln(\phi_{c}^{2}/\mu^{2})\ll1$
to obtain the accurate potential, it is just necessary that $g_{i}\ll1$. 

The Callan-Symanzik equation is a first-order partial differential equation. To solve it the method of characteristic curves is usually used. This method is based on restricting the partial differential  RG equation on curves, called characteristics, along which the eq. (\ref{eq:RGE-EP}) becomes an ordinary differential equations set. Suppose for a moment that a solution $V(\phi_{c})$ is known. In a renormalizable theory the effective potential must be of the form
\begin{eqnarray*}
V(\phi_{c})=\frac{1}{2}m^{2}(t)\phi_{c}^{2}(t)+\frac{1}{4}\lambda(t)\phi_{c}^{4}(t).
\end{eqnarray*}
As a consequence, the quadratic and the quartic term of the potential independently satisfies the 
Callan-Symanzik equation. The RG equation for the quartic term leads to:
\begin{eqnarray}
\left(\frac{\partial}{\partial t}+\beta_{g}\frac{\partial}{\partial g}+m^{2}\gamma_{m}\frac{\partial}{\partial m^{2}}+ \phi_{c} \gamma\frac{\partial}{\partial\phi_{c}}+4\gamma\right)\lambda(g,m^{2},\phi_{c}, t)=0. \label{RGforLambda}
\end{eqnarray}
We consider that the quartic coupling $\lambda=\lambda(g,m^{2},\phi_{c},t)$ is a surface in a five dimensional space. A normal vector of this surface would be
\begin{eqnarray}
\textbf{U}=\left(~\dfrac{\partial \lambda}{\partial g}~,~ \dfrac{\partial \lambda}{\partial m^{2}}~, ~\dfrac{\partial \lambda}{\partial \phi_{c}}~,~ \dfrac{\partial \lambda}{\partial t}~,~ 1 ~\right),
\end{eqnarray}
consequently, the vector field
\begin{eqnarray}
\textbf{V}=\left(~\beta_{g}~,~m^{2}\gamma_{m}~,~\phi_{c}\gamma~,~1~, ~ 4\gamma\lambda ~ \right)
\end{eqnarray}
is tangent to the surface $\lambda(g,m^{2},\phi_{c},t)$ at every point because (\ref{RGforLambda}) implies:
\begin{eqnarray}
{\bf U}\cdot {\bf V}= 0. 
\end{eqnarray}
Therefore, the graph of the surface must be a union of parametric curves of the vector field $\textbf{V}$, that represent a specific solution to the PDE (\ref{RGforLambda}) and are called characteristics. If a particular parametrization $t$ of the curves is fixed, the equations of the characteristic curves is the system of ordinary differential equations:
\begin{equation}
dt'={dg\over{\beta_{g}}}={dm^{2}\over{m^{2}\gamma_{m}}}={d\phi_{c}\over{\phi_{c}\gamma}}=-
{d\lambda\over 4\gamma\lambda}. \label{Eq:Curves-Char}
\end{equation}
Subject to initial conditions, we then have:
\begin{eqnarray}
\lambda(g_{i}(\Lambda), m^{2}(\Lambda),\phi_{c}(\Lambda)) =\lambda(\mu)exp\left\{-4\int^\Lambda_\mu dt'\gamma(g_{i}(t'), m^{2}(t'),\phi_{c}(t'))\right\},\label{SolLambda}
\end{eqnarray}
where $\Lambda$ is some energy scale much
larger than $\mu$, and $\mu$ is the scale that fixed the initial
conditions of the ODEs, it usually taken as the minimum of the classical
potential, $\mu\approx267$ GeV. The running parameters $g_{i}(\Lambda)$, $m^{2}(\Lambda)$ and $\phi_{c}(\Lambda)$ are solutions of the
ODEs (\ref{eq:MC-EP}). In particular the energy
dependent field $\phi_{c}(\Lambda)$ has the form:
\begin{eqnarray*}
\phi_{c}(\Lambda)=\phi_{c}(\mu)G(\Lambda) & ; & G(\Lambda)=exp\left\{ \int_{\mu}^{\Lambda}dt'\gamma(g_{i}(t'), m^{2}(t'),\phi_{c}(t'))\right\}.
\end{eqnarray*}
Therefore, the potential can be rewritten as
\begin{eqnarray*}
V(\phi_{c})= & \frac{1}{2}m^{2}(\Lambda)G^{2}(\Lambda)\phi_{c}^{2}(\mu)+\frac{1}{4}\lambda(\Lambda)G^{4}(\Lambda)\phi_{c}^{4}(\mu).
\end{eqnarray*}
To find the running coupling $\lambda(\Lambda)$ we need the explicit
form of the renormalization group equations (\ref{eq:MC-EP}) for
all couplings $g_{i}$ of the SM. This produces a system of coupled
differential equations that must be solved, involving numerical integration
of the coupled first-order equations. Once one set the appropriate boundary conditions,
the complete renormalization group improved (RGI) effective potential
is obtained. 

The expression of the RGI effective potential seems don't have any relation with the 1PI version obtained from the vacuum bubbles diagrams. However, one can show that the RGI potential, if one assume that $\beta_{\lambda}$, $\gamma_{m}$ and $\gamma$ are constants, is equal to the 1PI potential in the limit when $\phi_{c}$ is larger than the electro-weak scale. Let see this in more detail. If the beta and gamma functions are constants, the coupling $\lambda$, the parameter $m^{2}$ and the classical field $\phi_{c}$ have the approximate solutions:
\begin{eqnarray*}
\lambda(\Lambda)\approx\lambda(\mu)+\beta_{\lambda}ln\left(\frac{\Lambda^{2}}{\mu^{2}}\right),\\
m^{2}(\Lambda)\approx m^{2}(\mu)\left(1+\gamma_{m}ln\left(\frac{\Lambda^{2}}{\mu^{2}}\right)\right),\\
G^{n}(\Lambda)\approx1+n\gamma ln\left(\frac{\Lambda^{2}}{\mu^{2}}\right).
\end{eqnarray*}
These equations lead to the one-loop RGI effective potential 
\begin{eqnarray}
V_{RGI}(\phi_{c})\approx\frac{1}{2}m^{2}(\mu)\left[1+(\gamma_{m}+2\gamma)ln\left(\frac{\Lambda^{2}}{\mu^{2}}\right)\right]\phi_{c}^{2}(\mu) ~~~~~~~~~~~~~~~~~~~~~~ \nonumber \\ +\frac{1}{4}\left[\lambda(\mu)+(\beta_{\lambda}+4\lambda(\mu)\gamma)ln\left(\frac{\Lambda^{2}}{\mu^{2}}\right)\right]\phi_{c}^{4}(\mu).\label{eq:V-RGI}
\end{eqnarray}
Thus, if we want to find the one-loop improved potential, we need the
beta function $\beta_{\lambda}$, the anomalous dimension $\gamma$,
and the function $\gamma_{m}$ to one-loop order. All beta functions
and anomalous dimensions of the SM can be seen in the reference \cite{Ramond}.
Nevertheless, one can easily determine the beta function for $\lambda$
and the function $\gamma_{m}$ by looking at the 1PI effective
potential if one knows by other means the anomalous dimension. In
the Appendix~\ref{AppBetaF-EP} we show this computation. Here we only expose
the final results: 
\begin{eqnarray}
&\beta_{\lambda}=\dfrac{1}{16\pi^{2}}\left[12\lambda^{2}+\dfrac{3}{8}g^{4}+\dfrac{3}{16}(g^{2}+g'^{2})^{2}-3h_{t}^{4}-3\lambda g^{2}-\dfrac{3}{2}\lambda(g^{2}+g'^{2})+6\lambda h_{t}^{2}\right],\\ 
\nonumber \\
&\gamma=\dfrac{1}{64\pi^{2}}\left(\dfrac{9}{2}g^{2}+\dfrac{3}{2}g'^{2}-6h_{t}^{2}\right), ~~~~~~ \gamma_{m}=\dfrac{3\lambda}{8\pi^{2}}-2\gamma.
\end{eqnarray}
We now must introduce the above beta functions and anomalous dimension
into eq. (\ref{eq:V-RGI}) to get the renormalization
group improved potential. Defining $m^{2}=m^{2}(\mu)$, $g_{i}=g_{i}(\mu)$ and $\phi_{c}=\phi_{c}(\mu)$ we find the result:
\begin{eqnarray}
V_{RGI}(\phi_{c})\approx\frac{1}{2}m^{2}\left[1+\frac{12\lambda}{32\pi^{2}}ln\left(\frac{\Lambda^{2}}{\mu^{2}}\right)\right]\phi_{c}^{2} ~~~~~~~~~~~~~~~~~~~~~
~~~~~~~~~~~~~~~~~~~~~~~~~~~~~~~~~~ \label{V_RGI} \\ +\frac{1}{4}\left[\lambda+\frac{1}{16\pi^{2}}\left(12\lambda^{2}+\frac{3}{8}g^{4}+\frac{3}{16}(g^{2}+g'^{2})^{2}-3h_{t}^{4}\right)ln\left(\frac{\Lambda^{2}}{\mu^{2}}\right)\right]\phi_{c}^{4}. \nonumber 
\end{eqnarray}
The same outcome is obtained if one assumes that $\phi_{c}\sim\Lambda$ (in the limit $\Lambda\gg v$) over the one-loop 1PI potential
\begin{eqnarray}
V^{(1)}(\phi_{c})=\frac{1}{64\pi^{2}}\left[m_{H}^{4}\left(ln\frac{m_{H}^{2}}{\mu^{2}}-\frac{3}{2}\right)+3m_{G}^{4}\left(ln\frac{m_{G}^{2}}{\mu^{2}}-\frac{3}{2}\right)\right. ~~~~~~~~~~~~~~~~~~~~~\\ \nonumber \\
+\left.6 m_{W}^{4}\left(ln\frac{m_{H}^{2}}{\mu^{2}}-\frac{5}{6}\right)+3m_{Z}^{4}\left(ln\frac{m_{Z}^{2}}{\mu^{2}}-\frac{5}{6}\right)-12m_{t}^{4}\left(ln\frac{m_{t}^{2}}{\mu^{2}}-\frac{3}{2}\right)\right]. \nonumber \label{eq:EP-UP-1l}
\end{eqnarray}
In this case the scalar sector of the potential approximates to
\begin{eqnarray}
\dfrac{1}{64\pi^{2}} \left(12\lambda^{2}\phi_{c}^{4} + 12m^{2}\lambda \phi_{c}^{2}\right)ln\left(\dfrac{\Lambda^{2}}{\mu^{2}}\right),
\end{eqnarray}
the vector sector approximates to
\begin{eqnarray}
\dfrac{1}{64\pi^{2}} \left[\dfrac{3}{8}g^{4}+\dfrac{3}{16}(g^{2}+g'^{2})^{2}\right]\phi_{c}^{4}ln\left(\dfrac{\Lambda^{2}}{\mu^{2}}\right),
\end{eqnarray}
and the fermion sector to
\begin{eqnarray}
-\dfrac{1}{64\pi^{2}}3h_{t}^{4}\phi_{c}^{4}ln\left(\dfrac{\Lambda^{2}}{\mu^{2}}\right). 
\label{orig-inst}
\end{eqnarray}
Adding up the above approximations we recover the result (\ref{V_RGI}). This approximate result describes the correct behaviour of the effective potential for large values of the classical field, the unwanted logarithmic terms with $\phi_{c}$ as an argument have disappeared. It is usual to express this potential, preserving all terms proportional to $\phi_{c}^{4}$, in the form
\begin{eqnarray}
V(H) \approx \lambda_{eff}(H)\dfrac{H^4}{4},
\label{QUARTIC}
\end{eqnarray}
where the quadratic term of the potential has been ignored in the limit $H\gg v$, the reasons to make this approximation will be given in the next chapter. The new defined effective coupling $\lambda_{eff}(H)$ can be written at one-loop level as:
\begin{eqnarray}
\lambda_{eff}(H)=\dfrac{e^{4\Gamma (H)}}{(4\pi)^{2}} \left[\dfrac{3g^{2}}{8} \left( ln\dfrac{g^{2}}{4}-\dfrac{5}{6}+2\Gamma \right) + \dfrac{3}{16}(g^{2}+g'^{2})^{2} \left( ln\dfrac{g^{2}+g'^{2}}{4}-\dfrac{5}{6}+2\Gamma \right) \right. \nonumber \\
 \left. -3h_{t}^{4} \left( ln\dfrac{h_{t}^{2}}{2} - \dfrac{3}{2} + 2\Gamma \right)  + 
 3\lambda ^{2}\left(4ln\lambda -6+3ln3 + 8\Gamma \right) \right], \label{lambda-eff}
\end{eqnarray}
with 
\begin{eqnarray}
\Gamma(H) = \int _{\mu}^{H} \gamma (\bar{\mu})d ln(\bar{\mu}).
\label{GAMMA}
\end{eqnarray}
This expression is very useful in the study of the vacuum stability properties, we postpone its use to the Chapter \ref{cha:TwoLoopTadpoles}. where an effective quartic coupling are determined in the Sirlin-Zucchini renormalization scheme. 

\chapter{\noun{\label{cha:The Stability Problem of the SM}\index{Scale Dependent Properties of the SM}}Scale Dependent Properties of the SM}

\lettrine{B}{ efore} the discovery of the Higgs boson in the LHC by the ATLAS  \cite{ATLAS} and CMS  \cite{CMS} experiments, the Higgs boson's mass was the only unknown input parameter in the SM. Although its experimental value was unknown, it was always possible to obtain lower or higher bounds of the Higgs mass by imposing theoretical restrictions as the stability of the Higgs effective potential, the perturbativity of the self-coupling $\lambda$ (known as triviality) or some additional theoretical criterion. The stability analysis is the most stringent procedure to obtain a lower bound of the Higgs mass value. The requirement that $\lambda(\mu)$ has to be positive up to a given value of the energy scale $\mu=\Lambda_{I}$, translates into a lower limit of the initial condition $\lambda(\mu=v)$, or equivalently into a lower limit of $m_{H}=2\lambda(v)v^{2}$. The bound becomes more stringent if one requires the positivity of $\lambda(\mu)$ in a large range of $\mu$. With the new data for the Higgs mass, $m_{H}=125.09 \pm 0.21 ({\rm stat})\pm 0.11 ({\rm syst})$ GeV, arising from the recent combined ATLAS and CMS analysis \cite{comb}, the problem of the vacuum stability  of the SM has started to be considered by itself. \\
Following this guideline, in the section 3.1 we discuss the scale dependent properties of the Standard Model. We first look at the behaviour of the running parameters $m^{2}(\mu)$ and $\lambda(\mu)$, and we review the state of the art of the vacuum stability analysis, then we focus our attention on two very important issues, the triviality of SM and the hierarchy problem.  

\section{The Stability Problem}

\begin{figure}
\begin{center}
\scalebox{0.5}{
\fcolorbox{white}{white}{
  \begin{picture}(576,196) (133,-63)
    \SetWidth{1.0}
    \SetColor{Black}
    \Arc[arrow,arrowpos=0.5,arrowlength=5,arrowwidth=2,arrowinset=0.2](192,36)(57.689,326,686)
    \Arc[arrow,arrowpos=0.5,arrowlength=5,arrowwidth=2,arrowinset=0.2](426,34)(57.689,326,686)
    \Line[dash,dashsize=10,arrow,arrowpos=0.5,arrowlength=5,arrowwidth=2,arrowinset=0.2](372,34)(486,34)
    \Arc[arrow,arrowpos=0.5,arrowlength=5,arrowwidth=2,arrowinset=0.2](648,34)(57.689,326,686)
    \Text(186,112)[lb]{\Large{\Black{$t$}}}
    \Text(420,112)[lb]{\Large{\Black{$t$}}}
    \Text(426,-8)[lb]{\Large{\Black{$(t,~b)$}}}
    \Text(426,52)[lb]{\Large{\Black{$S$}}}
    \Text(642,106)[lb]{\Large{\Black{$t$}}}
    \Text(648,-8)[lb]{\Large{\Black{$(t,~b)$}}}
    \Text(648,52)[lb]{\Large{\Black{$V$}}}
    \Vertex(368,36){4}
    \Vertex(484,32){4}
    \Vertex(592,36){4}
    \Vertex(704,36){4}
    \Photon(592,36)(704,36){7.5}{6}
    \Text(196,-68)[lb]{\Large{\Black{$(a)$}}}
    \Text(432,-68)[lb]{\Large{\Black{$(b)$}}}
    \Text(656,-68)[lb]{\Large{\Black{$(c)$}}}
  \end{picture}
}}
\end{center}
\caption{\label{instability}Diagrams that generate the instability in the two-loop Higgs effective potential.}
\end{figure}
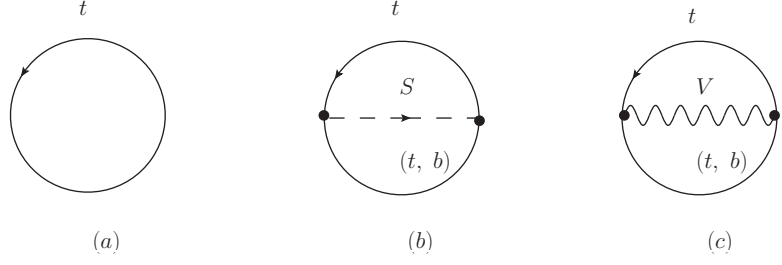

\begin{figure}
\begin{tabular}{c c}
\includegraphics[scale=0.28]{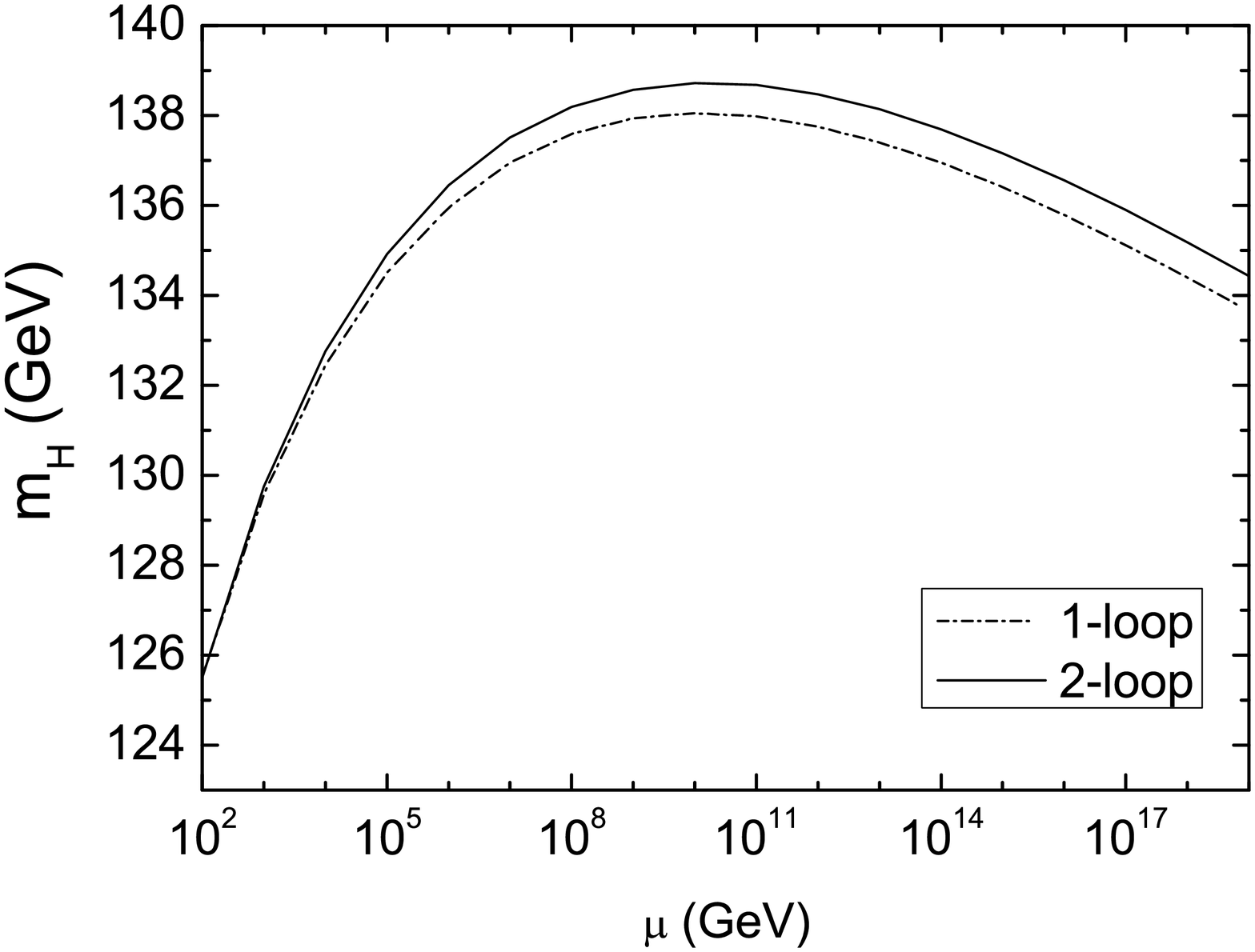}  & \includegraphics[scale=0.28]{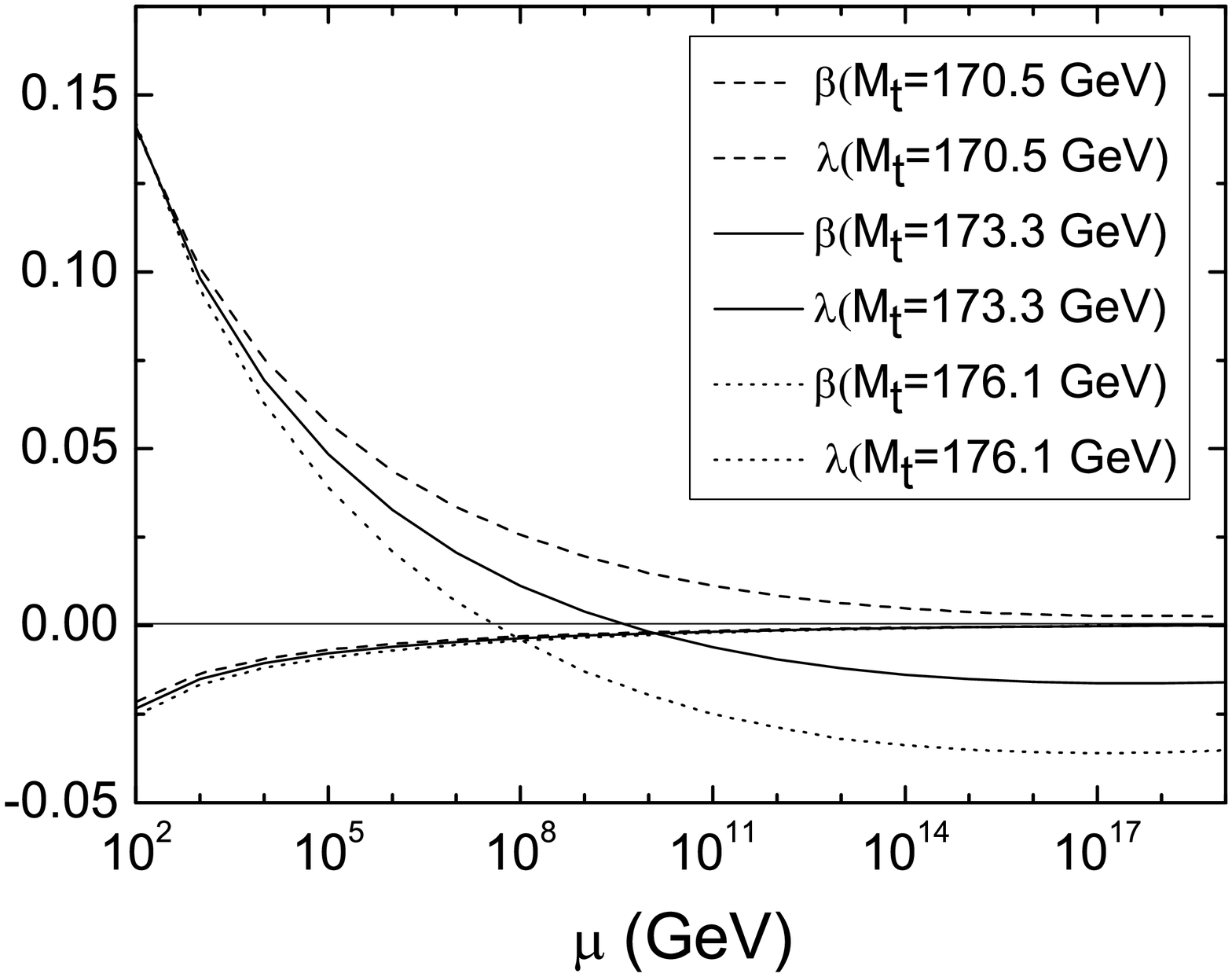}
\end{tabular}
\caption{\label{lambdarun} \textbf{Left:} Evolution of the mass term $m(\mu)$ up to one-loop and two-loop level. \textbf{Right:} RG evolution of the Higgs quartic coupling $\lambda$ up to the Planck scale at different values of the scale $\mu=M_t$. \cite{Ligong}}
\end{figure}

To study the stability of the SM vacuum, we need to look at the shape
of the Higgs effective potential. As mentioned in the previous chapter,
the $1PI$ effective potential is not reliable when we want to study
the shape of the potential at large values of the classical field
$\phi_{c}$. The analysis of the stability must be obtained from a
renormalization group improved version of the potential that requires
the study of the RGEs (Renomalization Group Equations) of all couplings ($\lambda,\: h_{t},\: g,\: g',\: g_{s},$
etc) in the SM and find the most precise relation between the running
coupling constants and the value of the SM observables at given scales, a sort of boundary conditions for those RGE. We first focus our attention in the running quartic Higgs
coupling $\lambda(\Lambda)$, and its threshold relation with the
Higgs physical mass
\begin{eqnarray}
\lambda(\Lambda)=\frac{G_{\mu}}{\sqrt{2}}m_{H}^{2}+\Delta\lambda(\Lambda). \label{runl}
\label{threSZ}
\end{eqnarray}
In (\ref{runl}) $G_{\mu}$ is the Fermi constant of the muon decay and $\Lambda$ represents the subtraction point. The equation  (\ref{threSZ}) will be analyzed in more detail in Chapter 4 where also the equation will be characteristic of the Sirlin-Zucchini renormalization scheme \cite{SirlinZucchini}.  
The RG evolution from the EW scale until the Planck scale implies
that $\lambda$ gets negative at the energy scale of $10^{10}$ GeV
as you can see in fig. \ref{lambdarun} - Right. The instability occurs due to the effects of the top quark corrections represented in fig. (\ref{instability}) up two-loop order.
For instance in the one-loop RGI potential (eq. \ref{orig-inst}) and therefore in $\lambda_{eff}(H)$ (eq. \ref{lambda-eff}) they contribute as $-\frac{3h_{t}^{4}}{16\pi^{2}}2\Gamma$, 
being $h_{t}$ the top Yukawa coupling. The troubles with $\lambda$
becoming negative is that it will cause an instability in the Higgs
potential, because at the very high values of the Higgs field the
effective potential is dominated by the quartic term. The fig. \ref{lambdarun} - Left, shows the running of $m(\mu)$, the quadratic coupling is always lower than $1$TeV, thus $m^{2}\ll\phi_{c}^{2}$ at the large energy scale \cite{Ligong}. Lets see this more detailed, consider the running of the parameter $m(\mu)$ obtained by solving the RGE (\ref{eq:MC-EP}) up to two-loop order\footnote{The three loop functions play a minor role in the vacuum stability analysis \cite{M.F.Zoller}} and plotted in fig. \ref{lambdarun} - Left. The boundary conditions on the SM running parameters are imposed from the experimental values of all physical masses in the SM, for instance the top Yukawa coupling is obtained from the top quark pole mass $m_{t}$ at $\Lambda=m_{t}^{{\rm pole}}= 173.36\pm 2.8~{\rm GeV}$ \cite{ADM}, while the mass term is normalised in such way that $m=m_{H}=125.66\pm 0.34~{\rm GeV}$ at tree level. For a more refined analysis it is necessary include the matching conditions and the threshold corrections to the relations between the couplings of the SM and the electro-weak observables. These relations will be
obtained in the on-shell renormalization scheme of Sirlin-Zucchini together with all running properties.\\
The dashed line in fig. \ref{lambdarun} - Left, represents the behaviour of the
parameter $m(\mu)$ when the RGEs are taken into account up to
one loop level, while the solid line includes the beta and gamma functions
up to two loop order. We can observe that the two-loop corrections
increases the value of $m(\mu)$ for any value of $\mu$,
the increase is greater about the scale $10^{10}$ GeV where it is well
know that the instability of the potential occurs. The most
appropriate is to use the function $\gamma_{m}$ at least up two-loop
order in eq. (\ref{eq:MC-EP}) when one studies the RG evolution
of the Higgs mass; the value of the Higgs mass is near-critical in the sense that it is very close to the division line between the metastability and absolute stability regions, the stability status of SM is thus very sensitive to the higher order radiative corrections. Now, the fig. \ref{lambdarun} - Left shows that $m$ stops increasing around the energy $\Lambda=10^{10}$ GeV and then damping
always when the scale $\mu$ grows. The value of $m$ remains positive
and lower than $140$ GeV $\sim 0.1$ TeV. As a consequence, when one is
interested in studying the shape of the potential at large values of
the classical field, $\phi_{c}\gg1$ TeV, the quadratic term of the
RGI potential can be ignored, and the potential approximates to
\begin{eqnarray}
V_{RGI}(\phi_{c})\approx\frac{\lambda(\Lambda)}{4}\phi_{c}^{4}(\Lambda).
\label{eq:RGI-Vapprox}
\end{eqnarray}
The stability of the potential translates to the study of the evolution
of the quartic coupling $\lambda(\Lambda)$, more precisely to its
positivity at large scales. The behaviour of the Higgs quartic coupling
is obtained from the beta functions of all couplings. The parameter $m^{2}$ does not enter into the beta function of $\lambda$ explicitly, as you can see at one-loop level in eq. (\ref{beta-lambda-1l}) and at two-loop in eq. (\ref{beta-lambda-2l}), therefore the RGE for the $m^{2}$ (eq. \ref{gammam2-1l} and \ref{gammam2-2l}) should not be considered. Solving the RGE flow numerically one obtains the fig. \ref{lambdarun} - Right \cite{Ligong}.\\ 
The dotted and the dashed lines correspond to the extreme values
of the top mass, $m_{t}=173.36+2.8$ GeV and $m_{t}=173.36-2.8$ GeV
respectively, the continuous line is the central value of the top mass,
$m_{t}=173.36$ GeV. The fig. \ref{lambdarun} - Right, shows that the scale
at which $\lambda(\mu)$ becomes negative, decreases when the value
of the top mas grows. For the central value of $m_{t}$ the coupling
$\lambda(\mu)$ damping to be negative from about $\Lambda_{I}=10^{10}$ GeV,
the subscript $I$ indicates that it is the critical energy scale
where the instability appears. The instability scale occurs at energies
much bigger than the EW scale ($v\sim1$TeV) thus, the approximation
in eq. (\ref{eq:RGI-Vapprox}) of neglecting $m$ with respect to
the value of the field $\phi_{c}$ is justified. For the extreme value of the top quark pole mass $m_{t}=173.36-2.8$ GeV, $\lambda(\mu)$ approaches zero around
the scale $10^{17}$ GeV, just below the Planck scale and where the
beta function of the coupling $\lambda$ also approaches zero (see
\cite{Degrassi}). The Higgs potential is stable and the vacuum stability
of the SM is achieved for $m_{t}<170.5$GeV \cite{Degrassi, LindnerSher}.

\subsection{Threshold Corrections in Stability Analysis}

The state of the art of the NNLO (next-to-next-to-leading order) stability analysis in the SM involves nowadays the RGEs of all coupling constants up to three loop level (see for instance \cite{Zoller}) and the threshold corrections at the weak scale up two-loop order, moreover the three-loop threshold corrections are not relevant. The reason for that is that we are working with the improved effective potential and as it was proved in \cite{kast},\cite{Bando} that the L-loop effective potential improved by (L+1)-loop RGE resums all the Lth-to-leading  logarithm contributions. Therefore for the vacuum stability analysis at L-loop the threshold value at L-loop also will be needed.
The first point is achieved by imposing the renormalization group equation to the effective potential in an appropriate choice of coordinates, defined in the space of fields and parameters. By imposing  the RG equation to the new parametrized version of the effective potential, one is able to reconstruct the full potential since the leading logarithms have coefficients determined by the tree-level potential and the one-loop result for the beta and gamma-functions. Analogous approach for the subleading logarithms. The crucial point is that the improved potential is obtained in terms of running parameters in the fields space.
For the last point  consider the improved effective potential, which for large fields and by dimensional analysis will have the shape $V(\phi_c)=\lambda_{\rm eff}\phi_c^4$, here $\lambda_{\rm eff}$ has the same order in loop of the effective potential the same must be for its threshold value.

For the NNLO vacuum stability the three loop RGEs have been completely computed. The three loop beta function of the gauge couplings $g_{s}$, $g$ and $g'$ can
be found in \cite{Mihaila1,Mihaila2}. The Yukawa coupling
of the top quark was fully computed in \cite{M.F.Zoller} and finally the
RGEs of the quartic coupling $\lambda$ and the quadratic parameter
$m^{2}$ up three loop level was obtained in \cite{Chetyrkin1,Bednyakov}. The two-loop threshold corrections of all relevant SM couplings are found in \cite{Degrassi}, where the NNLO corrections of $\lambda(\mu)$ was computed in the gauge-less limit: here the dynamics of the electro-weak bosons is not taken into account ($g=g'=0$). The full NNLO computation for $\lambda$, $m^{2}$ and $h_{t}$ is found in \cite{Degrassi2}, the
gauge contribution to the two-loop correction of $\lambda$ represents
around $0.6\%$ of the total value. The NNLO corrections of $g$ and
$g'$ have not yet fully calculated. We focus here our attention in the threshold
corrections of the quartic coupling $\lambda$. The calculation was
done in a renormalization scheme where the effects of the physics
at very large energy scales can be deduced from those of the theory
with the electro-weak scale, the so called Sirlin-Zucchini scheme.
\begin{figure}
\centering
\begin{tabular}{c c c}
\includegraphics[scale=0.7]{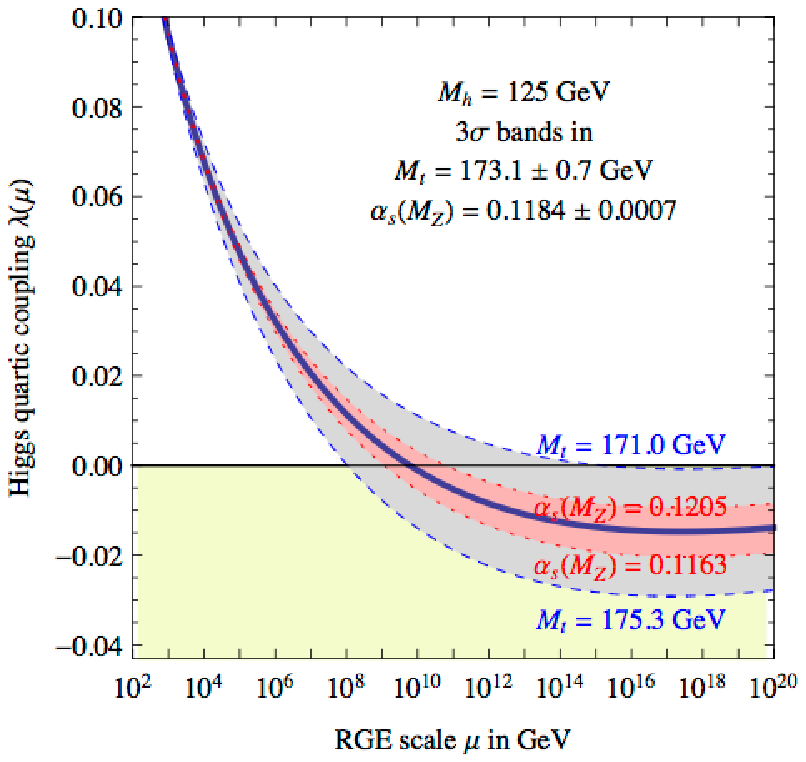}  & \;\;\; & \includegraphics[scale=0.67]{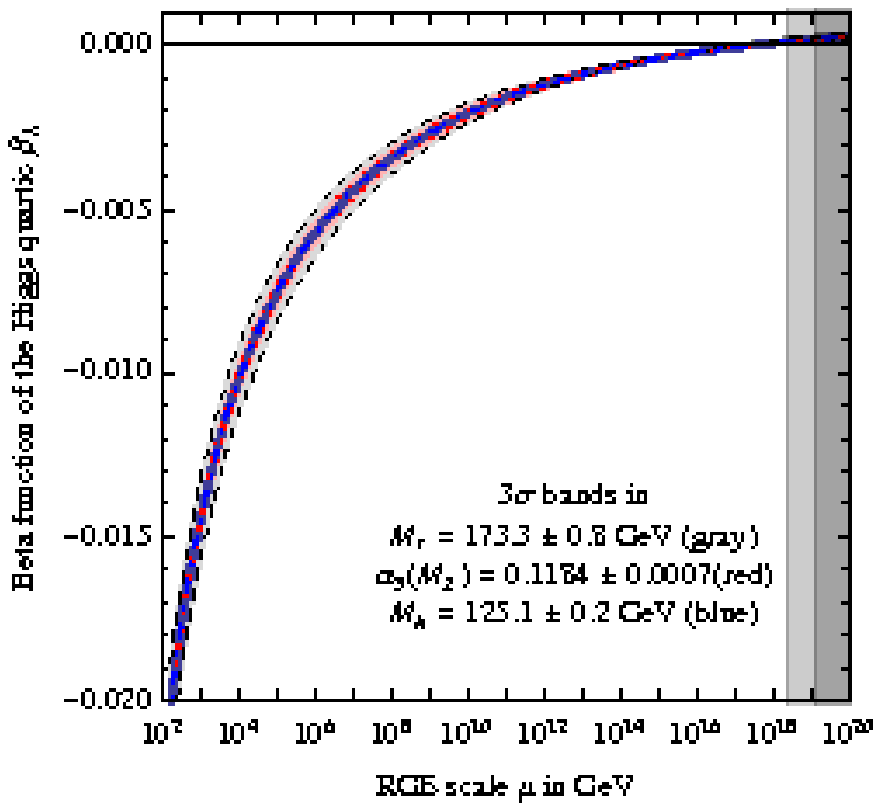}  
\end{tabular}
\caption{\label{BetaandLambda} \textbf{Left:} Evolution of the $\overline{MS}$ quartic coupling $\lambda(\mu)$. \textbf{Right:} Beta function of $\lambda$ varying $m_t$, $\alpha_{s}(m_Z)$ and $m_H$ by $\pm \sigma.$ \cite{Degrassi}}
\end{figure}
Thus, $\lambda(\mu)$ is computed in
terms of the Fermi coupling $G_{\mu}$, the pole masses{\footnote {This denomination applies to the K\" allen-Lehmann spectral decomposition of the propagator, for which the physical mass appears as pole of the propagator. Of course in the case of the top quark this is just a formal definition.}} $m_{H}$,
$m_{t}$, $m_{Z}$, $m_{W}$ and $\alpha_{s}(m_{Z})$. The threshold condition up two-loop order is given as

\begin{eqnarray}
\lambda(\mu)=\frac{G_{\mu}}{\sqrt{2}}m_{H}^{2}+\delta\lambda^{(1)}(\mu)+\delta\lambda^{(2)}(\mu).
\end{eqnarray}
The one-loop correction $\delta\lambda^{(1)}$ was computed analytically
by Sirlin and Zucchini in \cite{SirlinZucchini}, whereas the two-loop
correction \cite{Degrassi}
\begin{eqnarray}
\delta\lambda^{(2)}(\mu)=-\frac{G_{\mu}}{\sqrt{2}}m_{H}^{2}\left[-\frac{\Delta r^{(1)}}{m_{H}^{2}}\left(m_{H}^{2}\Delta r^{(1)}+\frac{3}{2}\frac{T^{(1)}}{v}+Re\Pi_{HH}^{(1)}(m_{H}^{2})\right)\right. \nonumber \\ 
\left.+\Delta r^{(2)}+\frac{1}{m_{H}^{2}}\left(\frac{T^{(2)}}{v}+Re\Pi_{HH}^{(2)}(m_{H}^{2})\right)\right]_{fin}+\Delta_{\lambda},
\end{eqnarray}
included the radiative correction ($\Delta r$) to the relation between the $SU(2)_{L}$ gauge coupling and the Fermi constant , eq. (\ref{muon-g}), the two-loop
tadpoles $T^{(2)}$, and the numerical evaluated Higgs self-energy
$\Pi_{HH}^{(2)}(m_{H}^{2})$ using the Martin's loop functions calculated in \cite{Martin1,Martin2003}. As it will be explained later $\Delta_\lambda$ is
the two-loop finite contribution that is obtained when the on-shell parameters of the Sirlin-Zucchini renormalization scheme, entering in the pole of the regularizator at one-loop are expressed in terms of MS quantities, that finite contribution comes from
the $O(\epsilon)$ part of the shifts.
The numerical result at $m_{t}$
is \cite{Degrassi}:
{\small
\begin{eqnarray}
\lambda(m_{t})=0.12710+0.00206\left(\dfrac{m_{H}}{GeV}-125.66\right)-0.00004\left(\dfrac{m_{t}}{GeV}-173.35\right)\pm0.00030_{th}. \label{runLatMt}
\end{eqnarray}}
The sign of $\lambda$ at the Planck scale $\Lambda_P\sim 10^{19}$ GeV depends primarily on the assumed value for $m_{t}$ and secondly of $m_{H}$. For its central
value $m_{t}=173.4$ GeV, $\lambda$ becomes negative starting from $\Lambda\sim10^{10}-10^{11}$ GeV.
The three loop corrections do not change the stability condition
substantially. The most relevant three-loop effect is due to the QCD
contributions, its effect is larger around Planck scale, see the red
bands in figure \ref{BetaandLambda}. Including the three loop QCD
corrections, the stability requires \cite{Degrassi2}
\begin{eqnarray}
m_{H} >  129.6 \pm 1.5~GeV  ~~~;~~~ m_t < 171.36  \pm 0.46~GeV. \label{StabCond}
\end{eqnarray}
The uncertainties include experimental and theoretical errors found in \cite{ADM}. The main source of uncertainty comes from the experimental error in $m_t$, its contribution to the uncertainty on $m_H$ is $\pm 1.4$ GeV. Along with the error on $m_H$ due to the combined theoretical and experimental uncertainty in $\alpha_{s}$ evaluated at the scale of the Z boson mass \cite{Nakamura} ($\Delta^{\alpha_{s}}~m_{H} \approx 1$ GeV) and the fairly small theoretical errors due to the threshold corrections ($\Delta^{th}~m_{H}\approx 0.7$ GeV) we obtain the total error on $m_H$ and $m_t$ of eq. (\ref{StabCond}). It is remarkable to note here that $\lambda$ never becomes too negative at $\Lambda_{P}$, in fact, $\lambda$ and $\beta_{\lambda}$ are very close to zero around $\Lambda_{P}$ (see fig.~\ref{BetaandLambda}), this situation is currently known as near-criticality, because the Higgs and top mass values of the eq. (\ref{StabCond}) put the SM very close to a critical line, determined by the condition $\lambda(\Lambda_{P})=\beta_{\lambda}(\Lambda_{P})=0$, that divides two stability phases: the metastability and the absolute stability. It will be developed in more detail in the next subsection.

\section{Metastability Scenario}

The behaviour of $\lambda(\Lambda)$ studied in the above section modified the shape of the Higgs effective potential at large energy scale values. The RGI effective potential bends down for values of the classical field $\phi_{c}$ much larger than the location of the electroweak minimum $v$, and develops a new minimum $v'$ at $\phi_{c}\gg v$. Depending on the SM parameters, in particular on the top and Higgs masses, the second minimum can be higher or lower than the electroweak one. In the first case, the electroweak vacuum is stable. In the second one, it is metastable and we have to consider the life-time of the false electroweak vacuum, $\tau$, and compare it with the age of the universe $\tau_{U}$.
The requirement of metastability under quantum fluctuations is given at zero
temperature. The only cosmological input required is the age of the universe $\tau_{U}$, but does not rest on any cosmological assumptions. The bound is formulated by requiring that the probability of quantum tunnelling out of the electroweak minimum be sufficiently small when integrated over this time interval. If $\tau$ turns out to be larger than $\tau_{U}$, even though the electroweak vacuum is not the absolute minimum of $V_{eff}(\phi_{c})$, our universe may well be sitting on a metastable vacuum, otherwise the vacuum is unstable. This is the well-known metastability scenario. Let see first the details of the computation of the SM vacuum lifetime and then its implications in the stability analysis. 

\subsection{Lifetime of the SM Vacuum}

The analysis of the above section shows that our universe is potentially unstable for the measured values of $m_{H}$ and $m_{t}$, and therefore that the electroweak vacuum is not the true vacuum of the theory. It is a problem that must be cured via the appearance of new physics only if the transition probability of the false vacuum to the true vacuum is smaller than the life of the universe. But, if $\lambda$ remains small in absolute magnitude, the SM vacuum is unstable but can be sufficiently long-lived compared to the age of the Universe. This is the condition to a metastable vacuum. The lifetime can be long because quantum tunnelling $\wp$ is an exponentially damped process. In semi-classical approximation, one can compute the quantum tunnelling probability as \cite{Coleman2}: 
\begin{eqnarray}
\wp \approx\left(\tau_{U}\Lambda_{B}\right)^{4}e^{-S(\Lambda_{B})},
\end{eqnarray}
where $S(\Lambda_{B})$ is the Euclidean action evaluated in the saddle-point approximation at the solution (the bounce $H=h(r)$) of the classical field equations \cite{Lee}
\begin{eqnarray}
-\partial_{\mu}\partial_{\mu}h+V'(h)=-\dfrac{d^{2}h}{dr^{2}}-\dfrac{3}{r}\dfrac{dh}{dr}+V'(h)=0, \label{eqofmot}
\end{eqnarray}
that interpolates between the false vacuum, $v$, and the opposite side of the barrier, the new minimum $v'$. Here the Higgs field depends only on the radial coordinate, $r$, $V'(h)$ represents the first derivative of the effective potential $V(h \gg v)\approx\frac{1}{4}\lambda(h)h^{4}$ with respect to $h$, and $\Lambda_{B}$ is the scale at which $\wp$ is maximized. The equation (\ref{eqofmot}) satisfies the boundary conditions:
\begin{eqnarray}
h'(0)=0 & ; & h(\infty)=v\rightarrow 0.
\end{eqnarray}
With this boundary conditions, the Euclidean equation of motion can be solved analytically, with negative $\lambda$, to obtain the tree-level bounce \cite{Lee} (always in Natural Units)
\begin{eqnarray}
h(r)=\sqrt{\dfrac{2}{|\lambda|}}\dfrac{2R}{r^{2}+R^{2}}, \label{bounce}
\end{eqnarray} 
where $R=\Lambda_{B}^{-1}$ is the size of the bounce. This leads to the action \cite{Arnold}
\begin{eqnarray}
S(\Lambda_{B})=\dfrac{8\pi^{2}}{3|\lambda(\Lambda_{B})|}. \label{actionSLB}
\end{eqnarray}
The scale $\Lambda_{B}$ and therefore $R$ are obtained when $\wp$ is maximized, according to eq. (\ref{actionSLB}) it is equivalent to minimize $\lambda(\Lambda_{B})$, which correspond to the condition $\beta_{\lambda}=0$. At tree level the quartic coupling $\lambda$ is scale-invariant and the size of the bounce $\Lambda_{B}^{-1}$ is arbitrary. This leads to a degenerated action with $R=\Lambda_{B}^{-1}$, however the degeneracy being lifted by quantum fluctuations, because the RG flow breaks scale invariance, as is shown at one-loop level in~\cite{Isidori}. Including one-loop corrections to the action, $\Delta S_{1}$, the one-loop tunnelling rate becomes
\begin{eqnarray}
\wp = \dfrac{1}{\tau_U}\left[\dfrac{S(\Lambda_{B})^{2}}{4\pi^{2}}\dfrac{\tau_{U}^{4}}{R^{4}}e^{-S(\Lambda_{B})} \right]\times e^{-\Delta S_{1}}.
\end{eqnarray}
The quantum corrections to the tunnelling rate and the explicit computation of the one-loop action, $\Delta S_{1}$, are found in \cite{Isidori}. Substituting the bounce (\ref{bounce}) in the above equation, the condition $\wp<1$ for a universe of about $\tau\approx10^{10}$ years old is equivalent to a lower limit over $\lambda$, that means $\lambda$ cannot take too large negative values. The bound on $\lambda$ can be translated into a lower bound on the Higgs mass, from the RG evolution of $\lambda(\mu)$, this lower bound can be approximated as \cite{Isidori}:
{\small
\begin{eqnarray}
m_H (GeV) > 117 + 2.9[m_t (GeV) - (175 \pm 2)] - 2.5\left[\dfrac{\alpha_{s}(m_Z)-0.118}{0.002} \right] + 0.1log\left(\dfrac{\tau _{U}}{10^{10}yr} \right).
\end{eqnarray}
}
This condition is known as the metastability condition, and it is useful to build the meta-stability diagram discussed in the next subsection.

\subsection{Meta-Stability and SM Phase Diagram}

\begin{figure}
\centerline{\includegraphics[width=0.45\linewidth]{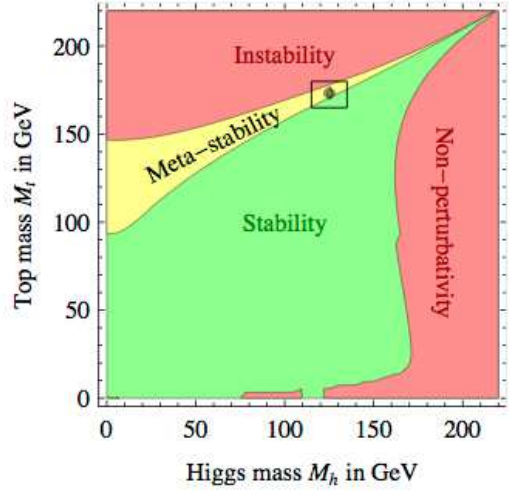}}
\caption[]{Stability phase diagram \cite{Degrassi}. The $m_H-m_t$ phase diagram is divided in three sectors, stability, instability, and metastability regions. The dot indicates $m_{H}=125.66 \pm 0.34$ GeV and $m_t\sim 173.1$ GeV with $3\sigma$ of accuracy.}\label{PhaseD}
\end{figure}
The stability analysis is generally presented with the help of a stability phase diagram in the $m_{H}-m_{t}$ plane. Neglecting the presence of new interactions up to the Planck scale, the zero-temperature meta-stability analysis provides the graph given by figure \ref{PhaseD}. This phase diagram is divided into three different sectors. An absolute stability region, the green region, where the effective potential evaluated in the electroweak minimum is lower than the effective potential evaluated in the new minimum $v'$, $V_{eff}(v)<V_{eff}(v')$. A meta-stability region, the yellow region, where the effective potential at the new minimum is lower than the effective potential at the electroweak minimum, $V_{eff}(v')<V_{eff}(v)$, but with the life-time of the electroweak vacuum larger than the age of the universe, $\tau > \tau_{U}$. Finally, an instability region, the red region, where $V_{eff}(v')<V_{eff}(v)$ but the life-time of the electroweak vacuum is lower than the age of the universe, $\tau < \tau_{U}$. The phase diagram also has two division lines. The stability line separates the stability and the meta-stability regions, this line is obtained when $V_{eff}(v)=V_{eff}(v')$. By other side, the instability line separates the metastability and the instability regions, this line is obtained when the life-time of the electroweak vacuum is equal to the age of the universe, $\tau=\tau_{U}$. \\
Given the value of the top mass $m_{t}=173.36\pm 2.8~{\rm GeV}$, and the Higgs mass value $m_{H}= 125.09 \pm 0.21 ({\rm stat})\pm 0.11 ({\rm syst})$~GeV, the analysis situate the SM inside a metastability region, very close to the stability line. When the experimental and perturbative errors in the determination of $m_{t}$ and $m_{H}$ are taken into account, represented with the $1\sigma$, $2\sigma$ and $3\sigma$ ellipses in fig. \ref{PhaseD}, the SM could be sitting on the stability region, i.e. it could reach and even cross the stability line. When it sits on the stability line, the coupling $\lambda(\Lambda)$ and the beta function $\beta_{\lambda}$ vanish at the Planck scale $\Lambda_{P}$, this scenario is known as the \textit{near-criticality}. The measured values of the top and Higgs masses suggest a near-critical electro-weak vacuum. Including the two-loop threshold corrections of all SM parameters \cite{Degrassi} Degrassi et. al. estimates an overall theory error on $m_{H} \pm 1.0$ GeV, that combined with the experimental errors on $m_{t}$ and $g_{s}$ gives as result that vacuum stability of the SM up to the Planck scale is excluded at $2.5\sigma$ (99.3\% C.L. one sided) for $ m_{H}< 126$ GeV.\\
The vanishing of $\beta_{\lambda}$ at $\Lambda_{P}$ is due to a cancellation between different large contributions, rather than an asymptotic approach to zero. For this reason the vanishing of $\lambda$ and $\beta_{\lambda}$ near to Planck scale looks like more as a coincidence that a deep consequence of the theory at large scales. However, the smallness of $\beta_{\lambda}$ and $\lambda$ at high energy implies that very small variations of the input values of the couplings at Planck scale lead to large fluctuations of the instability scale. The experimental Higgs mass value is so close to criticality that any more refined measurements, or more refined computations of theoretical and experimental errors, can be drastically change our conclusions about stability or metastability of the electro-weak vacuum. This is the main motivation to deepen our study about stability problem and its scale dependent properties. 

\section{Triviality Bound}

There are other scale-dependent properties weaker than the stability
condition that can give us information about the behaviour of the
SM at large energy scales and its theoretical restrictions. In particular important constraints can be put on the Higgs boson mass. Unfortunately, these constraints can always be evaded by postulating the existence of some unknown new physics which enters into the theory at a mass scale above the current experiments, but below the Planck scale. Nevertheless, in the SM there exists upper and lower bounds on the Higgs boson and heavy fermion masses \cite{Lindner}\cite{Casas}\cite{Maiani}. We are going to consider bounds
from triviality and from the naturalness, which, together with the previous bounds to stability and perturbativity, give limits on the Higgs mass as a function of the scale of new physics.  We first focus our attention in the triviality argument. This scenario is based on the requirement that the running self-coupling $\lambda(\mu)$ will stay finite at high energy scales, $\mu\gg v$ and that . From the renormalization group equation:
\begin{eqnarray*}
\frac{d\lambda}{d(log\Lambda^{2})}=\beta_{\lambda}\left(\lambda,g,g',h_{t},\dots\right),
\end{eqnarray*}
we can obtain the approximate solution
\begin{eqnarray}
\lambda(\Lambda)=\frac{\lambda(\mu_{0})}{1-\frac{\beta_{\lambda}}{\lambda(\mu_{0})}log\left(\frac{\Lambda^{2}}{\mu_{0}^{2}}\right)},\label{lambda-approx}
\end{eqnarray}
when the beta function is taken as constant. In the case of a theory with a spontaneous symmetry breaking there is a problem with
this solution, the running $\lambda(\Lambda)$ has a pole at some
value of $\Lambda=\Lambda_{L}$ given by 
\begin{eqnarray}
\Lambda_{L}=\mu_{0}e^{\frac{\lambda(\mu_{0})}{2\beta_{\lambda}}}. \label{LandauPole}
\end{eqnarray}
Around this particular value, $\lambda(\Lambda)$ grows up infinity
regardless of the value of the initial condition $\lambda(\mu_{0})$.
The energy scale $\Lambda_{L}$ is known as the Landau pole, and represents
the energy scale around which the theory cannot be any more described perturbatively. The general triviality argument \cite{Chivukula} establishes that Standard Model must be a perturbative theory at all energy scales, all coupling constants must be much less than unity up high energies, according to eq. (\ref{lambda-approx}) this is possible only
if $\lambda$ and $\beta_{\lambda}$ are zero. As $\beta_{\lambda}$
is an explicit function of all SM parameters, the perturbativity is
preserved only when all SM couplings vanishes, $g_{i}=0$, thus rendering
the theory non-interacting i.e. trivial. As a consequence the triviality implies
a massless Higgs boson, but the new results by ATLAS and CMS shows that
actually that is no case, the Higgs mass is different of zero, and, in fact, the Standard Model is not trivial, all coupling constants have non-zero
perturbative values up to the Planck scale.\\
\begin{figure}
\centerline{\includegraphics[bb=0bp 150bp 600bp 600bp,clip,scale=0.4]{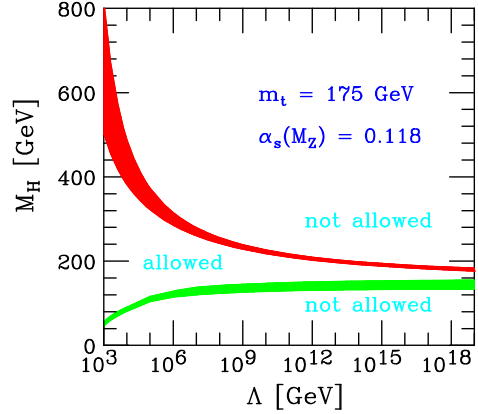}}
\caption[]{Triviality (Red) and Stability (Green) bounds as a function of the scale of new physics. \cite{Hambye}.}\label{triviality}
\end{figure}
However, we can see the triviality argument from another point of
view, we can use the solution of the RGE of $\lambda(\mu_{0})$ to
establish a valid energy region where the SM works, i.e. an energy
cut-off $\Lambda_{C}$ below which the self-coupling $\lambda$ remains
finite. If the cut-off is large, the coupling $\lambda$ must be small
to avoid the Landau pole, in turn, if the cut-off is small, $\lambda$
can be rather large according to:
\begin{eqnarray}
\frac{\beta_{\lambda}}{\lambda(\mu_{0})}log\left(\frac{\Lambda_{C}^{2}}{\mu_{0}^{2}}\right)\neq1.\label{eq:cotalambda}
\end{eqnarray}
Using the two-loop threshold relation (eq. \ref{runLatMt}) the value of the
Higgs mass can be obtained from the coupling $\lambda(\mu_{0})$ and
the central top quark mass. In this sense, the relation (\ref{eq:cotalambda})
provides an upper bound of $m_{H}$ inherited from $\lambda$. When the scale $\Lambda$ grows up, the permitted range for the Higgs mass value is reduced. For instance, if one assumes $\Lambda_{C}\sim10^{10}$ GeV the triviality imposes the upper bound $m_{H}\lesssim275$ GeV at $\mu_{0}=v$, whereas the stability analysis imposes the lower limit $m_{H}\gtrsim 150$ GeV. This range decreases with the scale of new physics $\Lambda_{C}$ as fig. \ref{triviality} shows.
We currently know the Higgs mass value at the electroweak scale,
$m_{H}\approx126$ GeV at $\mu_{0}=m_{t}$, this now imposes a new value of the Landau pole. From approximation (\ref{eq:cotalambda}) we obtain $\Lambda_{L}\gtrsim10^{38}$ at one-loop level. The Landau pole is then larger than $\Lambda_{P}$, therefore we have the freedom of choosing the energy scale where new physics can occur in any intermediate scale between the EW scale and the Planck scale or we can reasonably expect that no new physic occurs up to the Planck scale, the well known desert scenario. This issue together with the hierarchical problem and naturalness will be the subject of the following subsections.

\subsection{The Hierarchy Problem}

The observed Higgs mass leaves open the next question: why is the Higgs boson so light?
In renormalizing the Higgs mass, the radiative corrections are proportional to the mass of any particle which couples to the Higgs particle, allowing them in principle to be
as heavy as the Planck mass. For any other particle in the SM there is a symmetry protecting it to acquire a big mass by radiative corrections. For instance, the loop corrections of the electroweak vector boson masses of the theory are controlled by the three-level masses that are proportional to the vacuum expectation value (vev) according to the Spontaneous Symmetry Breaking (SSB) mechanism, ensuring that the $SU(2)_{L}\times U(1)_{Y}$ symmetry will be recovered in the limit where the masses vanish. There is no such mechanism for scalar particles, which get a mass contribution from every scale they are coupled to. A light Higgs implies a large accidental cancellation between different, in principle unrelated physical quantities, this very precise fine-tuning of the Higgs mass is known as the hierarchy or naturalness problem of the Standard Model \cite{tHooft2}. \\
To illustrate the hierarchy problem in gauge theories let us consider the radiative corrections to the Higgs propagator. The Higgs mass is an arbitrary parameter that emerges from the quadratic Higgs coupling $m^{2}$ after SSB, $m_H^{2}=2m^{2}$. The point $m^{2}=0$ is not protected by any symmetry, a non-zero quadratically UV divergent radiative corrections is induced by renormalization, this is the famous \textit{naturalness} or \textit{hierarchy problem}. The word \textit{naturalness} arises because the quadratic UV divergences make necessary a fine-tuning of the renormalized masses. All couplings of the SM are logarithmically divergent and don't require of a fine-tuning, in this sense the quadratic divergences that emerge in the masses are unnatural. Let see this in more detail. Take as an example the running Higgs mass in the broken phase, $m_{H}^2(\Lambda)=2\lambda (\Lambda) v^{2}(\Lambda)$. Since $\lambda(\Lambda)$ satisfies a RG evolution governed by logarithmic divergences only, all quadratic divergences must be generated by the renormalization of vev ($v$). More precisely, the quadratic divergences show up in the tadpole contributions to the self-energies, deeply related to the vev renormalization, or in any scalar momentum integral involved in self-energy computations that can be reduced to a superposition of the integral
\begin{eqnarray}
A_{0}(m)=\int \dfrac{d^{d}k}{(4\pi)^{2}} \dfrac{1}{k^{2}-m^{2}}= \dfrac{-i}{(4\pi)^{d/2}}\Gamma \left( 1- \dfrac{d}{2} \right) \left(m^{2} \right)^{\frac{d}{2}-1}, 
\end{eqnarray}
or its derivatives\footnote{For instance, the integral
{\footnotesize
\begin{eqnarray*}
\int \dfrac{d^{d}k}{(4\pi)^{2}} \dfrac{k_{\mu}k_{\nu}}{(k^{2}-m^{2})^{2}}= \dfrac{\partial}{\partial m^{2}} \left( \int \dfrac{d^{d}k}{(4\pi)^{2}} \dfrac{k_{\mu}k_{\nu}}{k^{2}-m^{2}}\right) = \dfrac{-ig_{\mu\nu}}{8\pi}\dfrac{1}{1-\frac{d}{2}}~ , ~~ {\rm at}~d=2. 
\end{eqnarray*}}} 
\begin{figure}
\begin{center}
\scalebox{0.5}{
\fcolorbox{white}{white}{
  \begin{picture}(683,170) (111,-15)
    \SetWidth{1.0}
    \SetColor{Black}
    \Line[dash,dashsize=10,arrow,arrowpos=0.5,arrowlength=5,arrowwidth=2,arrowinset=0.2](112,26)(304,26)
    \GOval(208,26)(40,40)(0){0.882}
    \Line[dash,dashsize=10](384,26)(544,26)
    \Line[dash,dashsize=10](464,26)(464,90)
    \GOval(464,122)(32,32)(0){0.882}
    \Line[dash,dashsize=10,arrow,arrowpos=0.5,arrowlength=5,arrowwidth=2,arrowinset=0.2](616,27)(776,27)
    \COval(693,27)(13.416,13.416)(153.43495){Black}{White}\Line(688.757,35.485)(697.243,18.515)\Line(701.485,31.243)(684.515,22.757)
    \Text(114,38)[lb]{\Large{\Black{$H$}}}
    \Text(300,38)[lb]{\Large{\Black{$H$}}}
    \Text(384,38)[lb]{\Large{\Black{$H$}}}
    \Text(528,38)[lb]{\Large{\Black{$H$}}}
    \Text(616,38)[lb]{\Large{\Black{$H$}}}
    \Text(759,38)[lb]{\Large{\Black{$H$}}}
    \Text(474,62)[lb]{\Large{\Black{$H$}}}
    \Text(204,86)[lb]{\Large{\Black{$(S,F,V)$}}}
    \Text(504,128)[lb]{\Large{\Black{$(S,F,V)$}}}
    \Text(693,60)[lb]{\Large{\Black{$\delta m_{H}^{2}$}}}
  \end{picture}
}}
\end{center}
\caption{\label{GI-H-Self-E} Gauge invariant Higss self energy corrections to the Higgs mass. The letters S, F and V represent all Scalar, Fermion and Vectors of SM respectively.}
\end{figure}
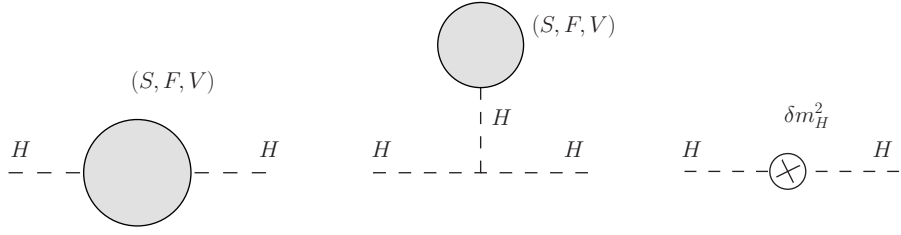
The above integral has two different poles, at $d=4$ or $d=2$. However, the dimensional regularization scheme hides the quadratic divergences ($d=2$) because the Laurent expansion is performed around $d=4$ . One must expand $d$ around $4$ so without meeting the pole at $d=2$ . Nevertheless, it is interesting to remark here, that the quadratic divergences are manifested in dimensional regularization as poles at $d=2$ on a complex plane. The one-loop counterterm $\delta m_{H}^{2}$ obtained from the self-energies diagrams of fig. \ref{GI-H-Self-E} amounts to
{\small
\begin{eqnarray}
\delta m_{H}^{2}=\frac{2}{(4\pi)}\frac{1}{1-d/2}\left[ 6\lambda-
6h^{2}_{t} +\frac{6g^{2}}{4}+3\frac{(g^{2}+g'^{2})}{4}\right] +\frac{1}{(4\pi)^2}
\frac{2}{4-d}\left(6\lambda+3 h^{2}_{t}
-\frac{9}{4}g^{2}-\frac{3}{4}g'^{2}\right). \label{eq:delta-mH2}
\end{eqnarray} }

If we take replacement 
\begin{eqnarray}
\dfrac{1}{1-d/2}\rightarrow \dfrac{\Lambda ^{2}}{(4\pi)},
\end{eqnarray}
and we take poles only at $d=2$, we find the common naturalness relation 
\begin{eqnarray}\label{eq:Naturalness}
(m^{0}_{H})^{2}=(m_{H})^{2}+\dfrac{2\Lambda^{2}}{(4\pi)^{2}v^{2}}
\left[ 3(m_{H})^{2}+3(m_{Z})^{2}+6(m_{W})^{2}-12(m_{t})^{2}\right],
\end{eqnarray}
derived usually with a cut-off method in four dimensions \cite{Veltman}. Now, why is unnatural this result?. The scale $\Lambda$ is considered as the scale where the SM is still valid, or the scale where new physics must appear.  Imagine this scale be the Planck scale $\Lambda_{P}$, the RG evolution of $m_{H}(\mu)$ leads to\footnote{Even when we use the DR scheme with $d=4$, this result is independent of the regularization procedure.}
\begin{eqnarray}
m_{H}^{2}(\Lambda_{EW})=m_{H}^{2}(\Lambda_{P}) - C\Lambda_{P}^{2}log\left(\dfrac{\Lambda_{P}}{\Lambda_{EW}} \right), \label{finetuning}
\end{eqnarray}  
where $\Lambda_{EW}$ is the low Electro-Weak scale, $\Lambda_{P}$ is the high Planck scale and $C$ is a coefficient that is a function of the couplings constants. As the scale $\Lambda_{P}$ is much higher than $m_{H}(\Lambda_{P})$, that according with the fig. \ref{lambdarun} is less than 1TeV, then the two contributions on the r.h.s. of eq. (\ref{finetuning}) have to balance out with a very high accuracy in order to generate a mass $m_{H}^{2}(\Lambda_{EW})\approx 126$ GeV much smaller than $\Lambda_{P}\approx 10^{18}$ GeV. Taking into account that the tuning is the precision to which the initial conditions at the high scale have to be given in order to have any parameter at the low scale, the quadratic divergences implies a really fine tuning ($\delta$) of the order
\begin{equation}
\delta \sim \dfrac{\Lambda_{P}^{2}}{m_{H}^{2}(\Lambda_{EW})} \sim 10^{34}.
\end{equation}
One needs a very fine arrangement of $34$ digits between the Higgs mass squared at $\Lambda_{P}$ and the radiative corrections to have a physical Higgs boson mass in the range of the EW scale. The above result can be explained in different ways, all of them are just speculations and some scenarios are currently discriminated by the new Higgs mass value. An example is the conformal conspiracy proposed by Veltman \cite{Veltman}, where the quadratic divergences can be absent if SM fermion contributions balance against the bosonic ones, this scenario requires a Higgs mass
$m_H \approx 314.92$ GeV in the one-loop approximation, or $m_{H}\approx 276.42$ GeV at two-loop order \cite{Hamada}, a numerical value far away from its currently established value. \\ An alternative possibility is to see the SM as a complete description of nature, no new physics appears at any energy, then the hierarchy problem could be viewed just as an aesthetic problem of the theory without a deep significance. The Higgs mass is not predicted in the SM, we only can obtain lower and upper bounds by imposing absolute vacuum stability or some another theoretical criterion, its renormalized value has to be determined by experiments, thus there is no reason to set the boundary condition for the RG evolution of $m_H(\mu)$ at some high scale. However, there are many indirect hints of new physics at high energies coming from dark matter observations, the oscillation of neutrinos, the Yukawa hierarchies in flavour physics, the baryon asymmetry in the Universe, etc. If one insists with naturalness and we consider the SM as an effective field theory (EFT), and the scale $\Lambda_{P}$ (or some intermediate scale) as the scale of new physics, then the coefficient $C$ includes the couplings of the new particles with the Higgs boson, the quadratic corrections at high energies could be present and the naturalness would represent a real problem of the theory. \\ Of course the coupling of the Higgs boson to all of those new degrees of freedom does not have to lead to the quadratic divergences. The Higgs mass could be protected from high energy effects through some unknown mechanism; any other phenomenon below the Planck scale can be the sufficiently decoupled from the Standard Model to make its correction irrelevant. Otherwise, the problem can be lessened when one considered that the new physics occurs around the TeV scale. For instance, in composite models the Higgs boson is a resonance of a new strongly interacting sector that reveals its true structure at the TeV scale. Thus, it makes no sense to speak about the Higgs mass at energies higher than the compositeness scale. \\ Finally, one has the possibility of ignoring the hierarchy problem and to accept a fine tuned Standard Model. In that case one needs to find a new approach, different of naturalness, to go beyond the SM. The work of Degrassi, Isidori et al. follows this direction. The near-criticality of the Higgs boson quartic coupling and of its beta-function, discussed in their papers, may be an intriguing possibility. The Higgs naturalness can be viewed as a problem of near-criticality between the broken and unbroken EW phases \cite{Giudice}. This argument is motivated by the observation that the measured value of $m_{H}$ also corresponds to a near-critical parameter separating two phases. In general these arguments use the multiverse hypothesis, leading to the speculation that within the multiverse critical points are attractors and the probability density in the space of parameters is peaked around the boundaries between different phases. In this picture generic universes are likely to live near critical lines, and the Higgs parameters found in our universe are not at all special, in fact, they correspond to the most likely occurrence in the multiverse.

\chapter{\noun{\label{cha:Sirlin-Zucchini}\index{Sirlin-Zucchini}}Sirlin-Zucchini Renormalization Scheme}

\lettrine{T}{ he} study performed in the previous chapter has shown that the NNLO vacuum stability analysis requires the three loop beta functions of the all Standard Model couplings \cite{Luminita} with particular reference to RG evolution of the Higgs coupling $\lambda$, the top Yukawa coupling $y_t$ available in  \cite{M.F.Zoller}. However, as remarked in the previous chapter, the most important NNLO pieces are the two-loop threshold corrections to $\lambda$ at the weak scale due to the QCD and the electroweak interactions. Those threshold corrections will be obtained from the relation discovered by Sirlin and Zucchini (SZ) in \cite{SirlinZucchini}, connecting $\lambda$ to the physical Higgs mass $m_H$ and to the Fermi coupling $G_\mu$, the precisely known muon decay constant:
\begin{equation}
\lambda(\mu)=\frac{G_\mu}{\sqrt{2}}m_H^2 +\Delta \lambda (\mu).
\label{SZrel}
\end{equation}   
This relation serves to define the SZ renormalization scheme in which the running coupling constants are expressed in terms of physical gauge invariant observables, which for the case of $\lambda$ would be the Higgs mass and the muon time decay. In such a scheme, therefore, the Higgs mass will not be a running parameter, as it happens in the $\overline{\rm MS}$ scheme but the RG invariant Higgs physical mass. Consequently in this renormalization scheme a vacuum stability analysis will be performed on all possible ranges of $m_H$ within the framework of perturbation theory, being still possible to extend the SZ scheme also the improved effective potential. The SZ scheme has an analogous relation of (\ref{SZrel}) including all other SM running coupling constants, generically called $g_i(\mu)$ and the other physical masses $m_i$ so that
\begin{equation}
g_i(\mu)=c_i\frac{G_F}{\sqrt{2}}m_i^2 +\Delta g_i(\mu)
\end{equation}   
being $c_i$ normalization constants. Naturally a question arises: could this scheme be suitable to compute the matching conditions also for the top quark?  The problem there is that free quarks are not observables in nature, their masses primarily are Lagrangian parameters related to the chiral symmetry breaking and deduced from the observed mass spectrum of the hadronic states which consists of permanently confined quarks and gluons. Therefore in the SZ scheme the role of the physical mass of the top quark one can use the mass pole or the running $\overline{\rm MS}$ mass, both are formal definition however accessible in perturbation theory. This chapter will be devoted to detail the characteristic features of the SM in the SZ renormalization scheme, stressing the aspects that we will need for the NNLO matching conditions to evaluate the SM vacuum stability.

\section{The On-Shell SZ Scheme\label{ONSHELL}}
Before entering in the details of the SZ renormalization scheme,
let's remember that to deal with ultraviolet divergences of four dimensional Feynman integrals in perturbation theory usually one first defines a regularized version
of the theory and then we give a procedure, known as renormalization
scheme, to remove the infinities.
Regularization consists replacing the theory by a slightly different
one, using some cut-off or modifying the dimension of the integration
space. In any case, by now there is a general agreement in using the dimensional
regularization as regularization scheme of gauge theories.
In dimensional regularization one computes the integrals in a $d$
-dimensional space-time, with $d$ chosen in such a way that the integral
converges, then one continues analytically the result in the complex
$d$ plane and one expresses the divergences as poles in $d-4$. \\
On the other side, the renormalization procedure consists of adding to
the Lagrangian extra terms, to cancel the regulator dependence
of the amplitudes. The extra terms added to the Lagrangian are the
so-called counter-terms.
As the Lagrangian of the SM has polynomial interactions, we may replace
the bare parameters\footnote{The bare parameters are the unrenormalized initial parameters of the Lagrangian that contain all UV divergences.} of the Lagrangian, $\{g_{0}\}$, and the unrenormalized fields, $\{\psi_{0}\}$,
by renormalized ones by multiplicative renormalization:
\begin{eqnarray}
g_{0}=Z_{g}g=g+\delta g, & \: & \psi_{0}=Z_{\psi}\psi,\label{eq:RenorDecomp}
\end{eqnarray}
with renormalization constants $Z_{i}$ different from $1$ when we
make radiative corrections. The counter-terms are chosen
to cancel the divergent part of the Feynman amplitudes, they
contain the dependence on the poles in $d-4$, and an arbitrary finite
contribution. The definition of the renormalized quantities in (\ref{eq:RenorDecomp}) is fixed
by imposing a finite set of renormalization conditions, meaning definition of the renormalized parameters at the so called subtraction points, which can be obtained, for instance, from some measurable observable at a given energy scale. 
The physical observables and the physical predictions are the same
in all renormalization schemes and the transformations making that invariance possible define the renormalization group. A change of a renormalization scheme can be compensated in the numerical values of the finite renormalized parameters of the theory \cite{collins}. \\
Now we discuss a renormalization scheme in SM proposed initially by
A. Sirlin \cite{ASirlin1} only for the electroweak sector, and then
expanded by A. Sirlin and R. Zucchini \cite{SirlinZucchini} to treat
the Higgs effective potential of the SM. That is, we will discuss
the generation of the principal counter-terms associated to the parameters
of the theory and we will see how we can fix their finite parts by fixing a set of the renormalization conditions.

\subsection{Renormalization of the gauge boson sector}

\begin{figure}
\begin{center}
\scalebox{0.5}{
\fcolorbox{white}{white}{
  \begin{picture}(514,195) (143,-31)
    \SetWidth{1.5}
    \SetColor{Black}
    \Line[dash,dashsize=10](400,35)(400,99)
    \SetWidth{1.0}
    \COval(400,131)(32,32)(0){Black}{White}
    \SetWidth{1.6}
    \Line[dash,dashsize=10](590,35)(590,125)
    \SetWidth{1.0}
    \COval(590,130)(11.18,11.18)(153.43495){Black}{White}\Line(586.464,137.071)(593.536,122.929)\Line(597.071,133.536)(582.929,126.464)
    \Vertex(590,35){5}
    \Vertex(400,35){5}
    \Text(209,-36)[lb]{\Large{\Black{$(a)$}}}
    \Text(396,-36)[lb]{\Large{\Black{$(b)$}}}
    \Text(594,-36)[lb]{\Large{\Black{$(c)$}}}
    \Photon(336,35)(464,35){7.5}{6}
    \Photon(528,35)(656,35){7.5}{6}
    \Photon(144,35)(272,35){7.5}{6}
    \COval(208,35)(32,32)(0){Black}{White}
  \end{picture}
}}
\end{center}
\caption{\label{selfen-VV} Feynman diagrams of the self-energies with two vector boson external lines. The dashed line represents the Higgs field.}
\end{figure}
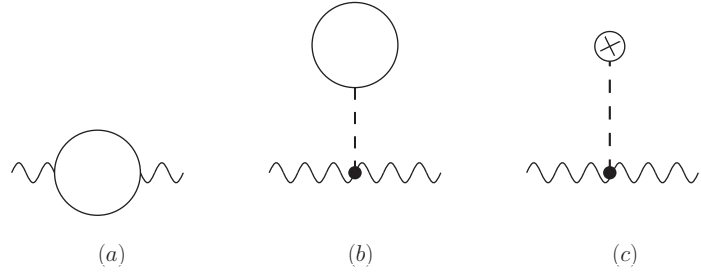

In the $SU(2)_{L}\times U(1)_{Y}$ sector of the SM, the mass matrix
of the vector boson (after SSB) is given by the Lagrangian 
\begin{eqnarray}
\mathcal{L}_{V}=\frac{v_{0}^{2}}{2}\left[\frac{g_{0}^{2}}{2}W_{\mu}^{+}W^{\mu-}+\frac{1}{4}\left(g'_{0}B_{\mu}-g_{0}W_{\mu}^{3}\right)^{2}\right],
\end{eqnarray}
where $v_{0}$ is the bare vacuum expectation value (vev), $g_{0}$
and $W_{\mu}^{a}$ are the unrenormalized coupling constant and bare
vector bosons associated with the $SU(2)_{L}$ group, and $g'_{0}$
and $B_{\mu}$ are the unrenormalized coupling and bare field associated
with the $U(1)$ group. 
By Weinberg rotating the bare fields 
\begin{eqnarray}
\left(\begin{array}{c}
W_{\mu}^{3}\\
B_{\mu}
\end{array}\right)=\left(\begin{array}{cc}
cos\theta_{W} & sin\theta_{W}\\
cos\theta_{W} & -sin\theta_{W}
\end{array}\right)\left(\begin{array}{c}
Z_{\mu}\\
A_{\mu}
\end{array}\right)\label{eq:RotationW}
\end{eqnarray}
the Lagrangian $\mathcal{L}_{V}$ takes the
form
\begin{eqnarray}
\mathcal{L}_{V}=\frac{v_{0}^{2}}{2}\left[\frac{g_{0}^{2}}{2}W_{\mu}^{+}W^{\mu-}+\frac{1}{4}\left(A_{\mu}(g'_{0}c-g_{0}s)-Z_{\mu}(g_{0}c+g'_{0}s)\right)^{2}\right],
\end{eqnarray}
where $c\equiv cos\theta_{W}$ and $s=sin\theta_{W}$. To generate
counterterms we define $g_{0}=g-\delta g$, $g'_{0}=g'-\delta g'$
and $v_{0}^{2}=v^{2}-\delta v^{2}$, with $g$, $g'$ and $v^{2}$
the renormalized parameters. 
In the Sirlin scheme the fields are not rescaled because for our analysis of the matching conditions we will be interested in scattering amplitudes and not in Green functions, for instance the quartic Higgs coupling constant can be related, for suitable choice of the subtraction point, to the four Higgs scattering amplitude \cite{SirlinZucchini}.  
The Weinberg rotation parameter $\theta_{W}$ is related with the renormalized couplings $g$ and $g'$ by the relation $gs=g'c$. This produces a mixing counterterm of the form  $Z_\mu A^\mu$. The counterterm $Z_\mu A^\mu$ is innocuous in the NNLO stability analysis, the renormalization of the running couplings are independent of the mixing counterterms. Besides, there is no counterterm proportional to $A_{\mu}A^{\mu}$, the field $A_{\mu}$ remains massless, and it is protected of radiative corrections. Without considering the mixing counterterm $Z_\mu A^\mu$, $\mathcal{L}_{V}$ can be written as:
\begin{eqnarray}
\mathcal{L}_{V}=\mathcal{L}_{V}^{r}+\Delta\mathcal{L}_{V},
\end{eqnarray}
where 
\begin{eqnarray}
\mathcal{L}_{V}^{r}=m_{W}^{2}W^{\mu+}W_{\mu}^{-}+\frac{m_{Z}^{2}}{2}Z_{\mu}Z^{\mu},
\end{eqnarray}
is the renormalized part of the Lagrangian, with
\begin{eqnarray}
m_{W}^{2}=\frac{g^{2}v^{2}}{4} & ; & m_{Z}^{2}=\frac{(g^{2}+g'^{2})v^{2}}{4},
\end{eqnarray}
and $\Delta\mathcal{L}_{V}$ is the counterterm
part
\begin{eqnarray}
\Delta\mathcal{L}_{V}=-\delta m_{W}^{2}W^{\mu+}W_{\mu}^{-}-\frac{\delta m_{Z}^{2}}{2}Z_{\mu}Z^{\mu},
\end{eqnarray}
with 
{\small
\begin{eqnarray}
&&\delta m_{W}^{2}=\dfrac{v^{2}\delta g^{2}+g^{2}\delta v^{2}}{4}-\dfrac{v^{2}(\delta g)^{2}+\delta g^{2}\delta v^{2}}{4},\label{countmw}\\
&&\delta m_{Z}^{2}=\dfrac{v^{2}\delta(g^{2}+g'^{2})+(g^{2}+g'^{2})\delta v^{2}}{4}-\dfrac{\delta v^{2}}{4}\left(\delta g^{2}+\delta g'^{2}\right)-\dfrac{v^{2}}{4}\left(c\delta g+s\delta g'\right),\label{countmZ}
\end{eqnarray}}
up two-loop order. The terms of order $\delta v^{2}(\delta g)^{2}$,
$\delta v^{2}(\delta g')^{2}$ have been neglected. The parameters $m_{W}^{2}$ and $m_{Z}^{2}$
are arbitrary finite parameters, the renormalization condition on them would identify them with the physical
masses of the $Z$ and $W$ particles, therefore
\begin{eqnarray}
&&\delta m_{W}^{2}=ReA_{WW}(m_{W}^{2})+t_{WW},\label{eq:Ren-mW}\\
&&\delta m_{Z}^{2}=ReA_{ZZ}(m_{Z}^{2})+t_{ZZ},\label{eq:Ren-mZ}
\end{eqnarray}
where $A_{VV}$ is the coefficient of the metric tensor in the unrenormalized
tensorial self-energy
\begin{eqnarray}
\Pi_{VV}^{\mu\nu}(q^{2})=A_{VV}(q^{2})g^{\mu\nu}+B_{VV}(q^{2})q^{\mu}q^{\nu},
\end{eqnarray}
with $\Pi_{VV}^{\mu\nu}(q^{2})$ defined as $-i$ times the vector
self-energy Feynman diagrams showed in fig. \ref{selfen-VV}~(a), those self-energies include tadpole diagrams of fig. \ref{selfen-VV} (b) whose corresponding Higgs tadpoles counterterms are in $t_{ZZ}$ and $t_{WW}$ and are depicted in \ref{selfen-VV} (c). 
The reason to remark the Higgs tadpole contribution is because they are deeply related to the SSB dynamics through the
renormalization of vev of the Higgs field. For instance,
if we require that $v$ be the exact vacuum expectation value of
he neutral Higgs field, we must impose $t_{WW}=t_{ZZ}=0$, as it will be seen in the section\ref{sub:Ren-Higgs-Pot} when we renormalize
the Higgs potential. 
We review now the renormalization of the interaction between quarks
and gauge bosons. This interaction is described by the bare Lagrangian:
{\small
\begin{eqnarray}
\mathcal{L}_{FV}=-\frac{1}{2}\sum_{i=i}^{3}\overline{N}_{iL}\gamma^{\mu}\left(g_{0}\overrightarrow{\tau}\centerdot\overrightarrow{W_{\mu}}+\frac{g'_{0}}{3}B_{\mu}\right)N_{iL}-\frac{g'_{0}}{3}\sum_{i=1}^{3}\left(2\overline{U}_{iR}\gamma^{\mu}U_{iR}-\overline{D}_{iR}\gamma^{\mu}D_{iR}\right)B_{\mu},\label{eq:lag-FV}
\end{eqnarray}}
where $N_{iL}$ are the isodoublets of the $SU(2)_{L}\equiv SU(2)_{ ew}$ group 
\begin{eqnarray*}
N_{1L}=\left(\begin{array}{c}
u\\
d'
\end{array}\right)_{L}, & N_{2L}=\left(\begin{array}{c}
c\\
s'
\end{array}\right)_{L} & N_{3L}=\left(\begin{array}{c}
t\\
b'
\end{array}\right)_{L},
\end{eqnarray*}
 and where the eigenstates $d'$, $s'$ and $b'$ are connected with
$d$, $s$ and $b$ through the Cabibbo-Kobayashi-Maskawa unitary matrix $A$ defined by
\begin{eqnarray}
\left(\begin{array}{c}
d'\\
s'\\
b'
\end{array}\right)_{L}=A\left(\begin{array}{c}
d\\
s\\
b
\end{array}\right)_{L}.
\end{eqnarray}
On the other side $U_{1R}=u_{R}$, $U_{2R}=c_{R}$, $U_{3R}=t_{R}$ and
$D_{1R}=d_{R}$, $D_{2R}=s_{R}$, $D_{3R}=b_{R}$ are right handed
isosinglets. In eq. (\ref{eq:lag-FV}) a summation over the quark
color indices are understood. \\
To generate the counterterms we use the definitions $g_{0}=g-\delta g$
and $g'_{0}=g'-\delta g'$, and we use again the rotation matrix \ref{eq:RotationW}
to express $\mathcal{L}_{FV}$ in terms of the fields $Z_{\mu}$ and
$A_{\mu}$. The gauge-quark interaction Lagrangian can be written
in the renormalized form:
\begin{eqnarray}
\mathcal{L}_{FV}=\mathcal{L}_{FV}^{r}+\Delta\mathcal{L}_{FV},
\end{eqnarray}
where
\begin{eqnarray}
\mathcal{L}_{FV}^{r}=-gsA_{\mu}J_{\gamma}^{\mu}-\frac{g}{c}Z_{\mu}J_{Z}^{\mu}-\frac{g}{\sqrt{2}}\left(W_{\mu}^{\dagger}J_{W}^{\mu}+h.c.\right),
\end{eqnarray}
and 
\begin{eqnarray}
\mathbf{\Delta\mathcal{L}}_{FV}=A_{\mu}J_{\gamma}^{\mu}\left(c^{3}\delta g'+s^{3}\delta g\right)+(c\delta g+s\delta g')Z_{\mu}J_{Z}^{\mu} ~~~~~~~~~~~~~~~~~~~~~~~~~~~~~~~~~~ \nonumber \\
~+~(s\delta g-c\delta g')A_{\mu}J_{Z}^{\mu}+\left(s\delta g-c\delta g'\right)scJ_{\gamma}^{\mu}Z_{\mu}+\frac{\delta g}{\sqrt{2}}\left(W_{\mu}^{\dagger}J_{W}^{\mu}+h.c.\right).
\end{eqnarray}
The currents $J$ appearing in the Lagrangian are defined as:
\begin{eqnarray}
&&J_{\gamma}^{\mu}=\overline{\psi}\gamma^{\mu}Q\psi,\\
&&J_{W}^{\mu}=\overline{\psi}\gamma^{\mu}a_{-}C_{-}\psi,\\
&&J_{Z}^{\mu}=\dfrac{1}{2}\overline{\psi}C_{3}\gamma^{\mu}a_{-}\psi-sin^{2}\theta_{W}\overline{\psi}\gamma^{\mu}Q\psi,
\end{eqnarray}
where the matrices are defined as
\begin{eqnarray*}
Q=\left(\begin{array}{cc}
\frac{2}{3}I\\
 & -\frac{1}{3}I
\end{array}\right); & C_{3}=\left(\begin{array}{cc}
I\\
 & -I
\end{array}\right); & C_{-}=\left(\begin{array}{cc}
0 & 0\\
A^{\dagger} & 0
\end{array}\right),
\end{eqnarray*}
and where $a_{-}=(1-\gamma^{5})/2$, $I$ is the $3\times3$ unit
matrix, and $\psi$ is a vector such that $\psi^{T}=(u,c,t,d,s,b)$.~Finally, analogous expressions are obtained in the leptonic sector if one
makes the substitutions: $\psi^{T}\rightarrow\psi_{L}^{T}=(\nu_{e},\nu_{\mu},\nu_{\tau},e,\mu,\tau)$,
$A\rightarrow A_{L}$ and $Q\rightarrow Q_{L}$, with 
\begin{eqnarray*}
Q_{L}=\left(\begin{array}{cc}
0\\
 & -I
\end{array}\right).
\end{eqnarray*}
If one assumes that the neutrinos are massless, then $A_{L}=I.$ 

\section{Determination of counterterms\label{detcount}}

The last section defined a set of renormalization conditions useful
to determine the counterterms $\delta g$, $\delta g'$ and $\delta v^{2}$.
The third counterterm, $\delta v^{2}$, need the renormalization of
the Higgs potential and will be analysed in the section \ref{sub:Ren-Higgs-Pot}.
We focus our attention here in the renormalization of the EW gauge
couplings. To obtain an expression of the EW counterterms, $\delta g$
and $\delta g'$, we need to study the renormalization of the electric
charge, identified in the SM by 
\begin{eqnarray}
e=gsin\theta_{W} & ; & tan\theta_{W}=\frac{g'}{g}.\label{eq:e-g-relation}
\end{eqnarray}
This definition determines $g$ and $g'$ as gauge-invariant parameters. Without going into the details and following
\cite{ASirlin2} the renormalization of the electric charge leads to
\begin{eqnarray}
&&\dfrac{\delta g}{g}=\dfrac{\delta e}{e}-\dfrac{c^{2}}{2s^{2}}Re\left[\dfrac{A_{ZZ}(m_{Z}^{2})}{m_{Z}^{2}}-\dfrac{A_{WW}(m_{W}^{2})}{m_{W}^{2}}\right],\label{ren-g}\\
&&\dfrac{\delta g'}{g'}=\dfrac{\delta e}{e}-\dfrac{1}{2}Re\left[\dfrac{A_{ZZ}(m_{Z}^{2})}{m_{Z}^{2}}-\dfrac{A_{WW}(m_{W}^{2})}{m_{W}^{2}}\right].\label{ren-g2}
\end{eqnarray} 

However, the introduction of the effective coupling constant $G_{\mu}$
that appears naturally in the study of muon decay, allows us to eliminate
the dependence of $\delta e$ from the above corrections avoiding the large fermionic corrections which arise
because $e^{2}$ is conventionally defined at null transferred
momentum \cite{ASirlin2}. Let's consider, in fact, the one-loop correction to $\Delta r$
\begin{eqnarray}
&&(\Delta r)_{\rm 1-loop}=\frac{ReA_{WW}(m_{W}^{2})-A_{WW}(0)}{m_{W}^{2}}-2\frac{\delta e}{e}+\frac{c^{2}}{s^{2}}Re\left[\frac{A_{ZZ}(m_{Z}^{2})}{m_{Z}^{2}}-\frac{A_{WW}(m_{W}^{2})}{m_{W}^{2}}\right]+E,\label{eq:one-loop-Dr}\nonumber\\
&&
\end{eqnarray}
with
\begin{eqnarray*}
E=\frac{g^{2}}{16\pi^{2}}\left[-8\left(\frac{1}{d-4}+\frac{\gamma-ln(4\pi)}{2}+ln\frac{m_{Z}}{\mu}\right)+\frac{lnc^{2}}{s^{2}}\left(\frac{7}{2}-6s^{2}\right)+6\right].
\end{eqnarray*}
The issues related to the one-loop muon decay are fully discussed in the references \cite{ASirlin3}\cite{ASirlin4}\cite{ASirlin5}. From equations (\ref{ren-g}) and (\ref{eq:one-loop-Dr}) we can rewrite $\delta g$ as:
\begin{eqnarray*}
\frac{\delta g}{g}=\frac{1}{2}\left[-\Delta r+\frac{ReA_{WW}(m_{W}^{2})-A_{WW}(0)}{m_{W}^{2}}+E\right].
\end{eqnarray*}
The radiative corrections to the electric charge are now contained
in $\Delta r$. If we relate $g$, defined in the Sirlin's scheme with $\bar{g}(\mu)$, the $SU(2)_{L}$
gauge coupling defined by modified minimal subtraction ($\overline{MS}$)
scheme at mass scale $\mu$, the correction $\Delta r$ can be eliminated
together with the counterterm of electric charge $\delta e$. The
connection between two schemes can be found by noting that the unrenormalized
$g_{0}$ is related to the renormalized couplings by
\begin{eqnarray}
g_{0}=g-\delta g=\bar{g}(\mu)-\delta\bar{g},\label{eq:con-schemes}
\end{eqnarray}
where $\delta\bar{g}$ is a counterterm that subtracts the terms proportional
to $(d-4)^{-1}+\frac{1}{2}(\gamma-ln(4\pi))$ in the one-loop correction
to the $SU(2)_{L}$ gauge coupling. The counterterms $\delta g$ and
$\delta\bar{g}$ have identical divergent parts, it allows to write
the eq. (\ref{eq:con-schemes}) as: 
\begin{eqnarray*}
\bar{g}(\mu)=g\left(1-\frac{(\delta g)_{fin}(\mu)}{g}\right),
\end{eqnarray*}
where
\begin{eqnarray}
(\delta g)_{fin}(\mu)=\frac{g}{2}\left[E-\Delta r+\frac{ReA_{WW}(m_{W}^{2})-A_{WW}(0)}{m_{W}^{2}}\right]_{fin}.\label{eq:delta-g-fin}
\end{eqnarray}
The subscript fin denotes the finite part of the counterterm $\delta g$,
obtained after subtracting the terms proportional to $(d-4)^{-1}+\frac{1}{2}(\gamma-ln(4\pi))$. Now, from eq. (\ref{muon-g}) we obtain:
\begin{eqnarray*}
\bar{g}(\mu)=2m_{W}\sqrt{2\frac{G_{\mu}}{\sqrt{2}}}\left(1-\Delta r\right)^{\frac{1}{2}}\left(1-\frac{(\delta g)_{fin}(\mu)}{g}\right)\\
\approx2m_{W}\sqrt{2\frac{G_{\mu}}{\sqrt{2}}}\left(1-\frac{(\delta g)_{fin}(\mu)}{g}-\frac{1}{2}\Delta r\right).
\end{eqnarray*}
Finally, using the eq. (\ref{eq:delta-g-fin}) we find:
\begin{eqnarray}
\bar{g}^{2}(\mu)=8m_{W}^{2}\frac{G_{\mu}}{\sqrt{2}}-8m_{W}^{2}\frac{G_{\mu}}{\sqrt{2}}\left(E+\frac{ReA_{WW}(m_{W}^{2})-A_{WW}(0)}{m_{W}^{2}}\right)_{fin}.\label{eq:MS-ren-g}
\end{eqnarray}
Following an analogous procedure we can easily obtain $\bar{g}'^{2}(\mu)$ at one-loop,
{\small
\begin{eqnarray}
\bar{g}'^{2}(\mu)=g'^{2}+\left(m_{Z}^{2}-m_{W}^{2}\right)8\frac{G_{\mu}}{\sqrt{2}}\left(-E+\frac{A_{WW}(0)}{m_{W}^{2}}-\frac{ReA_{ZZ}(m_{Z}^{2})-ReA_{WW}(m_{W}^{2})}{m_{Z}^{2}-m_{W}^{2}}\right)_{fin}.\label{eq:MS-ren-g2}
\end{eqnarray}}
The above expressions are independent of the renormalization of the
electric charge. In this sense, to obtain some quantity
in our on-shell scheme, as the 1PI effective potential, we
first make the computation in the $\overline{MS}$ scheme (with the
input parameters $\bar{g}$, $\bar{g}'$, $\bar{\lambda}$, $\bar{h_{t}}$,
$\bar{e}$, etc) and then we use matching conditions of the form (\ref{eq:MS-ren-g})
and (\ref{eq:MS-ren-g2}), where the input parameters are the physical
observables $m_{W}$, $m_{Z}$, $m_{H}$, $m_{t}$, $G_{\mu}$, etc.
There are a number of advantages in making this choice of the renormalized
parameters. First, as the $\overline{MS}$ definition is gauge invariant
and $G_{\mu}$ and the masses are physical observables, it follows
that all the renormalized parameters in equations (\ref{eq:MS-ren-g})
and (\ref{eq:MS-ren-g2}), are gauge-invariant quantities. Moreover,
the presence of $G_{\mu}$ instead of $e^{2}/(8sin^{2}\theta_{W}m_{W}^{2})$
avoids the big vacuum polarization effects associated with electric
charge renormalization at large momenta and finally $G_{\mu}$, $m_{W}$,
$m_{Z}$, $m_{H}$ and $m_{t}$ have a very well established physical
meaning from the phenomenological point of view.

\subsection{Corrections to the top Yukawa coupling}

The top Yukawa coupling ($h_{t}$) requires a special treatment. In the EW sector $h_{t}$ is fixed using the running mass $m_t(\mu)$ given by $m_t(\mu)=\frac{h_t(\mu)}{\sqrt{2}}v(\mu)$ and from the relation between $v$ and $G_\mu$,
\begin{eqnarray}
\dfrac{1}{2v^{2}}=\dfrac{G_{\mu}}{\sqrt{2}}(1-\Delta r),
\end{eqnarray} 
deduced from eq. (\ref{muon-g}). This leads to the one-loop relation between the $\overline{MS}$ Yukawa coupling $\bar{h}_{t}(\mu)$ and the top pole mass $m_{t(os)}$:
\begin{eqnarray}
\bar{h}_{t}(\mu)=2\left(\dfrac{G_{\mu}}{\sqrt{2}}m_{t(os)}^{2} \right)+2\left(\dfrac{G_{\mu}}{\sqrt{2}}m_{t(os)}^{2} \right)^{\frac{1}{2}}\left(\dfrac{\delta m_t}{m_{t(os)}} + \dfrac{1}{2}\left[E-\dfrac{A_{WW}(m_{W}^{2})}{m_{W}^{2}}\right] \right)_{fin}, \label{eq:1l-MS-ht}
\end{eqnarray}
where $\delta m_t$ refers to the one-loop correction of $m_{t(os)}$. R. Hempfling and B. Kniehl computed the full matching condition (\ref{eq:1l-MS-ht}) at one-loop level \cite{kniehl}. There is a theoretical problem when we use a renormalized mass $m_{t(os)}$ as an input parameter for theoretical predictions, because a rigorous relation of $m_{t(os)}$ with the currently LHC measurements of the top quark mass parameter is absent. The most precise measurement of the top-quark mass has been reported in \cite{ACCD} as the world
combination of the experiments ATLAS, CDF, CMS and D0. This combination of data from LHC and Tevatron obtained the result $m^{MC}_{t}=173.34 \pm 0.76$ GeV. The combination is based on determinations of top quark mass as a best fit to the mass parameter implemented in the
respective Monte Carlo (MC) program. However, the result obtained using MC output $m^{MC}_{t}$ cannot be used directly as an input for precise NLO or NNLO theoretical predictions because the measured quantity is the top mass parameter of the MC event generators which is not a renormalized field theory mass. Let see this in more detail.

Theoretically, the top quark mass is a renormalized quantity of the QCD Lagrangian. The renormalized mass is obtained from the top self energy Feynman diagrams. The finite contributions of the self energy can be absorbed in the renormalized mass and the UV divergences in a suitable counterterm, different choices for the finite contributions define different top mass schemes. In the $MS$ scheme, where only pure UV divergences are subtracted the top mass $m_{t}(\mu)$ is renormalization scale dependent. Physically, the $MS$ scheme is conceptually and numerically very far away from the notion of a physical particle mass. The parameter $m_{t}(\mu)$ should be thought more as a coupling for a heavy quark-antiquark correlation and is therefore a very good scheme for parametrizing the top Yukawa coupling $h_{t}$. The other very well known scheme is the top quark on-shell scheme, where all UV and finite
contributions of the self-energy are absorbed into the mass in the on-shell limit $q^{2}=m_{t(os)}^{(2)}$. The on-shell scheme would correspond to our intuitive notion of the top quark physical mass, but this notion and its connection to the top quark kinematic properties are physically limited because of confinement, free quarks are not observable in nature. The important conceptual issue in this context is that the on-shell mass scheme is based on the perception of the self energy diagram being a meaningful physical quantity. But, this is only for momenta above 1 GeV, which is the hadronization scale, because perturbation theory breaks down for energies below 1 GeV. In this sense, the Monte Carlo
top-quark mass parameter measured is not identical with the pole mass. However, the
measured values can be converted to the pole mass provided certain assumption on the relation of the MC mass to a short-distance mass at a low scale \cite{Hoang}. This conversion leads to an additional uncertainty of the order of 1 GeV \cite{Hoang, Moch}. 

To reduce the uncertainties related with the differences between 
Monte-Carlo and pole masses the mass of top-quark can be extracted directly
from a measurement of the total top-pair production cross section 
$\sigma_{\mbox{exp}}( p\bar p \to t\bar t\! + \!X)$. 
Such analysis performed in \cite{ADM} with NNLO accuracy
with inclusion of the full theoretical uncertainties gives rise to the following result, 
${ m_t^{\rm pole}  =  173.3 \pm 2.8~ {\rm GeV}} \;.$
The central value is very close to the MC value $m^{MC}_{t}=173.34 \pm 0.76$ 
but the theoretical uncertainty is much larger.
To improve the current precision of the top-mass determination from the total
cross section the higher order corrections are required. We need to consider not only QCD NNLO radiative corrections. The EW part as well as mixed EW $\times$ QCD corrections have to be included in a systematic way. In contrast to QCD, where the mass of a quark is the parameter of the Lagrangian, the notion of $\overline{MS}$ mass in EW theory is not determined entirely by the prescriptions of minimal subtraction. 
It depends on the value of vev $v(\mu^2)$ chosen as a parameter of the calculations
so that the running mass is $m_t(\mu) = h_t(\mu) v(\mu)/\sqrt{2}$. 

In one scheme with explicit inclusion of tadpoles the EW contribution is large and has opposite sign relative to the QCD contributions, so that the total SM correction is small and increases the theoretical error by $0.5~{\rm GeV}$ \cite{FredKniehl}.
If the scheme is defined in terms of the self-energy diagrams without including the tadpole contribution, it gives rise to a radiative corrections $\delta m_{t}$ that is gauge-dependent and, as a consequence, in this framework, the $\overline{MS}$ top mass is a gauge-
dependent quantity. However, a $\overline{MS}$ mass is not a physical quantity nor a Lagrangian parameter and therefore the requirement of gauge-invariance is not mandatory. With this choice the relation between the pole and $\overline{MS}$ masses of top quark not acquire the very large electroweak corrections.

\section{Renormalization of the Higgs Potential\label{sub:Ren-Higgs-Pot}}
To renormalize the self-interacting Higgs coupling $\lambda$ and the vacuum expectation value $v$ we need the renormalized version of the Higgs potential together with the renormalization conditions computed in the above section. We begin from the bare Higgs potential in the SM,
\begin{eqnarray*}
V_{0}=-m_{0}^{2}\Phi^{\dagger}\Phi+\lambda_{0}\left(\Phi^{\dagger}\Phi\right)^{2},
\end{eqnarray*}
where 
\begin{eqnarray*}
\Phi=\left(\begin{array}{c}
\phi^{\dagger}\\
\sqrt{\frac{1}{2}}\left(\phi_{1}+\imath\phi_{2}+v_{0}\right)
\end{array}\right).
\end{eqnarray*}
Defining the renormalized quantities
$\lambda$, $v$ and $m^{2}$ by 
\begin{eqnarray}
\lambda_{0}=\lambda-\delta\lambda, & v_{0}=v-\delta v, & m_{0}^{2}=m^{2}-\delta m^{2}, \label{ren-condit}
\end{eqnarray}
the Higgs potential is now given by
\begin{eqnarray}
V_{0}=V_{r}-\delta V,
\end{eqnarray}
where $V_{r}$ is the renormalized potential involving all the terms
of zeroth order in $\delta h$, $\delta v$ and $\delta m^{2}$. By
other side, $\delta V$ contains all the counterterms and UV divergences.
SZ set
\begin{eqnarray*}
m^{2}=\lambda v^{2},
\end{eqnarray*}
such that the term linear in $\phi_{1}$ vanishes in $V_{r}$ and $\phi_{1}$ has a zero vacuum expectation value
at tree-level, consequently the renormalized potential is up to constants
\begin{eqnarray}
V_{r}=\lambda\left[\phi^{\dagger}\phi\left(\phi^{\dagger}\phi+\phi_{1}^{2}
+\phi_{2}^{2}\right)+\frac{1}{4}\left(\phi_{1}^{2}+\phi_{2}^{2}\right)^{2}\right]
\nonumber \\
+\lambda v\phi_{1}\left[\phi_{1}^{2}+\phi_{2}^{2}+2\phi^{\dagger}\phi\right]+2\lambda v^{2}\frac{1}{2}\phi_{1}^{2}.\label{eq:ren-pot}
\end{eqnarray}
On the other
hand, after a bit of algebra, and considering only one-loop corrections,
so that only linear terms in $\delta\lambda$, $\delta m^{2}$ and
$\delta v$ will be retained%
\footnote{Terms of the form $(\delta v)^{2}$, $\delta m^{2}\delta v$, 
$\delta\lambda\delta v$
are ignored for the time being, but they are necessary when one requires
a two loop correction of the potential. Besides, we ignored the term
$\delta m^{2}v^{2}/2$ because it is independent of the fields.%
}, we obtain 
\begin{eqnarray}
\delta V=\delta\lambda\left[\phi^{\dagger}\phi\left(\phi^{\dagger}\phi +\phi_{1}^{2}+\phi_{2}^{2}\right)+\frac{1}{4}\left(\phi_{1}^{2}+\phi_{2}^{2}\right)^{2}\right] ~+~ \label{eq:1l-ct-pot} ~~~~~~~~~~~~~~~~~~~~~~~~~~~~~~~~~~~~~~ \\
\left[\lambda\delta v+v\delta\lambda\right]\phi_{1}\left[\phi^{\dagger}\phi+\phi_{1}^{2}+\phi_{2}^{2}\right]
+\delta\tau\left[\phi^{\dagger}\phi+\frac{1}{2}\phi_{2}^{2}\right] ~+~v\delta\tau\phi_{1}+\delta m_{H}^{2}\frac{1}{2}\phi_{1}^{2},\nonumber 
\end{eqnarray}
where we defined the new counter-terms:
\begin{eqnarray}
&&\delta\tau=v^{2}\delta\lambda+2\lambda v\delta v-\delta m^{2},\label{eq:deltatau}\\
&&\delta m_{H}^{2}=3v^{2}\delta\lambda+6\lambda v\delta v-\delta m^{2}.\label{eq:deltahiggs}
\end{eqnarray}
To determine the structure of $\delta V$ we need the counter-terms
$\delta\lambda$, $\delta m^{2}$ and $\delta v$, thus we need three
restrictions. The equations (\ref{eq:deltatau}) and (\ref{eq:deltahiggs})
are not sufficient so we use add the mass
counter-term (\ref{countmw})
\begin{eqnarray}
\delta m_{W}^{2}=\frac{1}{2}\left(v^{2}g\delta g+g^{2}v\delta v\right),\label{eq:deltaW}
\end{eqnarray}
where $g$ and $\delta g$ are the renormalized coupling of the $SU(2)_{L}$
sector and its counterterm respectively. The counter-term $\delta\tau$ will be fixed in a way that $v\delta\tau\phi_{1}$
in eq. (\ref{eq:1l-ct-pot}) exactly cancels the Higgs tadpoles, that
is, we demand that the vacuum expectation value of $\phi_{1}$ be
zero in presence of the radiative corrections,
\begin{eqnarray}
\left\langle \Omega\left|\phi_{1}\right|\Omega\right\rangle =0.
\end{eqnarray}
If we call $iT$ the sum of the tadpoles, with the external leg amputated,
we find the first restriction:
\begin{eqnarray}
iT+iv\delta\tau=0 & \Rightarrow & \delta\tau=-\frac{T}{v}.\label{eq:vev-ren}
\end{eqnarray}
This condition is equivalent to identify the renormalized vacuum with the minimum of the radiatively corrected potential. With this election, all mass counterterms $\delta m_i$ (see eqs. \ref{eq:Ren-mW} and \ref{eq:Ren-mZ}) are expressed in terms of self-energy diagrams without including the tadpole contribution. It gives rise to a $\delta m_i$ that is gauge-dependent and, as a consequence, the $\overline{MS}$ masses are gauge-dependent quantities. However, a $\overline{MS}$ mass is not a Lagrangian
parameter and therefore the requirement of gauge-invariance is not mandatory. The couplings computed in this chapter are parameters of the Lagrangian, and therefore are gauge-invariant quantities. The most important feature of this choice, according with the discussion of the above section, is that the relation between the pole and $\overline{MS}$ masses of top quark is free of the very large electroweak correction. 
\\ 
The second restriction is obtained imposing that the term proportional
to $\frac{1}{2}\phi_{1}^{2}$,
\begin{eqnarray*}
m_{H}^{2}=2\lambda v^{2},
\end{eqnarray*}
in eq. (\ref{eq:ren-pot}), corresponds to the physical Higgs boson
mass. According to equation (\ref{eq:vev-ren}), this implies 
\begin{eqnarray}
\delta m_{H}^{2}=Re\Pi_{HH}(m_{H}^{2}),\label{eq:2-cond}
\end{eqnarray}
where $-i\Pi_{HH}(p^{2})$ is the 1PI part of the Higgs self-energy. 
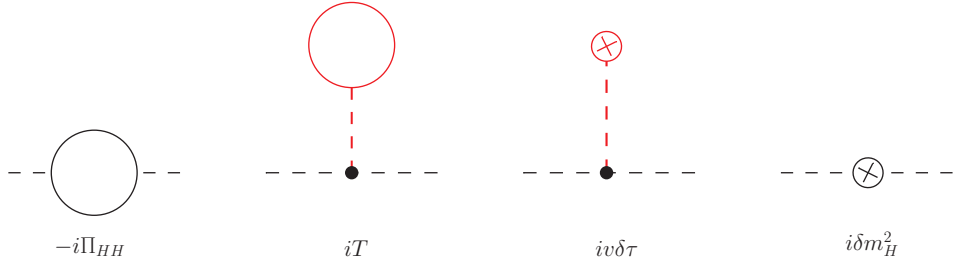
\begin{figure}
\begin{center}
\scalebox{0.5}{
\fcolorbox{white}{white}{
  \begin{picture}(706,195) (143,-31)
    \SetWidth{1.0}
    \SetColor{Black}
    \Line[dash,dashsize=10,arrow,arrowpos=0.5,arrowlength=5,arrowwidth=2,arrowinset=0.2](144,35)(272,35)
    \Line[dash,dashsize=10](336,35)(464,35)
    \Line[dash,dashsize=10](528,35)(656,35)
    \Line[dash,dashsize=10,arrow,arrowpos=0.5,arrowlength=5,arrowwidth=2,arrowinset=0.2](720,35)(848,35)
    \SetWidth{1.5}
    \SetColor{Red}
    \Line[dash,dashsize=10](400,35)(400,99)
    \SetWidth{1.0}
    \COval(400,131)(32,32)(0){Red}{White}
    \SetColor{Black}
    \COval(208,35)(32,32)(0){Black}{White}
    \SetWidth{1.6}
    \SetColor{Red}
    \Line[dash,dashsize=10](590,35)(590,125)
    \SetWidth{1.0}
    \COval(590,130)(11.18,11.18)(153.43495){Red}{White}\Line(586.464,137.071)(593.536,122.929)\Line(597.071,133.536)(582.929,126.464)
    \SetColor{Black}
    \COval(785,35)(11.18,11.18)(116.56505){Black}{White}\Line(777.929,38.536)(792.071,31.464)\Line(788.536,42.071)(781.464,27.929)
    \Vertex(590,35){5}
    \Vertex(400,35){5}
    \Text(179,-28)[lb]{\Large{\Black{$-i\Pi_{HH}$}}}
    \Text(396,-28)[lb]{\Large{\Black{$iT$}}}
    \Text(584,-28)[lb]{\Large{\Black{$iv\delta \tau$}}}
    \Text(771,-28)[lb]{\Large{\Black{$i\delta m_{H}^{2}$}}}
  \end{picture}
}}
\end{center}
\caption{\label{top-vev-ren}Renormalization of the vev. The sum of the topologies in red color vanishes.}
\end{figure}
The fig. (\ref{top-vev-ren}) clarifies the above condition. Finally,
the last condition is provided by the equation:
\begin{eqnarray*}
\delta m_{W}^{2}=ReA_{WW}(m_{W}^{2})+t_{WW},
\end{eqnarray*}
given in the above section. One more time, according to eq. (\ref{eq:vev-ren}),
the term $t_{WW}$ is absorbed by the counterterm $\delta\tau$, thus
the above relation reduces to
\begin{eqnarray}
\delta m_{W}^{2}=ReA_{WW}(m_{W}^{2}).\label{eq:3-cond}
\end{eqnarray}
Now, using the equations (\ref{eq:deltatau}-\ref{eq:deltaW}) in
(\ref{eq:vev-ren}-\ref{eq:3-cond}) we can obtain $\delta\lambda$,
$\delta m^{2}$ and $\delta v$, as a function of the tadpoles and
self-energies. For instance, to find $\delta m^{2}$ we need to make
the subtraction
\begin{eqnarray}
\delta m_{H}^{2}-3\delta\tau=2\delta m^{2}.
\end{eqnarray}
From this relation we deduce
\begin{eqnarray}
\delta m^{2}=\frac{1}{2}\left[Re\Pi_{HH}(m_{H}^{2})+3\frac{T}{v}\right].\label{eq:cond-ren-1}
\end{eqnarray}
By other side, if we make
\begin{eqnarray}
\frac{\delta m_{H}^{2}-\delta\tau}{m_{H}^{2}}=\frac{\delta\lambda}{\lambda}+2\frac{\delta v}{v} & \Rightarrow & \frac{\delta\lambda}{\lambda}=\frac{\left[\delta m_{H}^{2}-\delta\tau\right]}{m_{H}^{2}}-2\frac{\delta v}{v}
\end{eqnarray}
and 
\begin{eqnarray}
\frac{\delta m_{W}^{2}}{m_{W}^{2}}=2\frac{\delta g}{g}+2\frac{\delta v}{v} & \Rightarrow & 2\frac{\delta v}{v}=\frac{\delta m_{W}^{2}}{m_{W}^{2}}-2\frac{\delta g}{g},
\end{eqnarray}
therefore
\begin{eqnarray}
\frac{\delta v}{v}=\frac{ReA_{WW}(m_{W}^{2})}{2m_{W}^{2}}-\frac{\delta g}{g},\label{eq:cond-ren-2}
\end{eqnarray}

\begin{eqnarray}
\frac{\delta\lambda}{\lambda}=\frac{\left[Re\Pi_{HH}(m_{H}^{2})+T/v\right]}{m_{H}^{2}}-\frac{ReA_{WW}(m_{W}^{2})}{m_{W}^{2}}+2\frac{\delta g}{g}.\label{eq:cond-ren-3}
\end{eqnarray}
The counterterm $\delta g$ is obtained from the renormalization of
the electric charge, as we show in eq. (\ref{ren-g}). Combining eq. (\ref{eq:cond-ren-2})
and eq. (\ref{eq:cond-ren-3}) with the expression for $\delta g$,
and using the expression obtained for the quantum correction $\Delta r$ (eq. \ref{eq:one-loop-Dr}) to the relation between $G_{\mu}$ and $g$, we finally obtain
\begin{eqnarray}
\frac{\delta v}{v}=\frac{1}{2}\left[\frac{A_{WW}(0)}{m_{W}^{2}}+\Delta r-E\right]\label{eq:vev-final}
\end{eqnarray}
and
\begin{eqnarray}
\frac{\delta\lambda}{\lambda}=\frac{\left[Re\Pi_{HH}(m_{H}^{2})+T/v\right]}{m_{H}^{2}}-\frac{A_{WW}(0)}{m_{W}^{2}}-\Delta r+E.\label{eq:lambda-final}
\end{eqnarray}
Note that the above expressions are not computed in any particular
gauge, however one can easily prove that $\delta\lambda$ is a gauge
invariant quantity, although this is not evident from eq.~(\ref{eq:lambda-final}). But, if one starts from eq. (\ref{eq:cond-ren-3}), and writes it in the form:
\begin{eqnarray}
\frac{\delta\lambda}{\lambda}=\frac{\left[Re\Pi_{HH}(m_{H}^{2})+3T/v\right]}{m_{H}^{2}}-\left[\frac{ReA_{WW}(m_{W}^{2})}{m_{W}^{2}}+\dfrac{2T}{vm_{H}^ {2}}\right]+2\frac{\delta g}{g},\label{eq:cond-ren-3GI}
\end{eqnarray}
then $\delta\lambda$ is obviously a gauge-invariant quantity, because the Higgs or Vector self-energies including the tadpoles contributions are gauge-independent quantities. Besides, as pointed out in ref. \cite{ASirlin2} $\delta g/g$ is gauge invariant. As a consequence, we can compute all topologies involved in the matching condition (\ref{eq:lambda-final}) in any gauge. This result is valid too for all matching conditions computed in this chapter.

Finally, in the same way as the above section, the connection to one-loop order between $\lambda$ and the $\overline{MS}$ parameter $\bar{\lambda}$ leads to 
\begin{eqnarray}
\bar{\lambda}(\mu)=\dfrac{G_{\mu}}{\sqrt{2}}m_{H}^{2}\left[1 + \dfrac{A_{WW}(0)}{m_{W}^{2}}-\dfrac{Re\Pi_{HH}(m_{H}^{2})+T/v}{m_{H}^{2}} - E \right]_{fin}. \label{eq:ren-lambda}
\end{eqnarray} 

\section{Two-loop Radiative Corrections}

We look first the two loop renormalization of the Higgs sector of SM, and then we derive the two-loop relations between the $\overline{MS}$ couplings and the physical observables $G_{\mu}$, $m_{t}$, $m_{W}$, $m_{Z}$ and $m_{H}$, following the Sirlin-Zucchini scheme. We use the same unrenormalized Higgs potential written in terms of bare quantities, and the same renomalization conditions (\ref{ren-condit}), but this time we retain the two-loop terms, the correction at two-loop order obtained is:
\begin{eqnarray}
\delta V^{(2l)} &=& \delta \lambda \left[\phi ^{\dagger} \phi \left( \phi^{\dagger} 
\phi + \phi_{1}^2 + \phi_2 \right)^2 + 
                   \frac{1}{4} \left( \phi_{1}^2 + \phi_2 ^2\right)^2 \right] \nonumber \\
        & & + \left[ \lambda \left( \frac{\delta v^2}{2\,v} + 
               \frac{(\delta v^2)^2}{8\,v_{r}^{3}} \right) + v \,\delta \lambda 
            \left(1 -   \frac{\delta v^2}{2\,v^2}\right) \right]
           \phi_{1} \left[ \phi_{1}^2 + \phi_2 ^2 + 2\, \phi^{\dagger} \phi\right] \nonumber \\
        & & + \delta \tau \left( \frac{1}{2} \phi_2 ^2 + \phi^{\dagger} \phi \right) +
         \frac{1}{2} \delta m_H^2  \phi_{1}^2 + v \,\delta \tau 
        \left( 1 - \frac{\delta v^2}{2\, v^2} \right) \, \phi_{1}~,
\label{deltaV} 
\end{eqnarray}
where
\begin{eqnarray}
\delta m_h^2 &=& 3 \left[ \lambda \delta v^2 + v^{2} \delta \lambda 
 \left(   1 - \frac{\delta v^2}{v^{2}} \right) \right] - \delta m^{2}  
\label{dmh}\ ,\\
\delta \tau &=&  \lambda \delta v^2 + v^{2} \delta \lambda 
 \left(   1 - \frac{\delta v^2}{v^{2}} \right) - \delta m^{2} \ .
\label{dtau}
\end{eqnarray}
Now, the requirement of the tadpole cancellation are expressed by the condition:
\begin{eqnarray}
\delta \tau \left(1- \dfrac{\delta v^{2}}{2v^{2}}\right)=-\dfrac{T}{v} \label{vev2lc}
\end{eqnarray}
where $iT$ is the sum of the tadpole diagrams with the external leg truncated. Moreover if we uses the equations (\ref{dmh}), (\ref{dtau}) and (\ref{vev2lc}) we can find the two-loop corrections to the $\bar{\lambda}(\mu)$ coupling,
\begin{eqnarray}
\delta^{(2)} \lambda &=& \frac{G_\mu}{\sqrt{2}} m_H^2 \left\{\Delta r_{0}^{(2)}+ 
\frac{1}{m_H^2} \left[\frac{T^{(2)}}{v}
+\delta^{(2)} m_H^2 \right]+ \right. \nonumber \\ 
&& \left. -  \Delta r_{0}^{(1)} 
\left(\Delta r_{0}^{(1)}+ \frac{1}{m_{H}^{2}}\left[\frac{3\, T^{(1)}}{2\,v}
+\delta^{(1)} m_{H}^{2}  \right]\right)\right\} \,,
\label{eq:lh2}
\end{eqnarray}
where
\begin{eqnarray}
\Delta r_{0}^{(l)}=\dfrac{A_{WW}^{(l)}(m_{W}^{2})}{m_{W}^{2}}-E^{(l)}.
\end{eqnarray}
The superscript, $(l)=(1)$ or $(2)$ indicates the loop order. The correction $E$ is a sum of different contributions:
\begin{eqnarray}
E = V_{W} + 2v_{0}B_{W} + \xi + M,
\end{eqnarray}
where $V_{W}$ is the vertex contribution to the muon decay process, $B_{W}$ is the box contribution, $\xi$ is the term due to the renormalization of the external legs and $M$ is a two-loop mixed contribution due to product of different one-loop objects among $V_{W}$, $B_{W}$, $A_{WW}$ and $\xi$. We point out that in the above expression no tadpole contribution is included because of our choice of identifying the renormalized vacuum with the minimum of the radiatively corrected potential, as a consequence $\Delta r_0$ is a gauge-dependent quantity. Up two-loop level the correction $\Delta r_{0}$ has the form:
\begin{eqnarray}
&& \Delta r_{0}^{(1)}+\Delta r_{0}^{(2)} = \nonumber \\ && \nonumber \\ && \dfrac{A_{WW}^{(1)}(m_{W}^{2})}{m_{W}^{2}}-V_{W}^{(1)}-\dfrac{\sqrt{2}}{G_{\mu}}B_{W}^{(1)} - \xi^{(1)} + \dfrac{A_{WW}^{(2)}(m_{W}^{2})}{m_{W}^{2}} - V_{W}^{(2)}-\dfrac{\sqrt{2}}{G_{\mu}}B_{W}^{(2)} - \xi^{(2)} - M^{(2)} \nonumber \\ &&
~+~ \delta ^{(1)}m_{W}^{2}\dfrac{A_{WW}^{(1)}}{m_{W}^{4}}-\dfrac{\sqrt{2}}{G_{\mu}}B_{W}^{(1)}
\left( \dfrac{A_{WW}^{(1)}(m_{W}^{2})}{m_{W}^{2}}-V_{W}^{(1)}-\dfrac{\sqrt{2}}{G_{\mu}}B_{W}^{(1)} - \xi^{(1)} \right),
\end{eqnarray} 
with
\begin{eqnarray}
M^{(2)}= \dfrac{\sqrt{2}}{G_{\mu}}\xi^{(1)}B_{W}^{(1)}+\sum_{i<j}\xi_{i}^{1}\xi_{j}^{1}+\xi^{(1)}V^{(1)}-\left(
\xi^{(1)}+V^{(1)}\right)\dfrac{A_{WW}^{(1)}}{m_{W}^{2}}.
\end{eqnarray}
The indices $i,j$ in above equation label the different particles in the muon decay process: $\mu$, $e$, $\nu_{e}$ and $\nu_{\mu}$. We recall that $\Delta r_{0}$ is computed with all external four-momenta and all the light-fermion masses put to zero before the integration over the loop momenta \cite{Awramik}, such a procedure generates a lot of new infrared divergences in the calculations besides the ones contained in the photon diagrams of the topologies $B_{W}$ and $\xi$. The infrared divergences can be regularized by the dimensional regularization procedure and cancelled using the statement of the factorization theorem \cite{Awramik}. This cancellation is nontrivial, however, following this procedure the evaluation of $\Delta r_{0}$ is completely reduced to bubble diagrams with one or two loops and is an infrared (IR) safe quantity\footnote{The correction $\Delta r_0$ is however a UV divergent quantity.}. The analytical result is reduced to a superposition of the basis integrals $A_{1}^{(d)}$, $B_{11}^{(d)}$ and $K_{111}^{(d)}$ given in the Appendix \ref{AppTarcer}. 
\begin{figure}
\begin{center}
\scalebox{0.5}{
\fcolorbox{white}{white}{
  \begin{picture}(368,206) (139,-131)
    \SetWidth{1.0}
    \SetColor{Black}
    \Line[arrow,arrowpos=0.5,arrowlength=5,arrowwidth=2,arrowinset=0.2](140,-130)(224,-67)
    \Photon(224,-67)(224,3){5}{5}
    \Line[arrow,arrowpos=0.5,arrowlength=5,arrowwidth=2,arrowinset=0.2](140,73)(224,3)
    \Line[arrow,arrowpos=0.5,arrowlength=5,arrowwidth=2,arrowinset=0.2](224,3)(420,3)
    \Line[arrow,arrowpos=0.5,arrowlength=5,arrowwidth=2,arrowinset=0.2](224,-67)(420,-67)
    \Photon(322,3)(322,-67){5}{5}
    \Photon(420,2)(420,-68){5}{5}
    \Line[arrow,arrowpos=0.5,arrowlength=5,arrowwidth=2,arrowinset=0.2](420,-67)(497,-130)
    \Line[arrow,arrowpos=0.5,arrowlength=5,arrowwidth=2,arrowinset=0.2](420,3)(497,73)
    \Vertex(224,-66){4}
    \Vertex(224,2){4}
    \Vertex(324,2){4}
    \Vertex(320,-66){4}
    \Vertex(420,2){4}
    \Vertex(420,-66){4}
    \Text(156,-98)[lb]{\Large{\Black{$\mu$}}}
    \Text(188,54)[lb]{\Large{\Black{$\nu_{e}$}}}
    \Text(472,34)[lb]{\Large{\Black{$e$}}}
    \Text(440,-114)[lb]{\Large{\Black{$\nu_{\mu}$}}}
    \Text(268,18)[lb]{\Large{\Black{$f$}}}
    \Text(368,18)[lb]{\Large{\Black{$(f, \nu_{f})$}}}
    \Text(260,-96)[lb]{\Large{\Black{$f$}}}
    \Text(368,-96)[lb]{\Large{\Black{$(f, \nu_{f})$}}}
    \Text(240,-30)[lb]{\Large{\Black{$W$}}}
    \Text(336,-30)[lb]{\Large{\Black{$V$}}}
    \Text(432,-34)[lb]{\Large{\Black{$V$}}}
  \end{picture}
}}
\end{center}
\caption{\label{Box2l} A generic two-loop box diagram contributing to $\Delta r_{0}$.}
\end{figure}
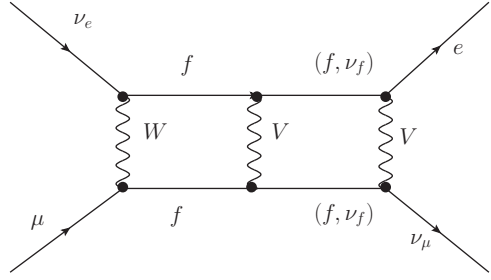
To visualize this, we consider the generic two-loop box diagram represented in fig. (\ref{Box2l}), the amplitude of this diagram can be written in the form:
{\small
\begin{eqnarray}
M_{WVV}=CM_{0} T_{\mu\nu\rho\sigma}(0,\{{\partial}_j\},{\bf d^+}){\rm TFI}[d,0,\{\{1,m_W\},\{1,m_V\},\{2,0\},\{2,0\},\{1,m_V\}\}],
\end{eqnarray}}
where $C$ is a coefficient of order $\sim g^{4}$, $M_{0}$ is the three-level muon decay amplitude, $T_{\mu\nu\rho\sigma}$ is the tensor operator defined in Chapter \ref{cha:TwoLoopCalc} - eq. (5.13) and TFI is the integral defined in Appendix \ref{AppTarcer} - eq. (E.3) with fermion masses and external momenta put to zero. The integral ${\rm TFI}[d,0,\{\{1,m_W\},\{1,m_V\},\{2,0\},\{2,0\},\{1,m_V\}\}]$ can be reduced to a superposition of vacuum bubbles integrals using the TARCER code, the result 
is:

{\small
\begin{eqnarray}
&& {\rm TFI}[d,0,\{\{1,m_W\},\{1,m_V\},\{2,0\},\{2,0\},\{1,m_V\}\}]=c_{1}\left(A_{\{1,m_Z\}}^{(d)}\right)^{2} + c_{2}A_{\{1,m_W\}}^{(d)}A_{\{1,m_V\}}^{(d)} \nonumber \\ 
&& + c_{3}B_{\{1,m_V\}\{1,0\}}^{(d)}A_{\{1,m_V\}}^{(d)} + c_{4}B_{\{1,m_W\}\{1,0\}}^{(d)}A_{\{1,m_V\}}^{(d)} + c_{5}A_{\{1,m_W\}}^{(d)} B_{\{1,m_V\}\{1,0\}}^{(d)}  \nonumber \\
&& + c_{6}F_{\{1,m_W\}\{1,m_V\}\{1,0\}\{1,0\}\{1,m_V\}}^{(d)} + c_{7}K_{\{1,m_V\}\{1,0\}\{1,0\}}^{(d)}+c_{8}K_{\{1,m_W\}\{1,m_V\}\{1,0\}}^{(d)}
 \nonumber \\ && +c_{9}K_{\{1,m_W\}\{1,m_V\}\{1,m_V\}}^{(d)} + c_{10}V_{\{1,m_V\}\{1,0\}\{1,m_V\}\{1,m_W\}}^{(d)} + c_{11}V_{\{1,m_V\}\{1,0\}\{1,m_W\}\{1,m_V\}}^{(d)}, \nonumber \\
\end{eqnarray}}
where the coefficients $c_{j}$ are functions of the masses $m_V$, $m_W$ and the space-time dimension $d$. The above expression can be reduced to a superposition of the vacuum bubbles integrals exposed in Chapter \ref{cha:TwoLoopCalc} - Section (5.2) by trivial algebraical manipulations. The functions $A$, $B$, $K$, $V$ and $F$ can be consulted in Appendix \ref{AppTarcer}. 

Similarly, one can find for the counterterm of the quadratic Higgs coupling in the potential
\begin{eqnarray}
\delta ^{(2)} m^{2} &=& 
3 \frac{T^{(2)}}{v}
+ \delta^{(2)} m_{h}^2  - \frac{3\, T^{(1)}}{2\,v}\Delta r_{0}^{(1)}\, ,
\label{eq:mh2}
\end{eqnarray}
and for the top Yukawa and gauge couplings,
\begin{eqnarray}
\delta^{(2)} h_{t} &=& 2 \left( \frac{G_{\mu}}{\sqrt{2}}  m_{t}^{2} \right)^{1/2}  
        \left( \frac{\delta^{(2)} m_t}{ m_t } + \frac{\Delta r_{0}^{(2)}}{2} -
 \frac{\Delta r_{0}^{(1)}}{2} \left[ \frac{\delta^{(1)} m_t}{ m_t} + 
   \frac{3\,\Delta r_{0}^{(1)}}{4} \right] \right),
\label{eq:yt2}
\end{eqnarray}
and
\begin{eqnarray}
\delta^{(2)} g & = &  \left( \sqrt{2}\,G_{\mu} \right)^{1/2} m_{W} \left(
\frac{\delta^{(2)} m_{W}^{2}}{{m}_{W}^{2}} + \Delta r_{0}^{(2)} +\right.\nonumber \\
&&-\left.\frac{\Delta r_{0}^{(1)}}{2}\left[\frac{\delta^{(1)} m_{W}^2}{m_{W}^2} +   \frac{3\Delta r_{0}^{(1)}}{2}\right]+\frac{1}{4}\left(\frac{\delta^{(1)} m_{W}^{2}}{m_{W}^{2}}\right)^{2}\right),
\label{eq:G22} 
\end{eqnarray}
for the $SU(2)_{L}$ gauge coupling. Finally for the hypercharge gauge coupling, one finds:
\begin{eqnarray}
\delta^{(2)} g' &=&  \left( \sqrt{2}\,G_{\mu} \right)^{1/2}  
\sqrt{m_{Z}^{2} -m^{2}_{W}}~ \left( 
\frac{\delta^{(2)} m_{Z}^{2} - \delta^{(2)} m_{W}^{2}}{m_{Z}^{2} -m^{2}_{W}} +  \Delta r_{0}^{(2)}+\right.\nonumber\\
&&\left.-\frac{\Delta r_{0}^{(1)}}{2}\left[\frac{\delta^{(1)} m_{Z}^{2} - \delta^{(1)} m_{W}^{2}}{m_{Z}^{2} -m^{2}_{W}} +  \frac{3\Delta r_{0}^{(1)}}{2}\right]+\frac{1}{4}\left(\frac{\delta^{(1)} m_{Z}^{2} - \delta^{(1)} m_{W}^{2}}{m_{Z}^{2} -m^{2}_{W}} \right)^{2}
\right)~. 
\label{eq:G12}
\end{eqnarray}

\subsection{SM couplings at the EW scale}
In this section we give numerical results for the SM  parameters ($g_i=\{\lambda, m^2, h_t, g, g'\}$) renormalised at the EW scale $\bar{\mu}=m_t$ in the $MS$ scheme, computed in terms of the observables $m_H,m_t,m_W,m_Z, G_\mu$ (see table \ref{tab:SMvalues}) and $\alpha_3(m_Z)$. Each $\overline{MS}$ parameter $g_i(\bar{\mu})$ is expanded in loops as
\begin{eqnarray} 
g_i = g_{i}^{(0)}+g_{i}^{(1)}+g_{i}^{(2)}+\cdots \label{eq:exp}
\end{eqnarray}
where the tree-level values (LO) $g_{i}^{(0)}$ are listed in the first column of table~\ref{tab:123}; the one-loop corrections $g_{i}^{(1)}$ are analytically given in appendix~\ref{app-counterterms} and the two-loop corrections $g_{i}^{(2)}$ can be consulted in \cite{Degrassi2}.

\begin{table}
$$ \begin{array}{c|ccccc}
 \bar{\mu}=m_t & \lambda  & h_t & g & g' & m/{\rm GeV} \\ \hline
 \text{LO} & 0.12917 & 0.99561 & 0.65294 & 0.34972 & 125.15 \\
 \text{NLO} & 0.12774 & 0.95113 & 0.64754 & 0.35940 & 132.37 \\
 \text{NNLO} & 0.12604 & 0.94018 & 0.64779 & 0.35830 & 131.55 \\
\end{array} $$
\caption{ Values of the fundamental SM parameters computed at tree level, one loop, two loops in the $\overline{MS}$ scheme and renormalised at $\bar{\mu}=m_t$ for the central values of the observables listed in table~\ref{tab:SMvalues}. \label{tab:123}}
\end{table}

\subsubsection{The Higgs quartic coupling}

For the Higgs quartic coupling we find:
\begin{eqnarray}
\lambda(\bar{\mu}=m_t) = 0.12604+0.00206\left( \frac{m_H}{{\rm GeV}}-125.15 \right) ~~~~~~~~~~~~~~~~ \nonumber \\  -0.00004 \left( \frac{m_t}{{\rm GeV}}-173.34 \right) \pm {0.00030}_{\rm th}~.  
\end{eqnarray}
The dependence on $m_t$ is  small because $\lambda$ is renormalised at $m_t$ itself.
Here and below the theoretical uncertainty is estimated from the dependence on $\bar{\mu}$ 
(varied around $m_t$ by one order of magnitude) of the higher-order unknown 3 loop corrections. Such dependence is extracted from the known SM RGE at 3 loops \cite{Degrassi2}.

\subsubsection{The Higgs mass term}

For the mass term of the Higgs doublet in the SM Lagrangian (normalised such that $m=m_H$ at tree level) one finds \cite{Degrassi2} 
\begin{eqnarray}
\frac{m(\bar{\mu}=m_t)}{{\rm GeV}}=131.55+0.94 \left( \frac{m_h}{{\rm GeV}} - 125.15 \right)+0.17\left( \frac{m_t}{{\rm GeV}}-173.34 \right) \pm 0.15_{\rm th}.
\end{eqnarray}
 
\subsubsection{The top Yukawa coupling}

For the top Yukawa coupling one get
\begin{eqnarray}
h_t(\bar{\mu}=m_t) &=& 0.93690 +0.00556 \left( \frac{m_t}{{\rm GeV}}-173.34 \right)+\\
&& -0.00042\dfrac{\alpha_{3}(m_Z)-0.1184}{0.0007} \pm {0.00050}_{\rm th}~.  \nonumber
\label{eq:ht_ew}
\end{eqnarray}
The central value differs from the NNLO value in table~\ref{tab:123} because is include here also the NNNLO (3 loop) pure QCD effect~\cite{Chetyrkin0}. The estimated theoretical uncertainty does not take into account the non-perturbative theoretical uncertainty of order $\Lambda_{\rm QCD}$ in the definition of $m_t$.

\subsubsection{The weak gauge couplings}

For the weak gauge couplings $g$ and $g'$ computed at NNLO accuracy in terms of $m_W$ and $m_Z$ one find
{\small
\begin{eqnarray}
g(\bar{\mu}=m_t) &=& 0.64779 +0.00004 \left( \frac{m_t}{{\rm GeV}}-173.34 \right)+ 0.00011 \dfrac{m_W-80.384{\rm GeV}}{0.014{\rm GeV}}, \\
g'(\bar{\mu}=m_t) &=& 0.35830 +0.00011 \left( \frac{m_t}{{\rm GeV}}-173.34 \right)- 0.00020 \dfrac{m_W-80.384{\rm GeV}}{0.014{\rm GeV}},
\end{eqnarray}}
where the adopted value for $m_W$ and its experimental error are reported in table~\ref{tab:SMvalues}.

\subsubsection{The strong gauge coupling}

\begin{figure}
\centerline{\includegraphics[width=0.45\linewidth]{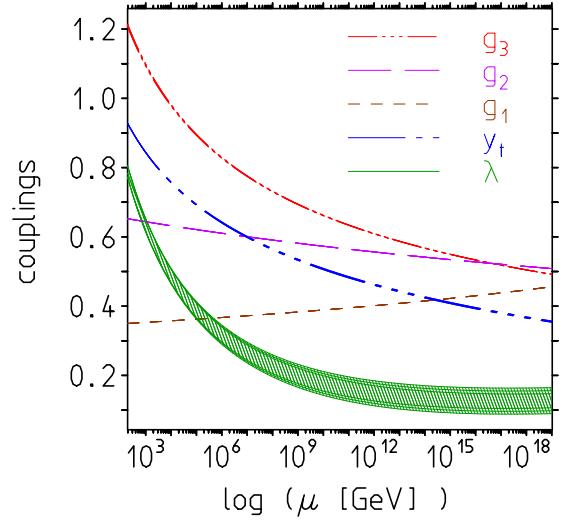}}
\caption[]{Renormalization of the SM couplings defined as: $g_1=\sqrt{5/3}g', ~g_{2}=g, ~g_{3}=g_s$, $y_{t}=h_t$ and of the Higgs selfcoupling $\lambda$ \cite{FredKniehl}. All parameters are defined in the $\overline{MS}$ scheme.}\label{Evolution}
\end{figure}

The central value of $\alpha_3(m_Z)$ ($\alpha_3(m_Z)=0.1184 \pm 0.0007$) is extracted from the global fit of~\cite{Siegfried} in the effective SM with 5 flavours. Including RG running from $m_Z$ to $m_t$ at 4 loops in QCD and at 2 loops in the electroweak gauge interactions, and 3 loop QCD matching at $m_t$ to the full SM with 6 flavours, one get
\begin{eqnarray}
g_s(\bar{\mu}=m_t)  = 1.1666 + 0.00314\dfrac{\alpha_{3}(m_Z)-0.1184}{0.0007} -0.00046 \left( \frac{m_t}{{\rm GeV}}-173.34 \right).
\end{eqnarray}
The SM parameters can be renormalised to any other desired energy by solving the SM renormalisation group equations. At EW scale the values of the SM parameters are summarized in table \ref{tab:123}, the evolution up to some large cut-off scale is shown in fig. \ref{Evolution}. At the Planck scale ($\Lambda_{P}$), the numerical values of the SM parameters are:
\begin{eqnarray}
g(\Lambda_{P}) &=& 0.6154+0.0003 \Mtdiff -0.0006 \MWdiff \\
g'(\Lambda_{P})&=& 0.5055 \\
g_s(\Lambda_{P})&=& 0.4873+ 0.0002 \asdiff \\
h_t(\Lambda_{P})&=& 0.3825+ 0.0051\Mtdiff - 0.0021\asdiff  \\
\label{eq:lammp}
\lambda (\Lambda_{P} ) &=&  -0.0143- 0.0066\Mtdiff +\\   \nonumber
&&+0.0018\asdiff +0.0029\Mhdiff \\
m(\Lambda_{P}) &=& 129.4{\rm GeV} +1.6{\rm GeV}\Mhdiff+
\\&&-0.25{\rm GeV}\Mtdiff + 0.05{\rm GeV}\asdiff \nonumber
\end{eqnarray}
All Yukawa couplings, other than the one of the top quark, are very small and can be disregarded in the stability analysis.

\chapter{\label{cha:TwoLoopCalc}The Two-Loop Calculations}

\lettrine{I}{ n} order to study the stability of the Electro-Weak vacuum, using the effective potential approach, we need a precise calculation of the Higgs quartic coupling $\lambda$. This is because due to its renormalization running the Higgs self-coupling can become large or develop new minima in the effective potential so to change the vacuum structure of the Standard Model. At present, the renormalized Higgs coupling $\lambda$ \cite{Degrassi} and  the Standard Model Higgs effective potential \cite{Ford-Jones} have been evaluated at two-loop order. The two-loop renormalization group determination of all physical parameters in the Standard Model require the evaluation of mass-dependent radiative corrections, therefore we need to compute in general $m$-point two-loop tensor integrals. Due to the complicated structure of integrals and the large number of diagrams, typically thousands, an efficient code-implemented organization of the calculation is needed. After generating Feynman diagrams and their corresponding integrands by the {\tt Mathematica} package {\tt FeynArts} \cite {FeynArts} we adopt a method of reduction of tensor integrals in terms of a combination of some basis of integrals, so called master integrals \cite{PaVe}. The process needs three steps: integrand tensor decomposition, reduction of scalar integrals to scalar master integrals and consequent evaluation of them. \\
In the first part of this chapter we will elucidate those steps by making explicit examples, in particular  concerning one-point and two-points two-loop diagrams. In calculating the different diagrams topologies with arbitrary masses we will follow the Tarasov's literature \cite{Tarasov1, Tarasov2}. In the second part, the Tarasov procedure is implemented through the {\tt Mathematica} code {\tt TARCER} \cite{Tarcer} to compute the 1PI effective potential in the $\overline{MS}$ renormalization scheme. The {\tt TARCER} code will be complemented with  the {\tt Mathametica} package {\tt FeynCalc} \cite{FeynCalc} for the algebra of the numerators. All of sectors of the potential are reduced using {\tt TARCER} in terms of two basis integrals whose explicit evaluation and their Laurent expansion around $d=4$ are computed with {\tt TARCER} too. We start the chapter by illustrating how to obtain an analytic form of the effective potential, its numerical evaluation will be provided. The implications of such a numerical evalution for what concerns the tadpole contribution to the threshold corrections of the input $\lambda(\mu)$ at two-loop level will be the subject of the next chapter.    

\section{\label{sec: TarasovAlg}Two-Loop Integrals and Tarasov Algorithm}
In this section we will consider only two-loop two-points (self-energy) dimensionally regulated diagrams with arbitrary masses. Nevertheless the analysis is straightforwardly extended to two-loop tadpoles and to two-loop vacuum bubbles. 
 As a sub-product of this analysis the one-loop integrals with one or two external legs will be obtained. 
The Tarasov's prescription \cite{Tarasov1, Tarasov2} consists in reducing any $m$-point two-loop integral with a tensorial structure to a superposition of scalar integrals through the application of a linear operator $T(q,\{{\partial}_j\},{\bf d^+})$ as
\begin{eqnarray}
\int \! \int \frac{d^dk_1 d^dk_2}{P_1^{\nu_1}P_2^{\nu_2} P_3^{\nu_3}P_4^{\nu_4}P_5^{\nu_5}}
  \prod_{n=1}^{r}~k_{1\mu_n}\prod_{m=1}^{s}~k_{2\lambda_m}=T_{\mu_1\ldots \lambda_s} (q,\{{\partial}_j\},{\bf d^+})\int \!\! \int \frac{d^dk_1 d^dk_2}{P_1^{\nu_1} P_2^{\nu_2} P_3^{\nu_3} P_4^{\nu_4} P_5^{\nu_5}}\,\,\, .\nonumber\\
\label{tensorial-int}
\end{eqnarray}
Here $ \partial_j=\frac{\partial}{\partial m_j^2}$, being $m_j$ the mass associated to the line $j$ in the diagram, ${\bf d^+}$ is the operator shifting the space-time dimensionality of a regularized integral by two-units $\left({\bf d^+} I^{(d)}=I^{(d+2)}\right)$ and
\begin{tabbing}
   $P_1=k_1^2-m_1^2+i\epsilon$,\qquad\=
   \qquad\=$P_3=(k_1-q)^2-m_3^2+i\epsilon$,
          \qquad\=   $P_5=(k_1-k_2)^2-m_5^2+i\epsilon$,\\
      $P_2=k_2^2-m_2^2+i\epsilon$, \>
    \> $P_4=(k_2-q)^2-m_4^2+i\epsilon.~~~~$ \> 
\end{tabbing}
On the right-hand side of (\ref{tensorial-int}) it is assumed that before differentiation the line $i$ has a non-zero mass $m_i$ to be sent to its physical (or field dependent) value after differentiation. The topology expressed by (\ref{tensorial-int}) is depicted in the figure~\ref{Two-Point-Two-Loop}.

To derive an explicit expression of the tensor operator $T_{\mu_1\ldots \lambda_s} (q,\{{\partial}_j\},{\bf d^+}) $ we need to introduce the independent auxiliary $d-$vectors $a_1, a_2$ of mass scaling $-1$ and the so-called $\alpha$-parametric
representation of the denominators $P_{i}$ \cite{BS}. The tensor structure of the integrand on the left-hand side of (\ref{tensorial-int}) can be written as
\begin{eqnarray}
\prod_{n=1}^{r}~k_{1\mu_n}\prod_{m=1}^{s}~k_{2\lambda_m}=
\left.\frac{1}{i^{r+s}}\left( \prod_{n=1}^{r}~\frac{\partial}{\partial a_{1}^{\mu_n}}\right) \left( \prod_{m=1}^{s}~
\frac{\partial}{\partial a_{2}^{\lambda_m}}\right) \exp \left[i (a_1k_1+a_2k_2)\right]
\right|_{ a_i=0 }.
\label{avectors}
\end{eqnarray}
The integral on the l.h.s. of (\ref{tensorial-int}) can be put therefore in the form
\begin{eqnarray*}
\left.\frac{1}{i^{r+s}}\left( \prod_{n=1}^{r}~\frac{\partial}{\partial a_{1}^{\mu_n}}\right) \left( \prod_{m=1}^{s} \frac{\partial}{\partial a_{2}^{\lambda_m}}\right) G^{(d)}(q^2,a_{1},a_{2}) \right|_{ a_i=0 },
\end{eqnarray*}
where
\begin{eqnarray}
G^{(d)}(q^2,a_{1},a_{2})=\int \! \int \frac{ d^dk_1 d^dk_2}
{P_1^{\nu_1}P_2^{\nu_2}P_3^{\nu_3}P_4^{\nu_4}P_5^{\nu_5}}
\exp \left[i (a_1k_1+a_2k_2) \right]. \label{Int-G}
\end{eqnarray}
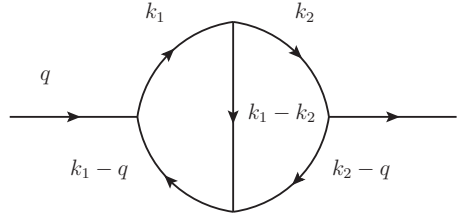
\begin{figure}
\begin{center}
\scalebox{0.7}{
\fcolorbox{white}{white}{
  \begin{picture}(240,124) (186,-149)
    \SetWidth{1.0}
    \SetColor{Black}
    \Line[arrow,arrowpos=0.5,arrowlength=5,arrowwidth=2,arrowinset=0.2](187,-97)(255,-97)
    \Arc[arrow,arrowpos=0.5,arrowlength=5,arrowwidth=2,arrowinset=0.2,clock](314.5,-105.5)(60.104,171.87,98.13)
    \Line[arrow,arrowpos=0.5,arrowlength=5,arrowwidth=2,arrowinset=0.2](306,-46)(306,-148)
    \Arc[arrow,arrowpos=0.5,arrowlength=5,arrowwidth=2,arrowinset=0.2,clock](314.5,-88.5)(60.104,-98.13,-171.87)
    \Arc[arrow,arrowpos=0.5,arrowlength=5,arrowwidth=2,arrowinset=0.2,clock](297.5,-105.5)(60.104,81.87,8.13)
    \Arc[arrow,arrowpos=0.5,arrowlength=5,arrowwidth=2,arrowinset=0.2,clock](297.5,-88.5)(60.104,-8.13,-81.87)
    \Line[arrow,arrowpos=0.5,arrowlength=5,arrowwidth=2,arrowinset=0.2](357,-97)(425,-97)
    \Text(204,-80)[lb]{{\Black{$q$}}}
    \Text(260,-46)[lb]{{\Black{$k_{1}$}}}
    \Text(340,-46)[lb]{{\Black{$k_{2}$}}}
    \Text(360,-131)[lb]{{\Black{$k_{2}- q$}}}
    \Text(220,-131)[lb]{{\Black{$k_{1} - q$}}}
    \Text(315,-100)[lb]{{\Black{$k_{1} - k_{2}$}}}
  \end{picture}
}}
\end{center}
\caption{\label{Two-Point-Two-Loop}{\small Topology for a generic two-loop two-points Feynman integrals.}}
\end{figure}
The $\alpha$-parametric representation of (\ref{Int-G}) is done in terms of the Schwinger parameters
\begin{eqnarray}
\frac{1}{(k_{i}^2-m_{i}^2+i\epsilon)^{\nu}}
 = \frac{i^{-\nu}}{ \Gamma(\nu)}\int_0^{\infty}
 d\alpha ~\alpha^{\nu-1} \exp\left[i\alpha(k_{i}^2-m_{i}^2+i\epsilon)\right],
\end{eqnarray}
and using the $d$- dimensional Gaussian integration  formula
\begin{eqnarray}
\int d^dk \exp \left[i(A k^2+ 2(pk))\right] =i
 \left( \frac{\pi}{i A} \right)^{\frac{d}{2}}
 \exp \left[ -\frac{ip^2}{A} \right] \label{GaussianI},
\end{eqnarray}
we can easily evaluate the integrals over loop momenta. The final result is:
\begin{eqnarray}
G^{(d)}(q^2,a_{1},a_{2})= i^2 \left ( \frac{\pi}{i} \right)^d
\prod^{5}_{j=1} \frac{i^{-\nu_j}}{\Gamma(\nu_j)} \int_0^{\infty} \frac{d \alpha_j \alpha^{\nu_j-1}_j}
     { [ D(\alpha) ]^{\frac{d}{2}}} ~~~~~~~~~~~~~~~~~~~~~~~~~~~~~~~~~~~~~~~~~~~~~~~ \nonumber \\         
         \times \exp \left[ i \left(\frac{Q(\alpha,a_1,a_2)}{D(\alpha)} - \sum_{l=1}^{5}\alpha_l(m_l^2-i\epsilon)\right) \right], \nonumber \\
         \label{represG1}
\end{eqnarray}
where
\begin{eqnarray}
\label{Dform}
&&D(\alpha)=\alpha_5(\alpha_1+\alpha_2+\alpha_3+\alpha_4)
  +(\alpha_1+\alpha_3)(\alpha_2+\alpha_4),    \\
&& \nonumber \\
&&Q(\alpha,a_1,a_2)=A(\alpha,a_1,a_2)q^2+B(\alpha,a_1,a_2),
\label{QI}
\end{eqnarray}
with 
\begin{eqnarray*}
&&A(\alpha,a_1,a_2)= (\alpha_1+\alpha_2)(\alpha_3+\alpha_4)
\alpha_5 +\alpha_1\alpha_2(\alpha_3+\alpha_4)
         +\alpha_3\alpha_4(\alpha_1+\alpha_2), \\
         && \nonumber \\
          && B(\alpha,a_1,a_2)=(qa_1)Q_1+(qa_2)Q_2+a_1^2Q_{11}
            +a_2^2Q_{22}+(a_1a_2)Q_{12}, \nonumber
\end{eqnarray*}
and
\begin{eqnarray}
Q_1&=&\alpha_3\alpha_5+\alpha_4\alpha_5+\alpha_2\alpha_3
+\alpha_3\alpha_4, \nonumber \\
Q_2&=&\alpha_4\alpha_5+\alpha_3\alpha_5+\alpha_1\alpha_4
+\alpha_3\alpha_4, \nonumber \\
-4 Q_{11}&=&\alpha_2+\alpha_4+\alpha_5,
\nonumber \\
-4Q_{22}&=&\alpha_1+\alpha_3+\alpha_5,
\nonumber \\
-2Q_{12}&=&\alpha_5.
\label{qbeta}
\end{eqnarray}
That result is proved in the Appendix \ref{AppDyQ}. If we make the substitution $\alpha_{j}\rightarrow i\partial_{j}$ in $B(\alpha,a_1,a_2)$ of (\ref{QI}),  we apply the resulting operator denoted as $B(i\partial,a_1,a_2)$ on the integral (\ref{Int-G}) in its $\alpha$-parametric representation (\ref{represG1}) and finally we evaluate it in $a_{1}=0$ and $a_{2}=0$ (when $Q(\alpha,a_1,a_2)=A(\alpha,a_1,a_2)q^{2}$) we obtain a $d-2$ dimensional integral:
\begin{eqnarray}
B(i\partial,a_1,a_2) G^{(d)}(q^2,0,0) = -\pi^{2}G^{(d-2)}(q^2,0,0)\dfrac{B(\alpha,a_1,a_2)}{D(\alpha)}.
\label{OperatorQ}
\end{eqnarray} 
The integral $G^{(d-2)}$ in the above formula must be understood as an operator acting over the ratio $\frac{B(\alpha,a_1,a_2)}{D(\alpha)}$, that is actually part of the integrand. If we define the operator $\rho$ by
\begin{eqnarray}
\rho= -\frac{i}{\pi^2}{\bf d^+}
\end{eqnarray}
and we use the equation (\ref{OperatorQ}), it is straightforward to obtain the following relation
\begin{eqnarray}
\sum_{n=0}^{\infty}\dfrac{(B(i\partial,a_1,a_2)\rho)^{n}}{n!} G^{(d)}(q^2,0,0) = G^{(d)}(q^2,0,0) \exp \left[i\dfrac{B(\alpha,a_1,a_2)}{D(\alpha)}\right]. \label{Tbuild}
\end{eqnarray}
The r.h.s of the above equation is just the integral $G^{(d)}(q^2,a_{1},a_{2})$ defined by (\ref{represG1}) and (\ref{Int-G}). As a consequence, we can construct the operator $T(q,\left\{\partial\right\},{\bf d^+})$ defined in (\ref{tensorial-int}) by applying the differential operator 
\begin{eqnarray*}
\frac{1}{i^{r+s}}\left( \prod_{n=1}^{r}~\frac{\partial}{\partial a_{1\mu_n}}\right) \left( \prod_{m=1}^{s} \frac{\partial}{\partial a_{2\lambda_m}}\right)
\end{eqnarray*}
over the l.h.s of equation (\ref{Tbuild}) and finally evaluating the obtained integral at $a_{i}=0$.  The resulting operator is: 
\begin{eqnarray}
&&T_{\mu_1 \ldots \lambda_s}(q,\left\{\partial\right\},{\bf d^+})
 =\frac{1}{i^{r+s}}\!\prod_{n=1}^{r} \frac{\partial}
 { \partial a_{1\mu_n}} \!\! \ldots \prod_{m=1}^{s}
 \frac{\partial}{ \partial a_{2 \lambda_m}}
 \nonumber \\
&&~~~\times \exp
 \left[\left(
 (qa_1)Q_1+(qa_2)Q_2+a_1^2Q_{11}+a_2^2Q_{22}+(a_1a_2)Q_{12}
 \right) \rho \right]
 \left|_{ a_j=0 \atop {\alpha_j=i \partial_j \atop
  \rho=-\frac{i}{\pi^2}{\bf d^+}} } \right. .
\label{Ttensor}
\end{eqnarray}
The tensor operator $T(q,\left\{\partial\right\},{\bf d^+})$ can be used for the direct evaluation of two-loop tensor integrals just by applying the formula (\ref{tensorial-int}). To understand its importance  we need to pay attention to the representation (\ref{represG1}). Any tensor integral in momentum space can be expressed as a sum over a set of tensors made of  external vectors and metric tensors, multiplied by a combination of scalar integrals with the shifted value of $d$ \cite{Davydychev4}. By making the differentiation of the integral (\ref{represG1}) with respect to vectors $a_{1}$ or $a_{2}$ this procedure generates external momenta $q_{\mu}$ and metric tensors $g_{\mu\nu}$ times some polynomials $P(\alpha)$ divided by $D(\alpha)$ to some power in the integrand of (\ref{represG1}). The polynomials $P(\alpha)$ are converted into operators $P(\partial)$ and the powers of $D(\alpha)$ are absorbed into the redefinition of space-time dimension $d$. Building the operator $T(q,\left\{\partial\right\},{\bf d^+})$ we can reproduce the above procedure automatically without direct application of the $\alpha$ -parametric representation. Once the operator is applied, the tensor integral is represented in terms of combinations of scalar integrals with the changed space-time dimension $d$ and with coefficients that are tensors made of external momenta and metric tensors. If we contract the resulting scalar integrals with the tensor $q_{\mu_{1}}\dots q_{v_{s}}$ of external momenta, this representation allows us to study Feynman integrals with irreducible scalar numerators and write them as a combination of scalar integrals having different values of $d$. Since $d$ is shifted, we also need an algorithm to obtain new generalized recurrence relations including integrals with different $d$ that leads to the solution of the problem of irreducible numerators in terms of the original space-time dimension. Those problems will be afforded and solved in the next sections.

\subsection{Non-Scalar Integrals and Irreducible Numerators}

In the previous section we mentioned that we can always contract a tensorial numerator with a proper projector in order to obtain a scalar numerator, with the aim of treating some specific integrals with irreducible scalar numerators. Let see this in more detail.
 
Integrals with scalar products in the numerator can be regarded as a contraction of the tensor integral (\ref{tensorial-int}) with the projector of external momenta $\prod _{n=1}^{r}q_{\mu _{n}}\prod _{m=1}^{s}q_{\lambda _{m}}$ or with the projector of loop momenta $\prod _{n=1}^{r} k_{1\mu _{n}}\prod _{m=1}^{s} k_{2\lambda _{m}}$. One commonly encounters integrals of the form: 
\begin{eqnarray}
I^{(d)}(q^2)=\frac{1}{\pi^d}
\int \!\!\int \frac{d^d k_1 d^dk_2}
 { P_1^{\nu_1} P_2^{\nu_2} P_3^{\nu_3} P_4^{\nu_4} P_5^{\nu_5}}
N(k_1^2,k_2^2,k_1q,k_2q,k_1k_2),
\label{Non-Scalar}
\end{eqnarray}
where $N(k_1^2,k_2^2,k_1q,k_2q,k_1k_2)$ is a polynomial of his arguments. 

If $N(k_1^2,k_2^2,k_1q,k_2q,k_1k_2)$ is a degree zero polynomial, then the integral (\ref{Non-Scalar}) is called "Scalar-Integral". If $N(k_1^2,k_2^2,k_1q,k_2q,k_1k_2)$ has non zero degree we call it "Non-Scalar Integral". The Tarasov algorithm to compute (\ref{Non-Scalar}) consists of three steps: the simplification of the numerator $N(k_1^2,k_2^2,k_1q,k_2q,k_1k_2)$ such that the integral $I^{(d)}(q^2)$ is expressed as a combination of scalar integrals with coefficients depending on $q$ and $m_i^2$, the reduction of these scalar integrals in terms of a set of irreducible integrals known as ``Master Integrals" and the evaluation of the basis integrals. 

In the specific case where the integral (\ref{Non-Scalar}) is a ``Non-Scalar" integral we need simplify the integrand as much as possible. The procedure to make the integrand into a most  simplified form is based on the repeated use of the identity: 
\begin{eqnarray}
\frac{(p_Ip)^{\beta}}{P_a^{\nu_a}}
&=& \frac{(p_Ip)^{\beta}}{P_a^{\nu_a}}\left[
\frac{p_I^2+p^2-(P_a+m_a^2)}{2~p_Ip}-\left( \dfrac{p_I^2+p^2-3(P_a+m_a^2)}{2~p_{I}p}\right)\delta_{p_Ip} \right]^{<\beta,\nu_a >},
\label{Substitution}
 \end{eqnarray}
where $p_I$ represents any internal momentum, $p$ represents any external ($q$) or internal ($k_{i}$) momentum and $P_{a}$ (with $a=1\ldots 5$) represents a denominator propagator that contains the momenta $p_{I}$ and $p$, moreover
$$
<\!\beta,\nu_a\!>={\rm min}\{\beta, \nu_a\}.
$$
For instance, if the numerator contains the scalar $(k_{1}q)^{\beta}$ then $p_{I}=k_{1}$, $p=q$, $\delta_{k_{1}q}=0$ and $a=3$ because $P_{3}=(k_{1}-q)^{2}-m_{3}^{2}$ is the propagator that contains the momentums $k_{1}$ and $q$. Therefore, we perform the substitution:
\begin{eqnarray*}
\frac{(k_1q)^{\beta}}{P_3^{\nu_3}}=\frac{ (k_1q)^{\beta}}{P_3^{\nu_3}}
\left(\frac{k_1^2+q^2-(P_3+m_3^2)}{2~k_1q}\right)
 ^{<\beta,\nu_3>}.
\end{eqnarray*}
After the repeated use of (\ref{Substitution}) into  (\ref{Non-Scalar}), that integral can be written as a superposition of scalar integrals, and non-scalar integrals with scalar products in the numerator. Nevertheless, the scalar products in the numerators remain only in the case when at least one internal line is eliminated ($\beta\rightarrow \beta-<\beta,\nu_a>$), take into account that in dimensional regularization $\int d^d p (p^2)^c=0$. If $k_{1}^{2}$ and $k_{2}^{2}$ remain in the numerator, the integral can be further simplified applying the relations 
\begin{eqnarray}
&&\frac{(k_1^2)^{\beta}}{P_3^{\nu_3}}
 =\frac{(k_1^2)^{\beta}}{P_3^{\nu_3}}
  ~\left(\frac{P_3+2k_1q-q^2+m_3^2}
     {k_1^2}\right)^{<\beta,\nu_3>},
\\
&& \nonumber \\
&&\frac{(k_2^2)^{\beta}}{P_4^{\nu_4}}
= \frac{(k_2^2)^{\beta}}{P_4^{\nu_4}}~
\left(\frac{P_4+2k_2q-q^2+m_4^2}{k_2^2}\right)^{<\beta, \nu_4>}
\end{eqnarray}
which still are particular cases of (\ref{Substitution}).
Moreover if the scalar product $(k_1k_2)^{\beta}$ remains in the numerator it is because the denominator $P_5$ has been cancelled i.e. the integral is the product of one-loop tensor integrals of the form:
\begin{eqnarray}
g_{\mu_{1}\nu_{1}}\dots g_{\mu_{\alpha}\nu_{\alpha}}\int d^dk_1 \dfrac{f_1(k_1,q,m_i)}{P_{1}^{\nu_{1}}P_{3}^{\nu_{3}}} k_{1_{\mu_{1}}}\dots k_{1_{\mu_{\alpha}}} \int d^dk_2 \dfrac{f_2(k_2,q,m_i)}{P_{2}^{\nu_{2}}P_{4}^{\nu_{4}}} k_{2_{\nu_{1}}}\dots k_{2_{\nu_{\alpha}}}.\nonumber \\
~ \label{k1k2}
\end{eqnarray}
In this case the substitution
\begin{eqnarray}
k_1k_2=k_{1\mu}\left(g_{\mu \nu}-\frac{q_{\mu}q_{\nu}}{q^2}\right)
k_{2\nu}+\frac{(k_1q)(k_2q)}{q^2}, \label{1lsum}
\end{eqnarray}
allows to transform the integral into a sum of products of one-loop scalar integrals \cite{PaVe}
\begin{eqnarray}
\sum_{n=0}^{\alpha}\dfrac{\alpha!}{n!(\alpha- n)!}\dfrac{1}{(q^{2})^{\alpha - n}}\int d^{d}k_{1} \dfrac{f_1(k_1,q,m_i)}{P_{1}^{\nu_{1}}P_{3}^{\nu_{3}}} (k_{1}q)^{\alpha - n}\int d^{d}k_{2} \dfrac{f_2(k_2,q,m_i)}{P_{2}^{\nu_{2}}P_{4}^{\nu_{4}}} (k_{2}q)^{\alpha - n} A^{n}(k_{1}, k_{2}), \nonumber \\
~~ \label{2lsum}
\end{eqnarray}
where $A(k_{1}, k_{2})$ comes from the transverse tensor in (\ref{1lsum})
\begin{eqnarray}
A(k_{1}, k_{2}) = k_{1\mu}\left(g_{\mu \nu}-\frac{q_{\mu}q_{\nu}}{q^2}\right)k_{2\nu}.
\end{eqnarray}
Integrals with odd powers of $A(k_{1}, k_{2})$ can be reduced to zero since the transverse tensor in $A(k_{1}, k_{2})$ will be always multiplied by an external momentum $q$ after going out of the integral. Integrals with even powers of $A(k_{1}, k_{2})$ can be completed decoupled using the relation 
\begin{eqnarray}
A^{2n}(k_{1}, k_{2}) = A^{n}(k_{1}, k_{1})A^{n}(k_{2}, k_{2}),
\end{eqnarray}
in the equation (\ref{2lsum}).

If in the numerator of (\ref{Non-Scalar}) the scalar products $(k_{1}q)^{\alpha}$ or $(k_{2}q)^{\beta}$ remain we can introduce the auxiliary scalar parameters $b_{1}$, $b_{2}$ of scaling mass $-2$ as
\begin{eqnarray}
(k_1q)^{r}(k_2q)^{s}=
\left( \frac{\partial}{(i\partial b_1)}\right)^{r}\left(\frac{\partial}{(i\partial b_2)}\right)^{s} 
\exp\left\{ i[b_1(k_1q)+b_2(k_2q)] \right\}
\left|_{b_i=0}\right. .
\label{Irs}
\end{eqnarray}
Following the same steps of the above section for the tensorial integral, we find the relation between the non-scalar and the scalar integral:
\begin{eqnarray}
\int \int \frac{d^dk_1 d^dk_2}{P_1^{\nu_1} P_2^{\nu_2} P_3^{\nu_3}P_4^{\nu_4} P_5^{\nu_5}}(k_1q)^{r}(k_2q)^{s} = T_{rs} (q,\{{\partial}_j\},{\bf d^+}) \int \int \frac{d^dk_1 d^dk_2}{P_1^{\nu_1} P_2^{\nu_2} P_3^{\nu_3} P_4^{\nu_4} P_5^{\nu_5}}
\label{form23}
\end{eqnarray}
where
\begin{eqnarray}
&&T_{r s}(q,\{{\partial}_j\},{\bf d^+})=
\frac{1}{i^{r+s}}
\left( \frac{\partial}{\partial b_1}\right)^{r}\left(\frac{\partial}{\partial b_2}\right)^{s}
\nonumber \\
&&~~~~~~~
 \times \exp\left\{i q^2 [
Q_{1}b_1+Q_{2}b_2
+Q_{11}b_1^2+Q_{22}b_2^2+Q_{12}b_1 b_2
      ]\rho\right\}
 \left|_{{b_i=0}
 \atop {\alpha_j=i\partial_j \atop
 \rho=-\frac{1}{\pi^2}  {\bf d^+}} }\right. ,
\label{Tbeta}
\end{eqnarray}
with $Q_i, Q_{ij}$  given in (\ref{qbeta}), compared with the operator in \ref{Ttensor} the operator $T_{r s}$ is not tensorial.

The evaluation of the scalar integrals in the form (\ref{form23}) will reduce any given integral with a generic numerator in terms of scalar integrals more efficiently than using of the tensorial reduction (\ref{tensorial-int}), but the scalar integrals have a different number of space-time dimensions, higher than $d$. For instance, if one considers the integral  (\ref{form23}) with $\nu_{1}=\nu_{4}=0$ and $r=s=1$, we have to set $\alpha_1=\alpha_4=0$ and thus obtain from~(\ref{Tbeta}) that: 
\begin{eqnarray}
T_{1 1}(q,\{{\partial}_j\},{\bf d^+})=
\frac{1}{i^{2}}
\left( \frac{\partial}{\partial b_1}\right)\left(\frac{\partial}{\partial b_2}\right)
~~~~~~~~~~~~~~~~~~~~~~~~~~~~~~~~~~~~~~~~~~~~~~~~~~\nonumber \\
 \times \exp\left\{i q^2 [
Q_{1}b_1+Q_{2}b_2
+Q_{12}b_1 b_2
      ]\rho\right\}
 \left|_{{b_i=0}
 \atop {\alpha_j=i\partial_j \atop
 \rho=-\frac{1}{\pi^2}  {\bf d^+}} }\right. .
\label{Tbeta11}
\end{eqnarray}
After applying the partial derivatives and evaluate at $b_{1}=b_{2}=0$, we obtain the operator
\begin{eqnarray}
T_{11}=\left. i\frac{q^2}{\pi^2}{\bf d^+} Q_{12}
+\frac{q^4}{\pi^4}({\bf d^+})^2 Q_{2}Q_{1}\right|_{\alpha_{j}=i\partial_{j}}  
\label{T11}
\end{eqnarray}
with
\begin{eqnarray*}
&& Q_{2} = \alpha _{3}\alpha _{5}, \\
&& Q_{1} = \alpha _{3}\alpha _{5} + \alpha _{2}\alpha_{3}, \\
&& Q_{12} = -\dfrac{\alpha_{5}}{2}.
\end{eqnarray*}
Therefore,
\begin{eqnarray*}
T_{11}=\frac{q^2}{2 \pi^2}{\bf d^+} \partial_5
+\frac{q^4}{\pi^4}({\bf d^+})^2\partial_3^2\partial_5
(\partial_2+\partial_5).
\end{eqnarray*}
With this operator is easily seen that:
\begin{eqnarray}
\label{exampleA}&&\int \!\! \int \!\!d^dk_1 d^dk_2
 \frac{(k_1q)(k_2q)}{P_2^{\nu_{2}}P_3^{\nu_{3}}P_5^{\nu_{5}}}\!= \frac{q^2}{2\pi^2} \nu_{5} \int \!\! \int \!\! \frac{d^{d+2}k_1 d^{d+2}k_2}{P_2^{\nu_{2}}P_3^{\nu_{3}}P_5^{\nu_{5}+1}}
  \\
&&\nonumber \\
&&~~~~~~~~~~~~~ \nonumber
\!
+\frac{q^4}{\pi^4}\int \!\! \int \!\! d^{d+4}k_1 d^{d+4}k_2\left[
\frac{\nu_{2}\nu_{5}\nu_{3}(\nu_{3}+1)}{P_2^{\nu_{2}+1}P_3^{\nu_{3}+2}P_5^{\nu_{5}+1}}
+\frac{\nu_{5}(\nu_{5}+1)\nu_{3}(\nu_{3}+1)}{P_2^{\nu_{2}}P_3^{\nu_{3}+2}P_5^{\nu_{5}+2}}            \right].
\end{eqnarray}
In this case, the resulting scalar integrals have dimensions $d+2$ and $d+4$. Nevertheless, is always possible reduce these integrals to ones in the generic dimension $d$ by using the appropriate recurrence relations, that will be given in the next section.

\subsection{\label{sc:RRandIBP} Scalar Integrals, Recurrence Relations and IBP's}

Once all irreducible numerators are eliminated and all integrals are expressed in terms of scalar integrals, without numerators and having different shifts of $d$, the next step is to repeatedly apply the appropriated recurrence relations that reduce the integrals to the generic dimension $d$ with the minimal exponent $\nu_i$ of the scalar propagators. The method of derivation of these recurrence relations was described by Tarasov in the reference \cite{Tarasov2}. One must proceed as follows. To obtain the recurrence relations, we start with the identity:
\begin{eqnarray}
\label{TarasovRecurrence}\prod_{i=1}^{L}\int d^{d}k_{i}\dfrac{\partial}{\partial k_{r}^{\mu}}\left[\left(\sum_{l}x_{l}\bar{k}_{l}^{\mu} \right)\prod_{j=1}^{N}P_{\bar{k}_{j},m_{j}}^{\nu_{j}}  \right] = 0, 
\end{eqnarray}
with
\begin{eqnarray}
P_{\bar{k}_{j},m_{j}}^{\nu_{j}}=\dfrac{1}{\bar{k}_{j}^{2}-m_{j}^{2}+i\varepsilon}, ~~~ \bar{k}_{j}^{\mu}=\sum_{n=1}^{L}\omega_{jn}k_{n}^{\mu} + \sum_{m=1}^{E}\eta_{jm}q_{m}^{\mu},
\label{PP}
\end{eqnarray}
where $x_{l}$ are arbitrary constants, $L$ is the number of the loops, $N$ is the number of propagators, $E$ is the number of external legs, $q_{m}$ are the external momenta, and $\omega$ and $\eta$ are matrices of incidences with the matrix elements being $\pm 1$ or $0$. 
The eq. (\ref{TarasovRecurrence}) is valid because in dimensional regularization the integral of a total derivative is zero \cite{Davydychev3}. After performing the differentiation one would usually express scalar products with integration momenta in terms of invariants that occur in the denominator of the propagators. In the present approach we write all or only some of the integrals containing scalar products involving loop momenta as a combination of integrals with changed~$d$ as for instance in (\ref{form23}) or just following the general strategy elucidated in section (\ref{sc:RRandIBP}). 
Choosing in a proper way the parameters $x_{l}$ in (\ref{TarasovRecurrence}) it is possible to find the most optimal set of relations for the reduction of the specific class of integrals to the set of some irreducible integrals which are also called "Master Integrals". This method is very similar to Integration By Parts method (IBP) \cite{Chetyrkin3} in which one imposes the relation
\begin{eqnarray}
\label{IBPrelation}\int \prod_{i}d^{d}k_{i}\dfrac{\partial f}{\partial k_{j}^{\mu}} = 0,
\end{eqnarray}
and writes down various equations for integrals of derivatives with respect to loop momenta and uses this set of relations between Feynman integrals in order to solve the reduction problem, i.e. to find out how a general Feynman integral of the given class can be expressed linearly in terms of some master integrals but without changing the dimension~$d$. In fact in the equation (\ref{IBPrelation}) $f$ depends on the loop and external momenta but, unlike equation (\ref{TarasovRecurrence}), $f$ is free of the external momenta ($q_{m}$) in his numerator. The Tarasov derivation is more general and, in fact, it includes the IBP method as a particular case just one needs to put $\eta_{jm}=0$ in (\ref{PP}). The IBP's recurrence relations correspond to some specific representation of scalar products in equation (\ref{TarasovRecurrence}).
\begin{figure}
\begin{center}
\scalebox{0.6}{
\fcolorbox{white}{white}{
  \begin{picture}(528,196) (147,-155)
    \SetWidth{1.0}
    \SetColor{Black}
    \Arc[arrow,arrowpos=0.5,arrowlength=5,arrowwidth=2,arrowinset=0.2](270,-49)(53.151,311,671)
    \Arc[arrow,arrowpos=0.5,arrowlength=5,arrowwidth=2,arrowinset=0.2](570,-49)(55,0,360)
    \Line[arrow,arrowpos=0.5,arrowlength=5,arrowwidth=2,arrowinset=0.2](568,4)(568,-104)
    \Text(490,-44)[lb]{\Black{$k_{2}$}}
    \Text(580,-50)[lb]{\Black{$k_{1}-k_{2}$}}
    \Text(640,-44)[lb]{\Black{$k_{1}$}}
    \Line[arrow,arrowpos=0.5,arrowlength=5,arrowwidth=2,arrowinset=0.2](324,-48)(216,-48)
    \Vertex(216,-48){4}
    \Vertex(324,-48){4}
    \Vertex(568,4){4}
    \Vertex(568,-104){4}
    \Text(270,15)[lb]{\Black{$k_{2}$}}
    \Text(260,-35)[lb]{\Black{$k_{1}-k_{2}$}}
    \Text(260,-90)[lb]{\Black{$k_{1}-q$}}
    \Line[arrow,arrowpos=0.5,arrowlength=5,arrowwidth=2,arrowinset=0.2](216,-48)(148,-48)
    \Line[arrow,arrowpos=0.5,arrowlength=5,arrowwidth=2,arrowinset=0.2](388,-48)(320,-48)
    \Text(156,-24)[lb]{\Black{$q$}}
    \Text(376,-24)[lb]{\Black{$q$}}
    \Text(264,-160)[lb]{\Large{\Black{$(a)$}}}
    \Text(564,-160)[lb]{\Large{\Black{$(b)$}}}
  \end{picture}
}}
\end{center}
\caption{\label{BubbleReduct}{\small (a) London Transport Topology $J_{111}(q^ 2)$. (b) Topology for a generic two-loops vacuum bubble $J_{111}(0)$.}}
\end{figure}
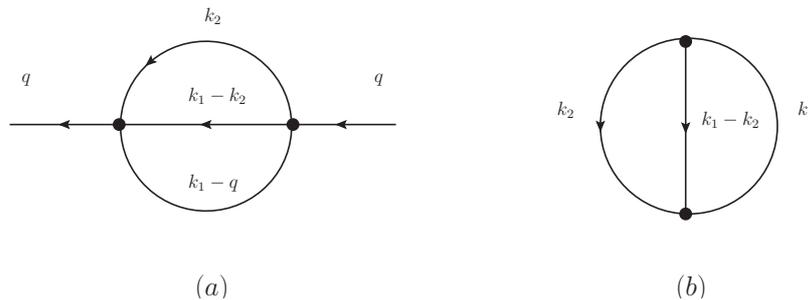

As an example  let us consider the two-loop integral in equation (\ref{exampleA}). In the particular case where $\nu_{2}=\nu_{3}=\nu_{5}=1$, this integral represents the two-loop self-energy given by the topology in figure (\ref{BubbleReduct}), known as the London Transport diagram or sunset or sunrise graph. By applying the operator $T_{11}$ of (\ref{T11}) on the London Transport diagram we got, from (\ref{exampleA}), a superposition of scalar integrals in ($d+2$) and ($d+4$) dimensions. Our task now is to express them in terms of scalar integrals in the original ($d$) dimensions. Just as an example we consider the dimensional reduction of 
\begin{eqnarray}
I^{(d+2)}_{\nu_{2}\nu_{3}\nu_{5+1}}(m_{1}^{2},m_{2}^{2},m_{5}^{2})=\int \!\! \int \!\! \frac{d^{d+2}k_1 d^{d+2}k_2}{P_2^{\nu_{2}}P_3^{\nu_{3}}P_5^{\nu_{5}+1}},
\label{ID}
\end{eqnarray}
to obtain a recurrence relation connecting integrals with different ($d$) using the prescription (\ref{TarasovRecurrence}). We start by the identity
\begin{eqnarray}
\int\int d^{d}k_{1}d^{d}k_{2} \dfrac{\partial}{\partial k_{2_{\mu}}} \left[ \dfrac{k_{2_{\mu}}}{P_{2}^{\nu_{2}}P_{3}^{\nu_{3}}P_{5}^{\nu_{5}}} \right] = 0. \label{identity}
\end{eqnarray}
From (\ref{identity}) we get a simple recurrence relation connecting integrals $I^{(d)}_{\nu_{2}\nu_{3}\nu_{5}}$ with different values of exponents $\nu_{i}$:
\begin{eqnarray}
\label{RR1} \nu_{5}\int\int d^{d}k_{1}d^{d}k_{2}\dfrac{2k_{1}k_{2}}{P_{2}^{\nu_{2}}P_{3}^{\nu_{3}}P_{5}^{\nu_{5}+1}}=
2\left(\nu_{2}-\dfrac{d}{2} \right)I^{(d)}_{\nu_{2}\nu_{3}\nu_{5}} + 2\nu_{2}m_{2}^{2}I^{(d)}_{\nu_{2}+1~\nu_{3}\nu_{5}}\\
 \nonumber \\
~~ + ~~ 2\nu_{5}I^{(d)}_{\nu_{2}-1~\nu_{3}~\nu_{5}+1} + 2\nu_{5}m_{5}^{2}I^{(d)}_{\nu_{2}\nu_{3}~\nu_{5}+1}. \nonumber 
\end{eqnarray}
The integral in the left hand side of equation (\ref{RR1}) can be reduced using equality:
\begin{eqnarray*}
2k_{1}k_{2}= P_{5} - P_{2} - P_{3} + q^{2} -2qk_{1} + m_{5}^{2}-m_{2}^{2}+m_{3}^{2},
\end{eqnarray*}
by replacing into (\ref{RR1}) and keeping the scalar product ($qk_{1}$) untouched we get 
\begin{eqnarray}
\label{RecReltwo} 2\nu_{5}\int\int d^{d}k_{1}d^{d}k_{2}\dfrac{qk_{1}}{P_{2}^{\nu_{2}}P_{3}^{\nu_{3}}P_{5}^{\nu_{5}+1}}~+~ 
\nu_{5}I^{(d)}_{\nu_{2}~\nu_{3}-1~\nu_{5}+1} ~+~ \left(2\nu_{2}-\nu_{5}-d\right)I^{(d)}_{\nu_{2}\nu_{3}\nu_{5}} \\ \nonumber
\\ \nonumber
~~+~~ 2\nu_{2}m_{2}^{2}I^{(d)}_{\nu_{2}+1~\nu_{3}\nu_{5}}
~ + ~ 3\nu_{5}I^{(d)}_{\nu_{2}-1~\nu_{3}~\nu_{5}+1} ~+~ \nu_{5}(m_{5}^{2}-m_{3}^{2}+m_{2}^{2}-q^{2})I^{(d)}_{\nu_{2}\nu_{3}~\nu_{5}+1} = 0. \nonumber
\end{eqnarray}
The first integral in equation (\ref{RecReltwo}) can be expressed in terms of integrals with shifted space-time dimension, by using  the formula (\ref{form23}) with $\alpha_{1}=\alpha_{4}=0$ and
\begin{eqnarray}
T_{10}=\dfrac{1}{i}\left(\dfrac{\partial}{\partial b_{1}} \right) exp \left\lbrace iq^{2}[ Q_{1}b_{1} + Q_{11}b_{1}^{2}]\rho \right\rbrace \left|_{{b_1=0}
 \atop {\alpha_j=i\partial_j \atop
 \rho=-\frac{1}{\pi^2}  {\bf d^+}} }\right. ,
\end{eqnarray}
where $Q_{1}=\alpha_{3}(\alpha_{2}+\alpha_{5})$  and $-4Q_{11}=(\alpha_{2}+\alpha_{5})$, therefore 
\begin{eqnarray}
\label{IntIrrOne}\int\int d^{d}k_{1}d^{d}k_{2}\dfrac{(qk_{1})}{P_{2}^{\nu_{2}}P_{3}^{\nu_{3}}P_{5}^{\nu_{5}+1}} = \dfrac{q^{2}}{\pi^{2}}\textbf{d}^{+}\partial_{3}(\partial_{2}+\partial_{5})\int\int \dfrac{d^{d}k_{1}d^{d}k_{2}}{P_{2}^{\nu_{2}}P_{3}^{\nu_{3}}P_{5}^{\nu_{5}+1}} ~~~~~~~ \\ \nonumber
~~~~~~~~~~ \\ \nonumber 
\\ \nonumber 
~~=~~\dfrac{q^{2}}{\pi^{2}}\left[ \nu_{2}\nu_{3}\int\int \dfrac{d^{d+2}k_{1}d^{d+2}k_{2}}{P_{2}^{\nu_{2}+1}P_{3}^{\nu_{3}+1}P_{5}^{\nu_{5}+1}}+(\nu_{5}+1)\nu_{3}\int\int \dfrac{d^{d+2}k_{1}d^{d+2}k_{2}}{P_{2}^{\nu_{2}}P_{3}^{\nu_{3}+1}P_{5}^{\nu_{5}+2}} \right]. 
\end{eqnarray}
We can find another relation between integrals in different dimensions $(d)$ introducing the polynomial differential operator $D(i\partial_{j})$, which is obtained from $D(\alpha)$ by~substituting~$\alpha_{j}\rightarrow i\partial_{j}$. The application of the operator $D(i\partial)$ on the integral (\ref{represG1}) is proportional to the same integral but in $d-2$ dimensions:
\begin{eqnarray}
D(i\partial)G^{(d)}(q^{2},a_{1},a_{2})=-\pi^{2}G^{(d-2)}(q^{2},a_{1},a_{2}).
\end{eqnarray}
The function $D(\alpha)$ for the integral (\ref{exampleA}) is $D(\alpha)=\alpha_{5}(\alpha_{2}+\alpha_{3})+\alpha_{3}\alpha_{2}$, and hence
\begin{eqnarray}
I^{(d-2)}_{\nu_{2}\nu_{3}\nu_{5}}=-\dfrac{1}{\pi^{2}}\left[i\partial_{5}(i\partial_{2}+i\partial_{3}) + i\partial_{3}i\partial_{2} \right] I^{(d)}_{\nu_{2}\nu_{3}\nu_{5}}.
\end{eqnarray}
where $ I^{(d)}_{\nu_{2}\nu_{3}\nu_{5}}$ is defined in (\ref{ID}).
From this relation we obtain a new recurrence relation in terms of the integral $I^{(d)}$:
\begin{eqnarray}
I^{(d-2)}_{\nu_{2}\nu_{3}\nu_{5}}=\dfrac{1}{\pi^{2}}\nu_{2}\nu_{5}I^{(d)}_{\nu_{2}+1~\nu_{3}~\nu_{5}+1}+\dfrac{1}{\pi^{2}}\nu_{3}\nu_{5}I^{(d)}_{\nu_{2}~\nu_{3}+1~\nu_{5}+1}+\dfrac{1}{\pi^{2}}\nu_{2}\nu_{3}I^{(d)}_{\nu_{2}+1~\nu_{3}+1~\nu_{5}}. \label{ReccRel3}
\end{eqnarray}
Making the change $\nu_{5}\rightarrow \nu_{5}+1$ and $d\rightarrow d+2$ we can put the recurrence relation (\ref{ReccRel3}) in the form:
{\small
\begin{eqnarray*}
I^{(d)}_{\nu_{2}\nu_{3}~\nu_{5}+1}-\dfrac{1}{\pi^{2}}\nu_{2}(\nu_{5}+1)I^{(d+2)}_{\nu_{2}+1~\nu_{3}~\nu_{5}+2}= \dfrac{1}{\pi^{2}}\left[\nu_{3}(\nu_{5}+1)I^{(d+2)}_{\nu_{2}~\nu_{3}+1~\nu_{5}+2}+\nu_{2}\nu_{3}I^{(d+2)}_{\nu_{2}+1~\nu_{3}+1~\nu_{5}+1}
\right]. 
\end{eqnarray*}}
Inserting the above expression into (\ref{IntIrrOne})
\begin{eqnarray}
\int\int d^{d}k_{1}d^{d}k_{2}\dfrac{qk_{1}}{P_{2}^{\nu_{2}}P_{3}^{\nu_{3}}P_{5}^{\nu_{5}+1}} = q^{2}\left(I^{(d)}_{\nu_{2}\nu_{3}~\nu_{5}+1}-\dfrac{1}{\pi^{2}}\nu_{2}(\nu_{5}+1)I^{(d+2)}_{\nu_{2}+1~\nu_{3}~\nu_{5}+2} \right). \label{ReccRel5}
\end{eqnarray}  
Finally, using the identity (\ref{ReccRel5}) with $\nu_{5}$ changed to $\nu_{5}-1$ and $\nu_{2}$ changed to $\nu_{2}-1$ in the equation (\ref{RecReltwo}),  we can write:
\begin{eqnarray}
&& 2\nu_{5}(\nu_{5}-1)(\nu_{2}-1)\dfrac{q^{2}}{\pi^{2}}\int\int \dfrac{d^{d+2}k_{1}d^{d+2}k_{2}}{P_{2}^{\nu_{2}}P_{3}^{\nu_{3}}P_{5}^{\nu_{5}+1}} = (\nu_{5}-1)(m_{5}^{2}-m_{3}^{2}+m_{2}^{2}+q^{2})I^{(d)}_{\nu_{2}-1~\nu_{3}~\nu_{5}} \nonumber \\  &&
~+~ (\nu_{5}-1)I^{(d)}_{\nu_{2}-1~\nu_{3}-1~\nu_{5}} ~~+~~ \left(2\nu_{2}-\nu_{5}-d-1 \right)I^{(d)}_{\nu_{2}-1~\nu_{3}~\nu_{5}-1} \nonumber \\ && \nonumber
\\ &&  ~~+~~ 2(\nu_{2}-1)m_{2}^{2}I^{(d)}_{\nu_{2}~\nu_{3}~\nu_{5}-1} ~~+~~ 3(\nu_{5}-1)I^{(d)}_{\nu_{2}-2~\nu_{3}~\nu_{5}}. 
\end{eqnarray}
Thus, the first integral in the right side of the equation (\ref{exampleA}) was reduced to a superposition of scalar integrals in ($d$) dimensions with coefficients that are functions of $q^2, \{m_i^2\}$ and $d$. Proceeding in a similar way with all two-points two-loops diagrams with arbitrary masses, Tarasov found, from the application of recurrence relations, a basis composed of 30 scalar integrals in $d$ dimensions
\begin{eqnarray}
I(q^2)=\sum_{j=1}^{30} R_j(q^2,\{m_i^2\},d) I_j^{(d)}(q^2), \label{Base}
\end{eqnarray}
with  $R_j(q^2,\{m_i^2\},d)$ being rational functions of $q^2, \{m_i^2\}$ and $d$ \cite{Tarasov1}. The two-loop integrals of the basis are expressed 
in terms of the following three two-loop integrals {\small$F^{(d)}_{11111}$}, {\small$V^{(d)}_{1111}$} and {\small$J^{(d)}_{111}(q^2)$}:
\begin{eqnarray}
&&  F^{(d)}_{\nu_1 \nu_2 \nu_3 \nu_4 \nu_5}=
\frac{1}{\pi^d}
\int\!\int
\frac{d^d k_1 d^dk_2}{[k_1^2-m_1^2]^{\nu_1}[k_2^2-m_2^2]^{\nu_2}}
\nonumber \\
&&~~~~~~~~~\times \frac{1}
{[(k_1-q)^2-m_3^2]^{\nu_3} [(k_2-q)^2-m_4^2]^{\nu_4}
[(k_1-k_2)^2-m_5^2]^{\nu_5}},
\nonumber \\
&& \nonumber \\
&&  V^{(d)}_{\nu_1 \nu_2 \nu_3 \nu_4}=
\frac{1}{\pi^d}
\int\!\! \int \frac{d^d k_1 d^dk_2}{[(k_1-k_2)^2-m_1^2]^{\nu_1}}
\nonumber \\
&&~~~~~~~~~\times\frac{1}{[k_2^2-m_2^2]^{\nu_2}
[(k_1-q)^2-m_3^2]^{\nu_3} [(k_2-q)^2-m_4^2]^{\nu_4}},
\nonumber \\
&&\nonumber \\
&&  J^{(d)}_{\nu_1 \nu_2 \nu_3} =
\frac{1}{\pi^d}
\int\!\! \int \frac{d^d k_1 d^dk_2}
{[k_1^2-m_1^2]^{\nu_1}[(k_1-k_2)^2-m_2^2]^{\nu_2}
[(k_2-q)^2-m_3^2]^{\nu_3}}. \label{JTopology}
\end{eqnarray}
All master integrals in (\ref{Base}) can be obtained from the previous ones by changing masses, external momenta or by differentiating. For instance, the two-loop bubble integrals represented by figure \ref{BubbleReduct} (b) are just the value of $J^{(d)}_{111}(q^2)$, represented by figure \ref{BubbleReduct} (a), at $q^2=0$. The application of recurrence relations to $F^{(d)}$ will produce $V^{(d)}$, $J^{(d)}$ and more simple one-loop integrals. In turn integrals $V^{(d)}$ will produce $J^{(d)}$ plus one-loop integrals. The one-loop integrals obtained will be denoted as
\begin{eqnarray}
&&B^{(d)}_{\nu_1 \nu_2 }=\frac{1}{\pi^{\frac{d}{2}} }
\int \frac{d^d k_1}
{[k_1^2-m_1^2]^{\nu_1}[(k_1-q)^2-m_2^2]^{\nu_2}}, \label{BTopology} \\
&& A^{(d)}_{\nu_1  }= \frac{1}{\pi^{\frac{d}{2}}}
\int \frac{d^d k_1} {[k_1^2-m_1^2]^{\nu_1}}. \label{ATopology}
\end{eqnarray}
Therefore, one should first apply the recurrence relations  to $F^{(d)}$, then to $V^{(d)}$  and finally to $J^{(d)}$. An example of how from the reduction of two-loop integrals, using recurrence relations, one obtains one-loop integrals is showed in Appendix \ref{AppIntegralJ}. The number of basic structures obtained after reduction strongly depends on the mass values. If some masses are equal to zero or there are equal masses then the number of basic structures substantially diminishes. ( For example, in the case of QED the number of relevant two-loop basic integrals for the photon propagator is 2). The evaluation of basis integrals $A^{(d)},~B^{(d)},~F^{(d)},~V^{(d)}$ and $J^{(d)}$ is a separate problem that will be studied in the next section.

Since it is not easy to extract a set of recurrence relations that reduce the complexity of the integrals at each step such that one finally arrives at only a small set of basic integrals, due to a vast number of interrelations between the integrals considered, we implemented the recurrence relations for reduce two-loop integrals using the Mathematica code TARCER \cite{Tarcer} that is part of the FEYNCALC package \cite{FeynCalc}. TARCER reduces two-loop propagator integrals with arbitrary masses to simpler basis integrals using the reduction algorithm proposed by Tarasov. For the reduction of scalar integrals TARCER contains the complete set of recurrence relations given in~\cite{Tarasov1} and some additions for particular parameter configurations. In some cases the number of basic integrals is reduced. 

The implementation and an example of these type of integrals is given in Appendix \ref{AppTarcer}. We also used the Mathematica package FEYNARTS \cite{FeynArts} for the generation and visualization of Feynman diagrams and amplitudes, its implementation and some examples are given in Appendix \ref{AppFeynArts}. 

\subsection{Master Integrals \label{sec:MI}}

We saw in the previous section that any non-scalar two-loop integral can be reduced to a superposition of scalar master integrals \footnote{The fact that in principle all integrals considered must be expressible in terms of a finite set of basic integrals may be formally seen in \cite{Lascoux}}, and that these master integrals can be actually computed in terms of only five integrals, the three two-loop integrals defined in (\ref{JTopology}) and the two one-loop integrals in equations (\ref{ATopology}) and (\ref{BTopology}). \\
The one-loop integrals are the scalar one-point and two-point Passarino-Veltman functions. The Laurent expansion around~$d~=~4$ of these functions include the divergent, finite and evanescent terms. The evanescent terms are the $\varepsilon$ -order terms to be included in  the two-loop calculations because we can have contributions to the finite part made of the product of the divergent and of the evanescent part  like for instance in diagrams in figure \ref{BubblesTop} (a). Those contributions are in fact very important in an on-shell renormalization scheme like the one we are working. That part is not present at two loop in the {\it MS} or $\overline{MS}$ scheme \cite{kleinert}. 
The Laurent expansion of $A_{1}^{(d)}$ is easily obtained but special treatments are required for $B_{11}^{(d)}$ \cite{Davydychev1}. Being $d=4-2\varepsilon$ we have
\begin{eqnarray}
A_{1}^{(d)} ~=~ i(m_{H})^{2}\Gamma(\varepsilon)\left(1+\dfrac{\varepsilon}{2}\right)\left(m_{H}^{2}\right)^{-\varepsilon} ~~~~~~~~~~~~~~~~~ \nonumber \\
\nonumber \\
~ = ~ m_{H}^{2}\left(m_{H}^{2}e^{\gamma_{_{E}}}\right)^{-\varepsilon}\left(\varepsilon\left(\dfrac{1}{12}i\pi^2 + i \right)+ \dfrac{i}{\varepsilon} + i\right).\label{LaurentA}
\end{eqnarray}  
Moreover, from Appendix \ref{App1lEpsExpansion} we get:
\begin{eqnarray}
&&B_{11}^{(d)} ~=~ \Gamma(\varepsilon)\int_{0}^{1}dx\left[ q^{2}x^{2}+ (-q^{2}+m_{2}^{2}-m_{1}^{2})x + m_{1}^{2} \right]^{-\varepsilon} \nonumber \\
&&\nonumber \\
&& ~~~~~~~ = ~ i\Gamma(\varepsilon)
\frac{m_0^{1-2\varepsilon}}{\sqrt{q^{2}}} \; 
\left\{ \Omega_1^{(2; 4-2\varepsilon)} 
+ \Omega_2^{(2; 4-2\varepsilon)} \right\} 
\end{eqnarray}
with
\begin{eqnarray}
\label{Omega_i}
\Omega_i^{(2; 4-2\varepsilon)} =
\int\limits_0^{\tau_{0i}} 
\frac{\mbox{d}\theta}{\cos^{2-2\varepsilon}\theta} .
\end{eqnarray}
Here it is assumed that $(m_1-m_2)^2\leq q^2 \leq (m_1+m_2)^2$. In equation (\ref{Omega_i}) the following notation was used \cite{Davydychev2}:
\begin{eqnarray}
\label{two-point}
\cos\tau_{12} = \frac{m_1^2+m_2^2-q^2}{2m_1 m_2}, \hspace{5mm}
m_0 = \frac{m_1 m_2 \sin\tau_{12}}{\sqrt{q^2}}, 
\hspace{5mm}
\cos\tau_{0i} = \frac{m_0}{m_i} \; .
\end{eqnarray}
Now, expanding (\ref{Omega_i}) in $\varepsilon$, we obtain
\begin{eqnarray}
\label{ep-exp}
\int_0^{\tau}
\frac{\mbox{d}\theta}{\cos^{2-2\varepsilon}\theta}
= \sum\limits_{j=0}^{\infty} \frac{(2\varepsilon)^j}{j!}
\int_0^{\tau}
\frac{\mbox{d}\theta}{\cos^2\theta}
\ln^j(\cos\theta) 
= \sum\limits_{j=0}^{\infty} \frac{(-2\varepsilon)^j}{j!}
f_j(\tau),
\end{eqnarray}
where
\begin{eqnarray}
\label{f_j}
f_j(\tau) \equiv (-1)^j
\int_0^{\tau}
\frac{\mbox{d}\theta}{\cos^2\theta}
\ln^j(\cos\theta) .
\end{eqnarray}
The lowest terms of the expansion are \cite{Davydychev2}
\begin{eqnarray}
f_0(\tau) &=& \tan\tau, \\
\label{f_1}
f_1(\tau) &=& -\tan\tau \ln(\cos\tau)-\tan\tau +\tau, \\
\label{f_2}
f_2(\tau) &=& 
\tan\tau \left[\ln^2(\cos\tau)\!+\!2\ln(\cos\tau)\!+\!2 \right]
-2\tau (1\!-\!\ln 2) - \mbox{Cl}_2\left(\pi\!-\!2\tau\right) ,
\end{eqnarray}
where we use that $\tau_{01}+\tau_{02} = \tau_{12}$ and
\begin{eqnarray}
\label{tan(tau0i)}
\tan\tau_{01}=\frac{m_1^2-m_2^2+q^2}{2m_0\sqrt{q^2}}, 
\hspace{5mm}
\tan\tau_{02}=\frac{m_2^2-m_1^2+q^2}{2m_0\sqrt{q^2}}.
\end{eqnarray}
The Clausen function of order 2\,\, $Cl_2(x)$ was defined in Appendix \ref{App1lEpsExpansion} eq. (\ref{Clausen}).

Taking into account the equations (\ref{f_1}) and (\ref{f_2}) the representation (\ref{Omega_i}) makes it possible to construct the well-known result \cite{NMB}
\begin{eqnarray}
B_{11}^{(4-2\varepsilon)} ~=~ e^{-\varepsilon \gamma_{_{E}}} \left( \dfrac{1}{\varepsilon}+B^{(fin)}+\varepsilon B^{(\varepsilon)} \right),
\end{eqnarray}
with
\begin{eqnarray}
B^{(fin)} = - ln(q^{2}) - \sum_{j=1}^{2}\left[ln(1-x_{j}) - x_{j}ln\dfrac{x_{j}-1}{x_{j}} - 1 \right],
\end{eqnarray}
and
\begin{eqnarray}
&& B^{(\varepsilon)} ~=~ \dfrac{\pi^{2}}{12} + \dfrac{1}{2}ln^{2}\left(q^{2} \right) - \left[ln\left(q^{2} \right) - 2 \right]\left[ B_{fin} + ln\left(q^{2} \right) \right] \nonumber \\
&& \nonumber \\
&& ~ +~ \sum_{j=1}^{2} \left\lbrace \dfrac{1}{2}(1-x_{j})ln^{2}(1-x_{j}) + \dfrac{1}{2}x_{j}ln^{2}(-x_{j})\right\rbrace + (1-x_{1})ln(1-x_{1})ln(1-x_{2})  \nonumber \\
&& \nonumber \\
&& ~+~ x_{1}ln(-x_{1})ln(-x_{2}) + (x_{1}-x_{2})\left[Li_{2}\left( \dfrac{x_{2}}{x_{2}-x_{1}}\right) - Li_{2}\left(\dfrac{x_{2}-1}{x_{2}-x_{1}} \right) \right. \nonumber \\
&& \nonumber \\
&& \left. + ln(x_{2}-x_{1})ln\left( \dfrac{x_{2}-1}{x_{2}}\right) \right], 
\end{eqnarray}
where $x_{1}$ and $x_{2}$ are defined by the relation 
\begin{eqnarray}
q^{2}x^{2}+ (-q^{2}+m_{2}^{2}-m_{1}^{2})x + m_{1}^{2} = q^{2}(x-x_{1})(x-x_{2}).
\end{eqnarray}
The corresponding result was obtained in Appendix (\ref{App1lEpsExpansion}). These expressions are valid for arbitrary masses and momenta. The dilogarithm function or Spence integral is defined as
\begin{equation}
Li_2(x)=-\int_0^x\frac{ln (1-t)}{t}dt.
\end{equation} 
We need now to compute the integrals in the equation (\ref{JTopology}). For arbitrary masses and external momenta only his divergent part is known analytically, for instance $V^{(d)}_{1111}$ is finite in four dimensions. The finite part are analytical only for particular values of masses and momenta. For the integral with three propagators $J^{(d)}_{111}(q^{2})$ the solution are related to the Lauricella functions \cite{BBBS}. Integrals with four propagators $V^{(d)}_{1111}$ can be written in terms of hypergeometric series \cite{BBBBW} and the integral with five propagators $F^{(d)}_{11111}$ have a representation for a general mass case given in \cite{BB} but its evaluation in terms of a similar series as for the other two integrals is not yet known. 

For the purposes of this thesis of computing the two-loop Higgs tadpoles as first derivative of the two-loop effective potential we specifically need only the two-loop integrals $V^{(d)}_{1111}(q=0)$ and $J^{(d)}_{111}(q=0)$ at vanishing external momentum. In this particular case the basis integrals can be rewrite in terms of the one-loop integrals computed above and the two-loop bubble integral $J^{(d)}_{111}(q=0)$ if we use the next relation between $V^{(d)}_{1111}(q=0)$ and $J^{(d)}_{111}(q=0)$:
\begin{eqnarray}
V^{(d)}_{1111}(q=0)= \dfrac{1}{m_{1}^{2}-m_{2}^{2}}\left( J_{111}^{(d)}(q=0, m_{1}^{2}, m_{3}^{2}, m_{4}^{2}) - J_{111}^{(d)}(q=0, m_{2}^{2}, m_{3}^{2}, m_{4}^{2}) \right) 
\end{eqnarray}
with
\begin{eqnarray}
J_{111}^{(d)}(q=0,  m_{1}^{2}, m_{2}^{2}, m_{3}^{2})=\int\int d^dk_{1}\,d^dk_{2}\,\left[{1\over{(k_{1}^2-m_{1}^{2})(k_{2}^2-m_{2}^{2})((k_{1}+k_{2})^2-m_{3}^{2})}}\right].
\label{Jaq2=0}
\end{eqnarray}
The finite parts of these integrals can be computed analytically by the standard methods shown in the Appendix \ref{App1lEpsExpansion} and Appendix \ref{AppIntegralJ}. 
In the next section we will adopt the methods exposed in this section to compute at two-loop the full effective potential in the $\overline{MS}$ and in the Sirlin-Zucchini's scheme.
All analytic results obtained here can be verified using the TARCER code that implement the Tarasov method.  

\section{The Two-Loop $\overline{MS}$ Effective Potential of SM}

\begin{figure}
\begin{center}
\scalebox{0.5}{
\fcolorbox{white}{white}{
  \begin{picture}(800,200)(0,0)
    \SetWidth{1.0}
    \SetColor{Black}
    \Arc[arrow,arrowpos=0.5,arrowlength=5,arrowwidth=2,arrowinset=0.2](256,97)(48,180,540)
    \Arc[arrow,arrowpos=0.5,arrowlength=5,arrowwidth=2,arrowinset=0.2](352,97)(48,0,360)

    \Arc[arrow,arrowpos=0.5,arrowlength=5,arrowwidth=2,arrowinset=0.2](560,99)(60,270,630)
    \GBox(561,40)(561,157){0.882}
    \Vertex(560,159){5}
    \Vertex(560,39){5}
    \Vertex(305,99){5}
    \Text(294,-5)[lb]{\Large{\Black{$(a)$}}}
    \Text(560,-5)[lb]{\Large{\Black{$(b)$}}}
  \end{picture}
}}
\end{center}
\caption{\label{BubblesTop}{\small 1PI Topologies contributing to two-loop effective potential}}
\end{figure}

In this section we calculate the two-loops effective potential using minimal subtraction, and we will use this result to obtain in the next chapter the potential in the Sirlin-Zucchini renormalization scheme. We split the two-loop potential in different sectors according to their diagrammatic origin. We started studying the scalar sector of $\overline{MS}$ effective potential. This sector contains the main ingredients to construct the full two-loop potential. In all other sectors we can reduce, by elementary manipulation, each individual Feynman diagram to a sum of scalar integrals of the form (\ref{JTopology}) or (\ref{ATopology}) with $\nu_{1}=\nu_{2}=\nu_{3}=1$ and $q^{2}\rightarrow 0$, which appear in scalar sector. 

There are only two types of 1PI topologies that contribute to the effective potential, as shown in figure \ref{BubblesTop}. When we insert the scalar fields to the topologies, arise nine different types of Feynman diagrams. Everyone of them can be expressed, after dimensional regularization, in terms of some of integrals: 
\begin{eqnarray}
 \label{IntegralI} I(m_{1}^{2},m_{2}^{2},m_{3}^{2})= \dfrac{\mu^{2(4-d)}}{(4\pi)^{d}}J_{111}^{(d)}(q^{2}=0) ~~~~~~~~~~~~~~~~~~~~~~~~~~~~~~~~~~~~\\
 \nonumber \\
 =  {(\mu^2)^{4-d}\over{(2\pi)^{2d}}}\int\int{d^dk_{1}\,d^dk_{2}\over
{(k_{1}^2-m_{1}^{2})(k_{2}^2-m_{2}^{2})((k_{1}+k_{2})^2-m_{3}^{2})}} \nonumber
\end{eqnarray}
or
\begin{eqnarray}
J(m_{1}^{2},m_{2}^{2})=J(m_{1}^{2})J(m_{2}^{2}) \label{IntegralJ} ~~;~~~~~~~~~~~~~~~~~~~~~~~~~~~~~~~~~~~~~~~~~~~~~~~~~~~~ \\
\nonumber  \\ 
J(m^{2})= \dfrac{\mu^{4-d}}{(4\pi)^{d/2}}A_{1}^{(d)}~=~{(\mu)^{4-d}\over{(2\pi)^d}}
\int{d^dk\over{k^2-m^{2}}}={(\mu)^{4-d}\over
{(4\pi)^{d\over2}}}i\Gamma(1-{d\over2})(m^{2})^{{d\over2}-1}. \nonumber
\end{eqnarray}
The evaluation of integral (\ref{IntegralI}) is non-trivial, and will be the subject of Appendix \ref{AppIntegralJ}. By other side, the expansion (\ref{IntegralJ}) does not include only the divergent and the finite parts, but includes also the first evanescent term, proportional to $(d-4)$, since we are dealing with products of two one-loop integrals, therefore, we could have products of evanescent and divergent parts that contribute to the finite part. \\
Let see the explicit computation of two-loop scalar effective potential. In terms of integrals (\ref{IntegralI}) and (\ref{IntegralJ}) we have the contributions of the nine diagrams written in a general covariant gauge:
\begin{center}
\scalebox{0.6}{
\fcolorbox{white}{white}{
  \begin{picture}(740,68) (200,-140)
    \SetWidth{1.0}
    \SetColor{Black}
    \Arc[dash,dashsize=10,arrow,arrowpos=0.5,arrowlength=5,arrowwidth=2,arrowinset=0.2](234,-106)(30,270,630)
    \Line[dash,dashsize=10](234,-76)(234,-136)
    \Vertex(234,-76){3}
    \Vertex(234,-136){3}
    \Text(183,-103)[lb]{\Black{$H$}}
    \Text(276,-103)[lb]{\Black{$H$}}
    \Text(243,-103)[lb]{\Black{$H$}}
    \Text(327,-112)[lb]{\Large{\Black{$~=~ -~3\lambda^{2}\phi_{c}^{2}I(m_{H}^{2},m_{H}^{2},m_{H}^{2})$}}}
  \end{picture}
}}

\scalebox{0.6}{
\fcolorbox{white}{white}{
  \begin{picture}(780,68) (189,-140)
    \SetWidth{1.0}
    \SetColor{Black}
    \Arc[dash,dashsize=10](234,-106)(30,270,630)
    \Line[dash,dashsize=10](234,-76)(234,-136)
    \Vertex(234,-76){3}
    \Vertex(234,-136){3}
    \Line[dash,dashsize=10,arrow,arrowpos=0.5,arrowlength=5,arrowwidth=2,arrowinset=0.2](372,-82)(372,-142)
    \Vertex(372,-136){3}
    \Vertex(372,-76){3}
    \Arc[dash,dashsize=10,arrow,arrowpos=0.2,arrowlength=5,arrowwidth=2,arrowinset=0.2](372,-106)(30,270,630)
    \Text(300,-113)[lb]{\Large{\Black{$+$}}}
    \Text(186,-103)[lb]{\Black{$H$}}
     \Text(324,-103)[lb]{\Black{$H$}}
    \Text(243,-103)[lb]{\Black{$G$}}
    \Text(267,-103)[lb]{\Black{$G$}}
    \Text(381,-103)[lb]{\Black{$G^{\pm}$}}
    \Text(408,-103)[lb]{\Black{$G^{\pm}$}}
    \Text(453,-112)[lb]{\Large{\Black{$~=~ -3\lambda^{2}\phi_{c}^{2}I(m_{H}^{2},m_{G}^{2},m_{G}^{2}) $}}}
  \end{picture}
}}

\scalebox{0.6}{
\fcolorbox{white}{white}{
  \begin{picture}(720,53) (173,-149)
    \SetWidth{1.0}
    \SetColor{Black}
    \Arc[dash,dashsize=10](198,-121)(24,180,540)
    \Arc[dash,dashsize=10](246,-121)(24,0,360)
    \Vertex(222,-121){3}
    \Text(160,-154)[lb]{\Black{$H$}}
    \Text(270,-154)[lb]{\Black{$H$}}
    \Text(321,-140)[lb]{\Large{\Black{$~=~ \dfrac{3}{4}\lambda J(m_{H}^{2},m_{H}^{2})$}}}
  \end{picture}
}}

\scalebox{0.6}{
\fcolorbox{white}{white}{
  \begin{picture}(720,56) (173,-149)
    \SetWidth{1.0}
    \SetColor{Black}
    \Arc[dash,dashsize=10](198,-118)(24,180,540)
    \Arc[dash,dashsize=5](246,-118)(24,0,360)
    \Vertex(222,-118){3}
    \Text(183,-154)[lb]{\Black{$H$}}
    \Text(255,-154)[lb]{\Black{$G$}}
    \Text(317,-125)[lb]{\Large{\Black{$+$}}}
    \Arc[dash,dashsize=10](397,-118)(24,180,540)
    \Arc[dash,dashsize=5,arrow,arrowpos=0.2,arrowlength=5,arrowwidth=2,arrowinset=0.2](445,-118)(24,0,360)
    \Vertex(421,-118){3}
    \Text(373,-154)[lb]{\Black{$H$}}
    \Text(457,-154)[lb]{\Black{$G^{\pm}$}}
    \Text(499,-140)[lb]{\Large{\Black{$ ~=~ \dfrac{3}{2}\lambda J(m_{H}^{2},m_{G}^{2})$}}}
  \end{picture}
}}

\scalebox{0.6}{
\fcolorbox{white}{white}{
  \begin{picture}(720,62) (173,-149)
    \SetWidth{1.0}
    \SetColor{Black}
    \Arc[dash,dashsize=5](198,-112)(24,180,540)
    \Arc[dash,dashsize=5](246,-112)(24,0,360)
    \Vertex(222,-112){3}
    \Arc[dash,dashsize=5](396,-112)(24,180,540)
    \Arc[dash,dashsize=5,arrow,arrowpos=0.2,arrowlength=5,arrowwidth=2,arrowinset=0.2](444,-112)(24,0,360)
    \Vertex(420,-112){3}
    \Arc[dash,dashsize=5,arrow,arrowpos=0.3,arrowlength=5,arrowwidth=2,arrowinset=0.2](582,-112)(24,0,360)
    \Arc[dash,dashsize=5,arrow,arrowpos=0.7,arrowlength=5,arrowwidth=2,arrowinset=0.2](630,-112)(24,180,540)
    \Text(174,-154)[lb]{\Black{$G$}}
    \Text(246,-154)[lb]{\Black{$G$}}
    \Text(378,-154)[lb]{\Black{$G$}}
    \Text(444,-154)[lb]{\Black{$G^{\pm}$}}
    \Text(570,-154)[lb]{\Black{$G^{\pm}$}}
    \Text(630,-154)[lb]{\Black{$G^{\pm}$}}
    \Vertex(606,-112){3}
    \Text(318,-112)[lb]{\Large{\Black{$+$}}}
    \Text(516,-115)[lb]{\Large{\Black{$+$}}}
    \Text(693,-130)[lb]{\Large{\Black{$~=~ \dfrac{15}{4}\lambda J(m_{G}^{2},m_{G}^{2})$}}}
  \end{picture}
}}
\end{center}
so that the two-loop effective potential of scalar sector is the sum of all above integrals:
\begin{eqnarray}
V_{S}^{(2l)}= - 3\lambda^{2}\phi_{c}^{2}\left( I(m_{H}^{2},m_{H}^{2},m_{H}^{2}) + I(m_{H}^{2},m_{G}^{2},m_{G}^{2})\right) ~~~~~~~~~~~~~~\nonumber \\
 ~~~~~~~~~~+~ \dfrac{3}{4}\lambda \left( J(m_{H}^{2},m_{H}^{2}) + 2J(m_{H}^{2}, m_{G}^{2}) + 5J(m_{G}^{2},m_{G}^{2})\right). \label{EffpotS}
\end{eqnarray} 
Nevertheless these quantities are divergent and we need to remove its divergences using a re-normalization prescription. From the results quoted in the Appendix \ref{AppIntegralJ} we have all ingredients to evaluate explicitly the scalar potential (\ref{EffpotS}). Using the equation (\ref{Jm00}) we find the Laurent expansion of the dimensional regulated integral 
\begin{eqnarray}
&&I(m_{H}^{2},0,0)=-\dfrac{(\mu^{2})^{4-d}}{(4\pi)^{d}}{\Gamma(2-{d\over2})
\Gamma(3-d)\Gamma({d\over2}-1)^2\over{\Gamma({d\over2})}}
\bigl({m_{H}^{2}}\bigr) ^{d-3}=\nonumber\\
&&\dfrac{m_{H}^{2}}{(4\pi)^{4}}\left(\dfrac{m_{H}^{2}}{4\pi\mu^{2}e^{-\gamma_{_{E}}}} \right)^{-2\varepsilon}\left(\dfrac{1}{2\varepsilon^{2}} + \dfrac{3}{2\varepsilon} + \dfrac{3\zeta(2)}{2} + \dfrac{7}{2}\right), \label{EP-SS1}
\end{eqnarray}
where $\zeta(2)=\pi^{2}/6$ is the Riemann zeta function evaluated at 2, and $\gamma_{E}$ is the Euler-Mascheroni constant.
From the equation (\ref{SolutionI7}) we obtain for $a^{2}= - \frac{3}{4}m_{H}^{4}< 0$,
\begin{eqnarray}
(4\pi)^{4}I(m_{H}^{2},m_{H}^{2},m_{H}^{2})=\left.{c\over {2(d/2 - 2)^2}}+{1\over{(d/2 - 2)}}\left(
{3c\over2}-L_1\right)+{1\over2}\Bigl[L_2-6L_1 \right. \nonumber \\
\nonumber \\ 
\left. ~+~ 3m_{H}^{2}(\overline{\ln}m_{H}^{2})^{2} + \xi (m_{H}^{2},m_{H}^{2},m_{H}^{2})+c(7+\zeta(2))\Bigr]\right. \label{ImH1}
\end{eqnarray}
with 
\begin{eqnarray*}
&& c = m_{H}^{2} + m_{H}^{2} + m_{H}^{2} ~~=~~ 3m_{H}^{2}, \\
&& \\
&& \overline{ln}(m_{H}^{2}) = ln\dfrac{m_{H}^{2}}{\mu^{2}} + \gamma_{E} - ln(4\pi), \\
&& \\
&& L_{1}= m_{H}^{2}\overline{ln}(m_{H}^{2}) + m_{H}^{2}\overline{ln}(m_{H}^{2}) + m_{H}^{2}\overline{ln}(m_{H}^{2}) ~~=~~ 3m_{H}^{2}\overline{ln}(m_{H}^{2}), \\
&& \\
&& L_{2}=m_{H}^{2}\overline{ln}^{2}(m_{H}^{2}) + m_{H}^{2}\overline{ln}^{2}(m_{H}^{2}) + m_{H}^{2}\overline{ln}^{2}(m_{H}^{2}) ~~=~~ 3m_{H}^{2}\overline{ln}^{2}(m_{H}^{2}), \\
&& \\
&&b= \sqrt{-a^{2}} ~~=~~ \dfrac{\sqrt{3}}{2}m_{H}^{2}, \\
&& \\
&&\theta_{H}=tan^{-1}\left(\dfrac{c/2 - m_{H}^{2}}{b} \right) = tan^{-1}\left(\dfrac{3/2m_{H}^{2} - m_{H}^{2}}{\sqrt{3}/2m_{H}^{2}} \right) = tan^{-1}\left(\dfrac{\sqrt{3}}{3} \right) ~=~ \dfrac{\pi}{6},\\
&& \\
&& L\left(\theta_{H}=\dfrac{\pi}{6}\right) = -\dfrac{1}{2}Cl_{2}\left(\dfrac{2\pi}{3}\right)+\dfrac{\pi}{6}ln2, \\
&& \\
&& Cl_{2}\left(\dfrac{2\pi}{3} \right) = \dfrac{1}{2}\left[Cl_{2}\left(\dfrac{\pi}{3}+\pi \right)+2Cl_{2}\left(\dfrac{\pi}{3}\right)\right] ~~;~~ {\small Cl_{2}\left(\dfrac{\pi}{3}+2m\pi \right) = 1.01494160\dots ~[m\in\mathbb{Z}]}, 
\end{eqnarray*}
and
\begin{eqnarray*}
\xi (m_{H}^{2},m_{H}^{2},m_{H}^{2}) = 8b \left[ 3L(\pi /6) - \dfrac{\pi}{2}ln(2) \right] = 4\sqrt{3}m_{H}^{2}\left[ 3L(\pi /6)-\dfrac{\pi}{2}ln(2) \right]. 
\end{eqnarray*}
The above special integrals and its numerical values can be consulted in \cite{Lobachevskiy}. Replacing all above identities in equation (\ref{ImH1}) we finally obtain 
\begin{eqnarray}
(4\pi)^{4}I(m_{H}^{2},m_{H}^{2},m_{H}^{2})= \dfrac{3}{2}m_{H}^{2}\dfrac{1}{\varepsilon^{2}}+\dfrac{1}{\varepsilon}\left( \dfrac{9}{2}m_{H}^{2} - 3m_{H}^{2}\overline{ln}(m_{H}^{2}) \right) ~~~~~~~~~~~~~~~~~~~~~~~~~ \nonumber \\ 
+ \dfrac{1}{2}\left[6m_{H}^{2}\overline{ln}^{2}(m_{H}^{2}) - 18m_{H}^{2}\overline{ln}m_{H}^{2} + \xi (m_{H}^{2},m_{H}^{2},m_{H}^{2}) + 3m_{H}^{2}(7 + \zeta(2)) \right]. \label{EP-SS2}
\end{eqnarray}
The last required integral $J(m_{H}^{2}, m_{H}^{2})$ can be easily computed, from equation (\ref{IntegralJ}): 
\begin{eqnarray}
J(m_{H}^{2},m_{H}^{2}) ~=~ (J(m_{H}^{2}))^{2} ~~~=~ \left(i{(m_{H})^{2}\over
{(4\pi)^{2}}}\Gamma(\varepsilon)\left(1+\dfrac{\varepsilon}{2}\right)\left(\dfrac{m_{H}^{2}}{4\pi\mu^{2}}\right)^{-\varepsilon} \right)^{2} ~~~~~~~~~~~~~~~~~~~~~~~ \nonumber \\
\nonumber \\
~ = ~ \left({m_{H}^{2}\over{(4\pi)^{2}}}\left(\dfrac{m_{H}^{2}}{4\pi\mu^{2}}\right)^{-\varepsilon}e^{-\varepsilon\gamma_{_{E}}}\left(\varepsilon\left(\dfrac{1}{2}i\zeta(2) + i \right)+ \dfrac{i}{\varepsilon} + i\right) \right)^{2} \nonumber \\
\nonumber \\
\nonumber \\ 
~ = ~ {m_{H}^{4}\over{(4\pi)^{4}}}\left(\dfrac{m_{H}^{2}}{4\pi\mu^{2}e^{-\gamma_{_{E}}}}\right)^{-2\varepsilon}\left(-\dfrac{1}{\varepsilon^{2}}-\dfrac{2}{\varepsilon}-\zeta(2)+\varepsilon\left(-2\zeta(2)+\dfrac{2\zeta(3)}{3} - 4\right) -3\right). \nonumber \\
~ \label{EP-SS3}
\end{eqnarray}  
For simplicity we make the calculation in the Landau gauge at the tree minimum of the potential, $\phi_{c}=v$. In this case $m_{H}^{2}=2\lambda v^{2}$ and $J(m_{H}^{2},m_{G}^{2})=J(m_{G}^{2},m_{G}^{2}) = 0$ because $m_{G}^{2}=0$ in the Landau Gauge and by dimensional regularization $\int \frac{d^dp}{p^2}=0$. Replacing the equations (\ref{EP-SS1}), (\ref{EP-SS2}) and (\ref{EP-SS3}) in the scalar potential $V_{S}^{(2l)}$ we obtain the expansion 
\begin{eqnarray}
V_{S}^{(2l)} = V_{S}^{(fin)} + \Delta V_{S}^{(2l)},
\end{eqnarray}
where
\begin{eqnarray}
&&V_{S}^{(fin)} = -3\lambda^{2}v^{2}\left[\dfrac{m_{H}^{2}}{(4\pi)^{4}}\left(\dfrac{3\zeta(2)}{2} + \dfrac{7}{2} - 3\overline{ln}(m_{H}^{2}) + \overline{ln}^{2}(m_{H}^{2}) \right. \right. \nonumber \\
&& ~~~~~~~~~~~~~~~~~~~~~ \left. \left. ~~ + ~~~3\overline{ln}^{2}(m_{H}^{2}) - 9\overline{ln}m_{H}^{2} + \dfrac{\xi (m_{H}^{2},m_{H}^{2},m_{H}^{2})}{2m_{H}^{2}} + \dfrac{3}{2}(7 + \zeta(2)) \right) \right]\nonumber \\
&& ~~~~~~~~~~~~~~~~~ ~~~~~~+~~~ \dfrac{3}{4}\lambda \left[ \dfrac{m_{H}^{4}}{(4\pi)^{4}}\left( - \zeta(2) - 3 + 4\overline{ln}(m_{H}^{2}) -2\overline{ln}^{2}(m_{H}^{2}) \right) \right]
\end{eqnarray}
is the finite part of the scalar two-loop potential, and
\begin{eqnarray}
&&\Delta V_{S}^{(2l)} =  -3\lambda^{2}v^{2}\left[\dfrac{m_{H}^{2}}{(4\pi)^{4}}\left( \dfrac{1}{2\varepsilon^{2}} + \dfrac{3}{2\varepsilon} -\dfrac{1}{\varepsilon}\overline{ln}(m_{H}^{2}) + \dfrac{3}{2}\dfrac{1}{\varepsilon^{2}} \right. \right. \nonumber \\
&& \nonumber \\
&& ~~~~~~~~~~~~~~~~~~~~~~~~ \left. \left. + ~~ \dfrac{1}{\varepsilon}\left( \dfrac{9}{2} - 3\overline{ln}(m_{H}^{2}) \right)\right) \right]  ~ + ~ \dfrac{3}{4}\lambda \left[\dfrac{m_{H}^{4}}{(4\pi)^{4}}\left( -\dfrac{1}{\varepsilon^{2}}-\dfrac{2}{\varepsilon} + \dfrac{2}{\varepsilon} \overline{ln}(m_{H}^{2})\right) \right] \nonumber \\
\end{eqnarray}
is the divergent part of the potential $V_{S}^{(2l)}$. The potential includes single and double poles in $\varepsilon$ apart local divergences, where a single pole is multiplied with a logarithmic term that is function of the field dependent mass $m_{H}^{2}$. The single and double poles in $\varepsilon$, are simply removed by two-loop counter-terms in modified minimal subtraction. Nevertheless to remove the sub-divergences
\[ -3\lambda^{2}v^{2}\left[\dfrac{m_{H}^{2}}{(4\pi)^{4}}\left(-\dfrac{1}{\varepsilon}\overline{ln}(m_{H}^{2}) - \dfrac{3}{\varepsilon}\overline{ln}(m_{H}^{2}) \right) \right] + \dfrac{3}{4}\lambda \left[\dfrac{m_{H}^{4}}{(4\pi)^{4}}\left( \dfrac{2}{\varepsilon} \overline{ln}(m_{H}^{2})\right) \right]=\dfrac{15}{2}\lambda\dfrac{m_{H}^{4}}{(4\pi)^{4}}\dfrac{1}{\varepsilon}\overline{ln}(m_{H}^{2}), \]
one must include counter-terms for the various one-loop divergent sub-diagrams. For the pure scalar sector the following diagram counterterm is needed
\begin{center}
\scalebox{0.6}{
\fcolorbox{white}{white}{
  \begin{picture}(230,82) (314,-179)
    \SetWidth{1.0}
    \SetColor{Black}
    \Arc[dash,dashsize=10,arrow,arrowpos=0.5,arrowlength=5,arrowwidth=2,arrowinset=0.2](355,-138)(40,180,540)
    \COval(395,-138)(7.071,7.071)(135.0){Black}{White}\Line(391.464,-134.464)(398.536,-141.536)\Line(398.536,-134.464)(391.464,-141.536)
    \Text(425,-138)[lb]{\Large{\Black{$~~=$}}}
    \Text(475,-150)[lb]{\Large{\Black{$i\dfrac{\delta m_{H}^{2}}{2}J(m_{H}^{2}).$}}}
  \end{picture}
}}
\end{center}
From eq. (\ref{eq:delta-mH2}), but without considering the tadpole contributions, we know that 
\begin{eqnarray}
\delta m_{H}^{2} ~~=~~ -\dfrac{3\lambda m_{H}^{2}}{(4\pi)^{2}}\dfrac{1}{\varepsilon} - \dfrac{9\lambda m_{H}^{2}}{(4\pi)^{2}}\dfrac{1}{\varepsilon} - \dfrac{\lambda m_{H}^{2}}{(4\pi)^{2}}\dfrac{1}{\varepsilon} - \dfrac{2\lambda m_{H}^{2}}{(4\pi)^{2}}\dfrac{1}{\varepsilon} \nonumber \\
\nonumber \\
~=~ -\dfrac{15 \lambda m_{H}^{2}}{(4\pi)^{2}}\dfrac{1}{\varepsilon}, ~~~~~~~~~~~~~~~~~~~~~~~~~~~~~~~~~~~~~~~~~~ 
\end{eqnarray} 
in the scalar sector. Therefore 
\begin{center}
\scalebox{0.6}{
\fcolorbox{white}{white}{
  \begin{picture}(720,82) (314,-179)
    \SetWidth{1.0}
    \SetColor{Black}
    \Arc[dash,dashsize=10,arrow,arrowpos=0.5,arrowlength=5,arrowwidth=2,arrowinset=0.2](355,-138)(40,180,540)
    \COval(395,-138)(7.071,7.071)(135.0){Black}{White}\Line(391.464,-134.464)(398.536,-141.536)\Line(398.536,-134.464)(391.464,-141.536)
    \Text(425,-144)[lb]{\Large{\Black{$~~=$}}}
    \Text(475,-160)[lb]{\Large{\Black{$\dfrac{15}{2}\lambda \dfrac{ m_{H}^{4}}{(4\pi)^{4}}\dfrac{1}{\varepsilon}\left[\dfrac{1}{\varepsilon} + (1 - \overline{ln}(m_{H}^{2})) + \varepsilon\left(\dfrac{1}{2}\zeta(2) + 1 -\overline{ln}(m_{H}^{2}) + \dfrac{1}{2}\overline{ln}^{2}(m_{H}^{2}) \right)\right].$}}}
  \end{picture}
}}
\end{center}
If we add this contribution to the potential $V_{S}^{(2l)}$, the sub-divergences are exactly cancelled and the finite and divergence terms are changed a bit. The resulting potential is:
\begin{eqnarray}
&&V_{S}^{(2l)} = -\dfrac{3}{4}\lambda\dfrac{m_{H}^{4}}{(4\pi)^{4}}\left[2\zeta(2) + 21 - 18\overline{ln}(m_{H}^{2}) + 5\overline{ln}^{2}(m_{H}^{2}) + \dfrac{\xi (m_{H}^{2},m_{H}^{2},m_{H}^{2})}{m_{H}^{2}} \right]\nonumber \\
&& \nonumber \\
&& ~~~~~~~~~~~~+~~\dfrac{3}{4}\lambda\dfrac{m_{H}^{4}}{(4\pi)^{4}}\left[\dfrac{5}{\varepsilon^{2}} - \dfrac{4}{\varepsilon} \right].\nonumber \\
\end{eqnarray}  
The divergent part in the above potential can be removed by the use of the renormalization constants introduced in renormalization procedure exposed in the Chapter \ref{cha:Effective Potential}. By the other hand, we note that the counterterm can be rewritten as
\begin{eqnarray}
&&-\dfrac{15}{2}\lambda \dfrac{m_{H}^{2}}{(4\pi)^{2}}\dfrac{i}{\varepsilon}J(m_{H}^{2}) ~=~ -\dfrac{3}{2}\lambda m_{H}^{2} \left[\dfrac{3i}{(4\pi)^{2}\varepsilon} + \dfrac{i}{(4\pi)^{2}\varepsilon}\right]J(m_{H}^{2})-\dfrac{3}{2}\lambda m_{H}^{2}\left(\dfrac{i}{(4\pi)^{2}\varepsilon} \right)J(m_{H}^{2}) \nonumber \\
&& \nonumber \\
&&  \nonumber \\
&& = -3\lambda^{2}v^{2}\left[\dfrac{i}{(4\pi)^{2}\varepsilon}\left(J(m_{H}^{2})+J(m_{H}^{2})+J(m_{H}^{2}) \right) + \dfrac{i}{(4\pi)^{2}\varepsilon}(J(m_{H}^{2})+J(m_{G}^{2})+J(m_{G}^{2})) \right] \nonumber\\
&& \nonumber \\
 && ~~~ + ~~~~ \dfrac{3}{4} \lambda \left[ \dfrac{-i}{(4\pi)^{2}\varepsilon}\left(m_{H}^{2}J(m_{H}^{2})+m_{H}^{2}J(m_{H}^{2})\right)-\dfrac{2i}{(4\pi)^{2}\varepsilon}\left(m_{G}^{2}J(m_{H}^{2})+m_{H}^{2}J(m_{G}^{2})\right) \right. \nonumber \\
 && \nonumber \\
&& \left. ~~~~~~~~~~~~~~~~~~~~~~~~~~~~ - ~~~ \dfrac{5i}{(4\pi)^{2}\varepsilon}\left(m_{G}^{2}J(m_{G}^{2})+m_{G}^{2}J(m_{G}^{2})\right) \right],
\end{eqnarray} 
therefore in the scalar potential (\ref{EffpotS}) the removing of the sub-divergences is equivalent to making the transformations \cite{Ford-Jack}
\begin{eqnarray}
&&I(x,y,z)\rightarrow \hat{I}(x,y,z) = I(x,y,z) + \dfrac{i}{(4\pi)^{2}\varepsilon}\left(J(x)+J(y)+J(z)\right), \nonumber \\
&& ~~J(x,y) \rightarrow \hat{J}(x,y) = J(x,y) - \dfrac{i}{(4\pi)^{2}\varepsilon}\left(yJ(x)+xJ(y)\right). \label{ContractedForm}
\end{eqnarray}
Summary $V_{S}^{(2l)}$ is then obtained by replacing the integrals $I$ and $J$ for its subtracted form $\hat{I}$ and $\hat{J}$ given in eq. (\ref{ContractedForm}) and discarding the $1/\varepsilon^{2}$ and $1/\varepsilon$ poles. For other sectors where no vector bosons are included in the Feynman diagrams, this procedure of renormalization is still valid. However when vector bosons are present, the algebra involved in reducing the sum of diagrams to dependence on $I$ and $J$ may produce explicit factors of $d$ and we will need the subtracted form of $dI$. The result depends on which sub-diagram yielded the factor of $d$ and therefore we need explicitly to evaluate the contribution of the subtractions to each diagram.

\subsection{The Scalar-Fermion Sector of the Effective Potential\label{sec:VSF}}

\begin{figure}
\begin{center}
\scalebox{0.5}{
\fcolorbox{white}{white}{
  \begin{picture}(522,204) (155,-123)
    \SetWidth{1.0}
    \SetColor{Black}
    \Arc[arrow,arrowpos=0.5,arrowlength=5,arrowwidth=2,arrowinset=0.2,clock](212,-12)(52,-180,-360)
    \Arc[dash,dashsize=10](212,-12)(52,-180,0)
    \Line[arrow,arrowpos=0.5,arrowlength=5,arrowwidth=2,arrowinset=0.2](264,-16)(160,-16)
    \Line[arrow,arrowpos=0.5,arrowlength=5,arrowwidth=2,arrowinset=0.2](472,-16)(368,-16)
    \Arc[arrow,arrowpos=0.5,arrowlength=5,arrowwidth=2,arrowinset=0.2,clock](420,-16)(52,-180,-360)
    \Arc[dash,dashsize=10](420,-16)(52,-180,0)
    \Line[arrow,arrowpos=0.5,arrowlength=5,arrowwidth=2,arrowinset=0.2](672,-16)(568,-16)
    \Arc[arrow,arrowpos=0.5,arrowlength=5,arrowwidth=2,arrowinset=0.2,clock](620,-16)(52,-180,-360)
    \Arc[dash,dashsize=10,arrow,arrowpos=0.5,arrowlength=5,arrowwidth=2,arrowinset=0.2](620,-16)(52,-180,0)
    \Text(204,55)[lb]{\Large{\Black{$f$}}}
    \Text(204,-4)[lb]{\Large{\Black{$f$}}}
    \Text(204,-90)[lb]{\Large{\Black{$H$}}}
    \Text(416,-90)[lb]{\Large{\Black{$G^{0}$}}}
    \Text(416,-4)[lb]{\Large{\Black{$f$}}}
    \Text(416,55)[lb]{\Large{\Black{$f$}}}
    \Text(610,55)[lb]{\Large{\Black{$(u_{i}, \nu_{i})$}}}
    \Text(610,-4)[lb]{\Large{\Black{$(d_{i},e_{i})$}}}
    \Text(610,-90)[lb]{\Large{\Black{$G^{\pm}$}}}
    \Text(212,-128)[lb]{\Large{\Black{$(a)$}}}
    \Text(420,-128)[lb]{\Large{\Black{$(b)$}}}
    \Text(620,-128)[lb]{\Large{\Black{$(c)$}}}
    \Vertex(160,-16){4}
    \Vertex(264,-16){4}
    \Vertex(368,-16){4}
    \Vertex(472,-16){4}
    \Vertex(568,-16){4}
    \Vertex(672,-16){4}
  \end{picture}
}}
\end{center}
\caption{\label{EP-SF-Sector}{\small Contributions to scalar-fermion sector of the two-loop effective potential.}}
\end{figure}
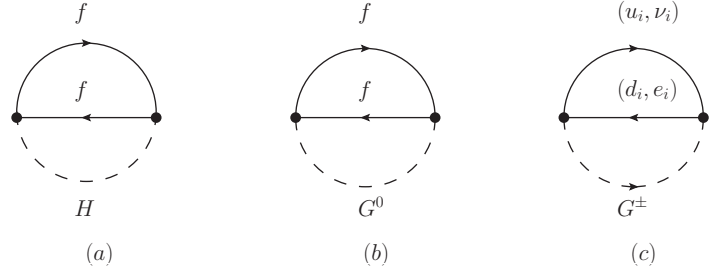

We give now all contributions to the effective potential $V^{(2l)}$. We divide the calculation into parts according its diagrammatic origin, thus
\[V^{(2l)}=V^{(2l)}_{S}+V^{(2l)}_{SF}+V^{(2l)}_{V}+V^{(2l)}_{VF}+V^{(2l)}_{VS}\]
where S, F and V denote scalar, fermion and vector fields respectively. The first contribution, $V_{S}^{(2l)}$, was computed in the above section. The second contribution, $V_{SF}^{(2l)}$, includes only the diagrams with scalar and fermion lines. The Figure \ref{EP-SF-Sector} shows the diagrams contributing to $V_{SF}^{(2l)}$. The line denoted with the letter $f$ represents all massive fermion fields of the SM: $e_{i}$, $u_{i}$ and $d_{i}$, where $e_{i}$ are the massive leptons, $u_{i}$ are the up-type quarks and $d_{i}$ are the down-type quarks. Here the $\nu_{i}$ represent the neutrinos. The individual diagram contribution to the potential $V_{SF}^{(2l)}$ can be written using the notation
\begin{eqnarray}
V_{SF}^{(2l)}(\phi_{c})=F_{ffH}(\phi_{c})+F_{ffG}(\phi_{c})+F_{udG}(\phi_{c}).
\end{eqnarray}
The diagram \ref{EP-SF-Sector} (a) represents the function $F_{ffH}(\phi_{c})$
given by
\begin{eqnarray}
F_{ffH}(\phi_{c})=-\frac{3}{2}\left(\frac{gm_{f}}{2m_{W}}\right)^{2}\int\frac{d^{4}p}{(2\pi)^{4}}\int\frac{d^{4}q}{(2\pi)^{4}}\dfrac{Tr\left[(\gamma^{\mu}q_{\mu}+m_{f})(\gamma^{\nu}(p+q)_{\nu}+m_{f})\right]}{(p^{2}-m_{H}^{2})(q^{2}-m_{f}^{2})((p+q)^{2}-m_{f}^{2})},
\end{eqnarray}
where
\begin{eqnarray}
Tr\left[(\gamma_{\mu}q^{\mu}+m_{f})(\gamma_{\nu}(p+q)^{\nu}+m_{f})\right]=4q\cdot p+4q^{2}+4m_{f}^{2} ~~~~~~~~~~~~~~~~~~~~~~~~~~~~~ \\ \nonumber \\
=2((q+p)^{2}-m_{f}^{2}+(q^{2}-m_{f}^{2})-(p^{2}-m_{H}^{2})+4m_{f}^{2}-m_{H}^{2}). \nonumber 
\end{eqnarray}
Therefore, in terms of the integrals $I$ and $J$, we obtain
\begin{eqnarray}
F_{ffH}(\phi_{c})=\sum_{f}\frac{3}{2}h_{f}^{2}\left[J(m_{f}^{2},m_{f}^{2})-2J(m_{f}^{2},m_{H}^{2})-(4m_{f}^{2}-m_{H}^{2})I(m_{f}^{2},m_{f}^{2},m_{H}^{2})\right],
\end{eqnarray}
where $h_{f}$ is the Yukawa coupling constant of the fermion $f$.
Moreover the contribution of the diagram \ref{EP-SF-Sector}
(b), \foreignlanguage{english}{$F_{ffG}(\phi_{c})$,} can be obtained
in the same way, just by changing the mass $m_{H}$ by $m_{G}$ and
using the result
\begin{eqnarray}
Tr\left[\gamma^{5}(\gamma_{\mu}q^{\mu}+m_{f})\gamma^{5}(\gamma_{\nu}(p+q)^{\nu}+m_{f})\right]=-4q\cdot p-4q^{2}+4m_{f}^{2},
\end{eqnarray}
giving
\begin{eqnarray}
F_{ffG}(\phi_{c})=\sum_{f}\frac{3}{2}h_{f}^{2}\left[J(m_{f}^{2},m_{f}^{2})-2J(m_{f}^{2},m_{G}^{2})+m_{G}^{2}I(m_{f}^{2},m_{f}^{2},m_{G}^{2})\right].
\end{eqnarray}
Finally, the diagram \ref{EP-SF-Sector} (c) requires the evaluation
of the integral
\begin{eqnarray}
F_{udG}(\phi_{c})=-3\left(\frac{g}{\sqrt{2}m_{W}}\right)^{2}\int\frac{d^{4}p}{(2\pi)^{4}}\int\frac{d^{4}q}{(2\pi)^{4}}\dfrac{1}{(p^{2}-m_{G}^{2})(q^{2}-m_{u}^{2})((p+q)^{2}-m_{d}^{2})}\times \nonumber \\ \nonumber \\
\;\;\;\quad ~~~~~~ Tr\left[(m_{d}\gamma_{+}-m_{u}\gamma_{-})(\gamma_{\mu}q^{\mu}+m_{u})(m_{d}\gamma_{+}-m_{u}\gamma_{-})(\gamma_{\nu}(p+q)^{\nu}+m_{d})\right], \nonumber \\ \nonumber \\
\end{eqnarray}
where $u(d)$ represents a generic fermionic field with weak isospin $T^{3}=\frac{1}{2}\left(-\frac{1}{2}\right)$ respectively, and $\gamma_{\pm}=1\pm\gamma^{5}/2$. The trace in the integrand amounts to
\begin{eqnarray}
&&Tr\left[(m_{d}\gamma_{+}-m_{u}\gamma_{-})(\gamma_{\mu}q^{\mu}+m_{u})(m_{d}\gamma_{+}-m_{u}\gamma_{-})(\gamma_{\nu}(p+q)^{\nu}+m_{d})\right] \nonumber \\
&&=4\left(\frac{m_{d}^{2}}{2}+\frac{m_{u}^{2}}{2}\right)(q\cdot p+q^{2})-4m_{u}^{2}m_{d}^{2} \\
&&=\left(m_{d}^{2}+m_{u}^{2}\right)\left((q+p)^{2}-m_{d}^{2}-(p^{2}-m_{G}^{2})+(q^{2}-m_{u}^{2})+m_{d}^{2}+m_{u}^{2}-m_{G}^{2}\right)-4m_{u}^{2}m_{d}^{2}. \nonumber 
\end{eqnarray}
Replacing this trace in the integral, we obtain in terms of $I$ and
$J$ the contribution
\begin{eqnarray}
F_{udG}(\phi_{c})=-3\left(\frac{g}{\sqrt{2}m_{W}}\right)^{2}\left\{ \left(m_{d}^{2}+m_{u}^{2}\right)\left[J(m_{G}^{2},m_{u}^{2})+J(m_{G}^{2},m_{d}^{2})-J(m_{u}^{2},m_{d}^{2})\right]~~~~~~~~~~~ \right.  \nonumber \\
\left.+\left[\left(m_{d}^{2}+m_{u}^{2}\right)\left(m_{d}^{2}+m_{u}^{2}-m_{G}^{2}\right)-4m_{u}^{2}m_{d}^{2}\right]I(m_{G}^{2},m_{u}^{2},m_{d}^{2})\right\}.	 \nonumber \\
\nonumber \\
\end{eqnarray}
The potential $V_{SF}^{(2l)}$ is the sum of these three contributions
above computed. The dominant contribution comes in the limit where
all fermion masses are put to zero except the top quark mass, $m_{t}$.
In this limit, the potential $V_{SF}^{(2l)}$ takes the form: 
\begin{eqnarray}
V_{SF}^{(2l)}=\frac{3}{2}h_{t}^{2}\left[J(m_{t}^{2},m_{t}^{2})-2J(m_{t}^{2},m_{H}^{2})-(4m_{t}^{2}-m_{H}^{2})I(m_{t}^{2},m_{t}^{2},m_{H}^{2})\right] ~~~~~~~~~~ \nonumber \\
+\frac{3}{2}h_{t}^{2}\left[J(m_{t}^{2},m_{t}^{2})-2J(m_{t}^{2},m_{G}^{2})+m_{G}^{2}I(m_{t}^{2},m_{t}^{2},
m_{G}^{2})\right] ~~~~~~ \nonumber \\
-3\left(\frac{gm_{t}}{\sqrt{2}m_{W}}\right)^{2}\left\{ J(m_{G}^{2},m_{t}^{2})+\left(m_{t}^{2}-m_{G}^{2}\right)I(m_{G}^{2},m_{t}^{2},0)\right\}. 
\end{eqnarray}
After a bit of algebra, we finally obtain:
\begin{eqnarray}
V_{SF}^{(2l)}=\frac{3}{2}h_{t}^{2}\left[2J(m_{t}^{2},m_{t}^{2})-2J(m_{t}^{2},m_{H}^{2})-4J(m_{t}^{2},m_{G}^{2})-(4m_{t}^{2}-m_{H}^{2})I(m_{t}^{2},m_{t}^{2},m_{H}^{2})\right. ~~~~~~ \nonumber \\
\left. ~~ + ~~~ m_{G}^{2}I(m_{t}^{2},m_{t}^{2},m_{G}^{2}) - 2\left(m_{t}^{2}-m_{G}^{2}\right)I(m_{G}^{2},m_{t}^{2},0)\right]. \nonumber \\ \nonumber \\
\end{eqnarray}
Since the coefficients of the integrals $I$ and $J$ are independent
of the space-time dimension $d$, this sector of the effective potential
can be renormalized in the same way that the scalar sector, making
the transformations (\ref{ContractedForm}) i.e. just by changing $I$
and $J$ by $\hat{I}$ and $\hat{J}$ respectively. Besides, is relevant note here that $V_{SF}^{(2l)}$ has terms of order $h_{t}^{6}$, those terms and the QCD contribution to $V_{FV}^{(2l)}$, computed later, represent the largest contributions of the potential.  
\begin{figure}
\begin{center}
\scalebox{0.4}{
\fcolorbox{white}{white}{
  \begin{picture}(549,366) (135,-29)
    \SetWidth{1.0}
    \SetColor{Black}
    \PhotonArc(180,240)(36.497,-171,189){7.5}{11}
    \Vertex(222,240){6}
    \PhotonArc(258,240)(36.497,-171,189){7.5}{11}
    \PhotonArc(440,241)(55,-180,180){7.5}{17}
    \Photon(380,236)(495,241){7.5}{6}
    \Vertex(380,236){5}
    \Vertex(495,241){5}
    \Vertex(590,241){5}
    \Vertex(675,241){5}
    \Line[dash,dashsize=2,arrow,arrowpos=0.5,arrowlength=5,arrowwidth=2,arrowinset=0.2](590,241)(675,241)
    \Arc[dash,dashsize=2,arrow,arrowpos=0.5,arrowlength=5,arrowwidth=2,arrowinset=0.2,clock](632.5,243.5)(42.573,-176.634,-363.366)
    \PhotonArc(632.5,238.5)(42.573,176.634,363.366){7.5}{7.5}
    \Line[dash,dashsize=2,arrow,arrowpos=0.5,arrowlength=5,arrowwidth=2,arrowinset=0.2](585,51)(670,51)
    \PhotonArc(627.5,48.5)(42.573,176.634,363.366){7.5}{7.5}
    \Arc[dash,dashsize=2,arrow,arrowpos=0.5,arrowlength=5,arrowwidth=2,arrowinset=0.2,clock](627.5,53.5)(42.573,-176.634,-363.366)
    \Vertex(670,51){5}
    \Vertex(585,51){5}
    \Line[dash,dashsize=2,arrow,arrowpos=0.5,arrowlength=5,arrowwidth=2,arrowinset=0.2](400,51)(485,51)
    \Arc[dash,dashsize=2,arrow,arrowpos=0.5,arrowlength=5,arrowwidth=2,arrowinset=0.2,clock](442.5,53.5)(42.573,-176.634,-363.366)
    \PhotonArc(442.5,48.5)(42.573,176.634,363.366){7.5}{7.5}
    \Vertex(400,51){5}
    \Vertex(485,51){5}
    \Line[dash,dashsize=2,arrow,arrowpos=0.5,arrowlength=5,arrowwidth=2,arrowinset=0.2](185,51)(270,51)
    \Arc[dash,dashsize=2,arrow,arrowpos=0.5,arrowlength=5,arrowwidth=2,arrowinset=0.2,clock](227.5,53.5)(42.573,-176.634,-363.366)
    \PhotonArc(227.5,48.5)(42.573,176.634,363.366){7.5}{7.5}
    \Vertex(270,51){5}
    \Vertex(185,51){5}
    \Text(215,141)[lb]{\Large{\Black{$(a)$}}}
    \Text(440,141)[lb]{\Large{\Black{$(b)$}}}
    \Text(625,141)[lb]{\Large{\Black{$(c)$}}}
    \Text(215,-34)[lb]{\Large{\Black{$(d)$}}}
    \Text(440,-34)[lb]{\Large{\Black{$(e)$}}}
    \Text(625,-34)[lb]{\Large{\Black{$(f)$}}}
    \Text(170,301)[lb]{\Large{\Black{$W$}}}
    \Text(230,301)[lb]{\Large{\Black{$(W, \gamma, Z)$}}}
    \Text(435,201)[lb]{\Large{\Black{$W$}}}
    \Text(435,251)[lb]{\Large{\Black{$W$}}}
    \Text(425,306)[lb]{\Large{\Black{$(\gamma, Z)$}}}
    \Text(630,206)[lb]{\Large{\Black{$\gamma$}}}
    \Text(610,246)[lb]{\Large{\Black{$(u_{+},u_{-})$}}}
    \Text(610,291)[lb]{\Large{\Black{$(u_{+},u_{-})$}}}
    \Text(225,16)[lb]{\Large{\Black{$Z$}}}
    \Text(205,56)[lb]{\Large{\Black{$(u_{+},u_{-})$}}}
    \Text(205,101)[lb]{\Large{\Black{$(u_{+},u_{-})$}}}
    \Text(440,16)[lb]{\Large{\Black{$W$}}}
    \Text(430,101)[lb]{\Large{\Black{$(u_{+},u_{-})$}}}
    \Text(440,56)[lb]{\Large{\Black{$u_{\gamma}$}}}
    \Text(620,16)[lb]{\Large{\Black{$W$}}}
    \Text(620,56)[lb]{\Large{\Black{$u_{Z}$}}}
    \Text(610,101)[lb]{\Large{\Black{$(u_{+},u_{-})$}}}
  \end{picture}
}}
\end{center}
\caption{\label{EP-V-Sector}{\small Contributions to vector sector of the two-loop effective potential.}}
\end{figure}
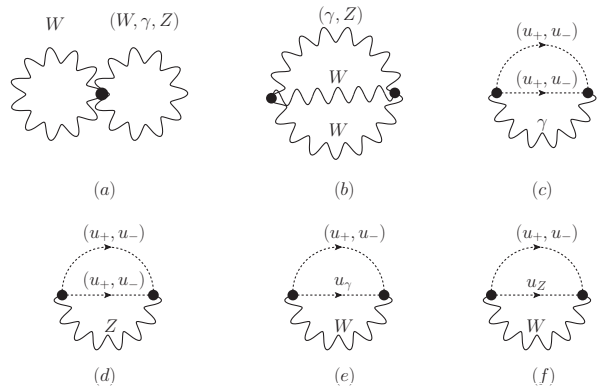

\subsection{The Pure Gauge Sector of the Effective Potential\label{sec:VGauge}}

Let's consider now the gauge sector of the effective potential. The thirteen
vacuum bubbles that contribute to the gauge sector are grouped in six typologies of diagrams showed in fig. \ref{EP-V-Sector}. This sector includes the contribution of the FP ghost fields represented with the letter $u_{i}$ (where $i=\gamma,\: Z,W_-,W_+$), see the diagrams \ref{EP-V-Sector} (c) - \ref{EP-V-Sector} (f).
Although these diagrams are computed in the Landau gauge, the diagrams
with ghost fields give non-zero contributions. This is a common mistake
when one computes the amplitudes with FeynArts. FeynArts contain all
vertices and propagators of the SM in the $R_{\zeta}$-gauge, where
each particle field have a gauge parameter associated $\zeta_{f}$\footnote{In the language of FeynArts each field have a gauge parameter represented
with the command {\tt GaugeXi[f]}, where {\tt f} denotes the specific
field.}. But, by default the option {\tt GaugeRules} sets all gauge parameters
to unity, or choose the Feynman gauge. If we want to enforce a particular
choice of gauge, we need specify the choice of gauge with the command: {\tt GaugeRules -> \_GaugeXi -> $\zeta$}.
This command changes the gauge parameters to be only three, denoted
by $\zeta_{W},\:\zeta_{Z},\:\zeta_{A}$, and puts the gauge parameters
in the propagators equal to our election $\zeta$. Nevertheless, it
does not change the gauge parameters in the vertices, and therefore
it produces some amplitudes proportional to the ratio between $\zeta$
and the new gauge parameters $\zeta_{W},\:\zeta_{Z},\:\zeta_{A}$.
For instance, if we compute the amplitude for the diagram \ref{EP-V-Sector}
(e), and we make the naive election $\zeta=0$, that is the condition
for the Landau gauge, we obtain with FeynArts an amplitude equal to
zero%
\footnote{In FeynArts this gauge election goes into the command {\tt CreateFeynAmp[]}
in the following way: {\tt CreateFeynAmp[\%, GaugeRules -> \_GaugeXi -> 0] }.%
}. The reason is because FeynArts produces an amplitude proportional
to the ratio $\frac{\zeta}{\zeta_{A}\zeta_{W}}$, that vanishes when
$\zeta$ goes to zero. In practice therefore, we must not make the gauge election over the integral produced by FeynArts, instead we must use a general non-zero gauge parameter $\zeta$, and once the amplitude is reduced in terms of $I$ and $J$ we put $\zeta=0$.\footnote{Recent conversations of our research group with Thomas Hahn about this concern led to the conclusion that FeynArts does not mistakenly fix the gauge rules. However, the use of the gauge rules for Landau and Unitary gauges will be clarified in the manual (reported in www.feynarts.de; Last update: 29 Oct 15).}

The thirteen diagrams in Fig. \ref{EP-V-Sector} can be put in the potential
$V_{V}^{(2l)}(\phi_{c})$ using the obvious notation 
\begin{eqnarray}
V_{V}^{(2l)}(\phi_{c})=F_{WV}(\phi_{c})+F_{WWV}(\phi_{c})+F_{u_{i}u_{i}V}(\phi_{c}).
\end{eqnarray}
The first contribution $F_{WV}(\phi_{c})$ represents the diagrams
\ref{EP-V-Sector} (a). In the Landau gauge we get:
\begin{eqnarray}
F_{WV}(\phi_{c})=\frac{1}{2}G_{WV}^{2}\int\frac{d^{4}q}{(2\pi)^{4}}\int\frac{d^{4}p}{(2\pi)^{4}}\left(g^{\mu\sigma}g^{\nu\rho}-2g^{\mu\rho}g^{\nu\sigma}+g^{\mu\nu}g^{\rho\sigma}\right)
\times ~~~~~~~~~~~ \nonumber \\
\left(\frac{-g_{\mu\nu}q_{\rho}p_{\sigma}}{(q^{2}-m_{V}^{2})(p^{2}-m_{W}^{2})(p^{2})}-\frac{-g_{\rho\sigma}q_{\mu}p_{\nu}}{(q^{2}-m_{V}^{2})(p^{2})(p^{2}-m_{W}^{2})}\right. ~~~~~~~ \nonumber \\
\left. + \frac{g_{\mu\nu}g_{\rho\sigma}}{(q^{2}-m_{V}^{2})(p^{2}-m_{W}^{2})}+\frac{q_{\mu}q_{\nu}p_{\rho}p_{\sigma}}{(q^{2}-m_{V}^{2})(q^{2})(p^{2}-m_{W}^{2})(p^{2})}\right),
\end{eqnarray}
where the coefficients $G_{WV}$ have the values 
\begin{eqnarray*}
G_{W\gamma}=e; & G_{WZ}=\dfrac{cos\theta_{W}}{sin\theta_{W}}e; & G_{WW}=\dfrac{e}{sin\theta_{W}}.
\end{eqnarray*}
After dimensional regularization and by expressing the integral in terms
of $I$ and $J$ we obtain:
\begin{eqnarray}
F_{WV}(\phi_{c})=\frac{1}{2}G_{WV}^{2}\left[\dfrac{~}{~}3(d-1)J(m_{V}^{2},m_{W}^{2})+d^{2}J(m_{V}^{2},m_{W}^{2})\right.\label{eq:FWV} ~~~~~~~~~~~~~~~ \nonumber\\
\left.-\int\frac{d^{d}q}{(2\pi)^{d}}\int\frac{d^{d}p}{(2\pi)^{d}}\frac{(q\cdot p)^{2}}{(q^{2}-m_{V}^{2})q^{2}(p^{2}-m_{W}^{2})p^{2}}\right].
\end{eqnarray}
Using the identity 
\begin{eqnarray*}
\frac{1}{q^{2}(q^{2}-m^{2})}=\frac{1}{m^{2}}\left(\frac{1}{q^{2}-m^{2}}-\frac{1}{q^{2}}\right),
\end{eqnarray*}
the last integral in eq. (\ref{eq:FWV}) can be reduced to
\begin{eqnarray*}
\int\frac{d^{d}q}{(2\pi)^{d}}\int\frac{d^{d}p}{(2\pi)^{d}}\frac{(q\cdot p)^{2}}{(q^{2}-m_{V}^{2})q^{2}(p^{2}-m_{W}^{2})p^{2}}=\frac{1}{m_{V}^{2}m_{W}^{2}}\int\frac{d^{d}q}{(2\pi)^{d}}\frac{q^{\mu}q^{\nu}}{(q^{2}-m_{V}^{2})}\int\frac{d^{d}p}{(2\pi)^{d}}\frac{p_{\mu}p_{\nu}}{(p^{2}-m_{W}^{2})}\\
=\frac{1}{m_{V}^{2}m_{W}^{2}}\left(\frac{1}{d}m_{V}^{2}J(m_{V}^{2})g^{\mu\nu}\right)\left(\frac{1}{d}m_{W}^{2}J(m_{W}^{2})g_{\mu\nu}\right).
\end{eqnarray*}
Adding all contributions we find:
\begin{eqnarray}
&&F_{WV}(\phi_{c})=\frac{1}{2}G_{WV}^{2}\left[\left(-3(d-1)+d^{2}-\frac{1}{d}\right)J(m_{V}^{2},m_{W}^{2})\right] \nonumber \\
&&~~~~~~~~~~~~=\frac{1}{2}G_{WV}^{2}\left[\frac{(d-1)^{3}}{d}J(m_{V}^{2},m_{W}^{2})\right]. \label{FWV}
\end{eqnarray}
The photon contribution $F_{W\gamma}(\phi_{c})$ evidently vanishes, we need only to compute the functions $F_{WW}(\phi_{c})$ and $F_{WZ}(\phi_{c})$. The renormalized version of $F_{WV}(\phi_{c})$ can be obtained making the transformations (\ref{ContractedForm}), but we must take into account
the terms proportional to the dimension $d$, this imply explicitly
evaluate the contribution of the subtractions to the function $F_{WV}(\phi_{c})$.
Expanding around $d=4$, the coefficient of ${J}$ take the form:
\begin{eqnarray*}
\frac{(d-1)^{3}}{d}=\frac{27}{4}\left(1-\frac{3}{2}\varepsilon+\frac{7\varepsilon^{2}}{12}+O(\varepsilon^{3})+\dots\right)=\frac{27}{4}+\left(\frac{(d-1)^{3}}{d}-\frac{27}{4}\right),
\end{eqnarray*}
therefore, the regularized function $F_{WV}(\phi_{c})$ has the explicit
form:
\begin{eqnarray}
F_{WV}(\phi_{c})=\frac{1}{2}G_{WV}^{2}\left[\frac{27}{4}\hat{J}(m_{V}^{2},m_{W}^{2})+\frac{27}{4}\left(-\frac{3}{2}\varepsilon+\frac{7\varepsilon^{2}}{12}+O(\varepsilon^{3})+\dots\right)\times \right. ~~~~~  \\
\left[J(m_{V}^{2},m_{W}^{2}) \left.+\frac{i}{(4\pi)^{2}\varepsilon}\left(m_{W}^{2}J(m_{V}^{2})+m_{V}^{2}J(m_{W}^{2})\right)\right]
\right],\nonumber
\end{eqnarray}
and its renormalized version is:
{\small
\begin{eqnarray}
&& F_{WV}(\phi_{c})= \\ && \frac{1}{2}G_{WV}^{2}\left[\frac{27}{4}\hat{J}(m_{V}^{2},m_{W}^{2})+\left(\frac{(d-1)^{3}}{d}-\frac{27}{4}\right)J(m_{V}^{2},m_{W}^{2}) -\dfrac{9}{2}\dfrac{i}{(4\pi)^{2}}(m_{V}^{2}J(m_{W}^{2})+m_{W}^{2}J(m_{V}^{2}))\right].
\nonumber
\end{eqnarray}}

The second contribution \foreignlanguage{english}{$F_{WWV}(\phi_{c})$}
represents the amplitude:
{\footnotesize
\begin{eqnarray}
&&F_{WWV}(\phi_{c})=\int\dfrac{d^d q}{(2\pi)^{4}}  \int \dfrac{d^d p}{(2\pi)^{4}} \left[\frac{G_{WWV}^{2}}{2}\left(g^{\mu\rho} (q-p)_{\alpha}+g^{\mu \alpha} (-2p-q)_{\rho}+g^{\rho\alpha}
(p+2q)_{\mu }\right)\times \right. \nonumber \\ 
&&\left. \left(g^{\nu\sigma} (-p+q)_{\beta}+g^{\nu\beta}
(2 p+q)_{\sigma}+g^{\sigma\beta} (-p-2 q)_{\nu}\right) \left(\frac{g^{\mu\nu}
(-p-q)_{\beta} \left(-q_{\rho}\right) \left(q_{\sigma}\right) (p+q)_{\alpha}}{\left(p^2-m_V^2\right)q^2\left(q^2-m_W^2\right)(p+q)^2\left((p+q)^2-m_W^2\right)} \right. \right. \nonumber \\ && \left. \left. +\frac{-g^{\rho\sigma}
p_{\mu} \left(p_{\nu}\right) (-p-q)_{\beta} (p+q)_{\alpha}}{p^2\left(p^2-m_V^2\right)\left(q^2-m_W^2\right)(p+q)^2\left((p+q)^2-m_W^2\right)}+\frac{g^{\mu\nu}
g^{\rho\sigma} (-p-q)_{\beta} (p+q)_{\alpha}}{\left(p^2-m_V^2\right)\left(q^2-m_W^2\right)(p+q)^2\left((p+q)^2-m_W^2\right)} \right. \right. \nonumber \\ && \left. \left. +\frac{-g^{\alpha\beta}
p_{\mu} \left(p_{\nu}\right) \left(-q_{\rho}\right) \left(q_{\sigma}\right)}{p^2\left(p^2-m_V^2\right)q^2\left(q^2-m_W^2\right)\left((p+q)^2-m_W^2\right)}+\frac{-g^{\mu\nu}
g^{\alpha\beta}q_{\rho} \left(q_{\sigma}\right)}{\left(p^2-m_V^2\right)q^2\left(q^2-m_W^2\right)\left((p+q)^2-m_W^2\right)} \right. \right. \nonumber \\ && \left. \left. +\frac{-g^{\rho\sigma}
g^{\alpha\beta}p_{\mu} \left(p_{\nu}\right)}{p^2\left(p^2-m_V^2\right)\left(q^2-m_W^2\right)\left((p+q)^2-m_W^2\right)}+\frac{g^{\mu\nu}
g^{\rho\sigma} g^{\alpha\beta}}{\left(p^2-m_V^2\right)\left(q^2-m_W^2\right)\left((p+q)^2-m_W^2\right)} \right. \right. \nonumber \\ && \left. \left. + \frac{-p_{\mu}
\left(p_{\nu}\right) (-p-q)_{\beta} \left(-q_{\rho}\right) \left(q_{\sigma}\right)
(p+q)_{\alpha}}{p^2 \left(p^2-m_V^2\right)q^2\left(q^2-m_W^2\right)(p+q)^2\left((p+q)^2-m_W^2\right)}\right)\right],
\end{eqnarray}
}
with $G_{WW\gamma}^{2}=g^{2}sin^{2}\theta_{W}$ and $G_{WWZ}^{2}=g^{2}cos^{2}\theta_{W}$. By the repeated use of the relations (\ref{Substitution}) and after a tedious algebra, the above integral can be expressed in terms of the renormalized master integrals $\hat{I}$ and $\hat{J}$. When the master integrals have coefficients that are functions of the dimension $d$ one additionally need evaluate the  explicit finite contributions due to the transformations (\ref{ContractedForm}). The final result is:
{\small
\begin{eqnarray}
&&F_{WWV}(\phi_{c})= \nonumber \\ && \nonumber  \\ && -{G_{WWV}^{2}\over {4m_{W}^{4}m_{V}^{2}}}\{(m_{V}^{8}-16m_{W}^{6}m_{V}^{2}-12m_{W}^{4}m_{V}^{4}+32a^2m_{W}^{4} + 64a^2m_{W}^{2}m_{V}^{2})\hat{I}(m_{W}^{2},m_{W}^{2},m_{V}^{2}) \nonumber \\ && \nonumber \\ && -2((m_{W}^{4}-m_{V}^{4})^2+8(m_{W}^{2}-m_{V}^{2})^2m_{W}^{2}m_{V}^{2})\hat{I}(m_{W}^{2},m_{V}^{2},0) +2m_{W}^{8}\hat{I}(m_{W}^{2},0,0)+m_{V}^{8}\hat{I}(m_{V}^{2},0,0) \nonumber \\ && \nonumber \\ && -m_{V}^{2}\hat{J}(m_{W}^{2},m_{W}^{2})(31m_{W}^{4}-18m_{W}^{2}m_{V}^{2}-m_{V}^{4}) - 2m_{W}^{2}\hat{J}(m_{W}^{2},m_{V}^{2})(9m_{V}^{4}+4m_{W}^{2}m_{V}^{2}-m_{W}^{4})\nonumber \\ && \nonumber \\ &&+({4\over d}-1)m_{W}^{4}m_{V}^{2}J(m_{W}^{2},m_{W}^{2}) +({4\over d}-1)m_{W}^{4}m_{V}^{2}J(m_{W}^{2},m_{V}^{2}) +({4\over d}-1)m_{V}^{2}m_{W}^{4}J(m_{V}^{2},m_{W}^{2}) \nonumber \\ && \nonumber \\ &&  + 16(m_{W}^{2}-m_{V}^{2})^2m_{W}^{2}m_{V}^{2}\varepsilon I(m_{W}^{2},m_{V}^{2},0) - 8m_{V}^{2}(2m_{W}^{2}m_{V}^{2}-3m_{W}^{4})\varepsilon J(m_{W}^{2},m_{W}^{2})\nonumber \\ && \nonumber \\ &&  + 16m_{W}^{2}m_{V}^{4}\varepsilon J(m_{W}^{2},m_{V}^{2}) - 8m_{W}^{4}m_{V}^{2}({25\over 3}m_{W}^{2}+6m_{W}^{2}+6m_{V}^{2})\dfrac{i}{(4\pi)^{2}}J(m_{W}^{2}) \nonumber \\ && \nonumber \\ && - 4m_{V}^{2}m_{W}^{4}({25\over 3}m_{V}^{2}+6m_{W}^{2}+6m_{W}^{2})\dfrac{i}{(4\pi)^{2}}J(m_{V}^{2})
-32a^2m_{W}^{4}\varepsilon I(m_{W}^{2},m_{W}^{2},m_{V}^{2})\nonumber \\ && \nonumber \\ &&  -64a^2m_{V}^{2}m_{W}^{2}\varepsilon I(m_{V}^{2},m_{W}^{2},m_{W}^{2})\},   \label{FWWV}     
\end{eqnarray}
}
where $a^{2}=\lambda(m_{W}^{2},m_{W}^{2},m_{V}^{2})/4$, with $\lambda(x,y,z)$ the Kallen's function given by:
\begin{eqnarray*}
\lambda(m_{W}^{2},m_{W}^{2},m_{V}^{2})=m_{W}^{4}+m_{W}^{4}+m_{V}^{4}-2(m_{W}^{2}m_{W}^{2}
+m_{W}^{2}m_{V}^{2}+m_{V}^{2}m_{W}^{2}) \\
= m_{V}^{4}-4m_{W}^{2}m_{V}^{2}. ~~~~~~~~~~~~~~~~~~~~~~~~~~~~~~~~~~~~~~~~~~~~~~
\end{eqnarray*}
The last contribution $F_{u_{i}u_{i}V}(\phi_{c})$ represents the
diagrams \ref{EP-V-Sector} (c-f). The eight diagrams can be explicitly
obtained from the general amplitude:
\begin{eqnarray}
F_{u_{i}u_{j}V}(\phi_{c})=\frac{G_{u_{i}u_{j}V}^{2}}{\zeta}\int\frac{d^{d}q}{(2\pi)^{d}}\int\frac{d^{d}p}{(2\pi)^{d}}q_{\mu}p_{\nu}\left(\frac{-g^{\mu\nu}}{(q^{2}-m_{u_{i}}^{2})(p^{2}-m_{u_{j}}^{2})((q+p)^{2}-m_{V}^{2})}\right. ~~~~ \nonumber \\
\left.+\frac{(1-\zeta)(q+p)^{\nu}(q+p)^{\nu}}{(q^{2}-m_{u_{i}}^{2})(p^{2}-m_{u_{j}}^{2})
((q+p)^{2}-m_{V}^{2})((q+p)^{2}-\zeta m_{V}^{2})}\right), \nonumber \\
\end{eqnarray}
where $m_{u-}=m_{u+}=\zeta m_{W}$, $m_{u_{Z}}=\zeta m_{Z}$ and $m_{u_{\gamma}}=m_{\gamma}=0$. The apparent singularity in the Landau gauge, $\zeta=0$, is spurious. We must first reduce the integral in terms of $I$ and $J$ and then make the gauge election. The coefficients of the basis integrals are proportional to $\zeta$, and therefore the gauge parameter in the denominator vanishes.  
The result in terms of the integrals $I$ and $J$ when the vector boson mass $m_{V}$ is non-zero is:
\begin{eqnarray}
F_{u_{i}u_{j}V}(\phi_{c})=-\frac{G_{u_{i}u_{j}V}^{2}}{4}\left[m_{V}^{2}I(0,0,m_{V}^{2})+\frac{2i}{3}\frac{m_{V}^{2}}{(4\pi)^{2}}J(m_{V}^{2})\right]. \label{FuuV}
\end{eqnarray}
Whereas, in the particular case where $m_{V}=0$, as for the photon,
the function $F_{u_{i}u_{j}V}(\phi_{c})$ vanishes. The coefficients
$G_{u_{i}u_{j}V}^{2}$ have the explicit values:
\begin{eqnarray*}
G_{u_{\pm}u_{\pm}Z}^{2}=\frac{g^{2}cos^{2}\theta_{W}}{2}, & G_{u_{\pm}u_{\gamma}W}^{2}=g^{2}sin^{2}\theta_{W}, & G_{u_{\pm}u_{Z}W}^{2}=g^{2}cos^{2}\theta_{W}.
\end{eqnarray*}
Therefore, the contributions, due to the ghost fields, are obtained adding the eight diagrams to produce:
\begin{eqnarray}
&&-\frac{g^{2}}{2}\left[m_{W}^{2}I(0,0,m_{W}^{2})+\frac{2i}{3}\frac{m_{W}^{2}}{(4\pi)^{2}}J(m_{W}^{2})\right]-\frac{g^{2}}{4}cos^{2}\theta_{W}\left[m_{Z}^{2}I(0,0,m_{Z}^{2})+\frac{2i}{3}\frac{m_{Z}^{2}}{(4\pi)^{2}}J(m_{Z}^{2})\right]. \nonumber \\ &&
\end{eqnarray} 
Finally we observe that the QCD contributions to the pure gauge sector of the potential, given by the diagrams 
\begin{center}
\scalebox{0.5}{
\fcolorbox{white}{white}{
  \begin{picture}(551,148) (88,-117)
    \SetWidth{1.0}
    \SetColor{Black}
    \GluonArc(135,-17)(38.079,-157,203){7.5}{19}
    \GluonArc(225,-17)(38.079,-157,203){7.5}{19}
    \Vertex(180,-17){5}
    \GluonArc(405,-17)(38.079,-157,203){7.5}{19}
    \Gluon(360,-17)(450,-17){7.5}{7}
    \Vertex(360,-17){5}
    \Vertex(450,-17){5}
    \Vertex(540,-17){5}
    \Vertex(630,-17){5}
    \GluonArc(585,-17)(45,-180,0){7.5}{11}
    \Text(165,-122)[lb]{\Large{\Black{$F_{gg}(\phi_{c})$}}}
    \Text(385,-122)[lb]{\Large{\Black{$F_{ggg}(\phi_{c})$}}}
    \Text(565,-122)[lb]{\Large{\Black{$F_{u_{g}u_{g}g}(\phi_{c})$}}}
    \Line[dash,dashsize=2,arrow,arrowpos=0.5,arrowlength=5,arrowwidth=2,arrowinset=0.2](540,-17)(630,-17)
    \Arc[dash,dashsize=2,arrow,arrowpos=0.5,arrowlength=5,arrowwidth=2,arrowinset=0.2,clock](585,-17)(45,-180,-360)
  \end{picture}
}}
\end{center}
are equal to zero, because the gluon boson ($g$) and its ghost field ($u_{g}$) are massless, as a consequence:
\begin{eqnarray}
F_{gg}(\phi_{c})=F_{ggg}(\phi_{c})=F_{u_{g}u_{g}g}(\phi_{c})=0. \label{FQCD}
\end{eqnarray} 
The integrals $F_{gg}$, $F_{ggg}$ and $F_{u_{g}u_{g}g}$ are computed in analogous way to (\ref{FWV}), (\ref{FWWV}) and (\ref{FuuV}) respectively. You just to need put all vector masses to zero. 

\subsection{The Vector-Fermion Sector of the Effective Potential}

\begin{figure}
\begin{center}
\scalebox{0.6}{
\fcolorbox{white}{white}{
  \begin{picture}(635,192) (91,-75)
    \SetWidth{1.0}
    \SetColor{Black}
    \Line[arrow,arrowpos=0.5,arrowlength=5,arrowwidth=2,arrowinset=0.2](96,32)(192,32)
    \Arc[arrow,arrowpos=0.5,arrowlength=5,arrowwidth=2,arrowinset=0.2](144,32)(48,-0,180)
    \SetWidth{1.1}
    \PhotonArc(144,32)(48,-180,0){3}{8.5}
    \SetWidth{1.0}
    \Line[arrow,arrowpos=0.5,arrowlength=5,arrowwidth=2,arrowinset=0.2](272,32)(368,32)
    \Arc[arrow,arrowpos=0.5,arrowlength=5,arrowwidth=2,arrowinset=0.2](320,32)(48,-0,180)
    \SetWidth{1.1}
    \PhotonArc(320,32)(48,-180,0){3}{8.5}
    \SetWidth{1.0}
    \Line[arrow,arrowpos=0.5,arrowlength=5,arrowwidth=2,arrowinset=0.2](448,32)(544,32)
    \SetWidth{1.1}
    \PhotonArc(496,32)(48,-180,0){3}{8.5}
    \SetWidth{1.0}
    \Arc[arrow,arrowpos=0.5,arrowlength=5,arrowwidth=2,arrowinset=0.2](496,32)(48,-0,180)
    \Line[arrow,arrowpos=0.5,arrowlength=5,arrowwidth=2,arrowinset=0.2](624,32)(720,32)
    \Arc[arrow,arrowpos=0.5,arrowlength=5,arrowwidth=2,arrowinset=0.2](672,32)(48,-0,180)
    \GluonArc(672,32)(48,-180,0){5}{10}
    \Text(144,-80)[lb]{\Large{\Black{$(a)$}}}
    \Text(320,-80)[lb]{\Large{\Black{$(b)$}}}
    \Text(496,-80)[lb]{\Large{\Black{$(c)$}}}
    \Text(672,-80)[lb]{\Large{\Black{$(d)$}}}
    \Text(124,96)[lb]{\Black{$(e_{i}, u_{i}, d_{i})$}}
    \Text(295,96)[lb]{\Black{$(\nu_{i}, e_{i}, u_{i}, d_{i})$}}
    \Text(476,96)[lb]{\Large{\Black{$(\nu_{i}, u_{i})$}}}
    \Text(652,96)[lb]{\Large{\Black{$(u_{i}, d_{i})$}}}
    \Text(124,48)[lb]{\Black{$(e_{i}, u_{i}, d_{i})$}}
    \Text(295,48)[lb]{\Black{$(\nu_{i}, e_{i}, u_{i}, d_{i})$}}
    \Text(476,48)[lb]{\Large{\Black{$(e_{i}, d_{i})$}}}
    \Text(652,48)[lb]{\Large{\Black{$(u_{i},d_{i})$}}}
    \Text(144,0)[lb]{\Large{\Black{$\gamma$}}}
    \Text(320,0)[lb]{\Large{\Black{$Z$}}}
    \Text(496,0)[lb]{\Large{\Black{$W$}}}
    \Text(672,0)[lb]{\Large{\Black{$g$}}}
  \end{picture}
}}
\end{center}
\caption{\label{EP-FV}{\small Contributions to vector-scalar sector of the two-loop effective potential. The graph~(d) is the QCD contribution to $V_{FV}^{(2l)}$ and represents the higher contribution to the two-loop effective potential, $V^{(2l)}$. }}
\end{figure}
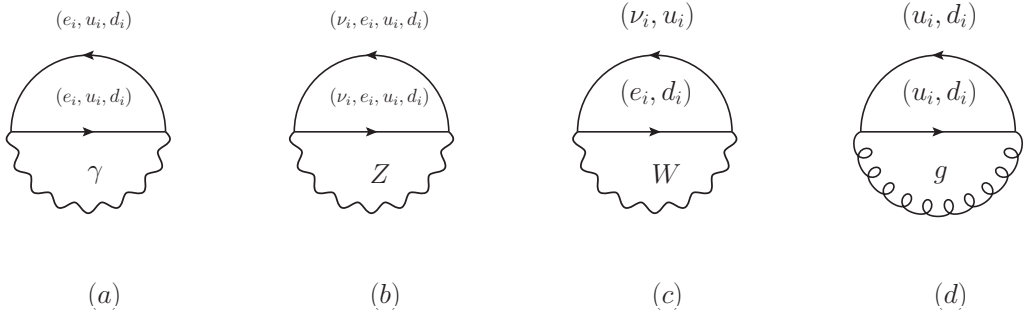

The last contribution to the effective potential comes from $V^{(2l)}_{FV}$. The diagrams contributing to this sector are represented in fig. \ref{EP-FV}. We write here only the explicit computation of the diagram of order $h_{t}^{4}g_{s}^{2}$ drawn in graph \ref{EP-FV} (d). The computation of all electroweak diagrams (fig. (a) - (c)) follows the same procedure seen in the QCD sector, without including the Casimir invariants and the colour algebra. For this reason, we just quote the final result. The diagram of order $h_{t}^{4}g_{s}^{2}$, contained in the QCD contribution to $V^{(2l)}_{FV}$, has the amplitude:
{\small
\begin{eqnarray}
V_{VF}^{(QCD)}=-\int\frac{d^{d}qd^{d}p}{512\pi^{8}}\left(\frac{g_{\mu\nu}}{(q^{2}-m_{t}^{2})(p^{2}-m_{t}^{2})(q+p)^{2}}-\frac{(1-\zeta)(q+p)_{\mu}(q+p)_{\nu}}{(q^{2}-m_{t}^{2})(p^{2}-m_{t}^{2})(q+p)^{2}(q+p)^{2}}\right) \nonumber \\
\times Tr\left[(m_{t}-\gamma^{\rho}q_{\rho})(-ig_{s}\gamma^{\nu}t_{ij}^{a})(m_{t}+\gamma^{\sigma}p_{\sigma})(-ig_{s}\gamma^{\mu}t_{ji}^{a})\right]. \nonumber \\
\end{eqnarray}}
Using the identity $t_{ij}^{a}t^{a}_{ji}=C_{2}(3)d(3)$ in $SU(3)$, the amplitude reduces to:
{\small
\begin{eqnarray*}
\frac{g_{s}^{2}C_{2}(3)d(3)}{512\pi^{8}}\int d^{d}qd^{d}p\left(\frac{g_{\mu\nu}}{(q^{2}-m_{t}^{2})(p^{2}-m_{t}^{2})(q+p)^{2}}-\frac{(1-\zeta)(q+p)_{\mu}(q+p)_{\nu}}{(q^{2}-m_{t}^{2})(p^{2}-m_{t}^{2})(q+p)^{2}(q+p)^{2}}\right)\\
\times Tr\left[(m_{t}-\gamma^{\rho}q_{\rho})\gamma^{\nu}(m_{t}+\gamma^{\sigma}p_{\sigma})\gamma^{\mu}\right].
\end{eqnarray*}}
Computing the trace 
\begin{eqnarray*}
Tr\left[(m_{t}-\gamma^{\rho}q_{\rho})\gamma^{\nu}(m_{t}+\gamma^{\sigma}p_{\sigma})\gamma^{\mu}\right]=4\left(m_{t}^{2}g^{\mu\nu}+g^{\mu\nu}p.q-q^{\mu}p^{\nu}-p^{\mu}q^{\nu}\right),
\end{eqnarray*}
we get:
\begin{eqnarray*}
V_{VF}^{(QCD)}=\frac{g_{s}^{2}C_{2}(3)d(3)}{512\pi^{8}}\int d^{d}qd^{d}p\left(\frac{4[dm_{t}^{2}+(d-2)p.q]}{(q^{2}-m_{t}^{2})(p^{2}-m_{t}^{2})(q+p)^{2}}\right. ~~~~~~~~~~~~~~~~~~~ \\ \left. -\frac{4[(q+p)^{2}m_{t}^{2}+(q+p)^{2}p.q-2(q+p).q(q+p).p]}{(q^{2}-m_{t}^{2})(p^{2}-m_{t}^{2})(q+p)^{2}(q+p)^{2}}\right).
\end{eqnarray*}
Applying the identity $p.q=\frac{1}{2}[(p+q)^{2}-p^{2}-q^{2}]$ and after a bit of algebra:
\begin{eqnarray*}
\frac{g_{s}^{2}C_{2}(3)d(3)}{2}4\left[(4\dfrac{~}{~}2\varepsilon)m_{t}^{2}I(m_{t}^{2},m_{t}^{2},0)+(1-\varepsilon)\left\{ J(m_{t}^{2},m_{t}^{2})-2m_{t}^{2}I(m_{t}^{2},m_{t}^{2},0)\right\} -\right.\\
\left.\left(m_{t}^{2}I(m_{t}^{2},m_{t}^{2},0)+\frac{1}{2}J(m_{t}^{2},m_{t}^{2})-m_{t}^{2}I(m_{t}^{2},m_{t}^{2},0)-\frac{1}{2}\left[J(m_{t}^{2},m_{t}^{2})\right]\right)\right].
\end{eqnarray*}
Finally, recalling that $C_{2}(3)d(3)=4$, we obtain:
\begin{eqnarray*}
V_{VF}^{(QCD)}=4g_{s}^{2}\left[4m_{t}^{2}I(m_{t}^{2},m_{t}^{2},0)+
2(1-\varepsilon)J(m_{t}^{2},m_{t}^{2})\right]. \label{EP-FV-QCD}
\end{eqnarray*}
The renormalized version is:
\begin{eqnarray}
V_{VF}^{(QCD)}=4g_{s}^{2}\left[4m_{t}^{2}\hat{I}(m_{t}^{2},m_{t}^{2},0)+2\hat{J}(m_{t}^{2},m_{t}^{2})-2\varepsilon J(m_{t}^{2},m_{t}^{2})\right].
\end{eqnarray}
The rest of the diagrams can be computed in analogous way. For instance, the three diagrams in fig. \ref{EP-FV} (a) gives the same contribution that $V_{FV}^{(QCD)}$ (eq. \ref{EP-FV-QCD}), but with a different coefficient that is function of the electric charge $e$. Taking into a account that the electric charge of the quarks type up ($u_i$) is $-\frac{2}{3}$, and the electric charge of the quarks type down ($d_i$) is $\frac{1}{3}$, one can easily find
\begin{eqnarray}
V_{VF}^{(\gamma)}=\dfrac{4}{3}e^{2}\left[4m_{t}^{2}\hat{I}(m_{t}^{2},m_{t}^{2},0)+2\hat{J}(m_{t}^{2},m_{t}^{2})-2\varepsilon J(m_{t}^{2},m_{t}^{2})\right],
\end{eqnarray}
here $V_{FV}^{(\gamma)}$ represents the sum of the three diagrams with a photonic internal line. The diagrams in the graph (b) are equal to:
\begin{eqnarray}
V_{VF}^{(Z)} = -3\sum_{f} (v_{f}^{2}+a_{f}^{2})F_f(m_f, m_f, m_Z)+(v_{f}^{2}-a_{f}^{2})G_f(m_f, m_f, m_Z), 
\end{eqnarray}
where the sum over $f$ is over all quarks and leptons, and $v_f$ and $a_f$ denote the vector
and axial couplings to the $Z$ boson. For the quarks type up and down, we have
\begin{eqnarray*}
v_{u_{i}}=\dfrac{g\left(1-\frac{8}{3}sin^{2}\theta \right)}{4cos\theta} ~,~ a_{u_{i}}=\dfrac{g}{4cos\theta} & ~;~ & v_{d_{i}}=\dfrac{g\left(1-\frac{4}{3}sin^{2}\theta \right)}{4cos\theta} ~,~ a_{d_{i}}=\dfrac{g}{4cos\theta}, 
\end{eqnarray*}
and for the leptons
\begin{eqnarray*}
 v_{e_{i}}=\dfrac{g\left(1-4sin^{2}\theta \right)}{4cos\theta}  ~,~ a_{e_{i}}=\dfrac{g}{4cos\theta} & ~;~ & v_{\nu _{i}}=\dfrac{g}{4cos\theta} ~,~ a_{\nu _{i}}=\dfrac{g}{4cos\theta}. 
\end{eqnarray*}
The new functions $F_f$ and $G_f$ has the explicit form:
\begin{eqnarray}
& F_{f}(m_{f}, m_{f}, m_{Z})= -\dfrac{1}{m_{Z}^{2}}\left\lbrace -(-2m_{Z}^{4}+2m_{f}^{2}m_{Z}^{2})\hat{I}(m_{f}^{2},m_{f}^{2},m_{Z}^{2})+2m_{Z}^{2}\hat{J}(m_{f}^{2},m_{f}^{2})\right. \nonumber \\ 
& -4m_{Z}^{2}\hat{J}(m_{f}^{2},m_{Z}^{2})+\dfrac{2}{3}m_{Z}^{2}(2m_{Z}^{2}-6m_{f}^{2})\dfrac{1}{(4\pi)^{2}}J(m_{Z}^{2}) \nonumber \\ & \left. -2\varepsilon m_{Z}^{2}\left[(m_{Z}^{2}-2m_{f}^{2})I(m_{f}^{2},m_{f}^{2},m_{Z}^{2})+J(m_{f}^{2},m_{f}^{2})-
2J(m_{f}^{2},m_{Z}^{2}  
) \right] \right\rbrace , 
\end{eqnarray}
and
\begin{eqnarray}
G_f(m_f, m_f, m_Z)=m_{f}^{2}\left(6\hat{I}(m_{f}^{2},m_{f}^{2},m_{Z}^{2})+4\dfrac{i}{(4\pi)^{2}}J(m_{Z}^{2})-4\varepsilon I(m_{f}^{2},m_{f}^{2},m_{Z}^{2}) \right).
\end{eqnarray}
Finally, the graph (c), in the limit where all fermion masses are zero except the top quark mass, gives the contribution:
\begin{eqnarray}
V_{VF}^{(W)}= -\dfrac{9}{2}g^{2}\left(-2m_{W}^{2}\hat{I}(0,0,m_{W}^{2}) - \dfrac{4i}{3(4\pi)^{2}}J(m_{W}^{2}) +2\varepsilon m_{W}^{2}I(0,0,m_{W}^{2}) \right) \nonumber \\
+\dfrac{3}{2}g^{2}\dfrac{1}{m_{t}^{2}} \left(-[m_{t}^{4}-2m_{W}^{4}+m_{t}^{2}m_{W}^{2}]\hat{I}(m_{t}^{2},0,m_{W}^{2})+m_{t}^{4}\hat{I}(m_{t}^{2},0,0) + \right. \nonumber \\ \left. (m_{t}^{2}-2m_{W}^{2})\hat{J}(m_{t}^{2},m_{W}^{2})+\dfrac{2}{3}m_{W}^{2}(2m_{W}^{2}-3m_{t}^{2})\dfrac{i}{(4\pi)^{2}}J(m_{W}^{2}) \right. \nonumber \\
\left. \dfrac{~}{~}2\varepsilon m_{W}^{2}[(m_{W}^{2}-m_{t}^{2})I(m_{t}^{2},0,m_{W}^{2})-J(m_{t}^{2},m_{W}^{2})] \right).
\end{eqnarray}
The fermion-vector contribution to the effective potential is the sum
\begin{eqnarray}
V_{VF}^{(2l)}=V_{VF}^{(\gamma)}+V_{VF}^{(Z)}+V_{VF}^{(W)}+V_{VF}^{(QCD)}.
\end{eqnarray}

\subsection{The Vector-Scalar Sector of the Effective Potential}

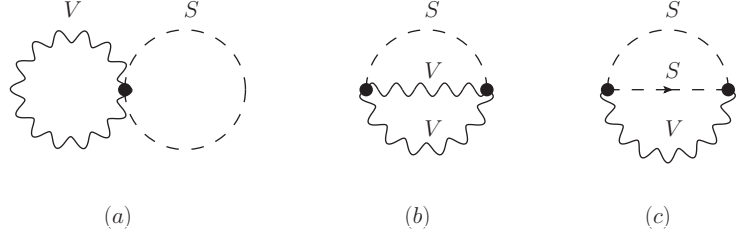
\begin{figure}
\begin{center}
\scalebox{0.5}{
\fcolorbox{white}{white}{
  \begin{picture}(544,186) (93,-79)
    \SetWidth{1.0}
    \SetColor{Black}
    \Vertex(180,21){5}
    \Vertex(360,21){5}
    \Vertex(450,21){5}
    \Vertex(540,21){5}
    \Vertex(630,21){5}
    \Text(165,-84)[lb]{\Large{\Black{$(a)$}}}
    \Text(390,-84)[lb]{\Large{\Black{$(b)$}}}
    \Text(570,-84)[lb]{\Large{\Black{$(c)$}}}
    \PhotonArc(140,21)(40.311,173,533){5}{14}
    \Arc[dash,dashsize=10,arrow,arrowpos=0.5,arrowlength=5,arrowwidth=2,arrowinset=0.2](225,21)(45,0,360)
    \Photon(360,21)(450,21){5}{5}
    \PhotonArc(405,21)(45,-180,0){5}{7.5}
    \Arc[dash,dashsize=10,clock](405,21)(45,-180,-360)
    \Line[dash,dashsize=10,arrow,arrowpos=0.5,arrowlength=5,arrowwidth=2,arrowinset=0.2](540,21)(630,21)
    \Arc[dash,dashsize=10,clock](585,21)(45,-180,-360)
    \PhotonArc(585,16.25)(45.25,173.974,366.026){5}{8.5}
    \Text(135,76)[lb]{\Large{\Black{$V$}}}
    \Text(225,76)[lb]{\Large{\Black{$S$}}}
    \Text(405,76)[lb]{\Large{\Black{$S$}}}
    \Text(585,76)[lb]{\Large{\Black{$S$}}}
    \Text(405,30)[lb]{\Large{\Black{$V$}}}
    \Text(405,-14)[lb]{\Large{\Black{$V$}}}
    \Text(585,30)[lb]{\Large{\Black{$S$}}}
    \Text(585,-14)[lb]{\Large{\Black{$V$}}}
  \end{picture}
}}
\end{center}
\caption{\label{EP-VS-Sector}{\small Contributions to vector-scalar sector of the two-loop effective potential.}}
\end{figure}

The last sector has the three contributions showed in fig. \ref{EP-VS-Sector}. The figure (a) represents the amplitude:
\begin{eqnarray}
F_{VS}(\phi_{c})=G_{VS}^{2}\int\dfrac{d^{d}p}{(2\pi)^{d}}\int\dfrac{d^{d}q}{(2\pi)^{d}}g_{\mu\nu}\left[\dfrac{g^{\mu\nu}}{(p^{2}-m_{S}^{2})(q^{2}-m_{V}^{2})}- \right.  ~~~~~~~~~~~~~~~~~~~~~~~\nonumber \\ \left. \dfrac{(1-\zeta)q^{\mu}q^{\nu}}{(p^{2}-m_{S}^{2})(q^{2}-m_{V}^{2})(q^{2}-\zeta m_{V}^{2})}\right].
\end{eqnarray}
This graph can be easily computed, and gives as result:
\begin{eqnarray}
F_{VS}(\phi_{c})=G_{VS}^{2}\left[3\hat{J}(m_{V}^{2},m_{S}^{2})-2\varepsilon J(m_{V}^{2},m_{S}^{2}) +2m_{V}^{2}\dfrac{i}{(4\pi)^{2}}J(m_{S}^{2})\right].
\end{eqnarray}
When one inserts the SM quartic vertices, the graph (a) generates seven Feynman diagrams. However, the diagram with an internal photonic line vanishes, the other six diagrams are computed in analogous way that $F_{VS}$, the main difference is the value of the coefficients $G_{VS}^{2}$. For the different contributions the coefficients are:
{\small
\begin{eqnarray*}
G_{ZH}^{2}=G_{ZG^{0}}^{2}=\dfrac{g^2}{8cos^{2}\theta} ~;~ G_{ZG^{\pm}}^{2}=\dfrac{g^2(1-2sin^{2}\theta)^{2}}{4cos^{2}\theta} ~;~ G_{WH}^{2}=G_{WG^{0}}^{2}=\dfrac{1}{4}g^2 ~;~ G_{WG^{\pm}}^{2}=\dfrac{1}{2}g^2. 
\end{eqnarray*}}
The diagram (b) represents the amplitude:
\begin{eqnarray}
&&F_{VV'S}(\phi_{c})=G_{VV'S}^{2}f_{VV'S} \nonumber \\
\nonumber \\
&&~~~~~~~~~~~~~ =G_{VV'S}^{2}\int\frac{d^{d}p}{(2\pi)^{d}}\int\frac{d^{d}q}{(2\pi)^{d}}g_{\mu\rho}g_{\nu\sigma}\left[\frac{g^{\mu\nu}g^{\rho\sigma}}{(p^{2}-m_{S}^{2})(q^{2}-m_{V}^{2})((p+q)^{2}-m_{V'}^{2})}\right. \nonumber \\
&&\left.-\frac{(1-\zeta)g^{\mu\nu}(q+p)^{\rho}(q+p)^{\sigma}}{(p^{2}-m_{S}^{2})(q^{2}-m_{V}^{2})((p+q)^{2}-m_{V'}^{2})((q+p)^{2}-\zeta m_{V'}^{2})}\right.\nonumber \\ && - \left. \frac{(1-\zeta)g^{\rho\sigma}q^{\mu}q^{\nu}}{(p^{2}-m_{S}^{2})(q^{2}-m_{V}^{2})((p+q)^{2}-m_{V'}^{2})(q^{2}-\zeta m_{V}^{2})}\right. \nonumber \\
&& \left.+\frac{(1-\zeta)q^{\mu}q^{\nu}(q+p)^{\rho}(q+p)^{\sigma}}{(p^{2}-m_{S}^{2})(q^{2}-m_{V}^{2})(q^{2}-\zeta m_{V}^{2})((p+q)^{2}-m_{V'}^{2})((q+p)^{2}-\zeta m_{V'}^{2})}\right].
\end{eqnarray}
Using the same techniques that the above sections we obtain $F_{VV'S}(\phi_{c})$ in terms of the basic integrals $I$ and $J$:
\begin{eqnarray*}
&&  f_{VV'S}(\phi_{c})= \\
&&{1\over {4m_{V}^{2}m_{V'}^{2}}}\{(10m_{V}^{2}m_{V'}^{2}+m_{S}^{4}+m_{V}^{4}+m_{V'}^{4}-2m_{V}^{2}m_{S}^{2}
-2m_{V'}^{2}m_{S}^{2}) \hat{I}(m_{V}^{2},m_{V'}^{2},m_{S}^{2}) \\ && ~+~(2m_{S}^{2}m_{V}^{2}-m_{V}^{4}-m_{S}^{4})\hat{I}(m_{V}^{2},m_{S}^{2},0)+ (2m_{S}^{2}m_{V'}^{2}-m_{V'}^{4}-m_{S}^{4})\hat{I}(m_{V'}^{2},m_{S}^{2},0) \\ && ~+~m_{S}^{4}\hat{I}(m_{S}^{2},0,0)-(m_{V}^{2}+m_{V'}^{2}-m_{S}^{2})\hat{J}(m_{V}^{2},m_{V'}^{2})+m_{V'}^{2}\hat{J}(m_{V}^{2},m_{S}^{2})+m_{V}^{2}\hat{J}(m_{V'}^{2},m_{S}^{2}) \\ && ~+~6m_{V}^{2}m_{V'}^{2}\dfrac{i}{(4\pi)^{2}}(J(m_{V}^{2})+J(m_{V'}^{2}))  -8m_{V}^{2}m_{V'}^{2}\varepsilon I(m_{V}^{2},m_{V'}^{2},m_{S}^{2})\}.
\end{eqnarray*}
Introducing the fields explicitly and evaluating the coefficients $G_{VV'S}^{2}$ one obtains:
{\small
\begin{eqnarray}
\sum_{VV'S}F_{VV'S}=g^{2}sin^{4}\theta m_{Z}^{2} f_{ZWG}+e^{2}m_{W}^{2} f_{W\gamma G}+\dfrac{1}{2}g^{2} m_{W}^{2} f_{WWH} + \dfrac{g^{2}}{4cos^{2}\theta}m_{Z}^{2}f_{ZZH}.
\end{eqnarray}}
The graph (c) represents the amplitude $F_{SS'V}(\phi_{c})=G_{VSS'}^{2}f_{SS'V}$, where:
{\small
\begin{eqnarray*}
f_{SS'V}(\phi_{c})=\int\dfrac{d^{d}p}{(2\pi)^{d}}\int\dfrac{d^{d}q}{(2\pi)^{d}}(p-q)_{\mu}(p-q)_{\nu}\left[\dfrac{-g^{\mu\nu}}{(p^{2}-m_{S}^{2})(q^{2}-m_{S'}^{2})((p+q)^{2}-m_{V}^{2})}\right. \nonumber \\
\left.-\dfrac{(1-\zeta)(q+p)^{\mu}(q+p)^{\nu}}{(p^{2}-m_{S}^{2})(q^{2}-m_{S'}^{2})((p+q)^{2}-m_{V}^{2})((q+p)^{2}-\zeta m_{V}^{2})}\right]. \nonumber
\end{eqnarray*}}
In terms of the basic integrals $I$ and $J$ we obtain:
{\small
\begin{eqnarray*}
f_{SS'V}={-1\over m_{V}^{2}}\{-4a^2\hat{I}(m_{S}^{2},m_{S'}^{2},m_{V}^{2})+(m_{S'}^{2}-m_{S}^{2})^2\hat{I}(m_{S}^{2},m_{S'}^{2},0)-(m_{S'}^{2}-m_{S}^{2}-m_{V}^{2})\hat{J}(m_{S}^{2},m_{V}^{2}) \\ - (m_{S}^{2}-m_{S'}^{2}-m_{V}^{2})\hat{J}(m_{S'}^{2},m_{V}^{2})-m_{V}^{2}\hat{J}(m_{S}^{2},m_{S'}^{2})
            +2m_{V}^{2}(m_{S}^{2}+m_{S'}^{2}-{1\over 3}m_{V}^{2})\dfrac{i}{(4\pi)^{2}}J(m_{V}^{2})\}.
\end{eqnarray*}}
Introducing the cubic vertices one produces five diagrams that can be reduced to:
{\small
\begin{eqnarray}
\sum_{VV'S} F_{VV'S}={g^2\over{8\cos^{2}\theta}}f_{HGZ}+{g^2(1-2
 \sin^{2}\theta)^2\over{8\cos^{2} \theta}}f_{GGZ}+{1\over 2}e^{2}f_{GG\gamma} + {1\over 4}g^2\left( f_{HGW} + f_{GGW}\right).
\end{eqnarray}}
Finally, the vector-scalar sector of the effective potential can be obtained as the sum of the terms:
\begin{eqnarray}
V_{VS}^{(2l)}=\sum_{SV}F_{SV} + \sum_{SS'V} F_{SS'V} + \sum_{VV'S} F_{VV'S}.
\end{eqnarray}

To obtain the $\overline{MS}$ effective potential we need sum all the above computed sectors. This potential is very important in our analysis. Its analytic expression is useful to compute the two-loop beta function of the coupling $\lambda$, the running $\lambda(\mu)$ from the RGEs, the effective coupling $\lambda_{eff}(\phi_{c})$ from the asymptotic behaviour of $V^{(2l)}(\phi_c)$ when $\phi_{c}\gg v$, and to obtain the tadpoles contribution to the threshold corrections to the boundary conditions over $\lambda$. However, the sum yields a very large analytic expression, therefore, is necessary the evaluation by an appropriate code to obtain a numerical result that is function of the renormalization scale $\bar{\mu}$ and of the classical field $\phi_c$. Due to all contributions to the potential $V^{(2l)}$ can be reduced to a superposition of the integrals $I(x,y,z)$ and $J(x,y)$, the potential has the following form:
\begin{eqnarray}
V^{(2l)} \approx c_{1}\phi_{c}^{4}+c_{2}^{j}\phi_{c}^{4}ln\left( \dfrac{m_{j}^{2}}{\bar{\mu}^{2}}\right)+c_{3}^{ij}\phi_{c}^{4}ln\left( \dfrac{m_{i}^{2}}{\bar{\mu}^{2}}\right)ln\left( \dfrac{m_{j}^{2}}{\bar{\mu}^{2}}\right),
\end{eqnarray}      
for high values of the Higgs field. We postpone the numerical details to the next chapter, where the potential and the tadpoles contribution are computed in the SZ renormalization scheme, and the numerical differences with the $\overline{MS}$ are presented.
  
\chapter{\label{cha:TwoLoopTadpoles} Tadpoles Contribution in SZ Scheme}

\lettrine{I}{ n} the first part of this chapter we compute the two-loop effective potential in the on-shell Sirlin Zucchini renormalization scheme, where the threshold corrections of the SM couplings are taken into account. A two-loop contribution can be deduced directly from the one-loop $\overline{MS}$ effective potential that we computed in the previous chapter, to get the latter we will perform the shifts over all couplings by imposing on them the matching conditions. In the second part we derive the two-loop tadpole contribution to the threshold relation between the Higgs quartic coupling $\lambda$ and the Higgs mass $m_{H}$, using two methods: from the first derivative of the Higgs effective potential with respect to the classical field and diagrammatically, using the Tarasov method automatized by the code TARCER.

\section{Matching Conditions and SZ Effective Potential}

\begin{figure}
\begin{center}
\scalebox{0.5}{
\fcolorbox{white}{white}{
  \begin{picture}(258,112) (255,-69)
    \SetWidth{1.0}
    \SetColor{Black}
    \Line[arrow,arrowpos=0.5,arrowlength=5,arrowwidth=2,arrowinset=0.2](256,-64)(512,-64)
    \GluonArc[clock](384,-64)(64,-180,-360){7.5}{16}
    \Vertex(318,-62){6}
    \Vertex(450,-62){6}
    \Text(384,22)[lb]{\Large{\Black{$g$}}}
    \Text(384,-50)[lb]{\Large{\Black{$t$}}}
  \end{picture}
}}
\end{center}
\caption{\label{QCDtopself} QCD contribution to the one-loop SZ effective potential.}
\end{figure}
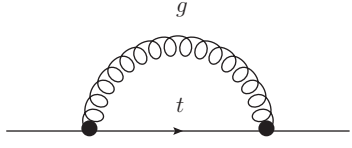

The effective potential computed in the Minimal Modified Subtraction Scheme $\overline{MS}$ is very well known up two-loop level \cite{Ford-Jones} \cite{Ford-Jack} \cite{Sher}, however the need of studying the vacuum stability as depending on the threshold corrections made the renormalization  scheme $\overline{MS}$ unsuitable and an on-shell scheme was preferred, the so called Sirlin-Zucchini (SZ) scheme. The two-loop effective potential in the SZ scheme can be obtained from the $\overline{MS}$ effective potential by making the appropriate shifts over the SM parameters, this procedure is straightforward but not trivial. The matching conditions obtained in Chapter \ref{cha:Sirlin-Zucchini}, useful in the evaluation of the boundary conditions of the RGI effective potential, can be applied to obtain these shifts. Let see how works. In the Landau gauge, the $\overline{MS}$ effective potential up two-loop level has the general form:
\begin{eqnarray}
V_{eff}^{\overline{MS}}=-\dfrac{1}{2}m^{2}(\bar{\mu})\phi_{c}^{2}+\dfrac{1}{4}\lambda(\bar{\mu}) \phi_{c}^{4}+V^{(1l)}(g_{i}(\bar{\mu}))+V^{(2l)}(g_{i}(\bar{\mu})),
\end{eqnarray}
where $V^{(1l)}$ is the one-loop $\overline{MS}$ potential given by eq. (\ref{eq:1l-SM-EP}), and $V^{(2l)}$ is the two-loop $\overline{MS}$ potential computed in the above chapter. To obtain the potential in the SZ scheme we need to apply the matching conditions like
\begin{eqnarray}
g_{i}(\overline{\mu})=g_{i(sz)}-\left. \delta ^{(1l)} g_{i(sz)}\right|_{fin} - \left. \delta ^{(2l)} g_{i(sz)}\right|_{fin} + \Delta _{g_{i}} \label{eq:match-cond}
\end{eqnarray}
over all $\overline{MS}$ parameters $g_{i}(\overline{\mu})$ contained in the potential $V_{eff}^{\overline{MS}}(\phi_{c})$. The additional term $\Delta _{g_{i}}$ is a two-loop
finite evanescent contribution coming from the $O(\varepsilon)$ part of the shifts when the SZ parameters entering the $1/\varepsilon$ in the SZ counterterm are expressed in terms of $\overline{MS}$ quantities. The subscript $fin$ denotes the finite part of the counterterm $\delta g_{i(sz)}$, obtained after subtracting the terms proportional to $(d-4)^{-1} + \frac{1}{2}(\gamma - ln(4\pi))$ i.e. after subtracting the $\overline{MS}$ counterterm $\delta g_{i}(\bar{\mu})$. Ignoring the L-loop terms, with $L\geq 3$, the contribution $V^{(2l)}(\phi_{c})$ remains equal to the computed in the above chapter, but with the couplings defined now in the SZ scheme:
\begin{eqnarray}
g_{i(sz)}=k_{i}\dfrac{G_{\mu}}{\sqrt{2}}m^{2}_{i},
\end{eqnarray}
where $m^{2}_i$ are the physical masses. Thus, in $V^{(2l)}(g_{i}(\bar{\mu}))$ we need to just relabel the couplings. From the potential $V^{(1l)}(g_{i}(\bar{\mu}))$ there is an additional two-loop contribution due to the one-loop correction to the coupling $\left. \delta ^{(1l)} g_{i(sz)}\right|_{fin}$ when one makes the shift of the parameters. The terms $\left. \delta ^{(2l)} g_{i(sz)}\right|_{fin}$  and $\Delta_{g_{i}}$ are disregarded at this order. To illustrate how the additional two-loop contribution emerges from the shift over the one-loop $\overline{MS}$ potential we begin by studying the QCD sector of $V^{(1l)}(\phi_{c})$, and then we generalize our results to the complete SM theory. The only QCD contribution to two-loop Higgs effective potential is through the Yukawa coupling constant $h_{t}(\overline{\mu})$ in the Yukawa sector potential:
\begin{eqnarray*}
V_{Y}^{(1)}(\phi_{c})=\frac{1}{64\pi^{2}}\left[-12T^{2}\left(log\frac{T}{\overline{\mu}^{2}}-\frac{3}{2}\right)\right] & ~;~ & T=\frac{1}{2}h_{t}^{2}(\overline{\mu})\phi^{2}_{c}.
\end{eqnarray*}
More precisely, through the gauge invariant one-loop correction to the $h_{t}(\overline{\mu})$ obtained from eq. (\ref{eq:match-cond})
\begin{eqnarray}
h_{t}^{2}(\overline{\mu})=h_{t(sz)}^{2}-\left.\delta^{(1l)}h_{t(sz)}^{2}\right|_{fin}+\dots , \label{YukawaCorr}
\end{eqnarray}
where $h_{t(sz)}$ and $h_{t}(\overline{\mu})$ are the renormalized Sirlin-Zucchini and $\overline{MS}$ parameters respectively. If we express the $\overline{MS}$ parameters in terms of the SZ parameters we obtain:
{\small
\begin{eqnarray*}
V_{Y}(\phi_{c})=\frac{1}{64\pi^{2}}\left[-12\left(\frac{1}{2}\left(h_{t(sz)}^{2}-\left.\delta^{(1l)}h_{t(sz)}^{2}\right|_{fin}\right)
\phi^{2}_{c}\right)^{2} ~~~~~~~~~~~~~~~~~~~~~~~~~~~~~~ \right.\\
\times \left. \left(log\frac{h_{t(sz)}^{2}\phi^{2}_{c}}{2\overline{\mu}^{2}}+log\left(1-\frac{\left.\delta^{(1l)}h_{t(sz)}^{2}\right|_{fin}}{h_{t(sz)}^{2}}\right)-\frac{3}{2}\right)\right],
\end{eqnarray*}
}
which in the perturbative regime is expanded as
\begin{eqnarray}
V_{Y}(\phi)=\frac{1}{64\pi^{2}}\left[-12\left(T_{(sz)}^{2}-\left[\delta^{(1)}T_{(sz)}^{2}\right]_{fin}+
\left(\left[\delta^{(1)}T_{(sz)}\right]_{fin}\right)^{2}\right)\right. \times ~~~~~~~~~~~~~~~~~
\nonumber \\
~~~~~~~~\left. \left(log\frac{T_{(sz)}}{\overline{\mu}^{2}}-\frac{3}{2}+\sum_{n=1}^{\infty}\frac{(-1)^{2n+1}}{n}\left(\frac{\left.\delta^{(1l)}h_{t(sz)}^{2}\right|_{fin}}{h_{t(sz)}^{2}}\right)^{n}\right)\right], \label{eq:V_Y-SZ}
\end{eqnarray}
where we have defined 
\begin{eqnarray}
\left[\delta^{(1)}T_{(sz)}^{2}\right]_{fin}=2T_{(sz)}\left[\delta^{(1)}T_{(sz)}\right]_{fin} & ; & \left[\delta^{(1)}T_{(sz)}\right]_{fin}=\frac{1}{2}\left.\delta^{(1l)}h_{t(sz)}^{2}\right|_{fin}\phi^{2}_{c}.
\end{eqnarray}
The validity of the perturbativity is up to the Planck scale because the results of recent analysis \cite{Jegerlehner} have confirmed that all coupling constants have smooth behaviour and are smaller than unity up to the Planck scale. Up two-loops level, the sum in eq. (\ref{eq:V_Y-SZ}) runs up $n=2$ and we have:
\begin{eqnarray}
V_{Y}(\phi)=\frac{1}{64\pi^{2}}\left[-12\left(T_{(sz)}^{2}\left(log\frac{T_{(sz)}}{\overline{\mu}^{2}}-\frac{3}{2}\right)+A\left[\delta^{(1)}T_{(sz)}^{2}\right]_{fin}+B\left(\left[
\delta^{(1)}T_{(sz)}\right]_{fin}\right)^{2}\right)\right], \nonumber
\end{eqnarray}
with 
\begin{eqnarray*}
A=-\frac{1}{2}-\left(log\frac{T_{(sz)}}{\overline{\mu}^{2}}-\frac{3}{2}\right) & {\rm and} &
B=\frac{3}{2}+\left(log\frac{T_{(sz)}}{\overline{\mu}^{2}}-\frac{3}{2}\right).
\end{eqnarray*}
The terms with the coefficients A and B are the additional two-loop contributions obtained from the one-loop potential $V^{(1l)}(g_{i}(\bar{\mu}))$ after shifting the $\overline{MS}$ parameters. According to equation (\ref{eq:1l-MS-ht}) the one-loop correction to the tree-level value of $h_{t}$ is given in terms of the W self-energy at zero momentum $A_{WW}(0)$, the contribution $E$ and the top mass counterterm $\delta m_{t}$, all of them to one-loop level. To determine only the QCD contribution, we must compute the top quark self-energy with a gluon internal line contained in $\delta m_{t}$ and depicted in fig. (\ref{QCDtopself}). 
The value of this diagram is:
\begin{eqnarray}
\Sigma^{QCD}(p)=  \int \dfrac{d^{4}q}{(2\pi)^{4}}(ig_{s})^{2}\gamma^{\mu}t^{a}\frac{i(q^{\sigma}\gamma_{\sigma}+m_{t})}{q^{2}-m_{t}^{2}}\gamma_{\mu}t^{a}\frac{-i}{(p+q)^{2}}.
\end{eqnarray}
The product of the SU(3) group matrices $t^{a}$ equals the quadratic Casimir operator, $t^{a}t^{a}=C_{2}(3)I$. The Dirac matrix structure can be reduced using the contraction identities in d dimensions: 
\begin{eqnarray}
\gamma^{\mu}(q^{\sigma}\gamma_{\sigma})\gamma_{\mu}=-(d-2)\gamma_{\sigma}q^{\sigma} & ; & \gamma^{\mu}\gamma_{\mu}=d.
\end{eqnarray}
Consequently, $\Sigma^{QCD}(p)$ can be reduced to
\begin{eqnarray}
\Sigma^{QCD}(p)= \mu^{4-d}g_{s}^{2}C_{2}(3) \int \dfrac{d^{d}q}{(2\pi)^{d}}\left[\frac{(d-2)q^{\sigma}\gamma_{\sigma}}{(q^{2}-m_{t}^{2})(p+q)^{2}}+\dfrac{dm_{t}}{(q^{2}-m_{t}^{2})(p+q)^{2}}\right].
\end{eqnarray}
The tensor and scalar two-point integrals can be reduced to linear combinations of Passarino-Veltman functions. Therefore, 
\begin{eqnarray}
\Sigma^{QCD}(p)=\frac{g_{s}^{2}C_{2}(3)}{(2\pi)^{d}}i\pi^{2} \times ~~~~~~~~~~~~~~~~~~~~~~~~~~~~~~
~~~~~~~~~~~~~~~~~~~~~~~~~~~~~~~~~~~~~~~~~~~~~~~~~~~ \\
\nonumber \\
\left[(d-2)\gamma_{\sigma}p^{\sigma}\left(\frac{A_{0}(m_{t}^{2})-(m_{t}^{2}+p^{2})B_{0}(p^{2},0,m_{t}^{2})}{2p^{2}}\right) + d\,m_{t}B_{0}(p^{2},0,m_{t}^{2})\right]. \nonumber 
\end{eqnarray}
The top quark mass counterterm defined in equation (\ref{eq:1l-MS-ht})
is expressed in terms of the self-energy diagrams $\delta^{(1)}m_{t}=Re\left[\Sigma_{1}(\widetilde{m_{t}})\right]$ with $\widetilde{m_{t}}=m_{t}-i\Gamma_{t}/2$, where $m_{t}$ is the pole mass of the unstable fermion and $\Gamma_{t}$ its width (a general discussion on the mass counterterm for unstable fermions is presented in \cite{kniehl}). Then, if we identify the position $\gamma_{\sigma}p^{\sigma}=m_{t}$ and if we introduce $d=4-\varepsilon$ and expand around $d=4$, the
following expression is derived:
{\scriptsize 
\begin{eqnarray*}
\Sigma^{QCD}(m_{t})=\frac{g_{s}^{2}C_{2}(3)}{(4\pi)^{2}}i\left[(2-\varepsilon)m_{t}\left(\frac{m_{t}^{2}\left(\frac{1}{\bar{\varepsilon}}+1-ln\frac{m_{t}^{2}}{\mu^{2}}\right)-2m_{t}^{2}\left(\frac{1}{\bar{\varepsilon}}+2-ln\frac{m_{t}^{2}}{\mu^{2}}\right)}{2m_{t}^{2}}\right)+(4-\varepsilon)m_{t}\left(\frac{1}{\bar{\varepsilon}}+2-ln\frac{m_{t}^{2}}{\mu^{2}}\right)\right],
\end{eqnarray*}
}
where $\frac{1}{\bar{\varepsilon}}=\frac{2}{\varepsilon}-\gamma+ln(4\pi)$. The last relation can be simplified to:
\begin{eqnarray}
\Sigma^{QCD}(m_{t})=\frac{g_{s}^{2}C_{2}(3)}{(4\pi)^{2}}im_{t}\left[3\left(\frac{1}{\bar{\varepsilon}}+1-ln\frac{m_{t}^{2}}{\mu^{2}}\right)+1+O(\varepsilon)\right].
\end{eqnarray}
Therefore, the QCD contribution to finite term $\left.\delta^{(1)}h_{t(os)}^{2}\right|_{fin}$, using $C_{2}(3)=\frac{4}{3}$, is:
\begin{eqnarray}
\left.\delta^{(1)}h_{t(os)}^{2}\right|_{fin}^{QCD} = 4\left(\frac{G_{\mu}}{\sqrt{2}}m_{t}^{2}\right)\left[\frac{\delta^{(1)}m_{t}^{2}}{m_{t}^{2}}\right]_{fin}^{QCD} = ~~~~~~~~~~~~~~~~~~~~~~~~~~~~~~~~~~~~~~~~~~~~~~~~~~
\\
\nonumber \\
\nonumber \\
4\left(\frac{G_{\mu}}{\sqrt{2}}\right)\frac{g_{s}^{2}m_{t}^{2}}{(4\pi)^{2}}\left[8\left(1-ln\frac{m_{t}^{2}}{\mu^{2}}\right)+\frac{8}{3}+O(\varepsilon)\right]. \nonumber
\end{eqnarray}
\begin{table}
\begin{center}
\begin{tabular}{ccc}
\hline 
$m_{W}=$ & $80.384\pm0.014$ GeV & Pole mass of the W boson\tabularnewline
$m_{Z}=$ & $91.1876\pm0.0021$ GeV & Pole mass of the Z boson\tabularnewline
$m_{H}=$ & $125.66\pm0.34$ GeV & Pole mass of the higgs\tabularnewline
$m_{t}=$ & $173.36\pm0.65\pm0.3$ GeV & Pole mass of the top quark\tabularnewline
$v=$ & $246.21971\pm0.00006$ GeV & EW vacuum expectation value\tabularnewline
$g_{s}(m_{Z})=$ & $0.1184\pm0.0007$  & $\overline{MS}$ gauge $SU(3)_{c}$ coupling\tabularnewline
\hline 
\end{tabular}
\\
\caption{Input values of the SM observables used to fix
the SM fundamental parameters. \label{tab:SMvalues}}
\end{center}
\end{table}
Note that the order of the above contribution is $\sim g_{s}^{2}h_{t}^{2}$, therefore the additional term $ \left[\delta^{(1)}T_{(sz)}^{2}\right]_{fin}=2T_{(sz)}
\left[\delta^{(1)}T_{(sz)}\right]_{fin} \sim g_{s}^{2}h_{t}^{4}$ has the same order of the QCD contributions to the two-loop potential $V_{FV}^{(2l)}$ computed in the above chapter (section 5.2.3). This confirms that the terms with coefficients A and B are actually two-loop terms. Moreover the Sirlin-Zucchini (SZ) top Yukawa coupling is fixed using the pole mass $(m_{t}=h_{t(sz)}v/\sqrt{2}=173.36\pm0.71~ GeV)$. Using the input values of the SM observables at $\mu=m_{t}$, listed in Table \ref{tab:SMvalues}, the QCD corrections give the value: {\small $\left[\delta^{(1)}h_{t(sz)}^{2}\right]_{fin}^{QCD}=0.091144$}. If one includes all contributions in eq. (\ref{eq:1l-MS-ht}) evaluated at $\mu=m_{t}$ one finds the value: {\small $\left.\delta^{(1)}h_{t(sz)}^{2}\right|_{fin}=0.089118$}, the QCD corrections above computed are the dominant two-loop contributions obtained from $V^{(1l)}$, besides has an opposite sign to the EW corrections. 

The same computations can be done for the most relevant SM parameters. If one applies the matching conditions over the one-loop $\overline{MS}$ effective potential, $V^{(1l)}$:
\begin{eqnarray*}
V_{sz}^{(1l)}(\phi_{c})=\frac{1}{64\pi^{2}}\sum_{j}(-1)^{2d_{j}}(2d_{j}+1)\left(a_{j}\left(m^{2}-\left.\delta^{(1l)}m^{2}\right|_{fin}\right)+b_{j}\left(g_{j}-\left.\delta^{(1l)}g_{j}\right|_{fin}\right)\phi_{c}^{2}\right)^{2}\\
\times\left[ln\frac{a_{j}m^{2}+b_{j}g_{j}\phi_{c}^{2}}{\mu^{2}}+ln\left[1-\frac{\left(a_{j}\left.\delta^{(1l)}m^{2}\right|_{fin}+b_{j}\left.\delta^{(1l)}g_{j}\right|_{fin}\phi_{c}^{2}\right)}{a_{j}m^{2}+b_{j}g_{j}\phi_{c}^{2}}\right]-c_{j}\right],
\end{eqnarray*}
where the couplings $g_{j}$, the coefficients $a_{j}$, $b_{j}$, $c_{j}$ and $d_{j}$ are defined in table \ref{table:def}. 
\begin{table}
\begin{center}
\scalebox{0.9}{
\begin{tabular}{|c|c|c|c|c|c|}
\hline 
$j$ & $g_{j}$ & $a_{j}$ & $b_{j}$ & $c_{j}$ & $d_{j}$\tabularnewline
\hline 
\hline 
$1$ & $\lambda$ & $1$ & $3$ & $3/2$ & $0$\tabularnewline
\hline 
$2$ & $\lambda$ & $1$ & $1$ & $3/2$ & $1$\tabularnewline
\hline 
$3$ & $g^{2}$ & $0$ & $1/4$ & $5/6$ & $5/2$\tabularnewline
\hline 
$4$ & $G^{2}$ & $0$ & $1/4$ & $5/6$ & $1$\tabularnewline
\hline 
$5$ & $h_{t}^{2}$ & $0$ & $1/2$ & $3/2$ & $11/2$\tabularnewline
\hline 
\end{tabular}
}
\end{center}
\caption{\label{table:def}Definitions of the coefficients $a_j$, $b_j$, $c_j$ and $s_j$.}
\end{table}
Here $G^{2}=(g^{2}+g'^{2})$, $d_{j}$ is a coefficient depending of the particle $j$, and $c_{j}$ are constants depending of the renormalization scheme. Using the definitions: $T_{j}=a_{j}m^{2}+b_{j}g_{j}\phi_{c}^{2}$,
\begin{eqnarray*}
\left.\delta^{(1l)}T_{j}^{2}\right|_{fin}=2T_{j}\left(\left.\delta^{(1l)}T_{j}\right|_{fin}\right) & {\rm and} & \left.\delta^{(1l)}T_{j}\right|_{fin}=a_{j}\left.\delta^{(1l)}m^{2}\right|_{fin}+b_{j}\left.\delta^{(1l)}g_{j}\right|_{fin}\phi_{c}^{2},
\end{eqnarray*}
and applying the matching conditions over the one-loop $\overline{MS}$ effective potential, the SZ potential $V_{sz}^{(1l)}(\phi_{c})$ can be written as:
\begin{eqnarray}
V^{(1l)}(g_{i(sz)})+V_{new}^{(2l)}=\frac{1}{64\pi^{2}}\sum_{j}(-1)^{2d_{j}}(2d_{j}+1)\left[T_{j}^{2}\left(ln\frac{a_{j}m^{2}+b_{j}g_{j}\phi_{c}^{2}}{\mu^{2}}-c_{j}\right) ~~~~~~~~~~~~ \right. \\ \left. ~+~ A_{j}\left(\left.\delta^{(1l)}T_{j}^{2}\right|_{fin}\right)+B_{j}\left(\left.\delta^{(1l)}T_{j}\right|_{fin}\right)^{2}\right] \nonumber
\end{eqnarray}
with the new constants:
\begin{eqnarray*}
A_{j}=-\frac{1}{2}-\left(ln\frac{a_{j}m^{2}+b_{j}g_{j}\phi_{c}^{2}}{\mu^{2}}-c_{j}\right) & {\rm and} & B_{j}=\frac{3}{2}+\left(ln\frac{a_{j}m^{2}+b_{j}g_{j}\phi_{c}^{2}}{\mu^{2}}-c_{j}\right).
\end{eqnarray*}
Using the complete analytical expressions for the one-loop corrections $\delta^{(1l)}g_{i(sz)}$ and $\delta^{(1l)}m^{2}$, exposed explicitly in Appendix \ref{app-counterterms}, we obtain the numerical results at $\bar{\mu}=m_{t}$ presented in table \ref{table:num}.
\begin{table}
\begin{center}
\scalebox{0.8}{
\begin{tabular}{|c|c|c|c|c|}
\hline 
$j$ & $g_{j}$ & $\left.\delta^{(1l)}g_{j}\right|_{fin}$ & $T_{j}/(GeV)^{2}$ & $\left.\delta^{(1l)}T_{j}\right|_{fin}/(GeV)^{2}$\tabularnewline
\hline 
\hline 
$1$ & $0.13023$ & $0.00144$ & $(125.66)^{2}+0.39069\phi_{c}^{2}$ & $-1806.99+0.00432\phi_{c}^{2}$\tabularnewline
\hline 
$2$ & $0.13023$ & $0.00144$ & $(125.66)^{2}+0.13023\phi_{c}^{2}$ & $-1806.99+0.00144\phi_{c}^{2}$\tabularnewline
\hline 
$3$ & $0.42633$ & $0.00705$ & $0.10658\phi_{c}^{2}$ & $0.00176\phi_{c}^{2}$\tabularnewline
\hline 
$4$ & $0.54863$ & $2.81\times10^{-4}$ & $0.13715\phi_{c}^{2}$ & $7.02932\times10^{-5}\phi_{c}^{2}$\tabularnewline
\hline 
$5$ & $0.99144$ & $0.08912$ & $0.49572\phi_{c}^{2}$ & $0.04456\phi_{c}^{2}$\tabularnewline
\hline 
\end{tabular}}
\caption{\label{table:num}Two-loop contribution obtained from the one-loop potential $V^{(1l)}$ by shifting the $\overline{MS}$ parameters at $\bar{\mu}=m_t$.}
\end{center}
\end{table}
The tree-level values of the couplings $g_{i}$ are obtained from the experimental values of the observables listed in table \ref{tab:SMvalues}.

Finally, from the three level potential we obtain two-loop contributions due to the corrections $\left.\delta^{(2l)}\lambda_{(sz)}\right|_{fin}$ and $\Delta_{\lambda}$, but the shift is trivial here. For high values of the Higgs field, we can ignore the quadratic terms in $\phi_{c}$ and the SZ potential has the following form
\begin{eqnarray}
V^{SZ}_{eff} \approx d_{1}\phi_{c}^{4}+d_{2}^{j}\phi_{c}^{4}ln\left( \dfrac{m_{j}^{2}}{\bar{\mu}^{2}}\right)+d_{3}^{ij}\phi_{c}^{4}ln\left( \dfrac{m_{i}^{2}}{\bar{\mu}^{2}}\right)ln\left( \dfrac{m_{j}^{2}}{\bar{\mu}^{2}}\right). \label{V2lsz}
\end{eqnarray}     
The coefficients $d_1$, $d_{2}^{j}$ and $d_{3}^{ij}$ have a simplified expression when we take into account only the dominant contributions from the strong and the top
Yukawa couplings at two-loop level. The potential reduces to
\begin{eqnarray}
V^{SZ}_{eff}\approx d_{1}\phi_{c}^{4}+d_{2}^{5}\phi_{c}^{4}ln\left( \dfrac{m_{5}^{2}}{\bar{\mu}^{2}}\right)+d_{3}^{55}\phi_{c}^{4}ln^{2}\left( \dfrac{m_{5}^{2}}{\bar{\mu}^{2}}\right),
\end{eqnarray}
where we use $m_{5}=m_{t}$ and
{\small
\begin{eqnarray}
 && d_1 = \lambda_{(sz)} - \frac{1}{64\pi^{2}}(-1)^{2s_{5}}(2s_{5}+1)\times \left[ b_{5}^{2}g_{5}^{2}c_{5} -\left(\dfrac{1}{2} + c_{5}\right)h_{t}^{2}\left[\delta^{(1l)}h_{t(sz)}^{2}\right]_{fin}^{QCD} \right. \\ && \left. ~~~~~~ + \left(\dfrac{3}{2}-c_{5}\right)\dfrac{1}{4}\left( \left[ \delta^{(1l)}h_{t(sz)}^{2}\right]_{fin}^{QCD}\right)^{2} \right]  + \dfrac{1}{(4\pi)^{4}}\left[72h_{t}^{4}g_{s}^{2}-\dfrac{3}{2}h_{t}^{6}\left(23+\dfrac{\pi^{2}}{3}\right) \right], \nonumber
\end{eqnarray}}
{\small
\begin{eqnarray}
~~~ d_{2}^{j=5}=(-1)^{2s_{5}}(2s_{5}+1)b_{5}^{2}g_{5}^{2}+ h_{t}^{2}\left[\delta^{(1l)}h_{t(sz)}^{2}\right]_{fin}^{QCD} +\dfrac{1}{4}\left( \left[ \delta^{(1l)}h_{t(sz)}^{2}\right]_{fin}^{QCD}\right)^{2} \\ +~\dfrac{1}{(4\pi)^{4}}\left[24h_{t}^{6}-64h_{t}^{4}g_{s}^{2} \right] \nonumber
\end{eqnarray}}
and
{\small
\begin{eqnarray}
d_{3}^{55}=\dfrac{1}{(4\pi)^{4}}\left[24g_{s}^{2}h_{t}^{4}-\dfrac{9}{2}h_{t}^{6}\right].
\end{eqnarray}}
From the above expressions we can easily see that the new two-loop terms obtained from $V^{(1l)}(g_{i}(\bar{\mu}))$ change the coefficients $c_1$ and $c_2$ of the $\overline{MS}$ effective potential $V_{eff}^{\overline{MS}}$. The complete two-loop SZ effective potential is useful for two reasons. From one side, one can obtain the instability condition by evaluating the positivity of the effective quartic coupling 
\begin{eqnarray}
\lambda_{eff}^{SZ}(\phi _{c})=4d_{1}+4d_{2}^{j}ln\left( \dfrac{m_{j}^{2}}{\bar{\mu}^{2}}\right)+4d_{3}^{ij}ln\left( \dfrac{m_{i}^{2}}{\bar{\mu}^{2}}\right)ln\left( \dfrac{m_{j}^{2}}{\bar{\mu}^{2}}\right),
\end{eqnarray}
obtained from the potential (\ref{V2lsz}) in the high energy limit $\phi_c \gg m_W$, where we can make the approximation:
\begin{eqnarray}
V_{eff}^{SZ}\approx \dfrac{\lambda_{eff}^{SZ}(\phi_c)}{4}\phi_{c}^{4}.
\end{eqnarray} 
By other side, from the potential is possible compute the tadpoles contribution, $T^{(2)}$, to the two-loop threshold corrections of the running coupling $\lambda(\bar{\mu})$.

\section{The Tadpoles Up Two-Loops Level}

The sum of the two-loop Tadpoles ($T^{(2)}$) is the most important computation in this work. The determination of $T^{(2)}$ plays a relevant role in the vacuum stability analysis. The major source of uncertainties in the (meta)stability conditions emerge from the theoretical uncertainty in the determination of the pole mass $m_{t}$, the translation of the Monte Carlo top-quark mass parameter to the pole mass scheme introduces an additional theoretical uncertainty of the order of 1 GeV. This uncertainty is obtained when only QCD NNLO radiative corrections are applied. In order to achieve percent level precision the EW part as well as mixed EW$\times$QCD corrections have to be included in a systematic way. The EW corrections depends of the vacuum expectation value $v(\mu)$, its renormalization requires the explicit inclusion of tadpoles in order to make large the EW contribution (the tadpoles represent the higher EW corrections). The EW corrections have opposite sign relative to the QCD contributions, in this sense, the inclusion of the tadpoles is relevant to make small the total SM correction. By other side, the tadpoles topologies together with the quantum correction $\Delta r$ are the only contributions to the threshold of $\lambda$ that can be analytically computed, however the larger radiative corrections comes from $T$ and the Higgs self-energy $\Pi_{HH}(m_{H}^{2})$ \cite{Degrassi}.\\ 
\begin{figure}
\begin{center}
\scalebox{0.5}{
\fcolorbox{white}{white}{
  \begin{picture}(714,106) (143,-107)
    \SetWidth{1.0}
    \SetColor{Black}
    \Line(144,-54)(208,-54)
    \Arc(256,-54)(48,0,360)
    \Arc(352,-54)(48,0,360)
    \Line(448,-54)(528,-54)
    \Arc[arrow,arrowpos=0.5,arrowlength=5,arrowwidth=2,arrowinset=0.2](576,-54)(48,0,360)
    \Line(576,-6)(576,-102)
    \Line(672,-54)(752,-54)
    \Arc(800,-54)(50.596,342,702)
    \Line(752,-54)(848,-54)
    \Vertex(208,-54){4}
    \Vertex(304,-54){4}
    \Vertex(528,-54){4}
    \Vertex(576,-6){4}
    \Vertex(576,-102){4}
    \Vertex(748,-54){4}
    \Vertex(852,-54){4}
  \end{picture}
}}
\end{center}
\caption{\label{T2topo} Topologies of the two-loop tadpoles.}
\end{figure}
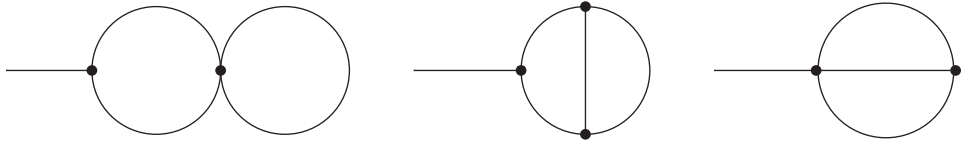

Let see now the calculation of $T^{(2)}$. In this section we decided to list in details the explicit computations due to the absence of analytic expressions in the literature. Without inserting the vertices and the fields of the SM, there are three types of 1PI topologies that contribute to the two-loop tadpoles $T^{(2)}$ shown in fig. (\ref{T2topo}). In a general $R_{\zeta}$ gauge, when we insert the vertices of the SM to the topologies (the SM vertices are shown in Appendix \ref{AppSM}), there are $184$ diagrams that contribute to $T^{(2)}$. These diagrams can be assorted in $25$ generic classes, where we have classified the fields as scalar (S), fermion (F), gauge (V) and ghost fields (G). In the language of FeynArts means without consider the mass and charge for each particle or the members of each class. We can divide these $25$ classes in $5$ sectors according to their diagrammatic origin, in the same way as for the effective potential:
\begin{eqnarray}
T^{(2)} = T_{S}^{(2)}+T_{SF}^{(2)}+T_{SV}^{(2)}+T_{FV}^{(2)}+T_{V}^{(2)}.
\end{eqnarray}
The tadpoles contribution can be computed from the first derivative of the effective potential with respect to the classical field, $\phi_{c}$:
\begin{eqnarray}
\left. \frac{\partial V_{eff}^{(2l)}}{\partial(\phi_{c})_{j}}\right|_{\phi_{c}=v}=-T_{j}^{(2l)}, \label{def:tadpole}
\end{eqnarray}
this derivative can be formally do it using the chain rule:
\begin{eqnarray}
T_{j}^{(2l)}=-\left. \frac{\partial V_{eff}^{(2l)}}{\partial m_{ik}^{2}}\frac{\partial m_{ik}^{2}}{\partial(\phi_{c})_{j}}\right|_{\phi_{c}=v}
\end{eqnarray}
over the potential, computed in terms of the master integrals $B^{(d)}_{11}(q=0)$, $A_{1}^{(d)}(q=0)$, $J^{(d)}_{111}(q=0)$ and $V_{1111}(q=0)$ exposed in section \ref{sec:MI}. This implies take the derivatives of the loop functions:
\begin{eqnarray}
\dfrac{\partial}{\partial (\phi_{c})_{j}}B(m_{i}^{2},m_{k}^{2})=A(m_{i}^{2},m_{l}^{2},m_{k}^{2})\left(\dfrac{\partial m_{il}^{2}}{\partial (\phi_{c})_{j}}+\dfrac{\partial m_{kl}^{2}}{\partial (\phi_{c})_{j}}\right),
\end{eqnarray}
\begin{eqnarray}
\dfrac{\partial}{\partial (\phi_{c})_{j}}J(m_{i}^{2},m_{k}^{2})=B(m_{i}^{2},m_{l}^{2})J(m_{k}^{2})\dfrac{\partial m_{il}^{2}}{\partial (\phi_{c})_{j}}+J(m_{i}^{2})B(m_{k}^{2},m_{l}^{2})\dfrac{\partial m_{kl}^{2}}{\partial (\phi_{c})_{j}},
\end{eqnarray}
and
\begin{eqnarray}
\dfrac{\partial}{\partial (\phi_{c})_{j}}I(m_{i}^{2},m_{k}^{2},m_{l}^{2})=V(m_{i}^{2},m_{s}^{2},m_{k}^{2},m_{l}^{2})\dfrac{\partial m_{is}^{2}}{\partial (\phi_{c})_{j}} ~~~~~~~~~~~~~~~~~~~~~~~~~~~~~~~~  \\ ~~+~~V(m_{k}^{2},m_{s}^{2},m_{i}^{2},m_{l}^{2})\dfrac{\partial m_{ks}^{2}}{\partial (\phi_{c})_{j}}+V(m_{l}^{2},m_{s}^{2},m_{i}^{2},m_{k}^{2})\dfrac{\partial m_{ls}^{2}}{\partial (\phi_{c})_{j}}, \nonumber
\end{eqnarray}
where $m_{ij}$ are the undiagonalized masses defined in Appendix \ref{AppSM}, $m_{j}=a_{j}m^{2}+b_{j}g_{j}\phi_{c}^{2}$ are the diagonalized field-dependent masses, and where we have defined the new functions
\begin{eqnarray}
&& B(x,y)=\dfrac{\mu^{4-d}}{(4\pi)^{d/2}}B_{11}^{(d)}(q=0)=\dfrac{1}{y-x}\left(J(x)-J(y)\right), \\
&& A(x,y,z)=\dfrac{B(x,z)-B(x,y)}{y-z}, ~~~~~ {\rm and } \\
&& V(x,y,z,w)=\dfrac{1}{y-x} \left(I(x,z,w) - I(y,z,w)\right).
\end{eqnarray} 
To obtain the above expressions we just need to use the trick 
\begin{eqnarray}
\dfrac{\partial}{\partial (\phi_{c})_{j}}\left(\dfrac{1}{k^{2}-m_{ik}^{2}}\right)=\left(\dfrac{1}{k^{2}-m_{il}^{2}}\right)\dfrac{\partial m^{2}_{ls}}{\partial (\phi_{c})_{j}}\left(\dfrac{1}{k^{2}-m_{sk}^{2}}\right)
\end{eqnarray}
over the propagators of the loop functions. Moreover, we use the explicit relations of the first derivatives regarding to the classical field $\phi_{c}$ of the loop functions $J(x)$, $\hat{J}(x,y)$, $\hat{I}(x,y,z)$, $\varepsilon J(x,y)$ and $\varepsilon I(x,y,z)$ to obtain a numerical evaluation of the renormalized quantity $T^{(2)}$ straightforwardly from the derivative of the 1PI potential $V^{(2)}$. For $J(x)$ we have:
\begin{eqnarray*}
\dfrac{\partial}{\partial\phi_{c}}J(x)=\dfrac{\partial}{\partial x}\left[i\dfrac{x}{(4\pi)^{2}}\left(\varepsilon \left(1+\dfrac{\zeta(2)}{2}\right) +\dfrac{1}{\varepsilon}+1-\overline{ln}(x)-\varepsilon\overline{ln}(x)\right)\right]\dfrac{\partial x}{\partial\phi_{c}}.
\end{eqnarray*}
Here $x,~y$ or $z$ represent the field-dependent masses $m_{j}^{2}=a_{j}m^{2}+b_{j}g_{j}\phi_{c}^{2}$. Solving the derivative:
\begin{eqnarray}
\dfrac{\partial}{\partial\phi_{c}}J(x)=\dfrac{i}{(4\pi)^{2}}\left[\left(\dfrac{1}{\varepsilon}-\overline{ln}(x)\right)-\varepsilon\left(\overline{ln}(x)-\dfrac{\zeta(2)}{2}\right)\right]\dfrac{\partial x}{\partial\phi_{c}},\label{eps0}
\end{eqnarray}
where we include the evanescent term proportional to $\varepsilon$ because is necessary to obtain the next derivative:
{\footnotesize
\begin{eqnarray*}
&&\dfrac{\partial J(x,y)}{\partial\phi_{c}}=\dfrac{\partial J(x)}{\partial\phi_{c}}J(y)+\dfrac{\partial J(y)}{\partial\phi_{c}}J(x), ~~~~~ {\rm with}\\ &&
\frac{\partial J(x)}{\partial\phi_{c}}J(y)=-\frac{y}{(4\pi)^{4}}\left[\frac{1}{\varepsilon}+\frac{1}{\varepsilon^{2}}-\frac{1}{\varepsilon}(\overline{ln}(x)+\overline{ln}(y))+\overline{ln}(x)\overline{ln}(y)-2\overline{ln}(x)-\overline{ln}(y)+\zeta(2)+1\right]\frac{\partial x}{\partial\phi_{c}},\\ &&
\frac{\partial J(y)}{\partial\phi_{c}}J(x)=-\frac{x}{(4\pi)^{4}}\left[\frac{1}{\varepsilon}+\frac{1}{\varepsilon^{2}}-\frac{1}{\varepsilon}(\overline{ln}(y)+\overline{ln}(x))+\overline{ln}(x)\overline{ln}(y)-2\overline{ln}(y)-\overline{ln}(x)+\zeta(2)+1\right]\frac{\partial y}{\partial\phi_{c}}.
\end{eqnarray*}}
From the above equation we can obtain the derivative of the renormalized quantity $\hat{J}(x,y)$:
\begin{eqnarray*}
\dfrac{\partial}{\partial\phi_{c}}\hat{J}(x,y)=\dfrac{\partial J(x,y)}{\partial\phi_{c}}-\dfrac{i}{(4\pi)^{2}\varepsilon}\left(\dfrac{\partial y}{\partial\phi_{c}}J(x)+y\dfrac{\partial J(x)}{\partial\phi_{c}}+\dfrac{\partial x}{\partial\phi_{c}}J(y)+x\dfrac{\partial J(y)}{\partial\phi_{c}}\right).
\end{eqnarray*}
Making the subtraction we find:
\begin{eqnarray}
\dfrac{\partial}{\partial\phi_{c}}\hat{J}(x,y)=\dfrac{y}{(4\pi)^{4}}\left[\dfrac{1}{\varepsilon^{2}}-\overline{ln}(x)\overline{ln}(y)+\overline{ln}(x)\right]\dfrac{\partial x}{\partial\phi_{c}} + (x\leftrightarrow y), \label{eps4} 
\end{eqnarray}
We need also the derivative of the term:
\begin{eqnarray}
\frac{\partial}{\partial\phi_{c}}\varepsilon J(x,y)=-\frac{y}{(4\pi)^{4}}\left[1+\frac{1}{\varepsilon}-(\overline{ln}(x)+\overline{ln}(y))\right]\frac{\partial x}{\partial\phi_{c}} + (x\leftrightarrow y), \label{eps3}
\end{eqnarray}
Let see now the explicit expressions that involve the derivative of 
{\small
\begin{eqnarray*}
\hat{I}(x,y,z)=\frac{1}{(4\pi)^{4}}\left[-\frac{c}{2\varepsilon^{2}}+\frac{1}{2\varepsilon}c+\frac{1}{2}\left(L_{2}-4L_{1}+(y+z-x)\overline{ln}(y)\overline{ln}(z) \right. \right. ~~~~~~~~~~~~~~~~~~ \\ \left. \left.  +(z+x-y)\overline{ln}(z)\overline{ln}(x)\right)+\frac{1}{2}\left((y+x-z)\overline{ln}(y)\overline{ln}(x)+\xi(x,y,z)+5c\right)\right],
\end{eqnarray*}}
where {\small$\xi(x,y,z)=8b\left[L(\theta_{x})+L(\theta_{y})+L(\theta_{z})-\frac{\pi}{2}ln2\right]$} with {\small$b=\frac{1}{2}(2xy+2yz+2zx-x^{2}-y^{2}-z^{2})^{1/2}$}. Besides
$c=x+y+z$ and {\small$L_{m}=x\overline{ln}^{m}x+y\overline{ln}^{m}y+z\overline{ln}^{m}z$}. The necessary derivatives are:
{\small
\begin{eqnarray}
\frac{\partial}{\partial\phi_{c}}\hat{I}(x,y,z)=\frac{1}{(4\pi)^{4}}\left[-\frac{1}{2\varepsilon^{2}}+\frac{1}{2\varepsilon}+\frac{1}{2}\left(\left(\overline{ln}(x)-1\right)^{2}+f(x,y,z)+\frac{\partial}{\partial x}\xi(x,y,z)\right)\right]\frac{\partial x}{\partial\phi_{c}}+c.p. \nonumber  \\ \label{eps1}
\end{eqnarray}}
with 
\begin{eqnarray*}
f(x,y,z)=\overline{ln}(x)\overline{ln}(y)+\overline{ln}(x)\overline{ln}(z)-\overline{ln}(y)\overline{ln}(z)+\overline{ln}(y)+\overline{ln}(z),
\end{eqnarray*}
and where $c.p.$ means cyclic permutations of the three indices $x$,
$y$ and $z$. The derivative of $\xi(x,y,z)$ amounts:
\begin{eqnarray*}
\frac{\partial}{\partial x}\xi(x,y,z)=\frac{2}{b}(y+z-x)\left[L(\theta_{x})+L(\theta_{y})+L(\theta_{z})-\frac{\pi}{2}ln2\right].
\end{eqnarray*}
Finally, we also need the derivative:
\begin{eqnarray}
\frac{\partial}{\partial\phi_{c}}\varepsilon I(x,y,z)=\frac{1}{(4\pi)^{4}}\frac{\partial}{\partial\phi_{c}}\left[\frac{c}{2\varepsilon}+\left(\frac{3c}{2}-L_{1}\right)\right],  ~~~~~~~~~~~~~~~~~~~~~~~ \label{eps2} \\
=\frac{1}{(4\pi)^{4}}\left[\frac{1}{2\varepsilon}+\frac{1}{2}-\overline{ln}x\right]\frac{\partial x}{\partial\phi_{c}}+(x\rightarrow y)+(x\rightarrow z). \nonumber
\end{eqnarray}
For the integrals with identical masses or with vanishes arguments, it is useful to have the finite terms of the next expressions:
{\small
\begin{eqnarray}
&&(4\pi)^{2}\hat{B}(x,x)= - \overline{ln}x, \\
&&(4\pi)^{4}\hat{I}(x,0,0)=\dfrac{1}{2}x\left( \overline{ln}x \right)^{2} - 2x\overline{ln}x + \dfrac{5}{2}x+\dfrac{\pi^{2}}{6}x, \\
&&(4\pi)^{4}\hat{I}(x,x,0)=x \left(5 - 4\overline{ln}x + \overline{ln}^{2}x \right), \\
&&(4\pi)^{4}\hat{I}(x,y,0)= (y-x)\left[Li_{2}\left( \dfrac{y}{x} \right) - ln\left( \dfrac{x}{y} \right)\overline{ln}(y-x)+\dfrac{1}{2}\left( \overline{ln}x \right)^{2} - \dfrac{\pi^2}{6}\right] \\ && ~~~~~~~~~~~~~~~~~~~~~ + \dfrac{5}{2}(x+y) - 2x\overline{ln}x - 2y\overline{ln}y + x\overline{ln}x\overline{ln}y, \nonumber 
\end{eqnarray}}
in addition to the trivial identities $J(0)=0$, $J(x,0)=0$ and $I(0,0,0)=0$.
We will verify analytically the results obtained by the diagrammatically computation of the tadpoles. For accomplish this check we use the TARCER code \cite{Tarcer}. Let us develop our method of calculation with a brief example, the scalar sector of the two-loop tadpoles. 

\subsection{The Pure Scalar Sector}
\begin{figure}
\begin{center}
\scalebox{0.4}{
\fcolorbox{white}{white}{
  \begin{picture}(777,201) (109,-134)
    \SetWidth{1.0}
    \SetColor{Black}
    \Line[dash,dashsize=10](110,-19)(175,-19)
    \Arc[dash,dashsize=10,arrow,arrowpos=0.5,arrowlength=5,arrowwidth=2,arrowinset=0.2](215,-19)(40,0,360)
    \Arc[dash,dashsize=10,arrow,arrowpos=0.5,arrowlength=5,arrowwidth=2,arrowinset=0.2](295,-19)(40,0,360)
    \Vertex(175,-19){5}
    \Vertex(255,-19){5}
    \Line[dash,dashsize=10](420,-19)(495,-19)
    \Arc[dash,dashsize=10,arrow,arrowpos=0.5,arrowlength=5,arrowwidth=2,arrowinset=0.2](540,-19)(50,0,360)
    \Line[dash,dashsize=10](540,31)(540,-69)
    \Vertex(490,-19){5}
    \Vertex(540,31){5}
    \Vertex(540,-69){5}
    \Line[dash,dashsize=10](680,-19)(875,-19)
    \Arc[dash,dashsize=10,arrow,arrowpos=0.5,arrowlength=5,arrowwidth=2,arrowinset=0.2](830,-19)(50,0,360)
    \Vertex(780,-19){5}
    \Vertex(880,-19){5}
    \Text(120,1)[lb]{\Large{\Black{$H$}}}
    \Text(425,1)[lb]{\Large{\Black{$H$}}}
    \Text(695,-4)[lb]{\Large{\Black{$H$}}}
    \Text(205,40)[lb]{\Large{\Black{$S$}}}
    \Text(295,40)[lb]{\Large{\Black{$S$}}}
    \Text(205,-85)[lb]{\Large{\Black{$S$}}}
    \Text(495,39)[lb]{\Large{\Black{$S$}}}
    \Text(495,-85)[lb]{\Large{\Black{$S$}}}
    \Text(515,-14)[lb]{\Large{\Black{$S$}}}
    \Text(600,-14)[lb]{\Large{\Black{$S$}}}
    \Text(830,46)[lb]{\Large{\Black{$S$}}}
    \Text(830,-4)[lb]{\Large{\Black{$S$}}}
    \Text(830,-49)[lb]{\Large{\Black{$S$}}}
    \Text(230,-139)[lb]{\Large{\Black{$(a)$}}}
    \Text(535,-139)[lb]{\Large{\Black{$(b)$}}}
    \Text(825,-139)[lb]{\Large{\Black{$(c)$}}}
  \end{picture}
}}
\end{center}
\caption{\label{SsectorT2} Scalar sector of the tadpoles topologies.}
\end{figure}
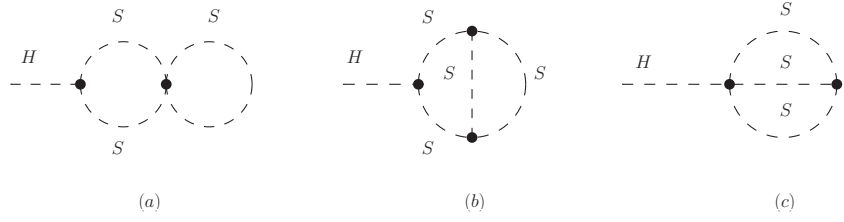

We begin studying the scalar sector of $T^{(2l)}$. We derive its computation by two different methods, we compute the first derivative of the scalar sector of the potential $V^{(2l)}_{S}$, see eq. (\ref{EffpotS}), and then we verify diagrammatically the expression obtained. In the Landau gauge, this implies compute the next derivatives:
\begin{eqnarray}
&& \frac{\partial I(m_{H}^{2},m_{H}^{2},m_{H}^{2})}{\partial\phi_{c}}=3\frac{\partial I(m_{H}^{2},m_{H}^{2},m_{H}^{2})}{\partial m_{H}^{2}}\frac{\partial m_{H}^{2}}{\partial\phi_{c}}, \label{dIdphi1} \\ && \frac{\partial I(m_{H}^{2},m_{G}^{2},m_{G}^{2})}{\partial\phi_{c}} = \frac{\partial I(m_{H}^{2},m_{G}^{2},m_{G}^{2})}{\partial m_{H}^{2}}\frac{\partial m_{H}^{2}}{\partial\phi_{c}} + 2\frac{\partial I(m_{H}^{2},m_{G}^{2},m_{G}^{2})}{\partial m_{G}^{2}}\frac{\partial m_{G}^{2}}{\partial\phi_{c}}, \label{dIdphi2}\\ && \frac{\partial J(m_{H}^{2},m_{H}^{2})}{\partial\phi_{c}}=2\frac{\partial J(m_{H}^{2},m_{H}^{2})}{\partial m_{H}^{2}}\frac{\partial m_{H}^{2}}{\partial\phi_{c}}. \label{dJdphi}
\end{eqnarray} 
To resolve the first two integrals (\ref{dIdphi1}) and (\ref{dIdphi2}) we need to use the recurrence relation 
\begin{eqnarray}
(d-3)I(x,y,y)=\dfrac{\partial J(y)}{\partial y}(J(x)-J(y)) ~~~~~~~~~~~ \label{recrel}\\ ~~~+~~~ 2x\dfrac{\partial}{\partial x}I(x,y,y) + x\dfrac{\partial}{\partial y}I(x,y,y), \nonumber 
\end{eqnarray}
and the explicit expressions of the derivative of the field-dependent masses:
\begin{eqnarray}
\dfrac{\partial m_{H}^{2}}{\partial \phi _{c}}=6\lambda \phi_{c} ~~~~~ {\rm and} ~~~~~ 
\dfrac{\partial m_{G}^{2}}{\partial \phi _{c}}=2\lambda \phi_{c}.
\end{eqnarray}
Using the eq. (\ref{recrel}) we obtain straightforwardly 
\begin{eqnarray}
\dfrac{\partial I(m_{H}^{2},m_{H}^{2},m_{H}^{2}) }{\partial \phi_{c}}= 3\times \dfrac{(d-3)}{3m_{H}^{2}}I(m_{H}^{2},m_{H}^{2},m_{H}^{2})\times 3\lambda \phi_{c}. \label{R1}
\end{eqnarray}
The second derivative (\ref{dIdphi2}) requires a bit more of algebra. From the relation (\ref{recrel}) we obtain:
\begin{eqnarray}
\frac{\partial I(m_{H}^{2},m_{G}^{2},m_{G}^{2})}{\partial m_{G}^{2}}=\dfrac{(d-3)}{m_{H}^{2}}I(m_{H}^{2},m_{G}^{2},m_{G}^{2})-2\frac{\partial I(m_{H}^{2},m_{G}^{2},m_{G}^{2})}{\partial m_{H}^{2}}.
\end{eqnarray}
Replacing the above equation in eq. (\ref{dIdphi2}), we can write:
\begin{eqnarray}
\dfrac{\partial I(m_{H}^{2},0,0) }{\partial \phi_{c}}=-2\lambda \phi _{c}\dfrac{\partial I(m_{H}^{2},0,0) }{\partial m_{H}^{2}}+\dfrac{(d-3)}{m_{H}^{2}}I(m_{H}^{2},0,0)4\lambda \phi_{c}.
\end{eqnarray}
Note that we first apply the derivative and then, when the derivative regard to $m_{G}^{2}$ is solved, we make the Landau gauge election. Computing explicitly the derivative $\partial I(m_{H}^{2}, 0, 0)/\partial m_{H}^{2}$~\footnote{We must be careful in don't use the recurrence relation (\ref{recrel}) because is true only for non-zero masses.}
\begin{eqnarray}
\dfrac{\partial I(m_{H}^{2},0,0) }{\partial m_{H}^{2}}=\dfrac{(d-3)}{m_{H}^{2}}I(m_{H}^{2},0,0),
\end{eqnarray}   
we obtain:
\begin{eqnarray}
\frac{\partial I(m_{H}^{2},0,0)}{\partial\phi_{c}}= \dfrac{(d-3)}{3m_{H}^{2}}I(m_{H}^{2},0,0)\times 6\lambda \phi_{c}. \label{R2}
\end{eqnarray}
The third derivative (\ref{dJdphi}) can be easily obtained by using the relation:
\begin{eqnarray}
\dfrac{\partial J(m^2)}{\partial m^{2}} = \dfrac{(d-2)}{2m^{2}}J(m^{2})
\end{eqnarray} 
The result is:
\begin{eqnarray}
\frac{\partial J(m_{H}^{2},m_{H}^{2})}{\partial\phi_{c}}= \dfrac{(d-2)}{2m_{H}^{2}}\left( J(m_{H}^{2}) \right)^{2} \times 12 \lambda \phi _{c}. \label{R3}
\end{eqnarray}
Finally to obtain the scalar sector of the tadpoles in the Landau gauge, $T^{(2)}_{S}$ we must to apply the equation (\ref{def:tadpole}) over the scalar potential $V^{(2)}_{S}$, that is:
\begin{eqnarray}
T^{(2l)}_{S}=-\dfrac{\partial }{\partial \phi_{c}}\left[ - 3\lambda^{2}\phi_{c}^{2}\left( I(m_{H}^{2},m_{H}^{2},m_{H}^{2}) + I(m_{H}^{2},m_{G}^{2},m_{G}^{2})\right) ~ + ~ \dfrac{3}{4}\lambda \left( J(m_{H}^{2},m_{H}^{2}) \right)\right]_ {\phi_{c}=v}
\end{eqnarray}
after to apply the derivative, and using the equations (\ref{R1}), (\ref{R2}) and (\ref{R3}), we obtain the expression, 
\begin{eqnarray}
T_{S}^{(2l)}=\dfrac{9}{64}\dfrac{g^{3}m_{H}^{2}}{m_{W}^{3}}(d-2)J(m_{H}^{2})^{2}+\dfrac{3}{16}\dfrac{g^{3}m_{H}^{4}}{m_{W}^{3}}I(m_{H}^{2},0,0) ~~~~~~~~~~~ \nonumber \\
+~~\dfrac{3}{32}\dfrac{g^{3}m_{H}^{4}}{m_{W}^{3}}(3d-7)I(m_{H}^{2},m_{H}^{2},m_{H}^{2}).
\end{eqnarray}
Diagrammatically the above expression can be verify using the {\tt TARCER} code. When one inserts the scalar fields over the topologies in fig. (\ref{SsectorT2}), one obtain 17 Feynman diagrams. This amplitudes are computed one by one in Appendix \ref{AppTarcer}. In the language of TARCER, the integral $I(m_{H}^{2},m_{H}^{2},m_{H}^{2})={\rm K}_{\{1,M_H\}\{1,0\}\{1,0\}}$.
\begin{figure}
\begin{center}
\scalebox{0.8}{
\fcolorbox{white}{white}{
  \begin{picture}(425,69) (157,-140)
    \SetWidth{1.0}
    \SetColor{Black}
    \Arc(208,-107)(32,0,360)
    \Arc[arrow,arrowpos=0.5,arrowlength=5,arrowwidth=2,arrowinset=0.2](384,-107)(32,0,360)
    \Arc(544,-107)(32,0,360)
    \Line[dash,dashsize=10](304,-107)(352,-107)
    \Line[dash,dashsize=10](464,-107)(512,-107)
    \COval(350,-106)(7,7)(90.0){Black}{White}\Line(345.05,-106)(354.95,-106)\Line(350,-101.05)(350,-110.95)
    \COval(574,-106)(7,7)(180.0){Black}{White}\Line(574,-101.05)(574,-110.95)\Line(578.95,-106)(569.05,-106)
    \Text(144,-120)[lb]{\Large{\Black{$\frac{\partial }{\partial \phi_{c}}$}}}
    \Text(259,-110)[lb]{\Large{\Black{$=$}}}
    \Text(434,-110)[lb]{\Large{\Black{$+$}}}
    \Text(322,-92)[lb]{\Large{\Black{$H$}}}
    \Text(490,-92)[lb]{\Large{\Black{$H$}}}
    \COval(238,-106)(7,7)(180.0){Black}{White}\Line(238,-101.05)(238,-110.95)\Line(242.95,-106)(233.05,-106)
  \end{picture}
}}
\end{center}
\caption{\label{ST2counter}Diagrams to remove the sub-divergences in $T^ {(2l)}$}
\end{figure}
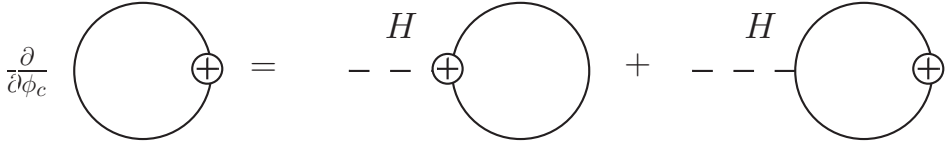

The above expression are not renormalized. To remove the divergences diagrammatically we must insert the one-loop sub-divergences depicted in fig. (\ref{ST2counter}) and then subtract the overall divergences with the standard $\overline{MS}$ procedure. We note that, the tadpoles counterterms can be obtained too, as the first derivative of the vacuum bubbles counterterms used to remove the sub-divergences in the effective potential, it is valid for all sectors of the Tadpoles, the continuous line in fig. (\ref{ST2counter}) can be replaced with any of the fields H, G, Z, W, F or U. However the sum of the all amplitudes obtained by replacing the fields in the diagrams (\ref{ST2counter}) must be distributed in some of the specific sectors of $T^{(2l)}$. For instance, to renormalize the sum $T_{S}^{(2)}+T_{SF}^{(2)}+T_{SV}^{(2)}$ we must insert the counterterms of fig. (\ref{ST2counter}) with the fermion (F), scalar (S) and vector (V) fields and choose the appropriate superposition of some of its terms. This is a nontrivial procedure, require of an exhaustive and tedious algebra and the use of some computer assistance. By other side, the diagrammatic relation shows in fig. (\ref{ST2counter}) have an enormous advantage, we verify with TARCER that the first derivative of the potential $V^{(2l)}_{eff}$ is actually the sum of the tadpoles $T^{(2l)}$, but TARCER computes $T^{(2l)}$ with the explicit Laurent expansion in terms of the poles $1/\varepsilon^{j}$, that is without removing the divergences. In this sense, the verification with TARCER of the relation (\ref{def:tadpole}) is incomplete. But this does not imply a problem because we can apply the first derivative with respect to the classical field over the renormalized quantities of the effective potential. Thus, the use of the diagrammatic relation of the fig. (\ref{ST2counter}) is equivalent to apply the relations (\ref{eps0}-\ref{eps2}). The first method is very efficient to remove the divergences. Using the relations (\ref{eps0}-\ref{eps2}) we obtain, after a bit of algebra, the explicit renormalized value of $T^ {(2l)}_{S}$ evaluated at the electroweak vacuum ($\phi_{c}=v$): 
{\small
\begin{eqnarray}
T_{S}^{(2l)}=\frac{3\lambda m_{H}^{2}}{2(4\pi)^{4}v}\left[\frac{41}{2}m_{H}^{2}\overline{ln}^{2}(m_{H}^{2})-14m_{H}^{2}\overline{ln}m_{H}^{2}+m_{H}^{2}\left(-11\sqrt{3}Cl_{2}\left(\frac{2\pi}{3}\right)+\frac{5}{6}\pi^{2}+26\right)\right].
\end{eqnarray}
}
We now give the explicit results obtained with our code for the other sectors of $T^{(2l)}$. The computation follows the methodology exposed above. This computation was done in the Landau gauge, for this reason all vertex couplings with one scalar and two ghost fields are zero. 

\subsection{The Yukawa Sector}
\begin{figure}
\begin{center}
\scalebox{0.4}{
\fcolorbox{white}{white}{
  \begin{picture}(665,330) (0,0)
    \SetWidth{1.0}
    \SetColor{Black}
    \Line[dash,dashsize=10](110,110)(175,110)
    \Vertex(175,110){5}
    \Line[dash,dashsize=10](435,110)(510,110)
    \Vertex(505,110){5}
    \Vertex(560,160){5}
    \Vertex(560,60){5}
    \Vertex(230,165){5}
    \Vertex(230,55){5}
    \Arc[dash,dashsize=10,arrow,arrowpos=0.5,arrowlength=5,arrowwidth=2,arrowinset=0.2](560,110)(50,0,360)
    \Arc[arrow,arrowpos=0.5,arrowlength=5,arrowwidth=2,arrowinset=0.2,clock](560,110)(50,90,-90)

    \Line[arrow,arrowpos=0.5,arrowlength=5,arrowwidth=2,arrowinset=0.2](560,160)(560,60)
    \Arc[arrow,arrowpos=0.5,arrowlength=5,arrowwidth=2,arrowinset=0.2](230.509,110.509)(55.511,91.558,269.475)
    \Line[arrow,arrowpos=0.5,arrowlength=5,arrowwidth=2,arrowinset=0.2](228,162)(228,54)
    \Arc[dash,dashsize=10,arrow,arrowpos=0.5,arrowlength=5,arrowwidth=2,arrowinset=0.2](230,110)(55.227,355,715)
    \Text(115,135)[lb]{\Large{\Black{$H$}}}
    \Text(440,130)[lb]{\Large{\Black{$H$}}}
    \Text(190,180)[lb]{\Large{\Black{$F$}}}
    \Text(190,40)[lb]{\Large{\Black{$F$}}}
    \Text(205,110)[lb]{\Large{\Black{$F$}}}
    \Text(295,110)[lb]{\Large{\Black{$S$}}}
    \Text(520,170)[lb]{\Large{\Black{$S$}}}
    \Text(520,40)[lb]{\Large{\Black{$S$}}}
    \Text(575,110)[lb]{\Large{\Black{$F$}}}
    \Text(630,110)[lb]{\Large{\Black{$F$}}}
    \Text(220,-5)[lb]{\Large{\Black{$(a)$}}}
    \Text(555,-5)[lb]{\Large{\Black{$(b)$}}}
  \end{picture}
}}
\end{center}
\caption{\label{F} Yukawa sector of the two-loop tadpoles.}
\end{figure}
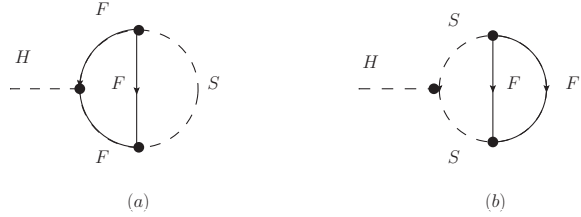
The sum of the individual diagrams contributing to $T_{SF}^{(2l)}$, represented schematically by fig. (\ref{F}), can be expressed as the sum of the two contributions with the obvious notation:
\begin{eqnarray}
T_{SF}^{(2l)} = T_{SS'FF'}^{(2l)} + T_{FF'\bar{F}S}^{(2l)}. 
\end{eqnarray}
Using the first derivative of the effective potential, the above two contributions to $T_{SF}^{(2l)}$ can be obtained as:
\begin{eqnarray}
T_{SS'FF'}^{(2l)}=\dfrac{\partial V_{SF}^{(2l)}}{\partial m_{S'}^{2}}\dfrac{\partial m_{S'}^{2}}{\partial \phi_{c}} & ; & T_{FF'\bar{F}S}^{(2l)}=\dfrac{\partial V_{SF}^{(2l)}}{\partial m_{F'}^{2}}\dfrac{\partial m_{F'}^{2}}{\partial \phi_{c}},
\end{eqnarray} 
where $V_{SF}^{(2l)}$ can be consulted in section \ref{sec:VSF} and where a sum over the repeated indices $S'$ and $F'$ are understood. The result for arbitrary fermion and scalar masses is:
\begin{eqnarray}
T_{SS'FF'}^{(2l)}=G_{SS'FF'}^{(1)}\left[ B(m^{2}_{S'},m^{2}_{S})J(m^{2}_{F})+B(m^{2}_{S'},m^{2}_{S})J(m^{2}_{F'})-I(m^{2}_{F},m^{2}_{F'},m^{2}_{S'})\right. ~~~~~ \nonumber \\ 
\left. - (m^{2}_{F}+m^{2}_{F'}-m^{2}_{S})V(m^{2}_{S},m^{2}_{S'},m^{2}_{F},m^{2}_{F'}) \right] - G_{SS'FF'}^{(2)}V(m^{2}_{S},m^{2}_{S'},m^{2}_{F},m^{2}_{F'}), \label{TSSFF}
\end{eqnarray}
and
\begin{eqnarray}
T_{FF'\bar{F}S}^{(2l)}= G_{FF'\bar{F}S}^{(1)}\left[ B(m^{2}_{F},m^{2}_{F'})J(m^{2}_{\bar{F}})+B(m^{2}_{F},m^{2}_{F'})J(m^{2}_{S})-I(m^{2}_{F},m^{2}_{\bar{F}},m^{2}_{S})\right. ~~~~~ \nonumber \\ 
\left. - (m^{2}_{F'}+m^{2}_{\bar{F}}-m^{2}_{S})V(m^{2}_{F},m^{2}_{F'},m^{2}_{\bar{F}},m^{2}_{S}) \right] + G_{FF'\bar{F}S}^{(2)}\left[I(m^{2}_{F},m^{2}_{\bar{F}},m^{2}_{S}) \right. \nonumber \\ \left. -m_{F}^{2}V(m^{2}_{F},m^{2}_{F'},m^{2}_{\bar{F}},m^{2}_{S})\right]-G_{FF'\bar{F}S}^{(3)}V(m^{2}_{F},m^{2}_{F'},m^{2}_{\bar{F}},m^{2}_{S}). \label{TFFFS}
\end{eqnarray}
Considering only the hard contribution, that is, the amplitude in the limit where the light fermion masses vanishes and with the external momenta put to zero, the non-zero coefficients have the values:
{\small
\begin{eqnarray*}
G_{H0tt}^{(1)}=\dfrac{3}{2}h_{t}^{2}6\lambda \phi_{c} ~~;~~G_{G0tt}^{(1)}=\dfrac{3}{2}h_{t}^{2}\lambda \phi_{c} & ; & G_{H0tt}^{(2)}=3h_{t}^{2}m_{t}^{2}6\lambda \phi_{c} ~~;~~G_{G0tt}^{(2)}=-3h_{t}^{2}m_{t}^{2}\lambda \phi_{c}, \\ 
G_{t0tS}^{(1)}=3\sqrt{2}h_{t}^{3}m_{t} ~~;~~ G_{t0tS}^{(2)}=3\sqrt{2}h_{t}^{3}m_{t} & ; & G_{t0tH}^{(3)}=3\sqrt{2}h_{t}^{3}m_{t}^{3} ~~;~~ G_{t0tG}^{(3)}=-3\sqrt{2}h_{t}^{3}m_{t}^{3}.
\end{eqnarray*}}
The diagrammatic computation can be easily reduced to the same superposition of the functions $I$, $J$, $B$ and $V$ given by eqs. (\ref{TSSFF}) and (\ref{TFFFS}) with TARCER code, the result is the same. 
The explicit renormalized value of $T_{SF}^{(2l)}$, derived from equations (\ref{eps0}-\ref{eps2}) and evaluated at the EW vacuum is:
{\footnotesize
\begin{eqnarray}
&& T_{SF}^{(2l)}=\frac{3}{2}\frac{h_{t}^{2}}{(4\pi)^{4}v}\left[\left(-66m_{t}^{4}+21m_{t}^{2}m_{H}^{2}-3m_{H}^{4}\right)
\overline{ln}^{2}(m_{t}^{2})+\left(3m_{H}^{4}-8m_{t}^{4}-7m_{t}^{2}m_{H}^{2}\right)\overline{ln}^{2}(m_{H}^{2})\right. \nonumber \\ &&
\left. +\left(3m_{H}^{4}+4m_{t}^{2}m_{H}^{2}-16m_{t}^{4}\right)\overline{ln}(m_{t}^{2})\overline{ln}(m_{H}^{2})+\left(76m_{t}^{4}-27m_{H}^{2}m_{t}^{2}\right)\overline{ln}(m_{t}^{2})\right. \nonumber \\ && \left. +\left(25m_{t}^{2}m_{H}^{2}-12m_{H}^{4}\right)\overline{ln}(m_{H}^{2}) +\left(\frac{34m_{t}^{4}m_{H}^{2}}{4m_{t}^{2}m_{H}^{2}-m_{H}^{4}}-\frac{16m_{t}^{6}}{4m_{t}^{2}m_{H}^{2}-m_{H}^{4}}-8m_{t}^{2}+3m_{H}^{2}\right)\xi(m_{t},m_{t},m_{H})
 \right. \nonumber \\ &&
 \left. + 15m_{H}^{4}-8m_{H}^{2}m_{t}^{2}-106m_{t}^{4}-\frac{2\pi^{2}}{3}m_{t}^{4}\right].
\end{eqnarray}}

\subsection{The Gauge Sector}
\begin{figure}
\begin{center}
\scalebox{0.4}{
\fcolorbox{white}{white}{
  \begin{picture}(814,216) (129,-124)
    \SetWidth{1.0}
    \SetColor{Black}
    \Line[dash,dashsize=10](130,-7)(195,-7)
    \SetWidth{0.0}
    \Vertex(199,-7){5.657}
    \Vertex(284,-11){5.657}
    \SetWidth{1.0}
    \Line[dash,dashsize=10](465,-6)(540,-6)
    \SetWidth{0.0}
    \Vertex(539,-7){5.657}
    \Vertex(585,37){5.657}
    \Vertex(587,-58){5.657}
    \SetWidth{1.0}
    \PhotonArc(588,-7)(47.17,148,508){7.5}{15}
    \Photon(586,40)(586,-60){7.5}{5}
    \PhotonArc(243,-6)(45.277,84,444){7.5}{14}
    \PhotonArc(334,-8)(45.277,174,534){7.5}{14}
    \Text(131,22)[lb]{\Large{\Black{$H$}}}
    \Text(464,24)[lb]{\Large{\Black{$H$}}}
    \Text(240,71)[lb]{\Large{\Black{$V$}}}
    \Text(239,-84)[lb]{\Large{\Black{$V$}}}
    \Text(398,-9)[lb]{\Large{\Black{$V$}}}
    \Text(540,50)[lb]{\Large{\Black{$V$}}}
    \Text(541,-66)[lb]{\Large{\Black{$V$}}}
    \Text(566,-1)[lb]{\Large{\Black{$V$}}}
    \Text(658,-1)[lb]{\Large{\Black{$V$}}}
    \Line[dash,dashsize=10](719,-8)(794,-8)
    \SetWidth{0.0}
    \Vertex(792,-8){5.657}
    \Vertex(844,-61){5.657}
    \Vertex(844,41){5.657}
    \SetWidth{1.0}
    \PhotonArc(843.5,-10.5)(51.502,90.556,270.556){7.5}{8.5}
    \Arc[dash,dashsize=2,arrow,arrowpos=0.5,arrowlength=5,arrowwidth=2,arrowinset=0.2,clock](844.48,-10.5)(50.523,88.276,-88.276)
    \Line[dash,dashsize=2,arrow,arrowpos=0.5,arrowlength=5,arrowwidth=2,arrowinset=0.2](843,41)(844,-62)
    \Text(720,19)[lb]{\Large{\Black{$H$}}}
    \Text(786,49)[lb]{\Large{\Black{$V$}}}
    \Text(783,-61)[lb]{\Large{\Black{$V$}}}
    \Text(853,1)[lb]{\Large{\Black{$u$}}}
    \Text(908,0)[lb]{\Large{\Black{$u$}}}
    \Text(269,-129)[lb]{\Large{\Black{$(a)$}}}
    \Text(589,-127)[lb]{\Large{\Black{$(b)$}}}
    \Text(846,-126)[lb]{\Large{\Black{$(c)$}}}
  \end{picture}
}}
\end{center}
\caption{\label{VT2} Vector sector of the two-loop tadpoles, $T_{V}^{(2l)}$.}
\end{figure}
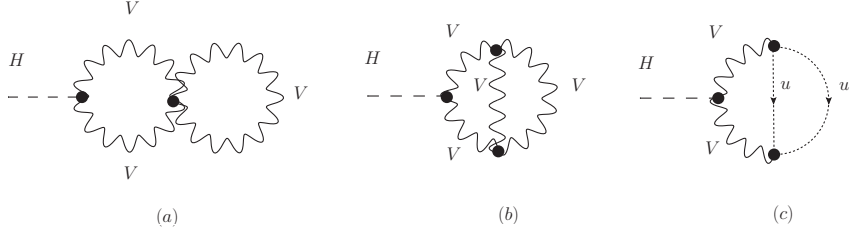

The diagrams contributing to the gauge sector are represented schematically by the fig. (\ref{VT2}).
The vector sector of $T^{(2l)}$ can be divided in three contributions:
\begin{eqnarray}
T_{V}^{(2l)}=T_{WVV'}+T_{WWVV'}+T_{uuVV'},
\end{eqnarray}
obtained from the the potential $V_{V}^{(2l)}(\phi_{c})$ derived
in section (\ref{sec:VGauge}) according with 
{\small
\begin{eqnarray*}
\left(T_{WVV'}\right)_{j}=\frac{\partial F_{WV}}{\partial m_{V'}^{2}}\frac{\partial m_{V'}^{2}}{\partial\left(\phi_{c}\right)_{j}}; & \left(T_{WWVV'}\right)_{j}=\dfrac{\partial F_{WWV}}{\partial m_{V'}^{2}}\dfrac{\partial m_{V'}^{2}}{\partial\left(\phi_{c}\right)_{j}}; & \left(T_{uuVV'}\right)_{j}=\frac{\partial F_{uuV}}{\partial m_{V'}^{2}}\frac{\partial m_{V'}^{2}}{\partial\left(\phi_{c}\right)_{j}},
\end{eqnarray*}}
where $\partial/\partial\left(\phi_{c}\right)_{j}$ denotes the derivative
with respect to the classical scalar field $j$. We are only interested
in the case where $j=H$. Note that makes the above derivatives is
equivalent to introduce a mass insertion to each propagator in the
vacuum bubble diagrams represented in fig. (\ref{EP-V-Sector}). When
the insertion is done over a ghost propagator, the tadpole obtained
is zero because the vertex coupling $G_{uuH}$ vanishes at the
Landau gauge. Including only the non-vanishing couplings the result
obtained is:
\begin{eqnarray}
\left(T_{V}^{(2l)}\right)_{j}=\sum_{a,b,c,d}G_{abc}G_{dbc}G_{adj}\frac{F_{V}(m_{a}^{2},m_{b}^{2},m_{c}^{2})-F_{V}(m_{d}^{2},m_{b}^{2},m_{c}^{2})}{m_{a}^{2}-m_{d}^{2}},\label{eq:TVSum}
\end{eqnarray}
where the indices $a,~b,~c,~d$ runs over the vector fields, whereas the
index $j$ run on the scalar fields as was defined in Appendix \ref{AppSM}.
The function $F_{V}$ has the value:
\begin{eqnarray*}
F_{V}(x,y,z)=-\frac{1}{16xyz}\left[4x^{3}yz+48xy^{2}z^{2}-\left(8x^{3}yz-22xy^{2}z-22xyz^{2}+\frac{40}{3}x^{2}yz\right)J(x)\right.\\
-\left(x^{2}-9y^{2}-9z^{2}+9xy+9xz+14yz\right)x\hat{J}(y,z)\\
+\left(z-y\right)^{2}\left(y^{2}+10yz+z^{2}\right)\hat{I}(0,y,z)+x^{2}\left(2yz-x^{2}\right)\hat{I}(x,0,0)\\
\left.+\left(-x^{4}-8x^{3}y-8x^{3}z+32x^{2}yz+18y^{2}z^{2}\right)\hat{I}(x,y,z)\right]+(x\leftrightarrow y)+(x\leftrightarrow z).
\end{eqnarray*}
To the sum (\ref{eq:TVSum}) contributes only the terms with the next
non-zero couplings:
\begin{eqnarray*}
G_{WWA}=gsin\theta_{W}~~;~~G_{WWZ}=gcos\theta_{W} & ; & G_{u\pm u\pm Z}=\frac{gcos\theta_{W}}{2}~~;~~G_{u\pm u_{\gamma}W}=gsin\theta_{W},\\
G_{u\pm u_{Z}W}=gcos\theta_{W}~~;~~G_{WWH}=gm_{W} & ; & G_{ZZH}=\frac{g}{cos\theta_{W}}m_{Z}.
\end{eqnarray*}
Finally we note that there is not QCD contributions to $T^{(2l)}$ because the functions $F_{gg}$, $F_{ggg}$ and $F_{u_{g}u_{g}g}$ are equal to zero as we proved in section \ref{sec:VGauge} - eq. (\ref{FQCD}). Diagrammatically this fact is evident because the gluon field $g$ and the ghost field $u_{g}$ are massless. 

\subsection{The Scalar-Gauge Sector}
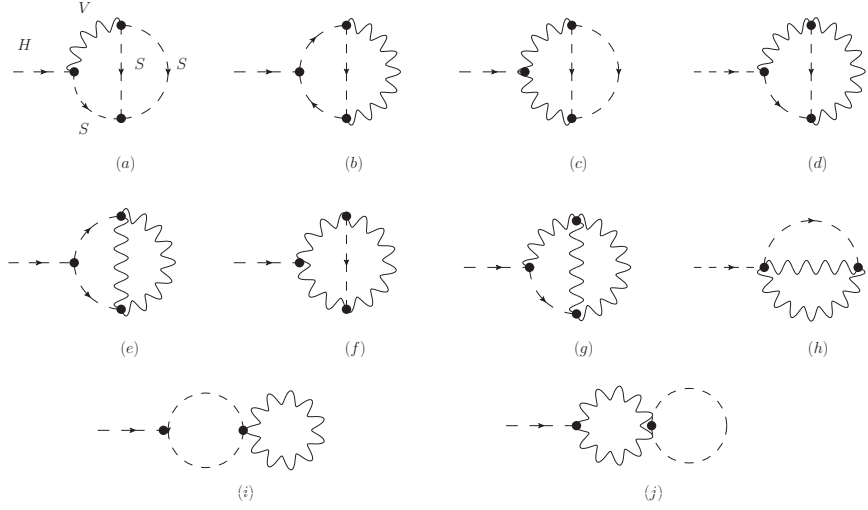
\begin{figure}
\begin{center}
\scalebox{0.35}{
\fcolorbox{white}{white}{
  \begin{picture}(915,536) (89,-44)
    \SetWidth{1.0}
    \SetColor{Black}
    \Line[dash,dashsize=10,arrow,arrowpos=0.5,arrowlength=5,arrowwidth=2,arrowinset=0.2](95,406)(160,406)
    \Vertex(160,406){5}
    \Line[dash,dashsize=10,arrow,arrowpos=0.5,arrowlength=5,arrowwidth=2,arrowinset=0.2](330,406)(405,406)
    \Vertex(400,406){5}
    \Line[dash,dashsize=10,arrow,arrowpos=0.5,arrowlength=5,arrowwidth=2,arrowinset=0.2](570,406)(645,406)
    \Vertex(640,406){5}
    \Line[dash,dashsize=10,arrow,arrowpos=0.5,arrowlength=5,arrowwidth=2,arrowinset=0.2](820,406)(900,406)
    \Vertex(895,406){5}
    \Line[dash,dashsize=10,arrow,arrowpos=0.5,arrowlength=5,arrowwidth=2,arrowinset=0.2](90,201)(155,201)
    \Vertex(160,201){5}
    \Line[dash,dashsize=10,arrow,arrowpos=0.5,arrowlength=5,arrowwidth=2,arrowinset=0.2](330,201)(405,201)
    \Vertex(400,201){5}
    \Line[dash,dashsize=10,arrow,arrowpos=0.5,arrowlength=5,arrowwidth=2,arrowinset=0.2](575,196)(650,196)
    \Vertex(645,196){5}
    \Line[dash,dashsize=10,arrow,arrowpos=0.5,arrowlength=5,arrowwidth=2,arrowinset=0.2](820,196)(900,196)
    \Vertex(895,196){5}
    \Vertex(210,356){5}
    \Vertex(210,456){5}
    \Arc[dash,dashsize=10,arrow,arrowpos=0.5,arrowlength=5,arrowwidth=2,arrowinset=0.2,clock](210,406)(50,90,-90)
    \Arc[dash,dashsize=10,arrow,arrowpos=0.5,arrowlength=5,arrowwidth=2,arrowinset=0.2](203.5,399.5)(43.983,171.501,278.499)
    \PhotonArc[clock](203.5,412.5)(43.983,-171.501,-278.499){7.5}{4.5}
    \Line[dash,dashsize=10,arrow,arrowpos=0.5,arrowlength=5,arrowwidth=2,arrowinset=0.2](210,456)(210,356)
    \Vertex(450,356){5}
    \Vertex(450,456){5}
    \PhotonArc[clock](450,406)(50,90,-90){7.5}{8.5}
    \Arc[dash,dashsize=10,arrow,arrowpos=0.5,arrowlength=5,arrowwidth=2,arrowinset=0.2,clock](451.25,404.75)(51.265,178.603,91.397)
    \Arc[dash,dashsize=10,arrow,arrowpos=0.5,arrowlength=5,arrowwidth=2,arrowinset=0.2,clock](451.25,407.25)(51.265,-91.397,-178.603)
    \Line[dash,dashsize=10,arrow,arrowpos=0.5,arrowlength=5,arrowwidth=2,arrowinset=0.2](450,456)(450,356)
    \Vertex(690,456){5}
    \Vertex(690,356){5}
    \PhotonArc(690,406)(50,90,270){7.5}{8.5}
    \Arc[dash,dashsize=10,arrow,arrowpos=0.5,arrowlength=5,arrowwidth=2,arrowinset=0.2,clock](690,406)(50,90,-90)
    \Line[dash,dashsize=10,arrow,arrowpos=0.5,arrowlength=5,arrowwidth=2,arrowinset=0.2](690,456)(690,356)
    \Vertex(945,456){5}
    \Vertex(945,356){5}
    \PhotonArc[clock](944.5,406.5)(50.502,89.433,-89.433){7.5}{8.5}
    \PhotonArc[clock](946.25,404.75)(51.265,178.603,91.397){7.5}{4.5}
    \Arc[dash,dashsize=10,arrow,arrowpos=0.5,arrowlength=5,arrowwidth=2,arrowinset=0.2](946.25,407.25)(51.265,-178.603,-91.397)
    \Line[dash,dashsize=10,arrow,arrowpos=0.5,arrowlength=5,arrowwidth=2,arrowinset=0.2](945,456)(945,356)
    \Vertex(210,251){5}
    \Vertex(210,151){5}

    \PhotonArc[clock](209.8,201.5)(50.5,89.773,-89.773){7.5}{8.5}
    \Arc[dash,dashsize=10,arrow,arrowpos=0.5,arrowlength=5,arrowwidth=2,arrowinset=0.2,clock](211.25,199.75)(51.265,178.603,91.397)
    \Arc[dash,dashsize=10,arrow,arrowpos=0.5,arrowlength=5,arrowwidth=2,arrowinset=0.2](211.25,202.25)(51.265,-178.603,-91.397)
    \Photon(210,251)(210,151){7.5}{5}
    \Vertex(450,251){5}
    \Vertex(450,151){5}
    \PhotonArc(450,201)(46.098,139,499){7.5}{14}
    \Line[dash,dashsize=10,arrow,arrowpos=0.5,arrowlength=5,arrowwidth=2,arrowinset=0.2](450,251)(450,151)
    \Vertex(695,246){5}
    \Vertex(695,146){5}
    \PhotonArc[clock](695.25,196)(50.001,90.286,-90.286){7.5}{8.5}
    \PhotonArc(696.25,194.75)(51.265,91.397,178.603){7.5}{4.5}
    \Arc[dash,dashsize=10,arrow,arrowpos=0.5,arrowlength=5,arrowwidth=2,arrowinset=0.2](696.25,197.25)(51.265,-178.603,-91.397)
    \Photon(695,246)(695,146){7.5}{5}
    \Vertex(995,196){5}
    \PhotonArc(945,195.75)(50.001,179.714,360.286){7.5}{8.5}
    \Arc[dash,dashsize=10,arrow,arrowpos=0.5,arrowlength=5,arrowwidth=2,arrowinset=0.2,clock](945,196.25)(50.001,-179.714,-360.286)
    \Photon(895,196)(995,196){7.5}{5}
    \Line[dash,dashsize=10,arrow,arrowpos=0.5,arrowlength=5,arrowwidth=2,arrowinset=0.2](185,21)(260,21)
    \Arc[dash,dashsize=10,arrow,arrowpos=0.5,arrowlength=5,arrowwidth=2,arrowinset=0.2](300,21)(40,0,360)
    \PhotonArc(385,21)(35.355,172,532){7.5}{11}
    \Vertex(340,21){5}
    \Vertex(255,21){5}
    \PhotonArc(735,26)(35.355,172,532){7.5}{11}
    \Arc[dash,dashsize=10,arrow,arrowpos=0.5,arrowlength=5,arrowwidth=2,arrowinset=0.2](815,26)(40,0,360)
    \Line[dash,dashsize=10,arrow,arrowpos=0.5,arrowlength=5,arrowwidth=2,arrowinset=0.2](620,26)(695,26)
    \Vertex(695,26){5}
    \Vertex(775,26){5}
    \Text(205,300)[lb]{\Large{\Black{$(a)$}}}
    \Text(450,300)[lb]{\Large{\Black{$(b)$}}}
    \Text(690,300)[lb]{\Large{\Black{$(c)$}}}
    \Text(945,300)[lb]{\Large{\Black{$(d)$}}}
    \Text(210,100)[lb]{\Large{\Black{$(e)$}}}
    \Text(450,100)[lb]{\Large{\Black{$(f)$}}}
    \Text(695,100)[lb]{\Large{\Black{$(g)$}}}
    \Text(945,100)[lb]{\Large{\Black{$(h)$}}}
    \Text(335,-55)[lb]{\Large{\Black{$(i)$}}}
    \Text(770,-55)[lb]{\Large{\Black{$(j)$}}}
    \Text(100,431)[lb]{\Large{\Black{$H$}}}
    \Text(165,471)[lb]{\Large{\Black{$V$}}}
    \Text(165,341)[lb]{\Large{\Black{$S$}}}
    \Text(225,411)[lb]{\Large{\Black{$S$}}}
    \Text(270,411)[lb]{\Large{\Black{$S$}}}
  \end{picture}
}}
\end{center}
\caption{\label{SVT2} Diagrams contributing to the Scalar-Vector sector of $T^{(2l)}$}
\end{figure}
The scalar - vector contribution to $T^{(2l)}$, represented schematically in fig. (\ref{SVT2}), can be obtained as the sum of the three terms:
\begin{eqnarray}
T_{SV}^{(2l)}=T_{SS'V}+T_{VV'S}+T_{SV},
\end{eqnarray}
where 
\begin{eqnarray*}
&& T_{SS'V}=\frac{\partial F_{SS'V}}{\partial m_{j}^{2}}\frac{\partial m_{j}^{2}}{\partial\phi_{c}}+\frac{\partial F_{SS'V}}{\partial m_{a}^{2}}\frac{\partial m_{a}^{2}}{\partial\phi_{c}}~~;~~ T_{VV'S}=\frac{\partial F_{VV'S}}{\partial m_{j}^{2}}\frac{\partial m_{j}^{2}}{\partial\phi_{c}}+\frac{\partial F_{VV'S}}{\partial m_{a}^{2}}\frac{\partial m_{a}^{2}}{\partial\phi_{c}}; \\ && T_{SV}=\frac{\partial F_{SV}}{\partial m_{j}^{2}}\frac{\partial m_{j}^{2}}{\partial\phi_{c}}+\frac{\partial F_{SV}}{\partial m_{a}^{2}}\frac{\partial m_{a}^{2}}{\partial\phi_{c}}.
\end{eqnarray*}
The index $a$ runs over the vector fields and $j$ runs over the scalar
fields. Diagrammatically, the contribution $T_{SS'V}$ is the sum
of the diagrams (a), (b), (c) and (d) of the fig. (\ref{SVT2}), the contribution $T_{VV'S}$ is the sum of the diagrams (e), (f),
(g) and (h),  whereas the contribution $T_{SV}$ is the sum of
the diagrams (i) and (j). These terms can be reduced to a superposition of the renormalized integrals $\hat{I}(x,y,z)$, $\hat{J}(x,y)$ and
$J(x)$ with TARCER. The results are:
\begin{eqnarray}
&& \left(T_{SS'V}\right)_{l}=\sum_{a,i,j,k}\frac{1}{2}G_{aij}G_{ajk}G_{ikl}\frac{f_{SS'V}(m_{i}^{2},m_{j}^{2},m_{a}^{2})-f_{SS'V}(m_{k}^{2},m_{j}^{2},m_{a}^{2})}{m_{i}^{2}-m_{k}^{2}} \\ && ~~~~~~~~~~~~~~~~+~~\sum_{a,b,i,j}\frac{1}{4}G_{aij}G_{bij}G_{abl}\frac{f_{SS'V}(m_{i}^{2},m_{j}^{2},m_{a}^{2})-f_{SS'V}(m_{i}^{2},m_{j}^{2},m_{b}^{2})}{m_{a}^{2}-m_{b}^{2}}, \nonumber
\end{eqnarray}
where the function $f_{SS'V}$ is defined as:
\begin{eqnarray*}
f_{SS'V}(x,y,z)=-\frac{1}{z}\left[-\left(x^{2}+y^{2}+z^{2}-2xy-2xz-2yz\right)\hat{I}(x,y,z)+(x-y)^{2}\hat{I}(x,y,0)\right.\\
\left.+(x+z-y)\hat{J}(x,z)+(y+z-x)\hat{J}(y,z)+z\hat{J}(x,y)\right]-2\left(x+y-\frac{z}{3}\right)J(z).
\end{eqnarray*}
By other side, the second contribution is:
\begin{eqnarray}
\left(T_{VV'S}\right)_{l}=\sum_{a,b,c,j}\frac{1}{2}G_{abj}G_{cbj}G_{acl}\frac{f_{VV'S}(m_{a}^{2},m_{b}^{2},m_{j}^{2})-f_{VV'S}(m_{c}^{2},m_{b}^{2},m_{j}^{2})}{m_{a}^{2}-m_{c}^{2}} \\ ~~~~~~~~~~~~~~~~~~~+~~\sum_{a,b,i,j}\frac{1}{4}G_{abi}G_{abj}G_{ijl}\frac{f_{VV'S}(m_{a}^{2},m_{b}^{2},m_{i}^{2})-f_{VV'S}(m_{a}^{2},m_{b}^{2},m_{j}^{2})}{m_{i}^{2}-m_{j}^{2}}, \nonumber
\end{eqnarray}
where $f_{VV'S}$ is:
\begin{eqnarray*}
f_{VV'S}(x,y,z)=\frac{1}{4xy}\left[-\left(x^{2}+y^{2}+z^{2}+10xy-2xz-2yz\right)\hat{I}(x,y,z)+(x-z)\hat{I}(x,0,z)\right.\\
\left.+(x-y)^{2}\hat{I}(0,y,z)-z^{2}\hat{I}(0,0,z)+(x+y-z)\hat{J}(x,y)-y\hat{J}(x,z)-x\hat{J}(y,z)\right]\\
+x+y+z+\frac{1}{2}J(x)+\frac{1}{2}J(y)+2J(z).
\end{eqnarray*}
To the terms $T_{SS'V}$ and $T_{VV'S}$, with $l=H$, only contributes
the couplings:
\begin{eqnarray*}
&& G_{WHG}=G_{WGG}=\frac{g}{2}~~;~~ G_{AGG}=gsin\theta_{W}~~;~~ G_{ZGG}=\frac{1}{2}\left(gcos\theta_{W}-g'sin\theta_{W}\right),\\
&& G_{HHH}=G_{GGH}=\lambda\phi_{c}~~~;~~ G_{WWH}=gm_{W}~~;~~ G_{ZZH}=\frac{g}{cos\theta_{W}}m_{Z}.
\end{eqnarray*}
Finally, the term $T_{SV}$ has the form:
\begin{eqnarray*}
\left(T_{SV}\right)_{l}=\sum_{a,b,i}\frac{1}{4}G_{abii}G_{abl}B(m_{a}^{2},m_{b}^{2})J(m_{i}^{2})+\sum_{a,i,j}\frac{1}{4}G_{aaij}G_{ijl}B(m_{a}^{2},m_{i}^{2})J(m_{j}^{2}),
\end{eqnarray*}
with the additional non-zero couplings:
\begin{eqnarray*}
&&G_{WWHH}=G_{WWGG}=\frac{g^{2}}{4}~~;~~  G_{AWGS}=\frac{1}{2}g^{2}sin\theta_{W}~~;~~G_{ZWGS}=\frac{1}{2}g^{2}sin\theta_{W}tan\theta_{W}; \\  && G_{ZZHH}=G_{ZZGG}=\frac{1}{8}\frac{g^{2}}{cos^{2}\theta_{W}}.
\end{eqnarray*}


\subsection{The Fermion-Gauge Sector}

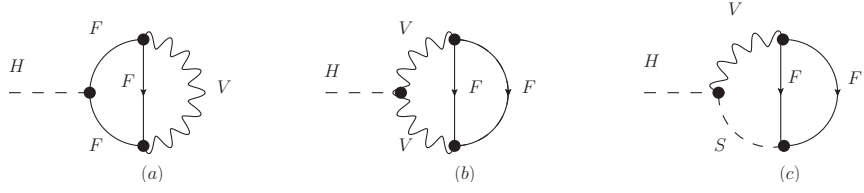
\begin{figure}
\begin{center}
\scalebox{0.4}{
\fcolorbox{white}{white}{
  \begin{picture}(821,176) (142,-66)
    \SetWidth{1.0}
    \SetColor{Black}
    \Line[dash,dashsize=10](143,16)(218,16)
    \SetWidth{0.0}
    \Vertex(218,16){5.657}
    \SetWidth{1.0}
    \Line[dash,dashsize=10](438,16)(513,16)
    \SetWidth{0.0}
    \Vertex(508,16){5.657}
    \Vertex(268,-34){5.657}
    \Vertex(268,66){5.657}
    \SetWidth{1.0}
    \PhotonArc[clock](268,16)(50,90,-90){7.5}{8.5}
    \SetWidth{0.0}
    \Vertex(558,66){5.657}
    \Vertex(558,-34){5.657}
    \SetWidth{1.0}
    \PhotonArc(558,16)(50,90,270){7.5}{8.5}
    \Arc[dash,dashsize=10,arrow,arrowpos=0.5,arrowlength=5,arrowwidth=2,arrowinset=0.2,clock](558,16)(50,90,-90)
    \Arc[arrow,arrowpos=0.5,arrowlength=5,arrowwidth=2,arrowinset=0.2,clock](558,16)(50,90,-90)
    \Line[arrow,arrowpos=0.5,arrowlength=5,arrowwidth=2,arrowinset=0.2](558,66)(558,-34)
    \Arc[arrow,arrowpos=0.5,arrowlength=5,arrowwidth=2,arrowinset=0.2](268,16)(50,90,270)
    \Line[arrow,arrowpos=0.5,arrowlength=5,arrowwidth=2,arrowinset=0.2](268,66)(268,-34)
    \Text(143,36)[lb]{\Large{\Black{$H$}}}
    \Text(218,71)[lb]{\Large{\Black{$F$}}}
    \Text(218,-39)[lb]{\Large{\Black{$F$}}}
    \Text(248,21)[lb]{\Large{\Black{$F$}}}
    \Text(338,16)[lb]{\Large{\Black{$V$}}}
    \Text(438,31)[lb]{\Large{\Black{$H$}}}
    \Text(508,71)[lb]{\Large{\Black{$V$}}}
    \Text(508,-39)[lb]{\Large{\Black{$V$}}}
    \Text(573,16)[lb]{\Large{\Black{$F$}}}
    \Text(623,16)[lb]{\Large{\Black{$F$}}}
    \Line[dash,dashsize=10](735,16)(810,16)
    \Line[arrow,arrowpos=0.5,arrowlength=5,arrowwidth=2,arrowinset=0.2](862,67)(862,-33)
    \Arc[arrow,arrowpos=0.5,arrowlength=5,arrowwidth=2,arrowinset=0.2,clock](865,16)(50,90,-90)
    \SetWidth{0.0}
    \Vertex(864,66){5.657}
    \Vertex(865,-34){5.657}
    \Vertex(804,16){5.657}
    \SetWidth{1.0}
    \PhotonArc[clock](854.336,15.561)(51.356,178.394,79.154){7.5}{4.5}
    \Arc[dash,dashsize=10](853.861,14.605)(49.865,-179.305,-77.093)
    \Text(737,40)[lb]{\Large{\Black{$H$}}}
    \Text(816,89)[lb]{\Large{\Black{$V$}}}
    \Text(872,25)[lb]{\Large{\Black{$F$}}}
    \Text(928,24)[lb]{\Large{\Black{$F$}}}
    \Text(803,-39)[lb]{\Large{\Black{$S$}}}
    \Text(267,-70)[lb]{\Large{\Black{$(a)$}}}
    \Text(560,-70)[lb]{\Large{\Black{$(b)$}}}
    \Text(864,-70)[lb]{\Large{\Black{$(c)$}}}
  \end{picture}
}} \\
\caption{Fermion - Vector sector of the two-loop tadpoles. \label{FVT2l}}
\end{center}
\end{figure}
The non-zero diagrams contributing to $T_{FV}^{(2l)}$ are drawn in
fig. \ref{FVT2l}. Its amplitude can be computed as the derivative:
\begin{eqnarray*}
T_{FV}^{(2l)}=\frac{\partial V_{FV}}{\partial m_{I}^{2}}\frac{\partial m_{I}^{2}}{\partial\phi_{c}}+\frac{\partial V_{FV}}{\partial m_{a}^{2}}\frac{\partial m_{a}^{2}}{\partial\phi_{c}}.
\end{eqnarray*}
The amplitude obtained is:
\begin{eqnarray*}
&& T_{FV}^{(2l)}=\sum_{I,J,L,a}2G_{I}^{aJ}G_{bJ}^{L}m_{LK}y^{KIi}\frac{f_{FV}(m_{I}^{2},m_{J}^{2},m_{a}^{2})-f_{FV}(m_{L}^{2},m_{J}^{2},m_{a}^{2})}{m_{I}^{2}-m_{L}^{2}} \\ && ~~~~~~~~~~~+~~ \frac{1}{2}\sum_{I,J,a,b}G_{I}^{aJ}G_{aJ}^{I}G^{abi}\frac{f_{FV}(m_{I}^{2},m_{J}^{2},m_{a}^{2})-f_{FV}(m_{L}^{2},m_{J}^{2},m_{b}^{2})}{m_{a}^{2}-m_{b}^{2}},
\end{eqnarray*}
where
\begin{eqnarray*}
f_{FV}(x,y,z)=-\frac{1}{z}\left[\left(x^{2}+y^{2}-2z^{2}-2xy+xz+yz\right)\hat{I}(x,y,z)-(x-y)^{2}\hat{I}(x,y,0)\right.\\
+\left(y+2z-x\right)\hat{J}(x,z)+\left(x+2z-y\right)\hat{J}(y,z)-2z\hat{J}(x,y)+2\left(-xz-yz+\frac{z^{2}}{3}\right)J(z)\\
\left.-2xzJ(x)-2yzJ(y)+(x+y)^{2}z-z^{3}\right].
\end{eqnarray*}
Introducing the SM couplings and considering only the more relevant contribution coming from the top quark fermion we can reduce the fermion-vector contribution to the next simple result:
\begin{eqnarray}
T_{FV}^{(2l)}\approx 4g^{2}d(3)C(3)4m_{t}^{3}h_{t}\left[6 - 5ln\left(\dfrac{m_{t}^{2}}{\bar{\mu}^{2}} \right)+3ln^{2}\left(\dfrac{m_{t}^{2}}{\bar{\mu}^{2}}\right) \right], \nonumber
\end{eqnarray}
where $C(3)=\dfrac{1}{2}$ and $d(3)=3$ are the Casimir operator and the dimension of the fundamental representation of the $SU(3)$ group.

\chapter{\noun{\label{cha:Future-Work}\index{Future Work}}Conclusions and Perspectives}

\lettrine{I}{n} this thesis we assumed that ATLAS and CMS discover the Higgs boson particle of the SM with a mass $m_{H}=125.09 \pm 0.21 ({\rm stat})\pm 0.11 ({\rm syst})$ since several of the properties of such a resonance do not show any deviation from the SM Higgs. The data obtained do not exclude any beyond Standard Model scenarios, but currently there is no signal of new physics at the TeV scale. This situation introduces us in a new precision era in which making new more precise measurements of the properties of the Standard Model particles and their interactions will be crucial and hopefully LHC will be answering to that. It is becoming a conventional wisdom that the SM is a valid just up to same scale $\Lambda_{P}$, and maybe $P$ stays for Planck. This is essentially the point of view adopted in these thesis where we do not give too much centrality to the naturalness. We remember here that by naturalness one must expect new physics already at few TeV. However can the pure SM be valid, in the weakly coupled regime up to the Planck scale? With the discovery of $m_{H}$ the Higgs quartic coupling $\lambda$ at EW scale is indirectly measured, summing it of SM type. To search some clue of what kind of new physics could emerge and at what scale of energy, it is necessary to make a very precise analysis of the extrapolation of the SM parameters up to the Planck scale. The study of the running of the SM parameters at NNLO is nowadays know in \cite{Degrassi2}, including the evolution of $\lambda$ up to the Planck scale. We can now review the  self-consistency of the SM by studying the usual scale-dependent properties of the theory: triviality, hierarchy problem and vacuum stability. 

The triviality constraint is related with the high energy behaviour of the running couplings, when the corresponding couplings have a beta function positive appears the famous Landau pole that spoils the perturbativity of the theory. The recent value obtained for the Higgs mass has confirmed that the SM is a self-consistent theory, all coupling constants are free from Landau singularities up to the Planck scale \cite{Degrassi2}\cite{Jegerlehner}. The SM parameters can be extrapolated from EW scale to $\Lambda_{P}$ since all couplings remain perturbative in that range. 

By other side, the hierarchy problem i.e. the problem of the quadratic divergences in the radiative corrections of the Higgs mass and its very high fine-tuning remains without solution in SM. In fact, if one considers that the new physics occurs only at $\Lambda_{P}$ there is a naturalness fine-tuning of 34 digits, that must be adjusted to obtain the squared Higgs mass value at EW scale from its value at $\Lambda_{P}$. This is due to the quadratic corrections in the energy cut-off supposed of the order of the Planck's mass. Of course the hierarchy problem can be just an aesthetic feature of the theory, but is too a good way to impose constraints over the scale where the new physics can be occur if one insists with naturalness. Supersymmetry is considered the way to avoid this large quadratic corrections since the quantum corrections do not renormalize the superpotential which contains the Higgs mass term. In the minimal supersymmetric version of the Standard Model (MSSM) there is no fine-tuning problem because supersymmetry is broken by soft terms \cite{Casas2}. 

The most intriguing perspectives come from the vacuum stability analysis. By studying the high energy properties of the RGI Higgs effective potential with the current precision in $m_{H}$, three-loop RG beta functions, two-loop matching conditions and as a function of the mass pole of the top quark the last published analysis for the the range of values of the Higgs mass allowing the vacuum stability until the Planck scale is \cite{Degrassi2}
\begin{equation}
M_H>\left[129.6+2.0\times \left(M_t^{\rm pole}-173.35 {\rm GeV}\right)-0.5\times\left(\frac{\alpha_s(M_Z)-0.1184}{0.0007}\right)\pm0.3\right]{\rm GeV}.
\label{est}
\end{equation}
The main error in (\ref{est}) is coming from the input value of the top-quark mass: $\Delta M_H\sim 2\Delta M_t^{\rm pole}$. Currently the most precise measurement of the top-quark mass has been reported as the world combination of the  ATLAS, CMS, CDF and D0 giving $M_t=173.34\pm0.27({\rm stat})\pm 0.71({\rm syst})$. With the previous date we would conclude that the EW vacuum is in a near-critical position between two phases: the metastability and absolute stability. In terms of the top-quark mass the stability bound is at $M_t<\left(171.36\pm 0.46\right){\rm GeV}$. The metastability, with a decay time much longer than the age of the universe, is now preferred at $99.3\%$ CL also if the metastability with a very long lifetime cannot be used as a motivation for a New Physics. The precision determinations of parameters and the study of methods to make higher order calculations are more important than ever and will be the challenges of LHC and ILC.

However the main goal of this thesis has been the study of the analytical methods of computation for the NNLO stability analysis. This includes the explicit computation of the 1PI Higgs effective potential in the $\overline{MS}$ scheme ($V^{(2l)}_{eff}(\phi_c)$) up two-loop level by the Tarasov's method and its implementation in Mathematica from TARCER code. Its analytic expression is useful to us for many reasons. We derived from $V^{(2l)}_{eff}(\phi_c)$ two computation that are not found in the literature. First, we computed the 1PI effective potential in the Sirlin Zucchini renormalization scheme at two-loop level. Using the matching conditions obtained from the threshold relations between the $\overline{MS}$ couplings and the physical observables of the on-shell SZ scheme we derived, as discussed deeply in Chapter 6, a new two-loop contribution coming from applying the shifts over the parameters in the one-loop $\overline{MS}$ effective potential. The higher contribution of this potential comes from the QCD corrections to the Yukawa sector, the dominant terms are of the order $\sim g_{s}^{2}h_{t}^{4}$. 

The second contribution of this work is the analytic computation of the sum of the tadpoles at two-loop level. We consider relevant its explicit computation because the tadpole topologies are deeply related with the renormalization of the EW vacuum expectation value ($v$) and it is a cornerstone in the determination of the top quark pole mass, the EW corrections due to the tadpoles are the higher EW contribution in the threshold corrections to the boundary conditions over $\lambda$ and therefore play an important role in the vacuum stability analysis. The tadpoles topologies can be obtained from the first derivative of the Higgs effective potential with respect to the classical field. This implies to obtain the derivative of the master integrals introduced in the computation of the effective potential i.e. in the reduction of the zero-point two-loop Feynman diagrams. We compute this expressions in order to obtain $T^{(2)}$ straightforwardly for the 1PI $\overline{MS}$ effective potential. As we don't have any reference of the analytic expression of $T^{(2)}$, we accomplish the check of our computation from the diagrammatic calculation of the tadpoles at two-loop level. The complete contribution of tadpoles require the evaluation of 161 diagrams in the $R_{\zeta}$ gauge, therefore we use computer codes to get the final expression; the code  FeynArts, to generate the amplitudes, and TARCER code to reduce the tensorial integrals as a superposition of some master integrals. We conclude that the result is the same by the two methods, however it is not evident. The analytic expressions obtained look different, thus it is necessary to use of appropriate recurrence relations over the result to verify its equality. The recurrence relations are obtained from the IBP (Integration by parts) method. A more technical issue is the renormalization of the tadpoles, its amplitudes has mixed non-local UV divergences i.e. we find in the negative order of the Laurent expansions terms of the form $ln^{i}(m_{i}^2)/\varepsilon^{j}$. As a consequence, we need first renormalize the sub-divergences using the results of one-loop diagrams, so by inserting the counter-terms at one-loop level, which will eliminate the mixed non-local terms, and the remaining overall divergences of the graph can then be subtracted by a suitable  genuine two-loop counterterms. This analytical procedure is not trivial, it requires large algebraic cancellations that must be done very carefully.  
In this sense, the use of the effective potential is more efficient. The process of removing divergences in the effective potential is equivalent to making suitable transformations over the master integrals, whose derivatives with respect to the classical field are straightforward. Moreover the number of diagrams to compute is less. The complete contribution of $V^{(2l)}_{eff}(\phi_c)$ requires the evaluation of 67 diagrams in the $R_{\zeta}$ gauge. 

We have some research perspectives for the future, motivated by the actual status of the vacuum stability analysis and recent discussions with the group of Prof. Gino Isidori at the University of Zurich, where the final stage of this work has been done, thanks to a grant provided by Universidad Nacional de Colombia. The EW vacuum of SM is very likely to be metastable, and we think that a more accurate measurement of the top quark mass and its uncertainty will lead to a confirmation of the meta-stability of the SM. If the model is extended with the inclusion of new physics at $10^{13}-10^{14}$ GeV range, an upper bound on the masses of the new degrees of freedom can be derived by the requirement that the electroweak vacuum has a lifetime longer than the age of the Universe. In particular, it is interesting to analyse the impact of the neutrinos models, Seesaw I - III, on the instability region of the Higgs sector. G. Isidori et. al. obtained upper bounds on the right handed neutrinos in a Seesaw-I scenario \cite{JJIsidori}. Their analysis was done with the inclusion of the one-loop beta function of the Yukawa coupling of the new right-handed neutrinos ($h_{\nu}$), and two-loop beta functions for all SM couplings ($g_i$). Kobakhidze and Spencer \cite{Kobakhidze} investigated vacuum stability within type-II seesaw and left-right symmetric models. We plan to extend the above mentioned analysis by including in the vacuum stability analysis the two-loop anomalous dimension for $h_{\nu}$, and three loop beta functions for all the couplings $g_i$. Moreover, there is no conclusive analysis in this direction for the three Seesaw model, their impact on the stability regions are unknown. Another direction of research would be the possibility that the Standard Model (SM) is valid up to the Planck scale, i.e. that new physics occurs only around $\Lambda_{P}$. For a metastable EW vacuum also in this case it would be interesting to study the impact of the new physics interactions on its lifetime. The stability phase diagram of the model must be independent on the new physics, this imposes constraints on the physics that can manifest at Planck scale \cite{Vincenzo}. Finally, the study of Higgs inflation scenarios, all based on results obtained neglecting new physics interactions, are worth to be explored \cite{Shaposhnikov}. 
 
\appendix

\chapter{The SM at Symmetric Phase}\label{AppSM}
\markboth{APPENDIX \ref{AppSM}}{APPENDIX \ref{AppSM}} 

In this appendix we present a brief introduction to the Standard Model (SM) of the strong and electroweak interactions before electroweak symmetry breaking. This will allow us to fix the notation which will be used later on. The GWS electroweak theory which describes the electromagnetic and weak interactions between quarks and leptons, is a Yang--Mills theory based on the symmetry group ${\rm SU(2)_L \times U(1)_Y}$. Combined with the ${\rm SU(3)_C}$ QCD gauge theory for the strong interactions between quarks, it provides a unified model of these three forces known as the Standard Model. The model at symmetric phase has two kinds of fields. 
There are first the matter fields, that is, the three generations of left-handed and right-handed chiral quarks and leptons, $f_{L,R} =\frac{1}{2}(1 \mp \gamma_5)f$. The left-handed fermions are in weak isodoublets, $T_{3}= \pm \frac{1}{2}$, while the right-handed fermions are in weak isosinglets, $T_{3}= 0$. We will assume that the neutrinos are massless fermions and therefore appear only with their left-handed components.
\begin{eqnarray}
\begin{array}{l} L_1= \left( \begin{array}{c} \nu_e \\ e^- \end{array} \right)_L, ~~~\ e_{R_1}=
e^-_R  \, , ~~~\ Q_1= \left( \begin{array}{c} u \\ d \end{array} \right)_L, ~~~\ 
u_{R_1}=u_R \, , ~~~\ d_{R_1}=d_R \\ 
L_2= \left( \begin{array}{c} \nu_\mu \\ \mu^- \end{array} \right)_L, ~~~\ 
e_{R_2} =\mu^-_R  \, , ~~~\ Q_2= \left( \begin{array}{c} c \\ s \end{array} 
\right)_L, ~~~\ u_{R_2}=c_R\ , ~~~\ d_{R_2} = s_R \\ 
L_3= \left( \begin{array}{c} \nu_\tau \\ \tau^- \end{array} \right)_L, ~~~\ 
e_{R_3}=\tau^-_R  \, , ~~~\ Q_3= \left( \begin{array}{c} t \\ b \end{array} \right)_L, ~~~\ u_{R_3}=t_R\ , ~~~\ d_{R_3}= b_R \\ 
\end{array}
\end{eqnarray}
Besides, there are the gauge fields corresponding to the spin-one bosons that mediate the interactions. In the electroweak sector, we have the field $B_\mu$ which corresponds to the generator $Y$ of the U(1)$_{\rm Y}$ group and the three fields $W^{1,2,3}_\mu$ which correspond to the generators $T^{a}$ (with {\small $a$=1,2,3}) of the SU(2)$_{\rm L}$ group. The matrices $T^{a}$ are in the representation of the multiplet the covariant derivative is acting on. When it acts on the gauge singlet $f_{R}$ we have $T^{a}\equiv 0$, and when it acts on the doublets $L_{i}$ or $Q_{i}$, $T^{a}$ are in fact equivalent  to half of the non-commuting $2 \times 2$ Pauli matrices:
\begin{eqnarray}
T^a= \frac{1}{2} \tau^a \, ; \quad 
\tau_1= \left( \begin{array}{cc} 0 & 1 \\ 1 & 0 \end{array} \right) \, , \ 
\tau_2= \left( \begin{array}{cc} 0 & -i \\ i & 0 \end{array} \right) \, , \ 
\tau_3= \left( \begin{array}{cc} 1 & 0 \\ 0 & -1 \end{array} \right)
\end{eqnarray}
with the commutation relations between these generators given by
\begin{eqnarray}
[T^a,T^b]=i\epsilon^{abc} T_c \ \ \ {\rm and} \ \ \  [Y, Y]=0 
\end{eqnarray}
where $\epsilon^{abc}$ is the antisymmetric tensor. $Y(\psi)$ is a quantum number, usually called the weak hypercharge, to be specified for each field $\psi$. Since the $SU(2)$ is a chiral group, it acts in a different way on left-handed and right-handed fermions, therefore, it is natural to allow for the possibility of assigning different hypercharge quantum numbers to the left and right components of the same fermion field. In general, for a generic fermion with charge $Q_f$, in units of the positron charge $+e$, and third component of the weak isospin $T_{3}$ ($1/2$ for $\nu_{e_{L}}$, $-1/2$ for $e_{L}$, $0$ for $\nu_{e_{R}}$ and $e_{R}$), the fermion hypercharge, defined in terms of the $T_{3}$ and the electric charge  is given by 
\begin{eqnarray}
Y_f=2(Q_f-T_{3}). 
\end{eqnarray}
Is custom, assign the quantum numbers $Y$ in such a way that the electromagnetic interaction term appear in  the SM Lagrangian. To do this, one performed the Weinberg rotation by an angle $\theta_{W}$ in the space of the two neutral gauge fields $W^{3}_{\mu}$, $B^{\mu}$:
\begin{eqnarray}
A^{\mu} = ~B^{\mu}cos\theta_{W} + W^{\mu}_{3}sin\theta_{W} \nonumber \\
~~~~~~ \nonumber \\
Z^{\mu} = -B^{\mu}sin\theta_{W} + W^{\mu}_{3}cos\theta_{W},
\end{eqnarray} 
and choose $Y(L)$, $Y(\nu_{e_{R}})$ and $Y(e_{R})$ so that $A^{\mu}$ couples to the electromagnetic current $J^{\mu}_{em}=-e(\bar{e}_{R}\gamma^{\mu}e_{R}+\bar{e}_{L}\gamma^{\mu}e_{L})$. The remaining terms of the Lagrangian will define the weak neutral current coupled to the other neutral vector boson $Z^{\mu}$. After some algebra, one find:
\begin{eqnarray}
Y(L)=-1, \ Y(e_{R})=-2, \  Y(\nu_{e_{R}})=0, \
Y(Q)=\frac{1}{3}, \ Y(u_{R})= \frac{4}{3}, \ Y(d_{R})= -\frac{2}{3}. 
\end{eqnarray}
Moreover, the quarks are triplets under the ${\rm SU(3)_C}$ group while leptons are color singlets, this leads to the relation 
\begin{eqnarray}
\sum_f Y_f\! =\! \sum_f Q_f\!= \!0
\end{eqnarray}
which ensures the cancellation of chiral anomalies within each generation, thus, preserving the renormalizability of the electroweak theory \cite{Anomaly}.

In the strong interaction sector, there is an octet of gluon fields $G_\mu^{1,\cdots,8}$ which correspond to the eight generators of the ${\rm SU(3)_C}$ group, equivalent to half of the eight $3\times 3$ anti-commuting Gell--Mann matrices, and which obey the relations
\begin{eqnarray}
[T^a,T^b]=if^{abc} T_c \ \ \ {\rm with} \ \ \  {\rm Tr}[T^a T^b]=
\frac12 \delta_{ab} 
\end{eqnarray}
where the tensor $f^{abc}$ is for the structure constants of the ${\rm SU(3)_C}$ group and where we have used the same notation as for the generators of SU(2) as little confusion should be possible. The field strengths are given by
\begin{eqnarray}
G_{\mu \nu}^a &=& \partial_\mu G_\nu^a -\partial_\nu G_\mu^a +g_s \, 
f^{abc} G^b_\mu G^c_\nu \nonumber \\
W_{\mu \nu}^a &=& \partial_\mu W_\nu^a -\partial_\nu W_\mu^a +g \, 
\epsilon^{abc} W^b_\mu W^c_\nu \nonumber \\ 
B_{\mu \nu} &=& \partial_\mu B_\nu -\partial_\nu B_\mu 
\end{eqnarray}
where $g_s$, $g$ and $g'$ are, respectively, the coupling constants of ${\rm SU(3)_C}$,  ${\rm SU(2)_L}$ and  ${\rm U(1)_Y}$. Because of the non-abelian nature of the SU(2) and SU(3) groups, there are self-interactions between their gauge fields, $V_\mu \equiv W_\mu $ or $G_\mu$, leading to triple and quartic gauge couplings 

\vspace*{-2mm}
\hspace*{0.5cm}
\SetWidth{1.}
\begin{picture}(300,100)(0,0)
\Photon(0,50)(50,50){4}{7}
\Photon(50,50)(100,75){-4}{8}
\Photon(50,50)(100,25){4}{8}
\Text(50,50)[]{{\black{\Large $\bullet$}}}
\Text(5,60)[]{$V_{\sigma}$}
\Text(110,75)[]{$V_\mu$}
\Text(110,25)[]{$V_\nu$}
\Text(260,50)[]{$~~~~~~~~~~~:~~~G_{VVV}\ =i g_i \, {\rm Tr} (\partial_\nu V_\mu 
- \partial_\mu V_\nu)[V_\mu,V_\nu]$ ,}
\end{picture}
\vspace*{-.5cm} 

\hspace*{0.5cm}
\begin{picture}(300,100)(0,0)
\Photon(0,25)(50,50){-4}{7}
\Photon(0,75)(50,50){4}{7}
\Photon(50,50)(100,75){-4}{7}
\Photon(50,50)(100,25){4}{7}
\Text(50,50)[]{{\black{\Large $\bullet$}}}
\Text(5,60)[]{$V_{\rho}$}
\Text(5,40)[]{$V_{\sigma}$}
\Text(110,75)[]{$V_\mu$}
\Text(110,25)[]{$V_\nu$}
\Text(260,50)[]{$:~~~G_{VVVV}\ =  \frac12 g_i^2 \, {\rm Tr}[V_\mu,V_\nu]^2$ .}
\end{picture}

The matter fields $\psi$ are minimally coupled to the gauge fields through the covariant derivative $D_\mu$ which, in the case of quarks, is defined as 
\begin{eqnarray}
D_{\mu} \psi = \left( \partial_\mu -ig_s T_a G^a_\mu -ig T_a W^a_\mu -i 
g' \frac{Y_q}{2} B_\mu \right) \psi  
\label{CovariantDerivative}
\end{eqnarray}
and which leads to an interaction Lagrangian between the fermion and gauge fields $V_\mu$ 
\vspace*{-2mm}
\hspace*{0.5cm}
\SetWidth{1.}
\begin{picture}(300,100)(0,0)
\ArrowLine(0,50)(50,50)
\ArrowLine(50,50)(100,75)
\Photon(50,50)(100,25){3}{7}
\Text(50,50)[]{{\black{\Large $\bullet$}}}
\Text(5,60)[]{$\bar{f}$}
\Text(110,75)[]{$f$}
\Text(110,25)[]{$V_{\mu}$}
\Text(260,50)[]{$~~~:~G_{ffV} = -g_i \overline \psi V_\mu \gamma^\mu \psi$ .}
\end{picture}

The SM Lagrangian at symmetric phase is then given by 
\begin{eqnarray}
\label{smlagrangian}
{\cal L}_{\rm SM}&=& -\frac{1}{4} G_{\mu \nu}^a G^{\mu \nu}_a 
-\frac{1}{4} W_{\mu \nu}^a W^{\mu \nu}_a -\frac{1}{4} 
B_{\mu \nu}B^{\mu \nu} \\ 
&& + \bar{L_i}\, i D_\mu \gamma^\mu \, L_i + \bar{e}_{Ri} \, i D_\mu 
\gamma^\mu \, e_{R_i} \ 
+ \bar{Q_i}\, i D_\mu \gamma^\mu \, Q_i + \bar{u}_{Ri} \, i D_\mu 
\gamma^\mu \, u_{R_i} \ + \bar{d}_{Ri} \, i D_\mu \gamma^\mu \, d_{R_i} \nonumber 
\end{eqnarray}
This Lagrangian is invariant under local ${\rm SU(3)_C \times SU(2)_L \times U(1)_Y}$ gauge transformations for fermion and gauge fields. In the case of the electroweak sector, for instance, one has  
\begin{eqnarray}
L(x) \to L'(x)=e^{i\alpha_a(x) T^a + i \beta(x)Y } L(x) \ \ , \ \
R(x) \to R'(x)=e^{i \beta (x) Y} R(x) \nonumber \\
\vec{W}_\mu (x) \to \vec{W_\mu}(x) -\frac{1}{g} \partial_\mu \vec{\alpha}(x)- 
\vec{\alpha}(x) \times \vec{W}_\mu(x) \ , \ B_\mu(x) \to B_\mu(x) -  \frac{1}
{g'} \partial_\mu \beta (x) 
\end{eqnarray}
At the symmetric phase the gauge fields and the fermions fields are massless. In the case of strong interactions, the gluons are indeed massless particles while mass terms of the form $m_q\overline{\psi}\psi$ can be generated for the colored quarks and for the leptons in an SU(3) gauge invariant way, however, in the case of electroweak interactions if we add mass terms the theory will have problems of renormalizability. 

\subsection*{The SM at Broken Phase}

In this section let us write the Lagrangian of the SM after spontaneous symmetry breaking. We write the quantum fields of the theory with the next general notation. For a set of real scalars we use the notation $\phi'_i$, for two-component
Weyl fermions $\psi'_I$, and for vector fields $A^{\prime\mu}_a$. The indices $i,j,k,\ldots$ represent scalar flavor indices; $I,J,K,\ldots$ are fermion flavor indices; and $a,b,c,\ldots$ run over the adjoint representation of the gauge group. Space-time vector indices 
are written as Greek letters $\mu,\nu,\rho,\ldots$, we use a metric
with signature ($-$ $+$ $+$ $+$). The conventions for the quantum fields follow
\cite{MartinNotation}. The primes are used to indicate that these fields are not
squared-mass eigenstates. The kinetic part of the undiagonalized Lagrangian includes
\begin{eqnarray}
- {\cal L} = {1\over 2} m^2_{ij} \phi' _i \phi' _j 
+ {1\over 2} (m^{IJ} \psi' _I \psi' _J + {\rm 
c.c.}) +
{1\over 2} m^2_{ab} A^{\prime \mu}_a A'_{\mu b} .
\label{undiagonalized}
\end{eqnarray}

The undiagonalized matrix $m^2_{ij}=(m^{2}+\lambda \omega^{2})\delta _{ij}+2\lambda\omega_{i}\omega_{j}$ is the real symmetric matrix described in Chapter \ref{cha:Effective Potential} for the scalar sector, in analogous way $m^2_{ab}$ is a real
symmetric matrix and $m^2_{IJ}$ is a Hermitian matrix that depend on the classical background scalar fields. In order to calculate the effective
potential, we need to diagonalize to squared-mass eigenstate
bases $\phi_i$, $\psi_I$, $A^\mu_a$. This can be done by using orthogonal rotation matrices $N^{(S)}$, $N^{(V)}$ for the scalar and vector degrees of freedom, and a unitary matrix $N^{(F)}$ for the fermion degrees of freedom. The rotations
\begin{eqnarray}
\phi' _i &=& N_{ji}^{(S)}\phi _j 
  ,\\
\psi'_I &=& N_{JI}^{(F)*} \psi_J 
  ,\\
A^{\mu\prime}_a &=& N_{ba}^{(V)} A^\mu_b ,
\end{eqnarray}
are chosen such that:
\begin{eqnarray}
N^{(S)}_{ik} m^2_{kl} N^{(S)}_{jl} &=& \delta_{ij} m^2_i 
\label{NS}
  ,\\
N^{(F)}_{IK} m^2_{KL} N^{(F)*}_{JL} &=& \delta_{IJ} m^2_I 
  ,\\
N^{(V)}_{ac} m^2_{cd} N^{(V)}_{bd} &=& \delta_{ab} m^2_a .
\label{NV}
\end{eqnarray}
Here $m^2_i$, $m^2_I$ and $m^2_a$ are respectively the
scalar, fermion, and vector squared-mass eigenvalues which will appear in
the effective potential. In the Standard Model, the diagonalizations just described can be done analytically. Now the interaction terms in a general renormalizable
theory can be written in terms of the squared-mass eigenstate fields
as
\begin{eqnarray}
{\cal L}_{\rm S} &=& - G_{ijk} \phi _i \phi _j \phi _k
- {1\over 4} G_{ijkl} \phi _i \phi _j \phi _k \phi _l
,  \label{LS}
\\
{\cal L}_{\rm SF} &=& - {1\over 2} y^{IJk} \psi _I \psi _J \phi _k + {\rm
c.c.}
, 
\\
{\cal L}_{\rm SV} &=& - {1\over 2} G_{abi} A_\mu^a A^{\mu b} \phi _i
-{1\over 4} G_{abij}  A_\mu^a A^{\mu b} \phi _i \phi _j
- G_{aij} A_{\mu}^{a} \phi _i \partial^{\mu} \phi _j 
, 
\label{LSV}
\\
{\cal L}_{\rm FV} &=& G^{aJ}_I A^a_\mu \psi^{\dagger I} {\overline
\sigma}^\mu \psi_J
,
\\
{\cal L}_{\rm gauge} &=& 
G_{abc} A^a_\mu A^b_\nu \partial^\mu A^{\nu c} 
- {1\over 4} G_{abe} g^{cde} A^{\mu a} A^{\nu b} A_\mu^c A_\nu^d
+ G_{abc} A^a_\mu u^b \partial^\mu \overline u^{c} ,
\label{Lgauge}
\end{eqnarray}
where $u^a$ and $\overline{u}^a$ are massless (in Landau gauge) 
ghost fields. This defines the field-dependent couplings
to be used in the two-loop effective potential calculation.
The scalar interaction couplings $G_{ijk}$ and $G_{ijkl}$
are each completely symmetric under interchange of indices, and real.
The Yukawa couplings $y^{IJk}$ are symmetric under interchange of
the fermion flavour indices $I,J$. The remaining couplings all have
their origins in gauge interactions. The vector-scalar-scalar coupling $G_{aij}$ is
antisymmetric under interchange of $i,j$. The pure gauge interaction
$G_{abc}$ is completely antisymmetric; it is determined by the 
original gauge coupling $g$, the antisymmetric
structure constants $f^{abc}$ of the gauge group, and $N^{(V)}$, according 
to
\begin{eqnarray}
G_{abc} = g f^{efg} 
N^{(V)}_{ae}
N^{(V)}_{bf}
N^{(V)}_{cg}.
\label{definegabc}
\end{eqnarray}
Similarly, if the fermions $\psi'_I$ transform under the gauge group with
representation matrices $(T^a)_I^J$, then the vector-fermion-fermion
couplings are
\begin{eqnarray}
G^{aJ}_I = g (T^b)^K_L N^{(F)*}_{JK} N^{(F)}_{IL} N^{(V)}_{ab} .
\label{definegapq}
\end{eqnarray}
Note that even the dimensionless couplings generically depend on the
classical scalar background fields $\phi_{c}$, through their dependence on the
rotation matrices $N^{(S)}$, $N^{(F)}$, and $N^{(V)}$.

Instead writing here the explicit expression of the SM Lagrangian and its interactions, we present all the vertices of the minimal SM in the $R_{\zeta}$ gauge. We use a few conventions: the gauge parameters in the $R_{\zeta}$ gauge are $\zeta$, $\zeta_{Z}$ and $\zeta_{A}$, we denote sine (cosine) of the weak mixing angle with $S_{\theta_{W}}$ ($C_{\theta_{W}}$). $Q_{f}$, $I_{f}^{(3)}$ denote the electric charge and the third component of isospin of a fermion. Finally we use the abbreviations $\gamma_{\pm}=1 \pm \gamma_{5}$, $v_{f}=I_{f}^{(3)}-2Q_{f}S^{2}_{\theta_{W}}$ and $a_{f}=I_{f}^{(3)}$.

\subsubsection*{Electroweak fermionic Feynman rules:}
\begin{center}
\scalebox{0.4}{
\fcolorbox{white}{white}{
  \begin{picture}(690,340) (77,-45)
    \SetWidth{1.0}
    \SetColor{Black}
    \Photon(78,214)(156,214){7.5}{4}
    \Vertex(156,214){6}
    \Line[arrow,arrowpos=0.5,arrowlength=5,arrowwidth=2,arrowinset=0.2](210,262)(156,214)
    \Line[arrow,arrowpos=0.5,arrowlength=5,arrowwidth=2,arrowinset=0.2](156,214)(210,166)
    \Photon(348,208)(426,208){7.5}{4}
    \Line[arrow,arrowpos=0.5,arrowlength=5,arrowwidth=2,arrowinset=0.2](480,256)(426,208)
    \Line[arrow,arrowpos=0.5,arrowlength=5,arrowwidth=2,arrowinset=0.2](426,208)(480,160)
    \Vertex(426,208){6}
    \Photon(594,208)(672,208){7.5}{4}
    \Line[arrow,arrowpos=0.5,arrowlength=5,arrowwidth=2,arrowinset=0.2](672,208)(726,160)
    \Line[arrow,arrowpos=0.5,arrowlength=5,arrowwidth=2,arrowinset=0.2](726,256)(672,208)
    \Vertex(672,208){6}
    \Line[arrow,arrowpos=0.5,arrowlength=5,arrowwidth=2,arrowinset=0.2](156,10)(210,-38)
    \Line[arrow,arrowpos=0.5,arrowlength=5,arrowwidth=2,arrowinset=0.2](210,58)(156,10)
    \Vertex(156,10){6}
    \Line[dash,dashsize=10](156,10)(78,10)
    \Line[dash,dashsize=10](420,10)(342,10)
    \Line[arrow,arrowpos=0.5,arrowlength=5,arrowwidth=2,arrowinset=0.2](480,64)(426,16)
    \Line[arrow,arrowpos=0.5,arrowlength=5,arrowwidth=2,arrowinset=0.2](420,10)(474,-38)
    \Vertex(420,10){6}
    \Line[dash,dashsize=10,arrow,arrowpos=0.5,arrowlength=5,arrowwidth=2,arrowinset=0.2](666,10)(588,10)
    \Line[arrow,arrowpos=0.5,arrowlength=5,arrowwidth=2,arrowinset=0.2](720,58)(666,10)
    \Line[arrow,arrowpos=0.5,arrowlength=5,arrowwidth=2,arrowinset=0.2](666,10)(720,-38)
    \Vertex(666,10){6}
    \Text(216,214)[lb]{\Large{\Black{$ieQ_{f}\gamma _{\mu}$}}}
    \Text(480,208)[lb]{\Large{\Black{$i\frac{g}{2\sqrt{2}}\gamma _{\mu} \gamma_{+}$}}}
    \Text(726,208)[lb]{\Large{\Black{$i\frac{g}{c\theta _{W}}\gamma _{\mu}(v_f + a_{f}\gamma _{5})$}}}
    \Text(210,10)[lb]{\Large{\Black{$-\frac{gm_{f}}{2m_{W}}$}}}
    \Text(480,10)[lb]{\Large{\Black{$igI_{f}^{(3)}\frac{m_{f}}{m_{W}}\gamma_{5}$}}}
    \Text(720,10)[lb]{\Large{\Black{$i\frac{g}{2\sqrt{2}}\left(\frac{m_d}{m_W}\gamma _{+}-\frac{m_u}{m_W}\gamma _{-} \right)$}}}
    \Text(78,232)[lb]{\Large{\Black{$A$}}}
    \Text(210,274)[lb]{\Large{\Black{$\overline{f}$}}}
    \Text(216,166)[lb]{\Large{\Black{$f$}}}
    \Text(354,226)[lb]{\Large{\Black{$W^{-}$}}}
    \Text(348,190)[lb]{\Large{\Black{$\mu$}}}
    \Text(486,274)[lb]{\Large{\Black{$\overline{u}$}}}
    \Text(486,154)[lb]{\Large{\Black{$d$}}}
    \Text(594,232)[lb]{\Large{\Black{$Z$}}}
    \Text(594,190)[lb]{\Large{\Black{$\mu$}}}
    \Text(78,196)[lb]{\Large{\Black{$\mu$}}}
    \Text(726,274)[lb]{\Large{\Black{$\overline{f}$}}}
    \Text(732,154)[lb]{\Large{\Black{$f$}}}
    \Text(84,28)[lb]{\Large{\Black{$H$}}}
    \Text(210,76)[lb]{\Large{\Black{$\overline{f}$}}}
    \Text(216,-44)[lb]{\Large{\Black{$f$}}}
    \Text(348,28)[lb]{\Large{\Black{$G$}}}
    \Text(486,82)[lb]{\Large{\Black{$\overline{f}$}}}
    \Text(480,-44)[lb]{\Large{\Black{$f$}}}
    \Text(594,28)[lb]{\Large{\Black{$G^{-}$}}}
    \Text(720,70)[lb]{\Large{\Black{$\overline{u}$}}}
    \Text(726,-50)[lb]{\Large{\Black{$d$}}}
  \end{picture}
}}
\end{center}

\subsubsection*{Electroweak trilinear vertices:}
\scalebox{0.4}{
\fcolorbox{white}{white}{
  \begin{picture}(982,522) (-8,-27)
    \SetWidth{1.0}
    \SetColor{Black}
    \Photon(97,394)(196,394){7.5}{5}
    \Vertex(193,394){4}
    \Photon(192,394)(255,458){7.5}{4}
    \Photon(193,392)(252,331){7.5}{4}
    \Text(99,420)[lb]{\Large{\Black{$(Z,~A)$}}}
    \Text(263,474)[lb]{\Large{\Black{$W^{+}$}}}
    \Text(259,321)[lb]{\Large{\Black{$W^{-}$}}}
    \Text(98,376)[lb]{\Large{\Black{$p,\mu$}}}
    \Text(247,430)[lb]{\Large{\Black{$k,\beta$}}}
    \Text(241,366)[lb]{\Large{\Black{$q,\alpha$}}}
    \Text(313,396)[lb]{\Large{\Black{$g(C_{\theta _{W}},~S_{\theta _{W}})\left[ g_{\mu \alpha}(p-q)_{\beta}+g_{\alpha \beta}(q-k)_{\mu}+g_{\mu \beta}(k-p)_{\alpha}\right]$}}}
    \Photon(99,210)(196,213){7.5}{5}
    \Vertex(195,211){5.099}
    \Photon(196,212)(253,150){7.5}{4}
    \Line[dash,dashsize=10,arrow,arrowpos=0.5,arrowlength=5,arrowwidth=2,arrowinset=0.2](196,211)(256,270)
    \Photon(455,212)(552,215){7.5}{5}
    \Photon(749,213)(846,216){7.5}{5}
    \Line[dash,dashsize=10](552,217)(612,276)
    \Photon(552,213)(609,151){7.5}{4}
    \Line[dash,dashsize=10](846,215)(906,274)
    \Photon(848,215)(905,153){7.5}{4}
    \Vertex(554,216){4}
    \Vertex(847,216){5.099}
    \Photon(754,396)(851,399){7.5}{5}
    \Line[dash,dashsize=10,arrow,arrowpos=0.5,arrowlength=5,arrowwidth=2,arrowinset=0.2](851,400)(911,459)
    \Photon(853,401)(910,339){7.5}{4}
    \Vertex(854,403){5}
    \Text(752,425)[lb]{\Large{\Black{$(Z,~A)$}}}
    \Text(754,380)[lb]{\Large{\Black{$\mu$}}}
    \Text(920,470)[lb]{\Large{\Black{$G^{+}$}}}
    \Text(919,339)[lb]{\Large{\Black{$W^{-}$}}}
    \Text(903,374)[lb]{\Large{\Black{$\nu$}}}
    \Text(939,406)[lb]{\Large{\Black{$ig\left(\frac{S^{2}_{\theta _{W}}}{C_{\theta _{W}}},~ -S_{\theta _{W}} \right)m_{W}g_{\mu\nu}$}}}
    \Text(102,238)[lb]{\Large{\Black{$(Z,~A)$}}}
    \Text(99,191)[lb]{\Large{\Black{$\mu$}}}
    \Text(264,284)[lb]{\Large{\Black{$G^{-}$}}}
    \Text(261,148)[lb]{\Large{\Black{$W^{+}$}}}
    \Text(244,190)[lb]{\Large{\Black{$\nu$}}}
    \Text(268,213)[lb]{\Large{\Black{$ig\left(-\frac{S^{2}_{\theta _{W}}}{C_{\theta _{W}}},~S_{\theta _{W}} \right) m_{W}g_{\mu\nu}$}}}
    \Text(616,294)[lb]{\Large{\Black{$H$}}}
    \Text(616,140)[lb]{\Large{\Black{$W^{+}$}}}
    \Text(457,240)[lb]{\Large{\Black{$W^{-}$}}}
    \Text(457,194)[lb]{\Large{\Black{$\mu$}}}
    \Text(606,183)[lb]{\Large{\Black{$\nu$}}}
    \Text(616,217)[lb]{\Large{\Black{$-gm_{W}g_{\mu\nu}$}}}
    \Text(750,239)[lb]{\Large{\Black{$Z$}}}
    \Text(752,193)[lb]{\Large{\Black{$\mu$}}}
    \Text(905,175)[lb]{\Large{\Black{$\nu$}}}
    \Text(910,143)[lb]{\Large{\Black{$Z$}}}
    \Text(910,291)[lb]{\Large{\Black{$H$}}}
    \Text(915,217)[lb]{\Large{\Black{$-g\frac{m_{W}}{C^{2}_{\theta_{W}}}g_{\mu\nu}$}}}
    \Photon(99,32)(196,35){7.5}{5}
    \Line[dash,dashsize=10,arrow,arrowpos=0.5,arrowlength=5,arrowwidth=2,arrowinset=0.2](255,95)(195,36)
    \Line[dash,dashsize=10,arrow,arrowpos=0.5,arrowlength=5,arrowwidth=2,arrowinset=0.2](197,36)(256,-25)
    \Vertex(197,37){5}
    \Photon(462,35)(559,38){7.5}{5}
    \Photon(751,34)(848,37){7.5}{5}
    \Line[dash,dashsize=10](618,95)(558,36)
    \Line[dash,dashsize=10,arrow,arrowpos=0.5,arrowlength=5,arrowwidth=2,arrowinset=0.2](908,95)(848,36)
    \Line[dash,dashsize=10,arrow,arrowpos=0.5,arrowlength=5,arrowwidth=2,arrowinset=0.2](560,36)(619,-25)
    \Line[dash,dashsize=10,arrow,arrowpos=0.5,arrowlength=5,arrowwidth=2,arrowinset=0.2](848,35)(907,-26)
    \Vertex(559,37){4.472}
    \Vertex(848,35){5}
    \Text(99,54)[lb]{\Large{\Black{$W^{-}$}}}
    \Text(99,12)[lb]{\Large{\Black{$\mu$}}}
    \Text(264,102)[lb]{\Large{\Black{$G^{-}$}}}
    \Text(263,-28)[lb]{\Large{\Black{$G^{+}$}}}
    \Text(247,1)[lb]{\Large{\Black{$q$}}}
    \Text(248,73)[lb]{\Large{\Black{$k$}}}
    \Text(259,39)[lb]{\Large{\Black{$\frac{1}{2}g(q-k)_{\mu}$}}}
    \Text(464,59)[lb]{\Large{\Black{$W^{+}$}}}
    \Text(466,17)[lb]{\Large{\Black{$\mu$}}}
    \Text(626,108)[lb]{\Large{\Black{$G^{0}$}}}
    \Text(632,-30)[lb]{\Large{\Black{$G^{-}$}}}
    \Text(614,-3)[lb]{\Large{\Black{$q$}}}
    \Text(613,74)[lb]{\Large{\Black{$k$}}}
    \Text(622,37)[lb]{\Large{\Black{$\frac{1}{2}g(q-k)_{\mu}$}}}
    \Text(754,57)[lb]{\Large{\Black{$(Z,~A)$}}}
    \Text(754,16)[lb]{\Large{\Black{$\mu$}}}
    \Text(915,105)[lb]{\Large{\Black{$G^{-}$}}}
    \Text(898,68)[lb]{\Large{\Black{$k$}}}
    \Text(898,-1)[lb]{\Large{\Black{$q$}}}
    \Text(913,-32)[lb]{\Large{\Black{$G^{+}$}}}
    \Text(907,37)[lb]{\Large{\Black{$g\left( \frac{C^{2}_{\theta_{W}}-S^{2}_{\theta_{W}}}{2C_{\theta_{W}}}, ~ S_{\theta_{W}}\right)(q-k)_{\mu}$}}}
  \end{picture}
}} \\
\begin{center}
\scalebox{0.4}{
\fcolorbox{white}{white}{
  \begin{picture}(848,355) (108,-8)
    \SetWidth{1.0}
    \SetColor{Black}
    \Photon(115,251)(212,254){7.5}{5}
    \Line[dash,dashsize=10](212,254)(272,313)
    \Line[dash,dashsize=10](212,252)(271,191)
    \Vertex(213,253){5}
    \Photon(455,252)(552,255){7.5}{5}
    \Photon(755,253)(852,256){7.5}{5}
    \Line[dash,dashsize=10](551,253)(611,312)
    \Line[dash,dashsize=10](852,255)(912,314)
    \Line[dash,dashsize=10,arrow,arrowpos=0.5,arrowlength=5,arrowwidth=2,arrowinset=0.2](553,253)(612,192)
    \Line[dash,dashsize=10,arrow,arrowpos=0.5,arrowlength=5,arrowwidth=2,arrowinset=0.2](852,255)(911,194)
    \Vertex(553,254){4.472}
    \Vertex(852,256){5}
    \Line[dash,dashsize=10](200,55)(260,114)
    \Line[dash,dashsize=10](199,54)(258,-7)
    \Line[dash,dashsize=10](561,56)(621,115)
    \Line[dash,dashsize=10,arrow,arrowpos=0.5,arrowlength=5,arrowwidth=2,arrowinset=0.2](561,57)(620,-4)
    \Line[dash,dashsize=10](845,59)(905,118)
    \Line[dash,dashsize=10](847,58)(906,-3)
    \Line[dash,dashsize=10](199,53)(101,54)
    \Line[dash,dashsize=10,arrow,arrowpos=0.5,arrowlength=5,arrowwidth=2,arrowinset=0.2](453,56)(562,56)
    \Line[dash,dashsize=10](845,58)(745,56)
    \Vertex(200,53){5.099}
    \Vertex(563,57){5}
    \Vertex(844,60){3.606}
    \Text(116,277)[lb]{\Large{\Black{$Z$}}}
    \Text(116,233)[lb]{\Large{\Black{$\mu$}}}
    \Text(283,324)[lb]{\Large{\Black{$H$}}}
    \Text(281,189)[lb]{\Large{\Black{$G^{0}$}}}
    \Text(266,220)[lb]{\Large{\Black{$q$}}}
    \Text(259,284)[lb]{\Large{\Black{$k$}}}
    \Text(278,253)[lb]{\Large{\Black{$\frac{i}{2}\frac{g}{C_{\theta_{W}}}(q-k)_{\mu}$}}}
    \Text(616,254)[lb]{\Large{\Black{$\frac{i}{2}g(q-k)_{\mu}$}}}
    \Text(918,256)[lb]{\Large{\Black{$\frac{i}{2}g(q-k)_{\mu}$}}}
    \Text(264,55)[lb]{\Large{\Black{$-\frac{i}{2}g\frac{m^{2}_{H}}{m_{W}}$}}}
    \Text(629,58)[lb]{\Large{\Black{$-\frac{i}{2}g\frac{m^{2}_{H}}{m_{W}}$}}}
    \Text(912,59)[lb]{\Large{\Black{$-\frac{3}{2}g\frac{m^{2}_{H}}{m_{W}}$}}}
    \Text(458,279)[lb]{\Large{\Black{$W^{-}$}}}
    \Text(461,232)[lb]{\Large{\Black{$\mu$}}}
    \Text(625,322)[lb]{\Large{\Black{$H$}}}
    \Text(620,187)[lb]{\Large{\Black{$G^{+}$}}}
    \Text(601,286)[lb]{\Large{\Black{$k$}}}
    \Text(607,218)[lb]{\Large{\Black{$q$}}}
    \Text(757,279)[lb]{\Large{\Black{$W^{+}$}}}
    \Text(759,232)[lb]{\Large{\Black{$\mu$}}}
    \Text(921,326)[lb]{\Large{\Black{$H$}}}
    \Text(919,191)[lb]{\Large{\Black{$G^{-}$}}}
    \Text(900,287)[lb]{\Large{\Black{$k$}}}
    \Text(902,223)[lb]{\Large{\Black{$q$}}}
    \Text(114,69)[lb]{\Large{\Black{$G^{0}$}}}
    \Text(266,123)[lb]{\Large{\Black{$H$}}}
    \Text(266,-9)[lb]{\Large{\Black{$G^{0}$}}}
    \Text(458,75)[lb]{\Large{\Black{$G^{-}$}}}
    \Text(625,-13)[lb]{\Large{\Black{$G^{+}$}}}
    \Text(628,129)[lb]{\Large{\Black{$H$}}}
    \Text(755,74)[lb]{\Large{\Black{$H$}}}
    \Text(912,131)[lb]{\Large{\Black{$H$}}}
    \Text(913,-11)[lb]{\Large{\Black{$H$}}}
  \end{picture}
}}
\end{center}

\subsubsection*{Electroweak quadri-linear vertices:}
\scalebox{0.45}{
\fcolorbox{white}{white}{
  \begin{picture}(736,543) (36,-5)
    \SetWidth{1.0}
    \SetColor{Black}
    \Photon(61,508)(214,377){7.5}{10}
    \Photon(60,376)(213,507){7.5}{10}
    \Vertex(137,442){6}
    \Photon(573,373)(726,504){7.5}{10}
    \Photon(572,505)(725,374){7.5}{10}
    \Photon(59,189)(212,320){7.5}{10}
    \Photon(60,320)(213,189){7.5}{10}
    \Photon(573,373)(726,504){7.5}{10}
    \Line[dash,dashsize=10,arrow,arrowpos=0.5,arrowlength=5,arrowwidth=2,arrowinset=0.2](722,321)(650,260)
    \Line[dash,dashsize=10,arrow,arrowpos=0.5,arrowlength=5,arrowwidth=2,arrowinset=0.2](648,258)(722,197)
    \Line[dash,dashsize=10,arrow,arrowpos=0.5,arrowlength=5,arrowwidth=2,arrowinset=0.2](209,126)(134,62)
    \Line[dash,dashsize=10,arrow,arrowpos=0.5,arrowlength=5,arrowwidth=2,arrowinset=0.2](134,62)(208,0)
    \Line[dash,dashsize=10](715,125)(644,62)
    \Line[dash,dashsize=10,arrow,arrowpos=0.5,arrowlength=5,arrowwidth=2,arrowinset=0.2](645,63)(717,-2)
    \Photon(60,127)(133,65){7.5}{5}
    \Photon(133,64)(59,0){7.5}{5}
    \Photon(570,127)(642,62){7.5}{5}
    \Photon(643,62)(567,-4){7.5}{5}
    \Photon(573,326)(647,260){7.5}{5}
    \Photon(648,261)(570,194){7.5}{5}
    \Vertex(651,440){5.099}
    \Vertex(649,259){4.472}
    \Vertex(645,62){5}
    \Vertex(135,62){4.472}
    \Vertex(136,255){4.472}
    \Text(78,517)[lb]{\Large{\Black{$\mu$}}}
    \Text(175,508)[lb]{\Large{\Black{$\beta$}}}
    \Text(83,376)[lb]{\Large{\Black{$\nu$}}}
    \Text(176,381)[lb]{\Large{\Black{$\alpha$}}}
    \Text(55,446)[lb]{\Large{\Black{$(Z,~A)$}}}
    \Text(221,514)[lb]{\Large{\Black{$W^{-}$}}}
    \Text(224,382)[lb]{\Large{\Black{$W^{+}$}}}
    \Text(203,446)[lb]{\Large{\Black{$-g^{2}\left(C^{2}_{\theta_{W}},~S^{2}_{\theta_{W}}  \right)\left[2g_{\mu\nu}g_{\alpha\beta}-g_{\mu\alpha}g_{\nu\beta}-g_{\mu\beta}g_{\nu\alpha} \right]$}}}
    \Text(721,440)[lb]{\Large{\Black{$-g^{2}C_{\theta_{W}}S_{\theta_{W}} \left[2g_{\mu\nu}g_{\alpha\beta}-g_{\mu\alpha}g_{\nu\beta}-g_{\mu\beta}g_{\nu\alpha} \right]$}}}
    \Text(551,512)[lb]{\Large{\Black{$A$}}}
    \Text(551,376)[lb]{\Large{\Black{$Z$}}}
    \Text(602,504)[lb]{\Large{\Black{$\mu$}}}
    \Text(608,378)[lb]{\Large{\Black{$\nu$}}}
    \Text(686,379)[lb]{\Large{\Black{$\alpha$}}}
    \Text(684,503)[lb]{\Large{\Black{$\beta$}}}
    \Text(737,505)[lb]{\Large{\Black{$W^{+}$}}}
    \Text(737,381)[lb]{\Large{\Black{$W^{-}$}}}
    \Text(196,254)[lb]{\Large{\Black{$-g^{2}\left[2g_{\mu\nu}g_{\alpha\beta}-g_{\mu\alpha}g_{\nu\beta}-g_{\mu\beta}g_{\nu\alpha} \right]$}}}
    \Text(35,324)[lb]{\Large{\Black{$W^{-}$}}}
    \Text(33,192)[lb]{\Large{\Black{$W^{-}$}}}
    \Text(223,331)[lb]{\Large{\Black{$W^{+}$}}}
    \Text(225,182)[lb]{\Large{\Black{$W^{+}$}}}
    \Text(92,323)[lb]{\Large{\Black{$\mu$}}}
    \Text(98,193)[lb]{\Large{\Black{$\nu$}}}
    \Text(162,313)[lb]{\Large{\Black{$\alpha$}}}
    \Text(168,197)[lb]{\Large{\Black{$\beta$}}}
    \Text(553,262)[lb]{\Large{\Black{$(Z,~A)$}}}
    \Text(594,335)[lb]{\Large{\Black{$\mu$}}}
    \Text(594,189)[lb]{\Large{\Black{$\nu$}}}
    \Text(736,332)[lb]{\Large{\Black{$G^{+}$}}}
    \Text(730,200)[lb]{\Large{\Black{$G^{-}$}}}
    \Text(726,260)[lb]{\Large{\Black{$g^{2}\left( -\frac{(C^{2}_{\theta_{W}}-S^{2}_{\theta_{W}})^{2}}{2C^{2}_{\theta_{W}}},~-2S^2_{\theta_{W}}\right)g_{\mu\nu}$}}}
    \Text(43,138)[lb]{\Large{\Black{$A$}}}
    \Text(42,-8)[lb]{\Large{\Black{$Z$}}}
    \Text(91,129)[lb]{\Large{\Black{$\mu$}}}
    \Text(105,16)[lb]{\Large{\Black{$\nu$}}}
    \Text(220,137)[lb]{\Large{\Black{$G^{+}$}}}
    \Text(213,2)[lb]{\Large{\Black{$G^{-}$}}}
    \Text(209,63)[lb]{\Large{\Black{$-g^{2}\frac{S_{\theta_{W}}}{C_{\theta_{W}}}(C^{2}_{\theta_{W}}-S^{2}_{\theta_{W}})g_{\mu\nu}$}}}
    \Text(554,138)[lb]{\Large{\Black{$A$}}}
    \Text(548,-6)[lb]{\Large{\Black{$W^{-}$}}}
    \Text(722,141)[lb]{\Large{\Black{$(G^{0},~H)$}}}
    \Text(722,-2)[lb]{\Large{\Black{$G^{\pm}$}}}
    \Text(609,119)[lb]{\Large{\Black{$\mu$}}}
    \Text(623,14)[lb]{\Large{\Black{$\nu$}}}
    \Text(721,62)[lb]{\Large{\Black{$\frac{1}{2}(1,~ \pm i)g^{2}S_{\theta_{W}}g_{\mu\nu}$}}}
  \end{picture}
}}
~~\\
~~\\
~~\\
~~\\
\scalebox{0.4}{
\fcolorbox{white}{white}{
  \begin{picture}(1007,657) (38,-14)
    \SetWidth{1.0}
    \SetColor{Black}
    \Photon(75,592)(160,517){7.5}{6}
    \Photon(70,442)(160,517){7.5}{6}
    \Line[dash,dashsize=10](250,592)(160,517)
    \Line[dash,dashsize=10](250,442)(160,517)
    \Vertex(160,517){5}
    \Vertex(540,522){5}
    \Vertex(920,522){5}
    \Line[dash,dashsize=10](625,592)(535,517)
    \Photon(455,602)(540,527){7.5}{6}
    \Line[dash,dashsize=10,arrow,arrowpos=0.5,arrowlength=5,arrowwidth=2,arrowinset=0.2](625,452)(535,527)
    \Photon(450,447)(540,522){7.5}{6}
    \Line[dash,dashsize=10,arrow,arrowpos=0.5,arrowlength=5,arrowwidth=2,arrowinset=0.2](1005,592)(915,517)
    \Photon(835,602)(920,527){7.5}{6}
    \Photon(825,447)(915,522){7.5}{6}
    \Vertex(145,287){5}
    \Vertex(535,287){5}
    \Vertex(920,287){5}
    \Line[dash,dashsize=10](230,357)(140,282)
    \Line[dash,dashsize=10](230,217)(140,292)
    \Line[dash,dashsize=10,arrow,arrowpos=0.5,arrowlength=5,arrowwidth=2,arrowinset=0.2](620,357)(530,282)
    \Line[dash,dashsize=10,arrow,arrowpos=0.5,arrowlength=5,arrowwidth=2,arrowinset=0.2](1000,357)(910,282)
    \Line[dash,dashsize=10,arrow,arrowpos=0.5,arrowlength=5,arrowwidth=2,arrowinset=0.2](915,292)(825,367)
    \Line[dash,dashsize=10](530,282)(440,207)
    \Line[dash,dashsize=10](530,292)(440,367)
    \Line[dash,dashsize=10](140,292)(50,367)
    \Line[dash,dashsize=10](135,282)(45,207)
    \Vertex(535,57){5}
    \Line[dash,dashsize=10](620,-13)(530,62)
    \Line[dash,dashsize=10](530,62)(440,137)
    \Line[dash,dashsize=10](535,57)(445,-18)
    \Line[dash,dashsize=10](625,132)(535,57)
    \Text(60,607)[lb]{\Large{\Black{$Z$}}}
    \Text(50,437)[lb]{\Large{\Black{$Z$}}}
    \Text(115,587)[lb]{\Large{\Black{$\mu$}}}
    \Text(110,452)[lb]{\Large{\Black{$\nu$}}}
    \Text(250,607)[lb]{\Large{\Black{$(G^{0},~H)$}}}
    \Text(260,437)[lb]{\Large{\Black{$(G^{0},~H)$}}}
    \Text(245,517)[lb]{\Large{\Black{$-\frac{1}{2}\frac{g^2}{C^{2}_{\theta_{W}}}g_{\mu\nu}$}}}
    \Text(440,622)[lb]{\Large{\Black{$Z$}}}
    \Text(430,430)[lb]{\Large{\Black{$(W^{-},~W^{+})$}}}
    \Text(630,612)[lb]{\Large{\Black{$\left((G^{0},H)~;~(G^{0},H)\right)$}}}
    \Text(635,432)[lb]{\Large{\Black{$(G^{+}~;~G^{-})$}}}
    \Text(495,597)[lb]{\Large{\Black{$\mu$}}}
    \Text(505,467)[lb]{\Large{\Black{$\nu$}}}
    \Text(625,522)[lb]{\Large{\Black{$-\frac{1}{2}\left((1,-i)~;~(1,i) \right)g^{2}\frac{S^{2}_{\theta_{W}}}{C_{\theta_{W}}}g_{\mu\nu}$}}}
    \Text(820,622)[lb]{\Large{\Black{$W^{-}$}}}
    \Text(810,430)[lb]{\Large{\Black{$W^{+}$}}}
    \Text(875,597)[lb]{\Large{\Black{$\mu$}}}
    \Text(870,462)[lb]{\Large{\Black{$\nu$}}}
    \Text(1005,612)[lb]{\Large{\Black{$(G^{+},~G^{0},~H)$}}}
    \Text(1005,430)[lb]{\Large{\Black{$(G^{-},~G^{0},~H)$}}}
    \Text(1000,527)[lb]{\Large{\Black{$-\frac{1}{2}(1,~1,~1)g^{2}g_{\mu\nu}$}}}
    \Text(40,377)[lb]{\Large{\Black{$G^{0}$}}}
    \Text(35,190)[lb]{\Large{\Black{$G^{0}$}}}
    \Text(230,377)[lb]{\Large{\Black{$(G^{0},~H)$}}}
    \Text(235,190)[lb]{\Large{\Black{$(G^{0},~H)$}}}
    \Text(225,287)[lb]{\Large{\Black{$-\frac{1}{4}(3,~1)g^{2}\frac{m^{2}_H}{m^{2}_{W}}$}}}
    \Line[dash,dashsize=10,arrow,arrowpos=0.5,arrowlength=5,arrowwidth=2,arrowinset=0.2](920,522)(1010,447)
    \Text(420,382)[lb]{\Large{\Black{$(G^{0},~H)$}}}
    \Text(420,186)[lb]{\Large{\Black{$(G^{0},~H)$}}}
    \Text(625,372)[lb]{\Large{\Black{$G^{+}$}}}
    \Text(630,190)[lb]{\Large{\Black{$G^{-}$}}}
    \Line[dash,dashsize=10,arrow,arrowpos=0.5,arrowlength=5,arrowwidth=2,arrowinset=0.2](535,287)(625,212)
    \Text(615,287)[lb]{\Large{\Black{$-\frac{1}{4}(1,~1)g^{2}\frac{m^{2}_{H}}{m^{2}_{W}}$}}}
    \Text(815,377)[lb]{\Large{\Black{$G^{-}$}}}
    \Text(820,192)[lb]{\Large{\Black{$G^{+}$}}}
    \Text(1000,372)[lb]{\Large{\Black{$G^{+}$}}}
    \Text(1010,192)[lb]{\Large{\Black{$G^{-}$}}}
    \Text(995,287)[lb]{\Large{\Black{$-\frac{1}{2}g^{2}\frac{m^{2}_{H}}{m^{2}_{W}}$}}}
    \Text(420,142)[lb]{\Large{\Black{$H$}}}
    \Text(420,-23)[lb]{\Large{\Black{$H$}}}
    \Text(635,142)[lb]{\Large{\Black{$H$}}}
    \Text(640,-23)[lb]{\Large{\Black{$H$}}}
    \Text(610,58)[lb]{\Large{\Black{$-\frac{3}{4}g^{2}\frac{m^{2}_{H}}{m^{2}_{W}}$}}}
    \Line[dash,dashsize=10,arrow,arrowpos=0.5,arrowlength=5,arrowwidth=2,arrowinset=0.2](920,287)(1010,212)
    \Line[dash,dashsize=10,arrow,arrowpos=0.5,arrowlength=5,arrowwidth=2,arrowinset=0.2](835,212)(920,287)
  \end{picture}
}}

\subsubsection*{Non-zero trilinear vertices involving FP ghosts in Landau gauge:}
\begin{center}
\scalebox{0.4}{
\fcolorbox{white}{white}{
  \begin{picture}(909,418) (80,-45)
    \SetWidth{1.0}
    \SetColor{Black}
    \Photon(81,264)(191,265){7.5}{6}
    \Line[dash,dashsize=2,arrow,arrowpos=0.5,arrowlength=5,arrowwidth=2,arrowinset=0.2](252,201)(193,266)
    \Line[dash,dashsize=2,arrow,arrowpos=0.5,arrowlength=5,arrowwidth=2,arrowinset=0.2](192,264)(254,335)
    \Vertex(193,265){6}
    \Photon(455,263)(565,264){7.5}{6}
    \Photon(773,266)(883,267){7.5}{6}
    \Photon(84,17)(194,18){7.5}{6}
    \Photon(462,19)(572,20){7.5}{6}
    \Photon(775,19)(885,20){7.5}{6}
    \Line[dash,dashsize=2,arrow,arrowpos=0.5,arrowlength=5,arrowwidth=2,arrowinset=0.2](566,265)(628,336)
    \Line[dash,dashsize=2,arrow,arrowpos=0.5,arrowlength=5,arrowwidth=2,arrowinset=0.2](624,204)(565,269)
    \Line[dash,dashsize=2,arrow,arrowpos=0.5,arrowlength=5,arrowwidth=2,arrowinset=0.2](884,267)(946,338)
    \Line[dash,dashsize=2,arrow,arrowpos=0.5,arrowlength=5,arrowwidth=2,arrowinset=0.2](941,207)(882,272)
    \Line[dash,dashsize=2,arrow,arrowpos=0.5,arrowlength=5,arrowwidth=2,arrowinset=0.2](251,-43)(192,22)
    \Line[dash,dashsize=2,arrow,arrowpos=0.5,arrowlength=5,arrowwidth=2,arrowinset=0.2](197,20)(259,91)
    \Line[dash,dashsize=2,arrow,arrowpos=0.5,arrowlength=5,arrowwidth=2,arrowinset=0.2](629,-41)(570,24)
    \Line[dash,dashsize=2,arrow,arrowpos=0.5,arrowlength=5,arrowwidth=2,arrowinset=0.2](572,19)(634,90)
    \Line[dash,dashsize=2,arrow,arrowpos=0.5,arrowlength=5,arrowwidth=2,arrowinset=0.2](942,-41)(883,24)
    \Line[dash,dashsize=2,arrow,arrowpos=0.5,arrowlength=5,arrowwidth=2,arrowinset=0.2](886,20)(948,91)
    \Vertex(569,265){6}
    \Vertex(888,268){6}
    \Vertex(887,19){6}
    \Vertex(574,18){6}
    \Vertex(196,17){6}
    \Text(84,292)[lb]{\Large{\Black{$(Z,~A)$}}}
    \Text(85,242)[lb]{\Large{\Black{$(p,~\mu)$}}}
    \Text(260,345)[lb]{\Large{\Black{$u^{-}$}}}
    \Text(259,193)[lb]{\Large{\Black{$u^{-}$}}}
    \Text(245,306)[lb]{\Large{\Black{$p$}}}
    \Text(244,267)[lb]{\small{\Black{$g\frac{1}{\zeta}\left( C_{\theta_{W}},~S_{\theta_{W}}\right)p_{\mu}$}}}
    \Text(632,268)[lb]{\small{\Black{$-g\frac{1}{\zeta}\left( C_{\theta_{W}},~S_{\theta_{W}}\right)p_{\mu}$}}}
    \Text(259,18)[lb]{\small{\Black{$-g\frac{1}{\zeta}\left( C_{\theta_{W}},~S_{\theta_{W}}\right)p_{\mu}$}}}
    \Text(634,18)[lb]{\small{\Black{$g\frac{1}{\zeta}\left( C_{\theta_{W}},~S_{\theta_{W}}\right)p_{\mu}$}}}
    \Text(946,270)[lb]{\small{\Black{$-g\left(\frac{C_{\theta_{W}}}{\zeta _{Z}} ,~\frac{S_{\theta_{W}}}{\zeta _{A}}\right)p_{\mu}$}}}
    \Text(954,21)[lb]{\small{\Black{$g\left(\frac{C_{\theta_{W}}}{\zeta _{Z}} ,~\frac{S_{\theta_{W}}}{\zeta _{A}}\right)p_{\mu}$}}}
    \Text(455,290)[lb]{\Large{\Black{$(Z,~A)$}}}
    \Text(457,242)[lb]{\Large{\Black{$\mu$}}}
    \Text(639,348)[lb]{\Large{\Black{$u^{+}$}}}
    \Text(638,199)[lb]{\Large{\Black{$u^{+}$}}}
    \Text(609,292)[lb]{\Large{\Black{$p$}}}
    \Text(777,290)[lb]{\Large{\Black{$W^{+}$}}}
    \Text(778,246)[lb]{\Large{\Black{$\mu$}}}
    \Text(954,352)[lb]{\Large{\Black{$u_{(Z,~A)}$}}}
    \Text(951,200)[lb]{\Large{\Black{$u^{-}$}}}
    \Text(925,295)[lb]{\Large{\Black{$p$}}}
    \Text(84,44)[lb]{\Large{\Black{$W^{-}$}}}
    \Text(90,-7)[lb]{\Large{\Black{$\mu$}}}
    \Text(272,110)[lb]{\Large{\Black{$u^{-}$}}}
    \Text(259,-49)[lb]{\Large{\Black{$u_{(Z,~A)}$}}}
    \Text(238,50)[lb]{\Large{\Black{$p$}}}
    \Text(463,47)[lb]{\Large{\Black{$W^{+}$}}}
    \Text(464,-4)[lb]{\Large{\Black{$\mu$}}}
    \Text(639,103)[lb]{\Large{\Black{$u^{+}$}}}
    \Text(637,-50)[lb]{\Large{\Black{$u_{(Z,~A)}$}}}
    \Text(609,44)[lb]{\Large{\Black{$p$}}}
    \Text(778,42)[lb]{\Large{\Black{$W^{-}$}}}
    \Text(953,109)[lb]{\Large{\Black{$u_{(Z,~A)}$}}}
    \Text(931,54)[lb]{\Large{\Black{$p$}}}
    \Text(952,-45)[lb]{\Large{\Black{$u^{+}$}}}
    \Text(783,-2)[lb]{\Large{\Black{$\mu$}}}
  \end{picture}
}}
\end{center}

\subsubsection*{QCD vertices:}
\begin{center}
\scalebox{0.5}{
\fcolorbox{white}{white}{
  \begin{picture}(766,588) (53,-29)
    \SetWidth{1.0}
    \SetColor{Black}
    \Vertex(160,448){5}
    \Vertex(145,218){5}
    \Text(60,538)[lb]{\Large{\Black{$g$}}}
    \Text(50,368)[lb]{\Large{\Black{$g$}}}
    \Text(115,518)[lb]{\Large{\Black{$a,\mu$}}}
    \Text(110,383)[lb]{\Large{\Black{$c,\rho$}}}
    \Text(250,538)[lb]{\Large{\Black{$g$}}}
    \Text(260,368)[lb]{\Large{\Black{$g$}}}
    \Text(245,448)[lb]{\Large{\Black{$-ig_{s}^{2}\left[ f^{abe}f^{cde}(g^{\mu\rho}g^{\nu\sigma}-g^{\mu\sigma}g^{\nu\rho})+f^{ace}f^{bde}(g^{\mu\nu}g^{\rho\sigma}-g^{\mu\sigma}g^{\nu\rho})+f^{ade}f^{bce}(g^{\mu\nu}g^{\rho\sigma}-g^{\mu\rho}g^{\nu\sigma})\right]$}}}
    \Gluon(80,518)(160,448){7.5}{9}
    \Gluon(245,518)(160,448){7.5}{9}
    \Gluon(75,378)(160,448){7.5}{9}
    \Gluon(255,373)(160,448){7.5}{10}
    \Text(195,518)[lb]{\Large{\Black{$b,~\nu$}}}
    \Text(186,383)[lb]{\Large{\Black{$d,~\sigma$}}}
    \Gluon(145,218)(145,313){7.5}{8}
    \Gluon(145,223)(65,158){7.5}{8}
    \Gluon(145,218)(220,153){7.5}{8}
    \Text(145,328)[lb]{\Large{\Black{$g$}}}
    \Text(225,143)[lb]{\Large{\Black{$g$}}}
    \Text(50,143)[lb]{\Large{\Black{$g$}}}
    \Line[arrow,arrowpos=0.5,arrowlength=5,arrowwidth=2,arrowinset=0.2](90,203)(110,218)
    \Line[arrow,arrowpos=0.5,arrowlength=5,arrowwidth=2,arrowinset=0.2](125,273)(125,248)
    \Line[arrow,arrowpos=0.5,arrowlength=5,arrowwidth=2,arrowinset=0.2](184,204)(172,216)
    \Text(88,224)[lb]{\Large{\Black{$p$}}}
    \Text(108,268)[lb]{\Large{\Black{$k$}}}
    \Text(180,224)[lb]{\Large{\Black{$q$}}}
    \Text(164,300)[lb]{\Large{\Black{$a,~\mu$}}}
    \Text(100,154)[lb]{\Large{\Black{$b,~\nu$}}}
    \Text(159,154)[lb]{\Large{\Black{$c,~\rho$}}}
    \Text(248,228)[lb]{\Large{\Black{$g_{s}f^{abc}\left[ g^{\mu\nu}(k-p)^{\rho}+g^{\nu\rho}(p-q)^{\mu}+g^{\rho\mu}(q-k)^{\nu} \right]$}}}
    \Vertex(144,36){4}
    \Gluon(144,108)(144,36){7.5}{6}
    \Line[arrow,arrowpos=0.5,arrowlength=5,arrowwidth=2,arrowinset=0.2](220,-28)(144,36)
    \Line[arrow,arrowpos=0.5,arrowlength=5,arrowwidth=2,arrowinset=0.2](144,36)(72,-28)
    \Vertex(668,36){4}
    \Line[dash,dashsize=2,arrow,arrowpos=0.5,arrowlength=5,arrowwidth=2,arrowinset=0.2](760,-28)(668,36)
    \Line[dash,dashsize=2,arrow,arrowpos=0.5,arrowlength=5,arrowwidth=2,arrowinset=0.2](668,36)(584,-24)
    \Gluon(668,108)(668,36){7.5}{6}
    \Text(144,112)[lb]{\Large{\Black{$g$}}}
    \Text(228,-28)[lb]{\Large{\Black{$f$}}}
    \Text(56,-28)[lb]{\Large{\Black{$\overline{f}$}}}
    \Text(772,-32)[lb]{\Large{\Black{$u_{g}$}}}
    \Text(568,-28)[lb]{\Large{\Black{$u_{g}$}}}
    \Text(668,112)[lb]{\Large{\Black{$g$}}}
    \Text(224,56)[lb]{\Large{\Black{$ig_{s}\gamma^{\mu}T^{a}$}}}
    \Text(160,88)[lb]{\Large{\Black{$a,~\mu$}}}
    \Text(684,88)[lb]{\Large{\Black{$b,~\mu$}}}
    \Text(612,16)[lb]{\Large{\Black{$a$}}}
    \Text(732,8)[lb]{\Large{\Black{$c$}}}
    \Text(636,0)[lb]{\Large{\Black{$p$}}}
    \Text(784,48)[lb]{\Large{\Black{$g_{s}f^{abc}p^{\mu}$}}}
  \end{picture}
}}
\end{center}

\chapter{Beta Functions and Anomalous Dimensions of SM \label{AppBetaF-EP}}
\markboth{APPENDIX \ref{AppBetaF-EP}}{APPENDIX \ref{AppBetaF-EP}}

As we mentioned in the last section of the Chapter \ref{cha:Effective Potential},
we can compute the beta function of the coupling $\lambda$ from the
non-improved effective potential (see eq. (\ref{eq:1l-SM-EP})) that,
up to one-loop order, have the explicit form:
\begin{eqnarray}
V(\phi_{c})=\frac{1}{2}m^{2}\phi_{c}^{2}+\frac{1}{4}\lambda\phi_{c}^{4} +\frac{1}{64\pi^{2}}\left[H^{2}\left(ln\frac{H}{\mu^{2}}-\frac{3}{2}\right)+3G^{2}\left(ln\frac{G}{\mu^{2}}-\frac{3}{2}\right)\right. ~~~~~~~~~~~~~~~~~~~~~\\ \nonumber \\
+\left.6W^{2}\left(ln\frac{W}{\mu^{2}}-\frac{5}{6}\right)+3Z^{2}\left(ln\frac{Z}{\mu^{2}}-\frac{5}{6}\right)-12T^{2}\left(ln\frac{T}{\mu^{2}}-\frac{3}{2}\right)\right], \nonumber \label{eq:EP-UP-1l}
\end{eqnarray}
where
\begin{eqnarray*}
H=m^{2}+3\lambda\phi_{c}^{2};\; G=m^{2}+\lambda\phi_{c}^{2};\; W=\frac{1}{4}g^{2}\phi_{c}^{2};\; Z=\frac{1}{4}(g^{2}+g'^{2})\phi_{c}^{2};\; T=\frac{1}{2}h_{t}^{2}\phi_{c}^{2}.
\end{eqnarray*}
The effective potential is invariant under renormalization scale transformations,
\begin{eqnarray*}
\frac{dV(\phi_{c})}{dt}=-\sum_{n}\frac{1}{n!}\left[\frac{d\tilde{\Gamma}_{n}(0)}{dt}\phi_{c}^{n}+\tilde{\Gamma}_{n}(0)\frac{d\phi_{c}^{n}}{dt}\right]=0,
\end{eqnarray*}
as a consequence of the Callan-Symanzik equation for the 1PI Green's
functions 
\begin{eqnarray}
\left(\frac{\partial}{\partial t}+\beta_{g_{i}}\frac{\partial}{\partial g_{i}}+m^{2}\gamma_{m}\frac{\partial}{\partial m^{2}}+n\gamma\right)\tilde{\Gamma}_{n}=0.\label{eq:C-S-eq-1PI}
\end{eqnarray}
Here $t=ln\mu^2$ represent the dependence of the energy scale, and
$\gamma$ is the anomalous dimension of the field $\phi_{c}$. The
first derivative of the potential $V(\phi_{c})$ with respect to $t$ can
be computed explicitly by differentiating eq. (\ref{eq:EP-UP-1l}).
Neglecting two-loops terms of order $O(g_{i}^{3}=\lambda^{3},\: g^{6},\: g^{4}g'^{2},\: g^{4}h_{t}^{2},\: etc.)$, we obtain
\begin{eqnarray*}
\frac{dV(\phi_{c})}{dt}= \frac{1}{2}m^{2}\phi_{c}^{2}\left[\gamma_{m}+2\gamma-\frac{12\lambda}{32\pi^{2}}\right] ~~~~~~~~~~~~~~~~~~~~~~~~~~~~~~~~~~~~~~~~~~~~~~~~~~~~~\\ + \frac{1}{4}\phi_{c}^{4}\left\{ \beta_{\lambda}+4\lambda\gamma-\frac{1}{16\pi^{2}}\left[12\lambda^{2}+\frac{3}{8}g^{4}+\frac{3}{16}(g^{2}+g'^{2})^{2}-3h_{t}^{^{4}}\right]\right\},
\end{eqnarray*}
and therefore
\begin{eqnarray}
&\beta_{\lambda}+4\lambda\gamma=\dfrac{1}{16\pi^{2}}\left[12\lambda^{2}+\dfrac{3}{8}g^{4}+\dfrac{3}{16}(g^{2}+g'^{2})^{2}-3h_{t}^{^{4}}\right],\nonumber \\ \nonumber \\
&\gamma_{m}+2\gamma=\dfrac{12\lambda}{32\pi^{2}}.\label{eq:anomalous-SE}
\end{eqnarray}
We have two equations for three variables: $\beta_{\lambda}$, $\gamma_{m}$
and $\gamma$. We need to compute explicitly one of them to obtain the
others two from the system of equations (\ref{eq:anomalous-SE});
we choose $\gamma$. The anomalous dimension $\gamma$ can easily
be determined from the Higgs self-energy. Let see this in more detail. 

When one makes a shift of the renormalization scale, in such way that:
\begin{eqnarray*}
\mu^{2}\rightarrow\mu^{2}+\delta\mu^{2},\\
g_{i}\rightarrow g_{i}+\delta g_{i},\\
m^{2}\rightarrow m^{2}+\delta m^{2},\\
H\rightarrow H'=(1+\delta\eta)H,
\end{eqnarray*}
then, the n-points 1PI Green's functions transforms as:
\begin{eqnarray*}
\tilde{\Gamma}_{n}\rightarrow\tilde{\Gamma}_{n}+\delta\tilde{\Gamma}_{n}=(1+n\delta\eta)\tilde{\Gamma}_{n}.
\end{eqnarray*}
Like $\tilde{\Gamma}_{n}$ is a function of $\mu^{2}$, $g_{i}$ and
$m^{2}$, we may use the chain rule to obtain
\begin{eqnarray*}
\delta\tilde{\Gamma}_{n}=\frac{\partial\tilde{\Gamma}_{n}}{\partial\mu^{2}}\delta\mu^{2}+\frac{\partial\tilde{\Gamma}_{n}}{\partial g_{i}}\delta g_{i}+\frac{\partial\tilde{\Gamma}_{n}}{\partial m^{2}}\delta m^{2}=n\delta\eta\tilde{\Gamma}_{n}.
\end{eqnarray*}
If we multiply by $\mu^{2}/\delta\mu^{2}$ on both side of the above
equation, we obtain the Callan-Symanzik equation given by the formula
(\ref{eq:C-S-eq-1PI}), where 
\begin{eqnarray}
\gamma=-\mu^{2}\frac{\delta\eta}{\delta\mu^{2}}.\label{eq:gammaexplicit}
\end{eqnarray}
To obtain an explicit result for $\gamma$, we need to give some meaning
to $\delta\eta$. We know that a shift in the scale $\mu$ change
the field $H$ by $H'$ in such way that
\begin{eqnarray*}
\delta\eta H=H'-H,
\end{eqnarray*}
where $H'$ is the field at $\mu^{2}+\delta\mu^{2}$ i.e. $H'=Z_{\phi}^{-1/2}(\mu^{2}+\delta\mu^{2})H_{0}$,
while $H$ is the field at $\mu^{2}$ i.e. $H=Z_{\phi}^{-1/2}(\mu^{2})H_{0}$.
The renormalization constant $Z_{\phi}$ renormalized the bare field
$H_{0}$ as is usual in the on-shell schemes. Therefore, we have the
next identity:
\begin{eqnarray}
\delta\eta=\frac{Z_{\phi}^{-1/2}(\mu^{2}+\delta\mu^{2})-Z_{\phi}^{-1/2}(\mu^{2})}{Z_{\phi}^{-1/2}(\mu^{2})}.\label{eq:variationscale}
\end{eqnarray}
Now, for $\delta\mu^{2}$ close to zero, and replacing the eq. (\ref{eq:variationscale})
in eq. (\ref{eq:gammaexplicit}), we obtain
\begin{eqnarray*}
\gamma=-\frac{\mu^{2}}{Z_{\phi}^{-1/2}(\mu^{2})}\frac{\partial Z_{\phi}^{-1/2}(\mu^{2})}{\partial\mu^{2}}=\frac{1}{2}\frac{\mu^{2}}{Z_{\phi}(\mu^{2})}\frac{\partial Z_{\phi}(\mu^{2})}{\partial\mu^{2}}.
\end{eqnarray*}
Is usually express the renormalization constant as $Z_{\phi}=1+\delta Z_{\phi}$.
Thus, neglecting two-loop contributions, we find 
\begin{eqnarray*}
\gamma=\frac{1}{2}\frac{\partial}{\partial (ln\mu^{2})}\delta Z_{\phi}.
\end{eqnarray*}
\begin{figure}
\begin{center}
\scalebox{0.6}{
\fcolorbox{white}{white}{
  \begin{picture}(528,52) (173,-169)
    \SetWidth{1.0}
    \SetColor{Black}
    \Line[dash,dashsize=10,arrow,arrowpos=0.5,arrowlength=5,arrowwidth=2,arrowinset=0.2](220,-143)(345,-143)
    \GOval(280,-143)(25,25)(0){0.882}
    \Text(365,-148)[lb]{\Large{\Black{$=$}}}
    \Line[dash,dashsize=10,arrow,arrowpos=0.5,arrowlength=5,arrowwidth=2,arrowinset=0.2](395,-143)(520,-143)
    \COval(460,-143)(25,25)(0){Black}{White}
    \Text(545,-148)[lb]{\Large{\Black{$+$}}}
    \Line[dash,dashsize=10,arrow,arrowpos=0.5,arrowlength=5,arrowwidth=2,arrowinset=0.2](575,-143)(700,-143)
    \COval(635,-143)(7.071,7.071)(-135.0){Black}{White}\Line(638.536,-139.464)(631.464,-146.536)\Line(638.536,-146.536)(631.464,-139.464)
    \Text(445,-148)[lb]{\Large{\Black{$1PI$}}}
    \Text(150,-152)[lb]{\Large{\Black{$\sum(p^2) =$}}}
  \end{picture}
}}
\end{center}
\caption{\label{HiggsS-E} The renormalized two-points Green function.}
\end{figure}

The counter-term $\delta Z_{\phi}$ can be computed from the two-points
Green's function $\Sigma(p^{2})$, represented schematically by Figure
\ref{HiggsS-E}. From the renormalized version of Higgs sector of the SM Lagrangian
and up one-loop level, is very easy to see that\footnote{The constant $Z_{\phi}$ appears in the kinetic term of the renormalized Higgs Lagrangian, 
\[\frac{Z_{\phi}}{2}\partial_{\mu}H\partial^{\mu}H + \frac{1}{2}(m_{H}^{2}+\delta m_{H}^{2})H^{2}.\]}
\begin{eqnarray*}
\Sigma(p^{2})=\Sigma(0)+\Sigma'(0)p^{2}+\widetilde{\Sigma}(p^{2})+i(\delta m_{H}^{2}-p^{2}\delta Z_{\phi}),
\end{eqnarray*}
where $\widetilde{\Sigma}(p^{2})$ is the finite part of the self-energy,
$\Sigma(0)$ and $\Sigma'(0)=\left.\frac{\partial\Sigma}{\partial p^{2}}\right|_{p^{2}=0}$
are the coefficients that contain the divergences, and $\delta m_{H}^{2}$
and $\delta Z_{\phi}$ are the counter-terms associated to the renormalization
scheme. To fix the finite part of the counter-terms we impose the
conditions:
\begin{eqnarray*}
\widetilde{\Sigma}(0)=0 & ; & \left.\frac{\partial\widetilde{\Sigma}(p^{2})}{\partial p^{2}}\right|_{p^{2}=0}=0.
\end{eqnarray*}
From the second condition we have
\begin{eqnarray*}
\delta Z_{\phi}=i\Sigma'(0).
\end{eqnarray*}
Explicitly, at the Landau gauge, the coefficient of $p^{2}$ in the
Higgs self-energy is
\begin{eqnarray*}
\Sigma'(0)=\frac{-i}{(4\pi)^{2}}\left(\frac{3}{4}G^{2}+\frac{3}{2}g^{2}-3h_{t}^{2}\right)\left(\frac{1}{\varepsilon}-ln\frac{1}{\mu^{2}}\right).
\end{eqnarray*}
With the above expression is easy to see that to one-loop level
\begin{eqnarray}
\gamma=\frac{1}{(4\pi)^{2}}\left(\frac{3}{8}G^{2}+\frac{3}{4}g^{2}-\frac{3}{2}h_{t}^{2}\right).\label{eq:gamma-1l}
\end{eqnarray}
Now, after a bit of algebra, we obtain from eq. (\ref{eq:anomalous-SE})
and eq. (\ref{eq:gamma-1l}) the beta function of $\lambda$ and the
function $\gamma_{m}$:
\begin{eqnarray}
&\beta_{\lambda}=\dfrac{1}{(4\pi)^{2}}\left(12\lambda^{2}+\dfrac{3}{8}g^{4}+\dfrac{3}{16}(g^{2}+g'^{2})^{2}-3h_{t}^{4}-3\lambda g^{2}-\dfrac{3}{2}\lambda(g^{2}+g'^{2})+6\lambda h_{t}^{2}\right), \label{beta-lambda-1l} \\ \nonumber \\
&\gamma_{m}=\dfrac{1}{(4\pi)^{2}}\left(6\lambda-\dfrac{3}{4}(g^{2}+g'^{2})-\dfrac{3}{2}g^{2}+3h_{t}^{2}\right). \label{gammam2-1l}
\end{eqnarray}
As was mentioned in the first chapter, the RGI potential is completely
determined if one find the running coupling $\lambda(\Lambda)$. This
required the knowledge of $\beta_{\lambda}$ together with the renormalization
group equations for the others coupling constants of the theory. For
this reason, we list here the beta functions for the gauge and Yukawa
couplings to one-loop level in the $\overline{MS}$ scheme \cite{Gross, Ramond}:
 \begin{eqnarray}
&\dfrac{dg}{dt}=\dfrac{1}{32\pi^{2}}\left(-\dfrac{19}{6}g^{3}\right),\\ \nonumber \\
&\dfrac{dg'}{dt}=\dfrac{1}{32\pi^{2}}\left(\dfrac{41}{6}g'^{3}\right),\\ \nonumber \\
&\dfrac{dg_{s}}{dt}=\dfrac{1}{32\pi^{2}}\left(-7g_{s}^{3}\right),\\ \nonumber \\
&\dfrac{dh_{t}}{dt}=\dfrac{1}{32\pi^{2}}\left[\dfrac{9}{2}h_{t}^{3}-\left(8g_{s}^{2}+\dfrac{9}{4}g^{2}+\dfrac{17}{12}g'^{2}\right)h_{t}\right].
\end{eqnarray}
The two-loop gauge, Yukawa and Higgs self-coupling beta functions are in \cite{beta-2l}, see also Ref. \cite{Ford-Jack}. Finally the corresponding gauge coupling beta function of the strong interaction part is known up to four-loop order \cite{Beta-QCD}.
The beta functions to two-loop order are:
 \begin{eqnarray}
 32\pi^{2}\beta^{(2)}_{\lambda}&=&-312\lambda^3-144 \lambda^2 h_t^2 + 36 \lambda^2 (3 g^2 + g'^2) -
 3 \lambda h_t^4 + \lambda h_t^2
 \left(80 g_s^2 + \frac{45}{2} g^2 + \frac{85}{6} g'^2\right)\nonumber \\
 &&\quad - \frac{73}{8} \lambda g^4 + \frac{39}{4} \lambda g^2 g'^2 +
 \frac{629}{24} \lambda g'^4 + 30 h_t^6 - 32 h_t^4 g_s^2 -
 \frac{8}{3} h_t^4 g'^2 - \frac{9}{4} h_t^2 g^4  \nonumber \\
 &&\quad+ \frac{21}{2} h_t^2 g^2 g'^2 - \frac{19}{4} h_t^2 g'^4
 + \frac{305}{16} g^6 - \frac{289}{48} g^4 g'^2 - \frac{559}{48} g^2 g'^4
 - \frac{379}{48} g'^6 , \label{beta-lambda-2l}\\
 32\pi^{2}\beta_{h_t}^{(2)}&=&h_t \left(-12 h_t^4 + h_t^2 \left(\frac{131}{16} g'^2
 + \frac{225}{16} g^2 + 36 g_s^2 - 12 \lambda\right) +\frac{1187}{216} g'^4\right.\nonumber \\
 &&\quad \left. - \frac{3}{4} g^2 g'^2 + \frac{19}{9} g'^2 g_s^2
 -\frac{23}{4} g^4 + 9 g^2 g_s^2 - 108 g_s^4 + 6 \lambda^2\right), \\
 32\pi^{2}\beta^{(2)}_{g'}&=&g'^3 \left(\frac{199}{18} g'^2 + \frac{9}{2} g^2
 + \frac{44}{3} g_s^2 - \frac{17}{6}h_t^2\right), \\
 32\pi^{2}\beta^{(2)}_{g}&=&g^3 \left(\frac{3}{2} g'^2 + \frac{35}{6} g^2
 + 12 g_s^2 - \frac{3}{2} h_t^2\right),\\
 32\pi^{2}\beta^{(2)}_{g_s}&=&g_s^3 \left(\frac{11}{6} g'^2 + \frac{9}{2} g^2
 - 26 g_s^2 -2 \lambda_t^2\right),
 \end{eqnarray}
and the function $\gamma_{m}$ to two-loop level is:
\begin{eqnarray}\label{eq:gammH2}
 \gamma^{(2)}_{m}&=&-\frac{1}{(16\pi^{2})^{2}}
 \bigg(-30\lambda^{2}
 -36\lambda h^{2}_{t}+12\lambda(3g^{2}+g'^{2})-\frac{27}{4}
 h^{4}_{t}+20g^{2}_{s}h^{2}_{t}+\frac{45}{8}g^{2}h^{2}_{t}
 \nonumber\\
 &&+\frac{85}{24}
 g'^{2}h^{2}_{t}-\frac{145}{32}g^{4}+\frac{15}{16}g'^{2}
 g^{2}
 +\frac{157}{96}g^{4}\bigg). \label{gammam2-2l}
 \end{eqnarray}
 
\chapter{FeynArts}\label{AppFeynArts}
\markboth{APPENDIX \ref{AppFeynArts}}{APPENDIX \ref{AppFeynArts}} 

FeynArts \cite{FeynArts} is a Mathematica package for the generation and visualization 
of Feynman diagrams and amplitudes. Mathematica is a basic programming language that makes it straightforward to use intermediate results of FeynArts in a variety of ways, for instance, the amplitudes computed can be exported to FeynCalc to make the Dirac algebra and then algebraical and numerical computations. This appendix gives an overview of the principal functions of FeynArts, useful to compute one-loop and two-loop Feynman amplitudes.

\section{Generation of Amplitudes}

The current version of FeynArts is \Var{FeynArts~3.5}, it uses almost 
the same syntax, though with many extensions of the launched version, so it is advisable to get the latest version. In any case, the generation of amplitudes is a 
three-steps process. In the first step, the distinct topologies for a given 
number of loops and external legs are produced using the code:

   {\tt ~~~ T1 = CreateTopologies[1, 2 \(\to\) 2]}

This process is a purely topological task and requires no quantum fields input. The internal algorithm starts from given zero-leg topologies of 
the requested loop order and successively adds legs.\\
In the second step, the fields are inserted over the topologies in all 
admissible ways according to a model's particle content that is read from a Model 
File. This step is implemented with the code:

   {\tt ~~~ F1 = InsertFields[T1, \{F[2,\{2\}], F[1,\{2\}]\} \(\to\) \{F[2,\{1\}], F[1,\{1\}]\}}

Finally, the Feynman rules are applied with

   {\tt ~~~ Amp1 = CreateFeynAmp[F1]}

The field labelling above is the one of the default model, 
\textit{SM.mod}, and corresponds to the muon decay $\mu ~ + ~\nu_{\mu} ~\to~ e ~+~ \bar{\nu}_{e}$, where $\mu$ 
and $e$ are the second and first members of the massive leptons class 
{\tt F[2]}, and {\tt F[1,\{~\}]} is the neutrinos class. The particle content of \textit{SM.mod} is summarized in Appendix B of ref. \cite{FeynArts}. This notation is part of the more general concept of field levels:
\begin{itemize}
\item
The Generic Level determines the space--time properties of a field, e.g. 
a fermion {\tt F}.  It also fixes the kinematic properties of the 
couplings.  For example, the {\tt FFS} coupling is of the form 
$G_{+} \omega_{+} + G_{-}\omega_{-}$, where $\omega_{\pm} = (1\pm\gamma_{5})/2$, 
with coefficients $G_{\pm}$ that depend on model parameters only.
\item
The Classes Level specifies the particle up to `simple' index 
substitutions, e.g. the down-type quark class {\tt F[4]} (where the 
generation index is not yet given).
\item
The Particles Level spells out any indices left unspecified, e.g. 
the bottom quark {\tt F[4,\{3\}]}.
\end{itemize}
The diagrams returned by {\tt CreateTopologies} and {\tt InsertFields} 
can be drawn with {\tt Paint}, with output as Mathematica Graphics 
object, PostScript, or \LaTeX.  \LaTeX\ code produced by {\tt Paint} 
can be post-processed with the FeynEdit editor \cite{FeynEdit} or with JaxoDraw \cite{JaxoDraw}. We prefer JaxoDraw 2.0 for draw the diagrams in publications. 
A diagram in the output is encoded as
{\tt FeynAmp[\Var{Id},\,\Var{Lm},\,\Var{Amp},\,\Var{Ins}]}.
For illustration, consider the diagram
\begin{center}
\vspace*{-7ex}
\begin{feynartspicture}(120,120)(1,1)
\FADiagram{}
\FAProp(0.,10.)(6.,10.)(0.,){Sine}{0}
\FALabel(3.,8.93)[t]{$\gamma$}
\FAProp(20.,10.)(14.,10.)(0.,){Sine}{0}
\FALabel(17.,11.07)[b]{$\gamma$}
\FAProp(6.,10.)(14.,10.)(0.8,){ScalarDash}{-1}
\FALabel(10.,5.73)[t]{$G^{+}$}
\FAProp(6.,10.)(14.,10.)(-0.8,){ScalarDash}{1}
\FALabel(10.,14.27)[b]{$G^{-}$}
\FAVert(6.,10.){0}
\FAVert(14.,10.){0}
\end{feynartspicture}
\vspace*{-6ex}
\end{center}
\begin{itemize}
\item
\Var{Id} is an identifier for bookkeeping, e.g.
{\tt GraphID[Topology == 1, Generic == 1]},

\item
\Var{Lm} identifies the loop momenta in the form
{\tt Integral[q1]},

\item
\Var{Amp} is the generic amplitude,

{\tt $\dfrac{I}{32~Pi^4}$~RelativeCF}~\ding{192} \\[1ex]
{\tt FeynAmpDenominator[$\dfrac{1}{q1^2 - Mass[S[Gen3]]^2},
  \dfrac{1}{(-p1 + q1)^2 - Mass[S[Gen4]]^2}]$}~\ding{193} \\[1ex]
{\tt (p1 - 2\,q1)[Lor1]~~(-p1 + 2\,q1)[Lor2]}~\ding{194} \\[1ex]
{\tt ep[V[1],p1,Lor1]~~ep$^*$[V[1],k1,Lor2]}~\ding{195} \\[1ex]
{\tt $G^{(0)}_{SSV}$[(Mom[1]-Mom[2])[KI1[3]]]~~
      $G^{(0)}_{SSV}$[(Mom[1]-Mom[2])[KI1[3]]]}~\ding{196}

where individual items can easily be identified: \ding{192} prefactor, 
\ding{193} loop denominators, \ding{194}~coupling structure, 
\ding{195} polarization vectors, \ding{196} coupling constants.

\item
\Var{Ins} is a list of rules substituting the unspecified items in the 
generic amplitude,

\{{\tt ~Mass[S[Gen3]],~Mass[S[Gen4]],} \\
{\tt $G^{(0)}_{SSV}$[(Mom[1]-Mom[2])[KI1[3]]],} \\
{\tt $G^{(0)}_{SSV}$[(Mom[1]-Mom[2])[KI1[3]]],~RelativeCF~~$\rightarrow$} \\
{\tt Insertions[Classes][MW,~MW,~I\ EL,~-I\ EL,~2 ] }\}
\end{itemize}

\subsection{Model Files}

The Model Files are ordinary Mathematica text files loaded by FeynArts 
during model initialization.  They supply certain objects, e.g.
{\tt M\$ClassesDescription}, the list of particles, and 
{\tt M\$CouplingMatrices}, the list of couplings.  Generic 
({\tt .gen}) and Classes ({\tt .mod}) Model Files store the kinematic 
and constant part of the coupling, respectively. \\
FeynArts further distinguishes Basic and Partial (Add-On) Model Files. 
Basic Model Files, such as {\tt SM.mod}, {\tt MSSM.mod}, can be 
modified by Add-On Model Files, as in

   {\tt ~~~ InsertFields[..., Model \(\to\) {"MSSMQCD", "FV"}]}

This loads the Basic Model File {\tt MSSMQCD.mod} and modifies it 
through the Add-On Model File {\tt FV.mod} (non-minimal flavour 
violation).  The brace notation works similarly for Generic Model 
files.  The Add-On Model File typically modifies (rather than
overwrites) its objects. FeynArts itself includes the ModelMaker tool which turns a suitably defined Lagrangian into a Model File.  For further details of model 
construction the reader is referred to the respective manuals.

\subsection{Enhanced Diagram Selection}

In recent FeynArts versions, many functions have been added or extended 
to ease diagram selection: {\tt DiagramSelect}, {\tt DiagramGrouping}, 
{\tt DiagramMap}, {\tt DiagramComplement}.
Also new or extended are many `filter functions' which simplify the 
construction of sophisticated filters for the selection functions above: 
{\tt Vertices}, {\tt FieldPoints}, {\tt FermionRouting}, 
{\tt FeynAmpCases}, {\tt FieldMatchQ}, {\tt FieldMemberQ}, 
{\tt FieldPointMatchQ}, {\tt FieldPointMemberQ}.

\def\diagding#1{\raise 2ex\hbox{#1}}
\def\diagyes{\diagding{\ding{52}}}
\def\diagno{\diagding{\ding{56}}}

To pick just two examples: the selection of wave-function corrections 
(WFc) has become more flexible.  The exclusion of WFc can be specified 
individually for every external leg,

   {\tt ~~~ CreateTopologies[..., ExcludeTopologies \(\to\) WFCorrections[1|3]]}

The filter function {\tt WFCorrectionFields} returns the in- and 
out-fields of the self-energy constituting the WFc.  It solves the 
problem of treating WFc with same outer particles (usually omitted)
and different particles (kept unless some on-shell scheme is employed) 
differently, e.g.
\begin{verbatim}
   DiagramSelect[..., UnsameQ@@ WFCorrectionFields[##] &]
\end{verbatim}

\begin{center}
\begin{feynartspicture}(180,90)(2,1)
\FADiagram{\kern -3em\diagyes}
\FAProp(0.,10.)(11.,10.)(0.,){ScalarDash}{1}
\FALabel(5.5,8.93)[t]{$H$}
\FAProp(20.,15.)(17.3,13.5)(0.,){Sine}{-1}
\FALabel(18.3773,15.1249)[br]{$W$}
\FAProp(20.,5.)(11.,10.)(0.,){ScalarDash}{0}
\FALabel(15.3487,6.8436)[tr]{$h^0$}
\FAProp(11.,10.)(13.7,11.5)(0.,){ScalarDash}{1}
\FALabel(12.0773,11.6249)[br]{$G$}
\FAProp(17.3,13.5)(13.7,11.5)(-0.8,){Straight}{1}
\FALabel(16.5727,10.1851)[tl]{$\nu_l$}
\FAProp(17.3,13.5)(13.7,11.5)(0.8,){Straight}{-1}
\FALabel(14.4273,14.8149)[br]{$e_l$}
\FAVert(11.,10.){0}
\FAVert(17.3,13.5){0}
\FAVert(13.7,11.5){0}

\FADiagram{\kern -1em\diagno}
\FAProp(0.,10.)(11.,10.)(0.,){ScalarDash}{1}
\FALabel(5.5,8.93)[t]{$H$}
\FAProp(20.,15.)(17.3,13.5)(0.,){Sine}{-1}
\FALabel(18.3773,15.1249)[br]{$W$}
\FAProp(20.,5.)(11.,10.)(0.,){ScalarDash}{0}
\FALabel(15.3487,6.8436)[tr]{$h^0$}
\FAProp(11.,10.)(13.7,11.5)(0.,){Sine}{1}
\FALabel(12.0773,11.6249)[br]{$W$}
\FAProp(17.3,13.5)(13.7,11.5)(-0.8,){Straight}{1}
\FALabel(16.5727,10.1851)[tl]{$\nu_l$}
\FAProp(17.3,13.5)(13.7,11.5)(0.8,){Straight}{-1}
\FALabel(14.4273,14.8149)[br]{$e_l$}
\FAVert(11.,10.){0}
\FAVert(17.3,13.5){0}
\FAVert(13.7,11.5){0}
\end{feynartspicture}
\end{center}
The new filter function {\tt FermionRouting} can be used to
select diagrams according to their fermion structure,  e.g.
\begin{verbatim}
   DiagramSelect[..., FermionRouting[##] === {1,3, 2,4} &]
\end{verbatim}
selects only diagrams where external legs 1--3 and 2--4 are connected
through fermion lines.
\begin{center}
\begin{feynartspicture}(200,100)(2,1)
\FADiagram{\diagyes}
\FAProp(6.5,13.5)(6.5,6.5)(0.,){ScalarDash}{1}
\FAProp(13.5,13.5)(13.5,6.5)(0.,){ScalarDash}{-1}
\FAProp(0.,15.)(6.5,13.5)(0.,){Straight}{1}
\FAProp(0.,5.)(6.5,6.5)(0.,){Straight}{-1}
\FAProp(20.,15.)(13.5,13.5)(0.,){Straight}{-1}
\FAProp(20.,5.)(13.5,6.5)(0.,){Straight}{1}
\FAProp(6.5,13.5)(13.5,13.5)(0.,){Straight}{1}
\FAProp(6.5,6.5)(13.5,6.5)(0.,){Straight}{-1}
\FAVert(6.5,13.5){0}
\FAVert(6.5,6.5){0}
\FAVert(13.5,13.5){0}
\FAVert(13.5,6.5){0}
\FALabel(3.59853,15.2803)[b]{\small $1$}
\FALabel(3.59853,4.71969)[t]{\small $2$}
\FALabel(16.4015,15.2803)[b]{\small $3$}
\FALabel(16.4015,4.71969)[t]{\small $4$}

\FADiagram{\diagno}
\FAProp(6.5,13.5)(13.5,13.5)(0.,){ScalarDash}{1}
\FAProp(6.5,6.5)(13.5,6.5)(0.,){ScalarDash}{-1}
\FAProp(0.,15.)(6.5,13.5)(0.,){Straight}{1}
\FAProp(0.,5.)(6.5,6.5)(0.,){Straight}{-1}
\FAProp(20.,15.)(13.5,13.5)(0.,){Straight}{-1}
\FAProp(20.,5.)(13.5,6.5)(0.,){Straight}{1}
\FAProp(6.5,13.5)(6.5,6.5)(0.,){Straight}{1}
\FAProp(13.5,13.5)(13.5,6.5)(0.,){Straight}{-1}
\FAVert(6.5,13.5){0}
\FAVert(6.5,6.5){0}
\FAVert(13.5,13.5){0}
\FAVert(13.5,6.5){0}
\FALabel(3.59853,15.2803)[b]{\small $1$}
\FALabel(3.59853,4.71969)[t]{\small $2$}
\FALabel(16.4015,15.2803)[b]{\small $3$}
\FALabel(16.4015,4.71969)[t]{\small $4$}
\end{feynartspicture}
\end{center}

\chapter{Weak Scale Thresholds at One-Loop\label{app-counterterms}}

In this appendix we summarise the one-loop corrections $\delta^{(1l)}g_{i}$ and $\delta^{(1l)}m^{2}$ to the SM parameters. The one-loop computations are in the Feynman gauge: $\zeta=1$. We recall that $\delta^{(1l)}g_{i}$ is a gauge-independent quantity.
Our expressions for $\delta^{(1l)}g_{i}$ are equivalent to the well known
expressions in the literature \cite{Degrassi2}.  We write $\delta^{(1l)}g_{i}$ in terms of
finite parts of the the Passarino-Veltman functions 
\begin{equation}
A_0 (M) =
M^2(1-\ln\frac{M^2}{\mu ^2})\ ,\qquad B_0(p;M_1,M_2) = -\int_0^1
\ln\frac{xM_1^2+(1-x) M_2^2-x(1-x)p^2}{\mu ^2}dx\ .  
\end{equation} 
Below we work in the limit where the fermion masses are zero excluding the top quark mass $m_{t}$.

\section{The Higgs Self-Coupling}
From the correction to the pole Higgs mass
\begin{eqnarray}
\delta^{(1)} m_H^2 + \frac{T^{(1)}}{v_{os}} &=&
\frac{1}{(4\pi v)^2}{\rm Re}\bigg[ 6m_t^2(4m_t^2-m_H^2) B_0(m_H;m_t,m_t)+ \nonumber\\
&& -(m_H^4-4m_H^2m_W^2+12m_W^4)B_0(m_H;m_W,m_W)+2m_H^2 A_0(m_W) +  \nonumber  \\
&&-\frac{1}{2}\left(m_H^4-4 m_H^2 m_Z^2+12 m_Z^4\right)B_0(m_H;m_Z,m_Z) +m_H^2  A_0(m_Z)+   \nonumber  \\
&&-\frac{9}{2} m_H^4 B_0(m_H;m_H,m_H) +4(2m_W^4+m_Z^4)- m_H^4 \bigg],
\label{d1Mh}
\end{eqnarray}
and from the one-loop correction to muon decay
{\small
\begin{eqnarray}
{\left. \Delta r^{(1)}_0 \right|_{\rm fin}}&=&\frac{1}{(4\pi V)^2}  
\bigg[3m_t^2 -m_W^2-\frac{m_Z^2}{2}-\frac{m_H^2}{2}+\frac{3m_W^2 A_0(m_H)}{m_H^2-m_W^2}+\frac{6m_W^2-3m_Z^2}{m_W^2-m_Z^2}A_0(m_Z)+ \nonumber \\
&&-6A_0(m_t) + \bigg(9-\frac{3m_H^2}{m_H^2-m_W^2}-\frac{3m_W^2}{m_W^2-m_Z^2}\bigg)A_0(m_W)+ 2A_0(m_W)+ A_0(m_Z) \bigg] \nonumber \\
\end{eqnarray}}
we obtain the one-loop correction $\delta^{(1l)}\lambda(\mu)$ according with eq.     (\ref{eq:ren-lambda}):
\begin{eqnarray}
\delta^{(1l)}\lambda(\mu) &=&
\frac{1}{(4\pi)^2v^4}{\rm Re}\bigg[  3m_t^2(m_H^2-4m_t^2) B_0(m_H;m_t,m_t)+3m_H^2 A_0(m_t)+ \\
&&+\frac{1}{4}\left(m_H^4-4 m_H^2 m_Z^2+12 m_Z^4\right)B_0(m_H;m_Z,m_Z) +\frac{m_H^2(7m_W^2-4 m_Z^2)}{2(m_Z^2-m_W^2)}  A_0(m_Z)+   \nonumber  \\
&& +\frac{1}{2}(m_H^4-4m_H^2m_W^2+12m_W^4)B_0(m_H;m_W,m_W)-\frac{3m_H^2 m_W^2}{2(m_H^2-m_W^2)} A_0(m_H)+ \nonumber \\
&&+\frac{m_H^2}{2}\left(-11 + \frac{3 m_H^2}{m_H^2-m_W^2} -\frac{3 m_W^2}{m_Z^2 - m_W^2}\right) A_0(m_W) +  \nonumber  \\
&&+\frac{9}{4} m_H^4 B_0(m_H;m_H,m_H) +\frac{1}{4}(m_H^4 +m_H^2(m_Z^2+2m_W^2-6 m_t^2)-8(m_Z^4+2m_W^4))
 \bigg] \ .\nonumber
\label{d1Mh}
\end{eqnarray}

\section{The Quadratic Higgs Term}
The correction is obtained from eq. (\ref{eq:cond-ren-1}):
\begin{eqnarray}
\delta^{(1l)} m^{2}(\mu) &=&
\frac{1}{(4\pi)^2 v^2} {\rm Re} \bigg[  6m_t^2(m_H^2-4m_t^2)B_0(m_H;m_t,m_t)+ 24 m_t^2 A_0(m_t) +\nonumber\\
&& +(m_H^4-4m_H^2m_W^2+12m_W^4)B_0(m_H;m_W,m_W)-2(m_H^2+ 6m_W^2) A_0(m_W) +  \nonumber  \\
&&+\frac{1}{2}\left(m_H^4-4 m_H^2 m_Z^2+12 m_Z^4\right)B_0(m_H;m_Z,m_Z) -(m_H^2+ 6m_Z^2) A_0(m_Z) +  \nonumber  \\
&&+\frac{9}{2} m_H^4 B_0(m_H;m_H,m_H)-3m_H^2 A_0(m_H) \bigg] \ .
\end{eqnarray}

\section{The top Yukawa Coupling}
The gauge-invariant one-loop correction to the top Yukawa coupling is obtained 
from eq. (\ref{eq:1l-MS-ht})
{\small
\begin{eqnarray}
\delta^{(1l)}h_t (\mu) &=&\frac{m_t}{\sqrt{2}v^3 (4\pi)^2}
{\rm Re}\bigg[
- \left(m_H^2-4 m_t^2\right) B_0\left(m_t;m_H,m_t\right) +
\nonumber \\ &&+  \nonumber
\frac{m_t^2 \left(80 m_W^2 m_Z^2-64 m_W^4-7
   m_Z^4\right)+40 m_W^2 m_Z^4-32 m_W^4 m_Z^2-17 m_Z^6 }{9  m_t^2
   m_Z^2}  B_0\left(m_t;m_t,m_Z\right)+
  \\ &&+ 
   \frac{\left(m_t^2 m_W^2+m_t^4-2 m_W^4\right)}{ m_t^2} B_0\left(m_t;0,m_W\right)+
   \\ &&+
  \left(\frac{3 m_H^2}{m_H^2-m_W^2}+\frac{2
   m_W^2}{m_t^2}+\frac{3 m_W^2}{m_W^2-m_Z^2}-10\right) A_0\left(m_W\right) +
    \left(\frac{3 m_W^2}{m_W^2-m_H^2}+1\right) A_0\left(m_H\right)+\nonumber
   \\ &&+\nonumber
   \frac{\left(36 m_t^2 m_Z^2-56 m_W^2 m_Z^2+64 m_W^4-17
   m_Z^4\right)}{9  m_t^2 m_Z^2} A_0\left(m_t\right)+
   \\ &&+\nonumber
   \left( \frac{3 m_W^2}{m_Z^2-m_W^2} + \frac{32 m_W^4 - 40 m_W^2 m_Z^2 +17 m_Z^4}{9 m_t^2 m_Z^2} -3\right) A_0\left(m_Z\right) +
   \\ &&+\nonumber
    \frac{m_H^2}{2} - 3 m_t^2 - 9 m_W^2 +\frac{7 m_Z^2}{18}+\frac{64 m_W^4}{9 m_Z^2} \bigg]
+ \frac{m_t}{\sqrt{2}v(4\pi)^2} g_3^2 \left(-\frac{8
   A_0\left(m_t\right)}{m_t^2}-\frac{8}{3}\right) \ .
\end{eqnarray}} 

 \section{The Weak Gauge Couplings}

The one-loop correction to the $SU(2)_L$ gauge coupling is obtained from
eq. (\ref{eq:MS-ren-g}): 
{\small
\begin{eqnarray}
\delta^{(1l)}g (\mu) &=& \frac{2m_W}{(4\pi)^2v^3} {\rm Re}\bigg[
\left(\frac{m_H^4}{6 m_W^2}-\frac{2 m_H^2}{3}+2 m_W^2\right) B_0\left(m_W,m_H,m_W\right)
+\nonumber
   \\ &&\nonumber
   +\left(-\frac{m_t^4}{m_W^2}-m_t^2+2
   m_W^2\right) B_0\left(m_W,0,m_t\right)+
      \\ &&
      +\frac{1}{6} \left(-\frac{48 m_W^4}{m_Z^2}+\frac{m_Z^4}{m_W^2}-68 m_W^2+16
   m_Z^2\right) B_0\left(m_W,m_W,m_Z\right)+
         \\ &&\nonumber
        + \frac{1}{6}\left(m_H^2
   \left(\frac{9}{m_H^2-m_W^2}+\frac{1}{m_W^2}\right)+\frac{m_Z^2}{m_W^2}+m_W^2
   \left(\frac{9}{m_W^2-m_Z^2}+\frac{48}{m_Z^2}\right)-27\right)  A_0\left(m_W\right) +
      \\ &&\nonumber
      + \left( 2- \frac{m_H^2 \left( m_H^2 + 8 m_W^2\right)}{6 m_W^2 \left(m_H^2 - m_W^2 \right)}\right) A_0\left(m_H\right) + \left(\frac{m_t^2}{m_W^2}+1\right) A_0\left(m_t\right)+
      \\ &&\nonumber
      +\frac{1}{6} \left(\frac{24 m_W^2}{m_Z^2} - \frac{m_Z^2}{m_W^2} + \frac{9 m_W^2}{m_Z^2 - m_W^2} - 17\right) A_0\left(m_Z\right) +
      \\ &&\nonumber
      +\frac{1}{36} \left(-3 m_H^2+18 m_t^2+\frac{288 m_W^4}{m_Z^2}-374 m_W^2-3 m_Z^2\right)\bigg] \ .
\end{eqnarray}}

The one-loop correction to the $U(1)_Y$ gauge coupling is obtained 
from eq. (\ref{eq:MS-ren-g2}): 
{\small
 \begin{eqnarray}
 \nonumber
 \delta^{(1l)}g'(\mu) &=& \frac{2 \sqrt{m_Z^2-m_W^2}}{(4\pi)^2 v^3}{\rm Re}\bigg[
   \left(\frac{88}{9} - \frac{124 m_W^2}{9 m_Z^2} + \frac{m_H^2+34 m_W^2}{6 (m_Z^2-m_W^2)} \right) A_0\left(m_Z\right)+
 \\ &&\nonumber+
    \frac{m_H^2-4 m_W^2}{2 (m_H^2-   m_W^2)}   A_0\left(m_H\right)  +
   \left(-\frac{7}{9} - \frac{m_t^2}{m_Z^2-m_W^2} +\frac{64 m_W^2}{9 m_Z^2} \right) A_0\left(m_t\right)+
    \\ &&\nonumber+
    \frac{m_H^4+ 2m_W^2 (m_W^2-15m_Z^2) +3m_H^2 (2 m_W^2 + 7 m_Z^2) }{6
   \left(m_H^2-m_W^2\right) \left(m_W^2-m_Z^2\right)}A_0\left(m_W\right)
  +  \\ &&\nonumber 
   -\frac{m_t^4+m_W^2 m_t^2-2 m_W^4   }{m_W^2-m_Z^2}B_0\left(m_W,0,m_t\right)
   -\frac{m_H^4-4 m_Z^2 m_H^2+12 m_Z^4 }{6 (m_W^2-m_Z^2)} B_0\left(m_Z,m_H,m_Z\right)+
    \\ &&+
    \frac{m_H^4-4 m_W^2 m_H^2+12 m_W^4 }{6(   m_W^2- m_Z^2)}B_0\left(m_W,m_H,m_W\right)+
    \\ &&\nonumber+
    \frac{m_Z^6-48 m_W^6-68 m_Z^2 m_W^4+16 m_Z^4 m_W^2 }{6 m_Z^2
   \left(m_W^2-m_Z^2\right)}B_0\left(m_W,m_W,m_Z\right)+
    \\ &&\nonumber +
    \frac{1}{9}\left(-23 m_W^2 + 7 m_t^2 +17 m_Z^2 -\frac{64 m_t^2 m_W^2}{m_Z^2} - \frac{9 m_W^2 (m_t^2-m_W^2)}{m_Z^2-m_W^2}\right) B_0\left(m_Z,m_t,m_t\right)+
    \\ &&\nonumber+
    \frac{m_Z^6 -48 m_W^6-68 m_Z^2 m_W^4+16 m_Z^4 m_W^2  }{6m_Z^2 \left(m_Z^2-m_W^2 \right)}B_0\left(m_Z,m_W,m_W\right)+ \\ &&\nonumber +
 \frac{1}{36} \left(\frac{576 m_W^4}{m_Z^2}-242 m_W^2-3 m_H^2 + 257
   m_Z^2 + \frac{36 m_W^4}{m_Z^2-m_W^2}+ m_t^2 \left(82-\frac{256 m_W^2}{m_Z^2}\right)\right)
   \bigg] \ .
\end{eqnarray}}

\section{Numerical Results at $\mu=m_t$}

All the $\delta^{(1l)}g_i$ have the numerical values for $\mu=m_t$:

  \begin{eqnarray}
     \begin{array}{ll}
 \delta^{(1l)}\lambda =-0.22716/(4\pi)^2, &
\delta^{(1l)} h_t = h_{t(os)}[0.4000-\frac{32}{3}g_s^2]/(4\pi)^2,\\
\delta^{(1l)} g = g_{(os)}[-2.611]/(4\pi)^2, \qquad &
 \delta^{(1l)} g' = g'_{(os)}[-0.0824]/(4\pi)^2.
  \end{array}
\end{eqnarray}

\chapter{Reduction of Two-Loop Integrals With TARCER \label{AppTarcer}}
\markboth{APPENDIX \ref{AppTarcer}}{APPENDIX \ref{AppTarcer}} 

TARCER is an implementation in Mathematica, as part of the code FEYNCALC, of the algorithm of O.V.~Tarasov for the reduction of two-loop integrals with arbitrary masses to a small group of master integrals using the complete set of recurrence relations, that contain as special case the IBP's method, given in \cite{Tarasov1}. His main purpose is reduce the general type of integrals (the Mathematica notation is listed first always)

\begin{eqnarray}
\lefteqn{
{\tt TFI[d, q^2, \Delta q, \{a, b\}, \{u, v, r, s, t\},
     \{ \{\nu_1,m_1\}, \{ \{\nu_2,m_2\}, \ldots, \{\nu_5,m_5\} \}] } 
= } 
\label{TFInotation1}
\nonumber \\
\nonumber \\
&& \frac{1}{\pi^d} \int\!\int
\frac{\ d^d k_1 d^dk_2 \ \ (\Delta k_1)^a \: (\Delta k_2)^b \:
 (k_1^2)^u \: (k_2^2)^v \: (q k_1)^r \: (q k_2)^s \: (k_1 k_2)^t}
{
[k_1^2-m_1^2]^{\nu_1}\ [k_2^2-m_2^2]^{\nu_2}\
[k_3^2-m_3^2]^{\nu_3}\ [k_4^2-m_4^2]^{\nu_4}\
[k_5^2-m_5^2]^{\nu_5}
} \; ,
\end{eqnarray}
to basic integrals. Where we use the abbreviations $k_3 = k_1-q,~k_4 = k_2-q$ and $k_5 = k_1 - k_2$.
The exponents $a, \ldots, t$ and the indices $\nu_1,\ldots,\nu_5$ are assumed to be nonnegative integers and $\Delta$ denotes a lightlike vector with $\Delta^2=0$.
If a mass vanishes the argument $\{\nu_j,0\}$ of TFI may be replaced by the index $\nu_j$ alone, and if some of the subsets $\{a, b\}$ or $\{u, v, r, s, t\}$ of exponents vanish, we can remove this argument of the Mathematica function TFI and use the reduced notation
\begin{eqnarray}
\lefteqn{
{\tt TFI[d, q^2, \Delta q, \{0, 0\}, \{u, v, r, s, t\},
     \{ \{\nu_1,0\}, \{\nu_2,m_2\}, \ldots, \{\nu_5,m_5\} \}] }}
\nonumber \\
 \nonumber \\
&&
~~~~~~~=~~~~~~~ {\tt TFI[d, q^2, \{u, v, r, s, t\},
     \{ \nu_1, \{\nu_2,m_2\}, \ldots, \{\nu_5,m_5\} \}] } \; .
\label{TFInotation}
\end{eqnarray}
In the course of reduction, all scalar products in the numerator will be eliminated. If only scalar integrals are present, we can express all basis integrals in terms of following input TARCER functions in {\tt Mathematica} notation (keep the notation in line with the Chapter~\ref{cha:TwoLoopCalc})~:
\begin{eqnarray}
\lefteqn{
{\tt TFI[d, q^2, \{ \{\nu_1,m_1\}, \{\nu_2,m_2\}, \ldots, 
\{\nu_5,m_5\} \}] }  
= } 
\nonumber \\
&&
F^{(d)}_{\nu_1 \nu_2 \nu_3 \nu_4 \nu_5} 
= 
\frac{1}{\pi^d}
\int\!\int
\frac{d^d k_1 d^dk_2}{
[k_1^2-m_1^2]^{\nu_1}\ [k_2^2-m_2^2]^{\nu_2}\
\cdots
[k_5^2-m_5^2]^{\nu_5}
} \; ,
\end{eqnarray}

\begin{eqnarray}
\lefteqn{
{\tt TVI[d, q^2, \{ \{\nu_1,m_1\},  \{\nu_2,m_2\}, 
\{\nu_3,m_3\}, \{\nu_4,m_4\} \}] }  
= }
\nonumber \\
&&
V^{(d)}_{\nu_1 \nu_2 \nu_3 \nu_4 } 
=
\frac{1}{\pi^d}
\int\!\int
\frac{d^d k_1 d^dk_2}{
[k_5^2-m_1^2]^{\nu_1}\ [k_2^2-m_2^2]^{\nu_2}\
[k_3^2-m_3^2]^{\nu_3}\ [k_4^2-m_4^2]^{\nu_4}
} \; ,
\end{eqnarray}

\begin{eqnarray}
\lefteqn{
{\tt TJI[d, q^2, \{ \{\nu_1,m_1\}, \{\nu_2,m_2\}, \{\nu_3,m_3\} \}] }  
= }
\nonumber \\
&&
J^{(d)}_{\nu_1 \nu_2 \nu_3 } 
=
\frac{1}{\pi^d}
\int\!\int
\frac{d^d k_1 d^dk_2}{
[k_1^2-m_1^2]^{\nu_1}\ [k_5^2-m_2^2]^{\nu_2}\ [k_4^2-m_3^2]^{\nu_3}
} \; ,
\end{eqnarray}

\begin{eqnarray}
\lefteqn{
{\tt TJI[d, 0, \{ \{\nu_1,m_1\}, \{\nu_2,m_2\}, \{\nu_3,m_3\} \}] }  
= }
\nonumber \\
&&
K^{(d)}_{\nu_1 \nu_2 \nu_3 } 
=
\frac{1}{\pi^d}
\int\!\int
\frac{d^d k_1 d^dk_2}{
[k_1^2-m_1^2]^{\nu_1}\ [k_5^2-m_2^2]^{\nu_2}\ [k_2^2-m_3^2]^{\nu_3}
}
\end{eqnarray}
and
\begin{eqnarray}
\lefteqn{
{\tt TBI[d, q^2, \{ \{\nu_1,m_1\}, \{\nu_2,m_2\} \}] }  
= } 
\nonumber \\
&& 
B^{(d)}_{\nu_1 \nu_2} =
\frac{1}{\pi^{d/2}}
\int\
\frac{d^d k_1}{[k_1^2-m_1^2]^{\nu_1}\ [k_3^2-m_2^2]^{\nu_2}} \; ,
\\ \nonumber \\
\lefteqn{
{\tt TAI[d, 0, \{ \{\nu_1,m_1\} \}] }  
= } 
\nonumber \\ 
&&
A^{(d)}_{\nu_1} =
\frac{1}{\pi^{d/2}}
\int\
\frac{d^d k_1}{[k_1^2-m_1^2]^{\nu_1}} \; .
\end{eqnarray}

To performs the complete reduction to the set of basic integrals, TARCER use the function {\tt TarcerRecurse}. It can be applied to any expression involving the functions {\tt TFI, TVI, TJI, TBI} and {\tt TAI} whose first argument is a symbol {\tt d}. Its usage is as follows. In the first step the numerator of the integrand in the {\tt TFI}-integrals is simplified as far as possible by standard manipulations until an irreducible numerator of the form $(\Delta k_1)^a \: (\Delta k_2)^b \: (p k_1)^r \: (p k_2)^s$ results. In the next step {\tt TarcerRecurse} rewrites the integrals containing irreducible numerators in terms of scalar integrals in higher space-time dimension by employing an operator $T$ that is a polynomial in the operator ${\bf d^+}$ representing a shift $d \rightarrow d+2$ in dimension and in the mass derivatives $\partial_j = \partial / \partial m_j^2$, and then, using the Tarasov recurrence relations, reduced again these integrals to scalar integrals in the original space-time dimension as was described in Chapter \ref{cha:TwoLoopCalc}. Once all irreducible numerators are eliminated and all integrals are reduced to original space-time dimension $d$, the next step taken by the function {\tt TarcerRecurse} is to repeatedly apply the recurrence relations that reduce the exponents of the scalar propagators in the integrals until no further reduction is possible. All recurrence relations explicitly or implicitly given by Tarasov are implemented.

For a simple example, consider the next two-loops tadpole in the Landau gauge:
\begin{center}
\begin{tabular}{|c c|}
\hline
 ~~~~~~~\scalebox{0.3}{
\fcolorbox{white}{white}{
  \begin{picture}(318,178) (149,-135)
    \SetWidth{1.0}
    \SetColor{Black}
    \Line[dash,dashsize=10,arrow,arrowpos=0.5,arrowlength=5,arrowwidth=2,arrowinset=0.2](150,-56)(240,-56)
    \Arc[dash,dashsize=10,arrow,arrowpos=0.5,arrowlength=5,arrowwidth=2,arrowinset=0.2](312,-56)(72.25,355,715)
    \Line[dash,dashsize=10,arrow,arrowpos=0.5,arrowlength=5,arrowwidth=2,arrowinset=0.2](312,16)(312,-128)
    \Vertex(240,-56){6}
    \Vertex(312,16){6}
    \Vertex(312,-128){6}
    \Text(150,-32)[lb]{\Large{\Black{$H$}}}
    \Text(246,16)[lb]{\Large{\Black{$H$}}}
    \Text(246,-140)[lb]{\Large{\Black{$H$}}}
    \Text(330,-56)[lb]{\Large{\Black{$G^{0}$}}}
    \Text(360,16)[lb]{\Large{\Black{$G^{0}$}}}
  \end{picture}
}} &  = ~~~~~$\dfrac{3g^{3}m_H^{6}}{32m_{W}^{3}}\int\int\dfrac{d^{d}k_{1}d^{d}k_{2}}{(2\pi)^{d}(2\pi)^{d}} \dfrac{1}{[k_{1}^{2}-m_H^{2}][k_{1}^{2}-m_H^{2}][k_{2}^{2}][(k_{2}-k_{1})^{2}]} $ \\
~~~~~~~~~~~~~&~~~~~~~~~~~~~~~~ \\
\hline
\end{tabular}
\end{center}
Application of {\tt TarcerRecurse} to the corresponding input form of above tadpole integral yields:
\begin{eqnarray}
{\tt T1 = (3e^{3}m_H^{6}/32m_{W}^{3}Sin^{3}\theta_{W})*TarcerRecurse[ TFI[d, 0, \{ \{2, m_H\}, \{1, 0\}, 0, 0, \{1,0\}\}]]}
\nonumber \\
\nonumber \\
\frac{ 3(d - 3)e^{3}m_{H}^{4}{\bf K}_{\{1,m_H\}, \{1,0\}, \{1,0\}}^{(d)}}
{32 m_{W}^{3} Sin^{3}\theta_{W}}~~~~~~~~~~~~~~~~~~~~~~~~~~~~~~~~~~~~~~~~~~~~~~~~~~~
\end{eqnarray}

The function {\tt TarcerExpand} inserts explicit results for some basis integrals as specified by the option {\tt TarcerReduce}. A second argument to  {\tt TarcerExpand} must be given in form of 
a rule, like ${\tt d \rightarrow 4 - \varepsilon}$. Then an expansion of the first argument of {\tt TarcerExpand} in the sole variable specified  $\varepsilon$ around $0$ will be performed. 

Applying {\tt TarcerExpand} to {\tt T1} yields:

\begin{eqnarray}
\lefteqn{
{\tt TarcerExpand[\mbox{{\tt T1}},~d \rightarrow 4 - \varepsilon ,~0]}
}
\nonumber \\ &&
\nonumber \\ &&
\frac{ e^{3}m_{H}^{6}}
{m_{W}^{3} Sin^{3}\theta_{W}}(m_{H}^2)^{-\varepsilon}S_{\varepsilon}^2 \left( - \frac{3}{16\varepsilon^{2}}- \frac{3}{32\varepsilon}-\frac{9\zeta(2)}{64}-\frac{3}{64}\right)
\end{eqnarray}

Here $S_{\varepsilon} = e^{\gamma_{E}\ (d-4)/2}$ is used, where $\gamma_{E}$ is the Euler-constant.

Let us see another example. The scalar sector of the two-loop tadpoles, this result is in agreement with the obtained in the Chapter \ref{cha:TwoLoopTadpoles}. The complete code to obtain the sum of the 17 diagrams of the scalar sector is: 

\noindent\(\pmb{\text{$\$$LoadPhi}=\text{False}}\)

\noindent\(\text{False}\)

\noindent\(\pmb{\text{$\$$LoadTARCER}=\text{True}}\)

\noindent\(\text{True}\)

\noindent\(\pmb{\text{$\$$LoadFeynArts} = \text{True}}\)

\noindent\(\text{True}\)

\noindent\(\pmb{<<\text{HighEnergyPhysics$\grave{ }$FeynCalc$\grave{ }$}}\)

FeynArts 3.9 patched for use with FeynCalc

\noindent\(\pmb{\text{tp2}= \text{CreateTopologies}[2,1\to  0, \text{ExcludeTopologies} \to  \text{Internal}]}\)

\noindent\(\pmb{\text{}}\\
\pmb{\text{SetOptions}[\text{FourVector},\text{FeynCalcInternal}\to  \text{False}]}\)

\noindent\(\{\text{Dimension}\to 4,\text{FeynCalcInternal}\to \text{False}\}\)

\noindent\(\pmb{\text{FDT}=\text{InsertFields}[\text{tp2}, S[1] \to  \{\}, \text{ExcludeParticles} \to  \{F,V,U\},\text{InsertionLevel} \to  \{\text{Classes}\}];}\)

\noindent\(\pmb{\text{AMP}=\text{CreateFeynAmp}[\text{FDT}] ;}\)

\noindent\(\text{creating amplitudes at level(s) }\{\text{Classes}\}\)

\noindent\(\text{$>$ Top. }1\text{: }\text{9 Classes amplitudes}\)

\noindent\(\text{$>$ Top. }2\text{: }\text{5 Classes amplitudes}\)

\noindent\(\text{$>$ Top. }3\text{: }\text{3 Classes amplitudes}\)

\noindent\(\text{in total: }\text{17 Classes amplitudes}\)

\noindent\(\pmb{\text{AMP}[[1]]}\)

{\small
\noindent\(\left( \text{Integral}[\text{q}1,\text{q}2],\frac{9
\text{EL}^3 \text{MH}^4 \text{FeynAmpDenominator}\left(\frac{1}{(\text{q}1)^2-\text{MH}^2},\frac{1}{(\text{q}1)^2-\text{MH}^2},\frac{1}{(\text{q}2)^2-\text{MH}^2}\right)}{8192
\pi ^8 \text{MW}^3 \text{SW}^3}\right)\)}

{\footnotesize
\noindent\(\pmb{\text{T1}=(9*\text{EL}{}^{\wedge}3*\text{MH}{}^{\wedge}4/(32*\text{MW}{}^{\wedge}3*\text{SW}{}^{\wedge}3))*\text{TarcerRecurse}[\text{TFI}[d,0,\{\{1,\text{MH}\},\{2,\text{MH}\},0,0,0\}]]}\)}

\noindent\(\frac{9 (d-2) \text{EL}^3 \text{MH}^2 \left(\pmb{A}_{\{1,\text{MH}\}}^{(d)}\right){}^2}{64 \text{MW}^3 \text{SW}^3}\)

\noindent\(\pmb{\text{T1F}=\text{TarcerExpand}[\text{T1}, d\text{-$>$}4-\varepsilon ,0] }\)

\noindent\(\frac{\text{EL}^3 \text{MH}^6 \left(\text{MH}^2\right)^{-\varepsilon } \pmb{S_{\varepsilon }}{}^2}{\text{MW}^3 \text{SW}^3}.\left(-\frac{9}{8
\varepsilon ^2}-\frac{9}{16 \varepsilon }-\frac{9 \zeta (2)}{32}-\frac{9}{32}\right)\)

\noindent\(\pmb{\text{AMP}[[2]]}\)

\noindent\(\left(\text{Integral}[\text{q}1,\text{q}2],\frac{3
\text{EL}^3 \text{MH}^4 \text{FeynAmpDenominator}\left(\frac{1}{(\text{q}1)^2-\text{MZ}^2},\frac{1}{(\text{q}1)^2-\text{MZ}^2},\frac{1}{(\text{q}2)^2-\text{MZ}^2}\right)}{8192
\pi ^8 \text{MW}^3 \text{SW}^3}\right)\)

{\footnotesize
\noindent\(\pmb{\text{T2}=(3*\text{EL}{}^{\wedge}3*\text{MH}{}^{\wedge}4/(32*\text{MW}{}^{\wedge}3*\text{SW}{}^{\wedge}3))*(1/(4*\text{Pi}){}^{\wedge}d)\text{TarcerRecurse}[\text{TFI}[d,0,\{\{1,0\},\{2,0\},0,0,0\}]]}\)}

\noindent\(0\)

\noindent\(\pmb{\text{AMP}[[3]]}\)

\noindent\(\left(\text{Integral}[\text{q}1,\text{q}2],\frac{3
\text{EL}^3 \text{MH}^4 \text{FeynAmpDenominator}\left(\frac{1}{(\text{q}1)^2-\text{MH}^2},\frac{1}{(\text{q}1)^2-\text{MH}^2},\frac{1}{(\text{q}2)^2-\text{MZ}^2}\right)}{8192
\pi ^8 \text{MW}^3 \text{SW}^3}\right)\)

{\footnotesize
\noindent\(\pmb{\text{T3}=(3*\text{EL}{}^{\wedge}3*\text{MH}{}^{\wedge}4/(32*\text{MW}{}^{\wedge}3*\text{SW}{}^{\wedge}3))*(1/(4*\text{Pi}){}^{\wedge}d)\text{TarcerRecurse}[\text{TFI}[d,0,\{\{1,0\},\{2,\text{MH}\},0,0,0\}]]}\)}

\noindent\(0\)

\noindent\(\pmb{\text{AMP}[[4]]}\)

\noindent\(\left(\text{Integral}[\text{q}1,\text{q}2],\frac{\text{EL}^3
\text{MH}^4 \text{FeynAmpDenominator}\left(\frac{1}{(\text{q}1)^2-\text{MZ}^2},\frac{1}{(\text{q}1)^2-\text{MZ}^2},\frac{1}{(\text{q}2)^2-\text{MH}^2}\right)}{8192
\pi ^8 \text{MW}^3 \text{SW}^3}\right)\)

{\footnotesize
\noindent\(\pmb{\text{T4}=(1*\text{EL}{}^{\wedge}3*\text{MH}{}^{\wedge}4/(32*\text{MW}{}^{\wedge}3*\text{SW}{}^{\wedge}3))*(1/(4*\text{Pi}){}^{\wedge}d)\text{TarcerRecurse}[\text{TFI}[d,0,\{\{1,\text{MH}\},\{2,0\},0,0,0\}]]}\)}

\noindent\(0\)

\noindent\(\pmb{\text{AMP}[[5]]}\)

\noindent\(\left(\text{Integral}[\text{q}1,\text{q}2],\frac{\text{EL}^3
\text{MH}^4 \text{FeynAmpDenominator}\left(\frac{1}{(\text{q}1)^2-\text{MW}^2},\frac{1}{(\text{q}1)^2-\text{MW}^2},\frac{1}{(\text{q}2)^2-\text{MW}^2}\right)}{1024
\pi ^8 \text{MW}^3 \text{SW}^3}\right)\)

{\footnotesize
\noindent\(\pmb{\text{T5}=(1*\text{EL}{}^{\wedge}3*\text{MH}{}^{\wedge}4/(4*\text{MW}{}^{\wedge}3*\text{SW}{}^{\wedge}3))*(1/(4*\text{Pi}){}^{\wedge}d)\text{TarcerRecurse}[\text{TFI}[d,0,\{\{1,0\},\{2,0\},0,0,0\}]]}\)}

\noindent\(0\)

\noindent\(\pmb{\text{AMP}[[6]]}\)

\noindent\(\left(\text{Integral}[\text{q}1,\text{q}2],\frac{3
\text{EL}^3 \text{MH}^4 \text{FeynAmpDenominator}\left(\frac{1}{(\text{q}1)^2-\text{MH}^2},\frac{1}{(\text{q}1)^2-\text{MH}^2},\frac{1}{(\text{q}2)^2-\text{MW}^2}\right)}{4096
\pi ^8 \text{MW}^3 \text{SW}^3}\right)\)

{\footnotesize
\noindent\(\pmb{\text{T6}=(3*\text{EL}{}^{\wedge}3*\text{MH}{}^{\wedge}4/(16*\text{MW}{}^{\wedge}3*\text{SW}{}^{\wedge}3))*(1/(4*\text{Pi}){}^{\wedge}d)\text{TarcerRecurse}[\text{TFI}[d,0,\{\{1,0\},\{2,\text{MH}\},0,0,0\}]]}\)}

\noindent\(0\)

\noindent\(\pmb{\text{AMP}[[7]]}\)

\noindent\(\left(\text{Integral}[\text{q}1,\text{q}2],\frac{\text{EL}^3
\text{MH}^4 \text{FeynAmpDenominator}\left(\frac{1}{(\text{q}1)^2-\text{MZ}^2},\frac{1}{(\text{q}1)^2-\text{MZ}^2},\frac{1}{(\text{q}2)^2-\text{MW}^2}\right)}{4096
\pi ^8 \text{MW}^3 \text{SW}^3}\right)\)

{\footnotesize
\noindent\(\pmb{\text{T7}=(1*\text{EL}{}^{\wedge}3*\text{MH}{}^{\wedge}4/(16*\text{MW}{}^{\wedge}3*\text{SW}{}^{\wedge}3))*(1/(4*\text{Pi}){}^{\wedge}d)\text{TarcerRecurse}[\text{TFI}[d,0,\{\{1,0\},\{2,0\},0,0,0\}]]}\)}

\noindent\(0\)

\noindent\(\pmb{\text{AMP}[[8]]}\)

\noindent\(\left(\text{Integral}[\text{q}1,\text{q}2],\frac{\text{EL}^3
\text{MH}^4 \text{FeynAmpDenominator}\left(\frac{1}{(\text{q}1)^2-\text{MW}^2},\frac{1}{(\text{q}1)^2-\text{MW}^2},\frac{1}{(\text{q}2)^2-\text{MH}^2}\right)}{4096
\pi ^8 \text{MW}^3 \text{SW}^3}\right)\)

{\footnotesize
\noindent\(\pmb{\text{T8}=(1*\text{EL}{}^{\wedge}3*\text{MH}{}^{\wedge}4/(16*\text{MW}{}^{\wedge}3*\text{SW}{}^{\wedge}3))*(1/(4*\text{Pi}){}^{\wedge}d)\text{TarcerRecurse}[\text{TFI}[d,0,\{\{1,\text{MH}\},\{2,0\},0,0,0\}]]}\)}

\noindent\(0\)

\noindent\(\pmb{\text{AMP}[[9]]}\)

\noindent\(\left(\text{Integral}[\text{q}1,\text{q}2],\frac{\text{EL}^3
\text{MH}^4 \text{FeynAmpDenominator}\left(\frac{1}{(\text{q}1)^2-\text{MW}^2},\frac{1}{(\text{q}1)^2-\text{MW}^2},\frac{1}{(\text{q}2)^2-\text{MZ}^2}\right)}{4096
\pi ^8 \text{MW}^3 \text{SW}^3}\right)\)

{\footnotesize
\noindent\(\pmb{\text{T9}=(1*\text{EL}{}^{\wedge}3*\text{MH}{}^{\wedge}4/(16*\text{MW}{}^{\wedge}3*\text{SW}{}^{\wedge}3))*(1/(4*\text{Pi}){}^{\wedge}d)\text{TarcerRecurse}[\text{TFI}[d,0,\{\{1,0\},\{2,0\},0,0,0\}]]}\)}

\noindent\(0\)

\noindent\(\pmb{\text{AMP}[[10]]}\)

\noindent\(\left(\text{Integral}[\text{q}1,\text{q}2],\frac{27
\text{EL}^3 \text{MH}^6 \text{FeynAmpDenominator}\left(\frac{1}{(\text{q}1)^2-\text{MH}^2},\frac{1}{(\text{q}1)^2-\text{MH}^2},\frac{1}{(\text{q}2)^2-\text{MH}^2},\frac{1}{(\text{q}2-\text{q}1)^2-\text{MH}^2}\right)}{8192
\pi ^8 \text{MW}^3 \text{SW}^3}\right)\)

{\footnotesize
\noindent\(\pmb{\text{T10}=(27*\text{EL}{}^{\wedge}3*\text{MH}{}^{\wedge}6/(32*\text{MW}{}^{\wedge}3*\text{SW}{}^{\wedge}3))*\text{TarcerRecurse}[\text{TFI}[d,0,\{\{2,\text{MH}\},\{1,\text{MH}\},0,0,\{1,\text{MH}\}\}]]}\)}

\noindent\(\frac{9 (d-3) \text{EL}^3 \text{MH}^4 \pmb{K}_{\{1,\text{MH}\}\{1,\text{MH}\}\{1,\text{MH}\}}^{(d)}}{32 \text{MW}^3 \text{SW}^3}\)

\noindent\(\pmb{\text{AMP}[[11]]}\)

\noindent\(\left(\text{Integral}[\text{q}1,\text{q}2],\frac{3
\text{EL}^3 \text{MH}^6 \text{FeynAmpDenominator}\left(\frac{1}{(\text{q}1)^2-\text{MH}^2},\frac{1}{(\text{q}1)^2-\text{MH}^2},\frac{1}{(\text{q}2)^2-\text{MZ}^2},\frac{1}{(\text{q}2-\text{q}1)^2-\text{MZ}^2}\right)}{8192
\pi ^8 \text{MW}^3 \text{SW}^3}\right)\)

{\footnotesize
\noindent\(\pmb{\text{T11}=(3*\text{EL}{}^{\wedge}3*\text{MH}{}^{\wedge}6/(32*\text{MW}{}^{\wedge}3*\text{SW}{}^{\wedge}3))*\text{TarcerRecurse}[\text{TFI}[d,0,\{\{2,\text{MH}\},\{1,0\},0,0,\{1,0\}\}]]}\)}

\noindent\(\frac{3 (d-3) \text{EL}^3 \text{MH}^4 \pmb{K}_{\{1,\text{MH}\}\{1,0\}\{1,0\}}^{(d)}}{32 \text{MW}^3 \text{SW}^3}\)

\noindent\(\pmb{\text{T11F}=\text{TarcerExpand}[\text{T11}, d\text{-$>$}4-\varepsilon ,0]}\)

\noindent\(\frac{\text{EL}^3 \text{MH}^6 \pmb{S_{\text{Global$\grave{ }\varepsilon $}}}{}^2 \left(\text{MH}^2\right)^{-\varepsilon }}{\text{MW}^3
\text{SW}^3}.\left(-\frac{3}{16 \varepsilon ^2}-\frac{3}{32 \varepsilon }-\frac{9 \zeta (2)}{64}-\frac{3}{64}\right)\)

\noindent\(\pmb{\text{AMP}[[12]]}\)

\noindent\(\left(\text{Integral}[\text{q}1,\text{q}2],\frac{\text{EL}^3
\text{MH}^6 \text{FeynAmpDenominator}\left(\frac{1}{(\text{q}1)^2-\text{MZ}^2},\frac{1}{(\text{q}1)^2-\text{MZ}^2},\frac{1}{(\text{q}2)^2-\text{MH}^2},\frac{1}{(\text{q}2-\text{q}1)^2-\text{MZ}^2}\right)}{4096
\pi ^8 \text{MW}^3 \text{SW}^3}\right)\)

{\footnotesize
\noindent\(\pmb{\text{T12}=(1*\text{EL}{}^{\wedge}3*\text{MH}{}^{\wedge}6/(16*\text{MW}{}^{\wedge}3*\text{SW}{}^{\wedge}3))*\text{TarcerRecurse}[\text{TFI}[d,0,\{\{2,0\},\{1,\text{MH}\},0,0,\{1,0\}\}]]}\)}

\noindent\(-\frac{(d-3) \text{EL}^3 \text{MH}^4 \pmb{K}_{\{1,\text{MH}\}\{1,0\}\{1,0\}}^{(d)}}{16 \text{MW}^3 \text{SW}^3}\)

\noindent\(\pmb{\text{T12F}= \text{TarcerExpand}[\text{T12}, d\text{-$>$}4-\varepsilon ,0]}\)

\noindent\(\frac{\text{EL}^3 \text{MH}^6 \left(\text{MH}^2\right)^{-\varepsilon } \pmb{S_{\varepsilon }}{}^2}{\text{MW}^3 \text{SW}^3}.\left(\frac{1}{8
\varepsilon ^2}+\frac{1}{16 \varepsilon }+\frac{3 \zeta (2)}{32}+\frac{1}{32}\right)\)

\noindent\(\pmb{\text{AMP}[[13]]}\)

\noindent\(\left(\text{Integral}[\text{q}1,\text{q}2],\frac{3
\text{EL}^3 \text{MH}^6 \text{FeynAmpDenominator}\left(\frac{1}{(\text{q}1)^2-\text{MH}^2},\frac{1}{(\text{q}1)^2-\text{MH}^2},\frac{1}{(\text{q}2)^2-\text{MW}^2},\frac{1}{(\text{q}2-\text{q}1)^2-\text{MW}^2}\right)}{4096
\pi ^8 \text{MW}^3 \text{SW}^3}\right)\)

{\footnotesize
\noindent\(\pmb{\text{T13}=(3*\text{EL}{}^{\wedge}3*\text{MH}{}^{\wedge}6/(16*\text{MW}{}^{\wedge}3*\text{SW}{}^{\wedge}3))*\text{TarcerRecurse}[\text{TFI}[d,0,\{\{2,\text{MH}\},\{1,0\},0,0,\{1,0\}\}]]}\)}

\noindent\(\frac{3 (d-3) \text{EL}^3 \text{MH}^4 \pmb{K}_{\{1,\text{MH}\}\{1,0\}\{1,0\}}^{(d)}}{16 \text{MW}^3 \text{SW}^3}\)

\noindent\(\pmb{\text{T13F}= \text{TarcerExpand}[\text{T13}, d\text{-$>$}4-\varepsilon ,0]]}\)

\noindent\(\frac{\text{EL}^3 \text{MH}^6 \left(\text{MH}^2\right)^{-\varepsilon } \pmb{S_{\varepsilon }}{}^2}{\text{MW}^3 \text{SW}^3}.\left(-\frac{3}{8
\varepsilon ^2}-\frac{3}{16 \varepsilon }-\frac{9 \zeta (2)}{32}-\frac{3}{32}\right)\)

\noindent\(\pmb{\text{AMP}[[14]]}\)

\noindent\(\left(\text{Integral}[\text{q}1,\text{q}2],\frac{\text{EL}^3
\text{MH}^6 \text{FeynAmpDenominator}\left(\frac{1}{(\text{q}1)^2-\text{MW}^2},\frac{1}{(\text{q}1)^2-\text{MW}^2},\frac{1}{(\text{q}2)^2-\text{MH}^2},\frac{1}{(\text{q}2-\text{q}1)^2-\text{MW}^2}\right)}{2048
\pi ^8 \text{MW}^3 \text{SW}^3}\right)\)

{\footnotesize
\noindent\(\pmb{\text{T14}=(1*\text{EL}{}^{\wedge}3*\text{MH}{}^{\wedge}6/(8*\text{MW}{}^{\wedge}3*\text{SW}{}^{\wedge}3))*\text{TarcerRecurse}[\text{TFI}[d,0,\{\{2,0\},\{1,\text{MH}\},0,0,\{1,0\}\}]]}\)}

\noindent\(-\frac{(d-3) \text{EL}^3 \text{MH}^4 \pmb{K}_{\{1,\text{MH}\}\{1,0\}\{1,0\}}^{(d)}}{8 \text{MW}^3 \text{SW}^3}\)

\noindent\(\pmb{\text{T14F}= \text{TarcerExpand}[\text{T14}, d\text{-$>$}4-\varepsilon ,0]}\)

\noindent\(\frac{\text{EL}^3 \text{MH}^6 \left(\text{MH}^2\right)^{-\varepsilon } \pmb{S_{\varepsilon }}{}^2}{\text{MW}^3 \text{SW}^3}.\left(\frac{1}{4
\varepsilon ^2}+\frac{1}{8 \varepsilon }+\frac{3 \zeta (2)}{16}+\frac{1}{16}\right)\)

\noindent\(\pmb{\text{AMP}[[15]]}\)

\noindent\(\left(\text{Integral}[\text{q}1,\text{q}2],\frac{3
\text{EL}^3 \text{MH}^4 \text{FeynAmpDenominator}\left(\frac{1}{(\text{q}1)^2-\text{MH}^2},\frac{1}{(\text{q}2)^2-\text{MH}^2},\frac{1}{(\text{q}1+\text{q}2)^2-\text{MH}^2}\right)}{4096
\pi ^8 \text{MW}^3 \text{SW}^3}\right)\)

{\footnotesize
\noindent\(\pmb{\text{T15}=(3*\text{EL}{}^{\wedge}3*\text{MH}{}^{\wedge}4/(16*\text{MW}{}^{\wedge}3*\text{SW}{}^{\wedge}3))*\text{TarcerRecurse}[\text{TFI}[d,0,\{\{1,\text{MH}\},\{1,\text{MH}\},0,0,\{1,\text{MH}\}\}]]}\)}

\noindent\(\frac{3 \text{EL}^3 \text{MH}^4 \pmb{K}_{\{1,\text{MH}\}\{1,\text{MH}\}\{1,\text{MH}\}}^{(d)}}{16 \text{MW}^3 \text{SW}^3}\)

\noindent\(\pmb{\text{AMP}[[16]]}\)

\noindent\(\left(\text{Integral}[\text{q}1,\text{q}2],\frac{\text{EL}^3
\text{MH}^4 \text{FeynAmpDenominator}\left(\frac{1}{(\text{q}1)^2-\text{MH}^2},\frac{1}{(\text{q}2)^2-\text{MZ}^2},\frac{1}{(\text{q}1+\text{q}2)^2-\text{MZ}^2}\right)}{4096
\pi ^8 \text{MW}^3 \text{SW}^3}\right)\)

{\footnotesize
\noindent\(\pmb{\text{T16}=(1*\text{EL}{}^{\wedge}3*\text{MH}{}^{\wedge}4/(16*\text{MW}{}^{\wedge}3*\text{SW}{}^{\wedge}3))*\text{TarcerRecurse}[\text{TFI}[d,0,\{\{1,\text{MH}\},\{1,0\},0,0,\{1,0\}\}]]}\)}

\noindent\(\frac{\text{EL}^3 \text{MH}^4 \pmb{K}_{\{1,\text{MH}\}\{1,0\}\{1,0\}}^{(d)}}{16 \text{MW}^3 \text{SW}^3}\)

\noindent\(\pmb{\text{T16F}= \text{TarcerExpand}[\text{T16}, d\text{-$>$}4-\varepsilon ,0]}\)

\noindent\(\frac{\text{EL}^3 \text{MH}^6 \left(\text{MH}^2\right)^{-\varepsilon } \pmb{S_{\varepsilon }}{}^2}{\text{MW}^3 \text{SW}^3}.\left(-\frac{1}{8
\varepsilon ^2}-\frac{3}{16 \varepsilon }-\frac{3 \zeta (2)}{32}-\frac{7}{32}\right)\)

\noindent\(\pmb{\text{AMP}[[17]]}\)

\noindent\(\left(\text{Integral}[\text{q}1,\text{q}2],\frac{\text{EL}^3
\text{MH}^4 \text{FeynAmpDenominator}\left(\frac{1}{(\text{q}1)^2-\text{MH}^2},\frac{1}{(\text{q}2)^2-\text{MW}^2},\frac{1}{(\text{q}1+\text{q}2)^2-\text{MW}^2}\right)}{2048
\pi ^8 \text{MW}^3 \text{SW}^3}\right)\)

{\footnotesize
\noindent\(\pmb{\text{T17}=(1*\text{EL}{}^{\wedge}3*\text{MH}{}^{\wedge}4/(8*\text{MW}{}^{\wedge}3*\text{SW}{}^{\wedge}3))*\text{TarcerRecurse}[\text{TFI}[d,0,\{\{1,\text{MH}\},\{1,0\},0,0,\{1,0\}\}]]}\)}

\noindent\(\frac{\text{EL}^3 \text{MH}^4 \pmb{K}_{\{1,\text{MH}\}\{1,0\}\{1,0\}}^{(d)}}{8 \text{MW}^3 \text{SW}^3}\)

\noindent\(\pmb{\text{T17F}= \text{TarcerExpand}[\text{T17}, d\text{-$>$}4-\varepsilon ,0]}\)

\noindent\(\frac{\text{EL}^3 \text{MH}^6 \left(\text{MH}^2\right)^{-\varepsilon } \pmb{S_{\varepsilon }}{}^2}{\text{MW}^3 \text{SW}^3}.\left(-\frac{1}{4
\varepsilon ^2}-\frac{3}{8 \varepsilon }-\frac{3 \zeta (2)}{16}-\frac{7}{16}\right)\)

\noindent\(\pmb{T=\text{T1}+\text{T11}+\text{T12}+\text{T13}+\text{T14}+\text{T16}+\text{T17} }\)

\noindent\(\frac{9 (d-2) \text{EL}^3 \text{MH}^2 \left(\pmb{A}_{\{1,\text{MH}\}}^{(d)}\right){}^2}{64 \text{MW}^3 \text{SW}^3}+\frac{3 (d-3) \text{EL}^3
\text{MH}^4 \pmb{K}_{\{1,\text{MH}\}\{1,0\}\{1,0\}}^{(d)}}{32 \text{MW}^3 \text{SW}^3}+\frac{3 \text{EL}^3 \text{MH}^4 \pmb{K}_{\{1,\text{MH}\}\{1,0\}\{1,0\}}^{(d)}}{16
\text{MW}^3 \text{SW}^3}\)

\noindent\(\pmb{\text{TFP}= \text{TarcerExpand}[T, d\text{-$>$}4-\varepsilon ,0]+\text{FullSimplify}[\text{T10}+\text{T15}] }\)

\noindent\(\frac{3 (3 d-7) \text{EL}^3 \text{MH}^4 \pmb{K}_{\{1,\text{MH}\}\{1,\text{MH}\}\{1,\text{MH}\}}^{(d)}}{32 \text{MW}^3 \text{SW}^3}+\frac{\text{EL}^3
\text{MH}^6 \left(\text{MH}^2\right)^{-\varepsilon } \pmb{S_{\varepsilon }}{}^2}{\text{MW}^3 \text{SW}^3}.\left(-\frac{27}{16 \varepsilon ^2}-\frac{39}{32
\varepsilon }-\frac{45 \zeta (2)}{64}-\frac{63}{64}\right)\)

\chapter{Computation of $D(\alpha)$ and $Q(\alpha, a_{1}, a_{2})$}\label{AppDyQ}
\markboth{APPENDIX \ref{AppDyQ}}{APPENDIX \ref{AppDyQ}}
In this appendix we are going to prove that
\begin{eqnarray}
G^{(d)}(q^2,a_{1},a_{2})=i^2\left ( \frac{\pi}{i} \right)^d \prod^{5}_{j=1} \frac{i^{-\nu_j}}{\Gamma(\nu_j)}\int_0^{\infty} \frac{d \alpha_j \alpha^{\nu_j-1}_j}{ [ D(\alpha) ]^{\frac{d}{2}}} ~~~~~~~~~~~~~~~~~~~~~~~~~~~~~~~~~~~~~~~~~~~~~~~~~ \nonumber \\
 \times \exp \left[ i \left(\frac{Q(\alpha,a_1,a_2)}{D(\alpha)}-\sum_{l=1}^{5}\alpha_l(m_l^2-i\epsilon)
                \right)\right], \nonumber \\         
         \label{represG}
\end{eqnarray}
where
\begin{eqnarray}
\label{Dform}
&&D(\alpha)=\alpha_5(\alpha_1+\alpha_2+\alpha_3+\alpha_4)
  +(\alpha_1+\alpha_3)(\alpha_2+\alpha_4),    \\
&& \nonumber \\
&&Q(\alpha,a_1,a_2)~~=~~ A(\alpha,a_1,a_2)q^2+B(\alpha,a_1,a_2),
\label{Q}
\end{eqnarray}
with 
\begin{eqnarray*}
&&A(\alpha,a_1,a_2)= (\alpha_1+\alpha_2)(\alpha_3+\alpha_4)
\alpha_5 +\alpha_1\alpha_2(\alpha_3+\alpha_4)
         +\alpha_3\alpha_4(\alpha_1+\alpha_2) \\
         && \nonumber \\
          && B(\alpha,a_1,a_2)=(qa_1)Q_1+(qa_2)Q_2+a_1^2Q_{11}
            +a_2^2Q_{22}+(a_1a_2)Q_{12} \nonumber
\end{eqnarray*}
and
\begin{eqnarray}
Q_1&=&\alpha_3\alpha_5+\alpha_4\alpha_5+\alpha_2\alpha_3
+\alpha_3\alpha_4, \nonumber \\
Q_2&=&\alpha_4\alpha_5+\alpha_3\alpha_5+\alpha_1\alpha_4
+\alpha_3\alpha_4, \nonumber \\
-4 Q_{11}&=&\alpha_2+\alpha_4+\alpha_5,
\nonumber \\
-4Q_{22}&=&\alpha_1+\alpha_3+\alpha_5,
\nonumber \\
-2Q_{12}&=&\alpha_5.
\end{eqnarray}
To compute the explicit expressions (\ref{Dform}) and (\ref{Q}), we begin with the $\alpha$ -representation of the integral (\ref{Int-G})
\begin{eqnarray}
&&G^{(d)}(q^2,a_{1},a_{2})\!= \int\int d^{d}k_{1}d^{d}k_{2} \! \prod^{5}_{j=1} \frac{i^{-\nu_j}}{\Gamma(\nu_j)}
\! \int_0^{\infty} d \alpha_j \alpha^{\nu_j-1}_j e^{i\alpha_{j}(k_{j}^{2}-m_{j}^{2}+i\varepsilon)}
       \exp \left[ia_{1}k_{1} + ia_{2}k_{2} \right]. \nonumber \\
        && \label{RepG1}
\end{eqnarray} 
We can rewrite the above expression as
\begin{eqnarray}
 G^{(d)}(q^2,a_{1},a_{2})\!= \! \prod^{5}_{j=1} \frac{i^{-\nu_j}}{\Gamma(\nu_j)}
\! \int_0^{\infty} d \alpha_j \alpha^{\nu_j-1}_j e^{i\sum_{i=1}^{5}\alpha_{i}(-m_{i}^{2}+i\varepsilon)} ~~~~~~~~~~~~~~~~~~~~~~~~~~~~~~~ \nonumber\\
\nonumber \\
    ~~~~\times ~~~~ \int\int d^{d}k_{1}d^{d}k_{2}\exp \left[i\sum_{i=1}^{5}\alpha_{i}k_{i}^{2} + ia_{1}k_{1} + ia_{2}k_{2} \right]. 
         \label{RepG2}
\end{eqnarray}
In (\ref{RepG2}) consider the sub-integral:
\begin{eqnarray}
\int\int d^{d}k_{1}d^{d}k_{2}\exp \left[i\sum_{i=1}^{5}\alpha_{i}k_{i}^{2} + ia_{1}k_{1} + ia_{2}k_{2} \right] ~~ = ~~~~~~~~~~~~~~~~~~~~~~~~~~~~~~~~~~~~~~~~~~ \\
\nonumber \\
\nonumber \\
 \int d^{d}k_{1} e^{\left[i\alpha_{1}k_{1}^{2} + i\alpha_{3}(k_{1}-q)^{2} + ia_{1}k_{1}\right]} \int d^{d}k_{2} e^{\left[i\alpha_{2}k_{2}^{2} + \alpha_{4}(k_{2}-q)^{2} + \alpha_{5}(k_{1}-k_{2})^{2} + ia_{2}k_{2} \right]}, \nonumber
         \label{RepG3}
\end{eqnarray}  
where we use that $k_{3}=k_{1}-q$, $k_{4}=k_{2}-q$  ~and~  $k_{5}=k_{1}-k_{2}$. The integral on the momentum $k_{2}$ is a Gaussian integral that, using the $d$ -dimensional Gaussian integration formula (\ref{GaussianI}), can be put in the form:
\begin{eqnarray}
\int d^{d}k_{2} e^{i\left(M k_{2}^{2} + 2pk_{2}+\alpha_{4}q^{2} + \alpha_{5}k_{1}^{2} \right)} = i\left(\dfrac{\pi}{iM} \right)^{d/2} \exp \left\lbrace -i \dfrac{p^{2}}{M}\right\rbrace   e^{i\left(\alpha_{4}q^{2} + \alpha_{5}k_{1}^{2}\right)} \label{RepG4}
\end{eqnarray}
with
\begin{eqnarray*}
 M = \alpha_{2}+\alpha_{4}+\alpha_{5} & ; & p = \dfrac{a_{2}}{2} - \alpha_{5}k_{1} - \alpha_{4}q.
\end{eqnarray*}
Therefore, 
\begin{eqnarray}
\int\int d^{d}k_{1}d^{d}k_{2}\exp \left[i\sum_{i=1}^{5}\alpha_{i}k_{i}^{2} + ia_{1}k_{1} + ia_{2}k_{2} \right] ~~ = ~~~~~~~~~~~~~~~~~~~~~~~~~~~~~~~~~~~~~~~\nonumber \\
\nonumber \\
\nonumber \\
i\left(\dfrac{\pi}{iM} \right)^{d/2} \int d^{d}k_{1} e^{\left[i\alpha_{1}k_{1}^{2} + i\alpha_{3}(k_{1}-q)^{2} + ia_{1}k_{1} + \alpha_{5}k_{1}^{2} + \alpha_{4}q^{2} \right]} \exp \left\lbrace -i \dfrac{p^{2}}{M}\right\rbrace.  \label{RepG5}
\end{eqnarray}
After some algebra and using (\ref{GaussianI}) we obtain 
\begin{eqnarray}
\int\int d^{d}k_{1}d^{d}k_{2}\exp \left[i\sum_{i=1}^{5}\alpha_{i}k_{i}^{2} + ia_{1}k_{1} + ia_{2}k_{2} \right] ~~ = ~~~~~~~~~~~~~~~~~~~~~~~~~~~~~~~~~~~~~~~\nonumber \\
\nonumber \\
\nonumber \\
i^{2}\left(\dfrac{\pi}{iM} \right)^{d/2} \left(\dfrac{\pi}{iN} \right)^{d/2} \exp \left\lbrace -i \dfrac{t^{2}}{N}\right\rbrace  e^{\left[i(\alpha_{4}+\alpha_{3}-\alpha_{4}^{2}/M)q^{2} + i(\alpha_{4}\alpha_{2}/M)q - ia_{2}^{2}/4M \right]}, \label{RepG6}
\end{eqnarray}
with
\begin{eqnarray*}
N = \alpha_{1}+\alpha_{3}+\alpha_{5} -\dfrac{\alpha_{5}^{2}}{M} & ; & t = \dfrac{1}{2}\left( a_{1} + \dfrac{\alpha_{5}a_{2}}{M} - 2\alpha_{3}q - \dfrac{2\alpha_{4}\alpha_{5}}{M}q\right).
\end{eqnarray*}
By replacing (\ref{RepG6}) into (\ref{RepG2}) we find: 
\begin{eqnarray}
G^{(d)}(q^2,a_{1},a_{2}) ~=~ i^{2}\left(\dfrac{\pi}{i} \right)^{d} \prod^{5}_{j=1} \frac{i^{-\nu_j}}{\Gamma(\nu_j)} \! \int_0^{\infty} d \alpha_j \alpha^{\nu_j-1}_j e^{i\sum_{i=1}^{5}\alpha_{i}(-m_{i}^{2}+i\varepsilon)} ~~~~~~~~~~~~~~~ \nonumber\\
\nonumber \\
~~~ \times ~~~ \left(\dfrac{1}{MN} \right)^{d/2} \exp \left\lbrace i \left(c_{0} + c_{1}q + c_{2}q^{2} \right)\right \rbrace, \label{RepG7}
\end{eqnarray}  
where
\begin{eqnarray*}
&& c_{0}= a_{1}^{2}\left(-\dfrac{1}{4N} \right) + a_{2}^{2}\left(-\dfrac{1}{4M} - \dfrac{\alpha_{5}^{2}}{4M^{2}N} \right) + a_{1}a_{2}\left(-\dfrac{\alpha_{5}}{2MN} \right),\\
&&\\
&&\\
&& c_{1}= a_{1}\left(\dfrac{\alpha_{3}}{N} + \dfrac{\alpha_{4}\alpha_{5}}{MN} \right) + a_{2}\left( \dfrac{\alpha_{4}}{M} + \dfrac{\alpha_{3}\alpha_{5}}{MN} + \dfrac{\alpha_{4}\alpha_{5}^{2}}{M^{2}N}\right), ~~ {\rm and} \\
&& \\
&& \\
&& c_{2} = \alpha_{4}+\alpha_{3}-\dfrac{\alpha_{4}^{2}}{M}-\dfrac{\alpha_{4}^{2}\alpha_{5}^{2}}{M^{2}N} - \dfrac{\alpha_{3}^{2}}{N} + \dfrac{2\alpha_{3}\alpha_{4}\alpha_{5}}{MN}.
\end{eqnarray*}
Defining the function $D(\alpha)$ by:
\begin{eqnarray}
D(\alpha) ~=~ MN ~=~  \left(\alpha_{2}+\alpha_{4}+\alpha_{5} \right)\left( \alpha_{1}+\alpha_{3}+\alpha_{5} -\dfrac{\alpha_{5}^{2}}{M} \right) ~~~~~~ ~~~\nonumber \\
\nonumber \\
~~=~~ \alpha_{5}(\alpha_{1} + \alpha_{2} + \alpha_{3}+ \alpha_{4}) + (\alpha_{2} + \alpha_{4})(\alpha_{1}+\alpha_{3}),
\end{eqnarray}
and the function $Q(\alpha, a_{1}, a_{2})$ by:
\begin{eqnarray}
Q(\alpha, a_{1}, a_{2}) ~=~ D(\alpha)(c_{0} + c_{1}q + c_{2}q^{2}) ~~~~~~~~~~~~~~~~~~~~~~~~~~~~~~~~~~~~~~~~~~~~~~~~~~~~~~~~~~ \nonumber \\
\nonumber \\ 
=~ a_{1}^{2}\left(-\dfrac{M}{4} \right) + a_{2}^{2}\left(-\dfrac{N}{4} - \dfrac{\alpha_{5}^{2}}{4M} \right) + a_{1}a_{2}\left(-\dfrac{\alpha_{5}}{2} \right)  ~~ + ~~ ~~~~~~~~~~~~~~~~~~~  \nonumber \\
\nonumber \\
qa_{1}\left(\alpha_{3}M + \alpha_{4}\alpha_{5}\right) + qa_{2}\left( \alpha_{4}N + \alpha_{3}\alpha_{5} + \dfrac{\alpha_{4}\alpha_{5}^{2}}{M}\right) ~~ + ~~ ~~~~~~~~~~\nonumber \\
\nonumber \\
~~~~~~ q^{2}\left(\alpha_{4}MN+\alpha_{3}MN-\alpha_{4}^{2}N-\dfrac{\alpha_{4}^{2}\alpha_{5}^{2}}{M} - \alpha_{3}^{2}M + 2\alpha_{3}\alpha_{4}\alpha_{5}\right),
\end{eqnarray}
it is easy to recognize that we obtained the claimed statement (\ref{represG}). In fact
\begin{eqnarray}
&&G^{(d)}(q^2,a_{1},a_{2})\!=i^2\!\left ( \frac{\pi}{i} \right)^d
\! \prod^{5}_{j=1} \frac{i^{-\nu_j}}{\Gamma(\nu_j)}
\! \int_0^{\infty} \!\!\!\! \ldots \!\! \int_0^{\infty}
\frac{d \alpha_j \alpha^{\nu_j-1}_j}
     { [ D(\alpha) ]^{\frac{d}{2}}} \nonumber \\
 && ~~~~~~~~~~~~~~~~~~~~~~~~~~~~~~~~~~~~~~~~~~ \times \exp \left[
              i \left(\frac{Q(\alpha,a_1,a_2)}
	                   {D(\alpha)}
             \!-\!\sum_{l=1}^{5}\alpha_l(m_l^2\!-\!i\epsilon)
                \right)
         \right], \nonumber \\
         && 
        \end{eqnarray}
where
\begin{eqnarray}
&&Q(\alpha,a_1,a_2)=[(\alpha_1+\alpha_2)(\alpha_3+\alpha_4)
\alpha_5 +\alpha_1\alpha_2(\alpha_3+\alpha_4)
         +\alpha_3\alpha_4(\alpha_1+\alpha_2)]q^2 ~~~\nonumber \\ 
         &&\nonumber \\
&&~~~~~~~~~~~~~~~~~~~~~~~~~~~~~~~~~~~+(qa_1)Q_1+(qa_2)Q_2+a_1^2Q_{11}
            +a_2^2Q_{22}+(a_1a_2)Q_{12}, \nonumber \\
            &&
\end{eqnarray}
and
\begin{eqnarray}
&&~~~~~Q_1 ~=~ \alpha_{3}M + \alpha_{4}\alpha_{5} ~=~ \alpha_3\alpha_5+\alpha_4\alpha_5+\alpha_2\alpha_3
+\alpha_3\alpha_4, \nonumber \\
&&~~~~~Q_2 ~=~  \alpha_{4}N + \alpha_{3}\alpha_{5} + \dfrac{\alpha_{4}\alpha_{5}^{2}}{M} ~=~ \alpha_4\alpha_5+\alpha_3\alpha_5+\alpha_1\alpha_4
+\alpha_3\alpha_4, \nonumber \\
&&-4 Q_{11} ~=~ M ~=~ \alpha_2+\alpha_4+\alpha_5, \nonumber \\
&&-4Q_{22} ~=~ N + \dfrac{\alpha_{5}^{2}}{M} ~=~ \alpha_1 + \alpha_3 + \alpha_5 , \nonumber \\
&&-2Q_{12} ~=~\alpha_5.
\end{eqnarray}

\chapter{$\varepsilon$-Expansion of One-Loop Two-Point Feynman Diagrams}\label{App1lEpsExpansion}
\markboth{APPENDIX \ref{App1lEpsExpansion}}{APPENDIX \ref{App1lEpsExpansion}} 
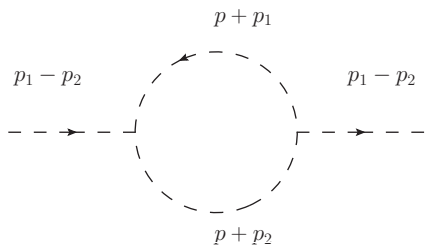
\begin{figure}
\begin{center}
\scalebox{0.5}{
\fcolorbox{white}{white}{
  \begin{picture}(307,191) (214,-74)
    \SetWidth{1.0}
    \SetColor{Black}
    \Arc[dash,dashsize=10,arrow,arrowpos=0.5,arrowlength=5,arrowwidth=2,arrowinset=0.2](370,6)(60.415,294,654)
    \Line[dash,dashsize=10,arrow,arrowpos=0.5,arrowlength=5,arrowwidth=2,arrowinset=0.2](215,6)(310,6)
    \Line[dash,dashsize=10,arrow,arrowpos=0.5,arrowlength=5,arrowwidth=2,arrowinset=0.2](430,6)(525,6)
    \Text(370,86)[lb]{\Large{\Black{$p + p_{1}$}}}
    \Text(370,-79)[lb]{\Large{\Black{$p + p_{2}$}}}
    \Text(220,41)[lb]{\Large{\Black{$p_{1}- p_{2}$}}}
    \Text(470,41)[lb]{\Large{\Black{$p_{1} - p_{2}$}}}
  \end{picture}
}}
\end{center}
\caption{\label{1l2PFunction} Topology for two-point one-loop scalar integral.}
\end{figure}

In this appendix we will expose a geometrical way to calculate the dimensionally regulated two-point one-loop integral (Figure \ref{1l2PFunction}) 
\begin{eqnarray}
B_{11}^{(d)} = \int \dfrac{d^{d}p}{\pi^{d/2}} \dfrac{1}{((p+p_{1})^{2}-m_{1}^{2})((p+p_{2})^{2}-m_{2}^{2})}
\end{eqnarray}
using the Laurent expansion up to first order in $\varepsilon$. We begin with the usual Feynman parametric representation for the propagators
\begin{eqnarray}
B_{11}^{(d)} = \int \dfrac{d^{d}p}{\pi^{d/2}} \int_{0}^{1} \int_{0}^{1} dx_{1}dx_{2}\dfrac{\delta(x_{1}+x_{2}-1)}{(x_{1}A_{1}+x_{2}A_{2})^{2}}, \label{1l2PF-1}
\end{eqnarray}  
where
\begin{eqnarray*}
A_{1}= (p+p_{1})^{2} - m_{1}^{2}   &~;~&  A_{2}=(p+p_{2})^{2} - m_{2}^{2}.
\end{eqnarray*}
Therefore, the denominator can be written as
\begin{eqnarray}
x_{1}A_{1}+x_{2}A_{2} = x_{1}((p+p_{1})^{2} - m_{1}^{2})+x_{2}((p+p_{2})^{2} - m_{2}^{2}) \nonumber \\
~ =  ~ x_{1}(p+p_{1})^{2} + x_{2}(p+p_{2})^{2} - \sum_{i=1}^{2}m_{i}^{2}x_{i}. ~~~
\end{eqnarray} 
After some trivial algebra and using that $x_{1}+x_{2}=1$ we can put the denominator in the form:
\begin{eqnarray}
x_{1}A_{1}+x_{2}A_{2} = l^{2} + \Delta ,
\end{eqnarray}
where
\begin{eqnarray*}
&& l = (p + p_{2}) + x_{1}(p_{1}-p_{2}), \\
&& \\
&& \Delta = - x_{1}^{2}(p_{1}-p_{2})^{2} + x_{1}p_{1}^{2} + x_{1}p_{2}^{2} - 2x_{1}p_{2}p_{1} - \sum_{i=1}^{2}m_{i}^{2}x_{i},\\
&& ~~~= x_{1}x_{2}(p_{1}-p_{2})^{2} - \sum_{i=1}^{2}m_{i}^{2}x_{i}.
\end{eqnarray*}
Defining $k_{12}=(p_{1}-p_{2})$ we have for $\Delta$:
\begin{eqnarray}
\Delta = \sum_{j=1, j<l}^{2}~\sum_{l=1}^{2}x_{j}x_{l}k_{jl}^{2} - \sum_{i=1}^{2} x_{i}m_{i}^{2}.
\end{eqnarray}
In this way, the integral (\ref{1l2PF-1}) can be put in the form:
\begin{eqnarray}
&&B_{11}^{(d)} = \int_{0}^{1} \int_{0}^{1} dx_{1}dx_{2}\int \dfrac{d^{d}l}{\pi^{d/2}}\dfrac{\delta(x_{1}+x_{2}-1)}{[l^{2}+\Delta]^{2}}, \nonumber \\
&&\nonumber \\
&& ~~~~~~=~\Gamma\left(2 - \dfrac{d}{2} \right) \int_{0}^{1}\int_{0}^{1}  \prod_{i=1}^{2} dx_{i}\dfrac{\delta(\sum_{j=1}^{2} x_{j}-1)}{\Delta ^{2-d/2}}. \label{1l2PF-2}
\end{eqnarray}
Now, using $\sum_{i=1}^{2}x_{i}=1$ can be rewriting $\Delta$ in a homogeneous form,
\begin{eqnarray}
&&\Delta = x_{1}x_{2}k_{12}^{2} - (x_{1}+x_{2})(x_{1}m_{1}+x_{2}m_{2}) \nonumber \\
&& ~~~ = -(x_{1}^{2}m_{1}^{2}+x_{2}^{2}m_{2}^{2}) - 2x_{1}x_{2}\dfrac{(m_{1}^{2}+m_{2}^{2}-k_{12}^{2})}{2m_{1}m_{2}}m_{1}m_{2} \nonumber \\
&& ~~~ = - \left[ \sum_{i=1}^{2}x_{i}^{2}m_{i}^{2} + 2\sum_{j=1, j<l}^{2}~\sum_{l=1}^{2}x_{j}x_{l}m_{j}m_{l}c_{jl}\right],
\end{eqnarray} 
where $c_{jl}=\dfrac{m_{j}^{2}+m_{l}^{2}-k_{jl}^{2}}{2m_{j}m_{l}}$, i.e. 
\begin{eqnarray}
k_{jl}^{2} = m_{j}^{2}+m_{l}^{2} - 2m_{j}m_{l}c_{jl}. \label{CosinesLaw}
\end{eqnarray} 

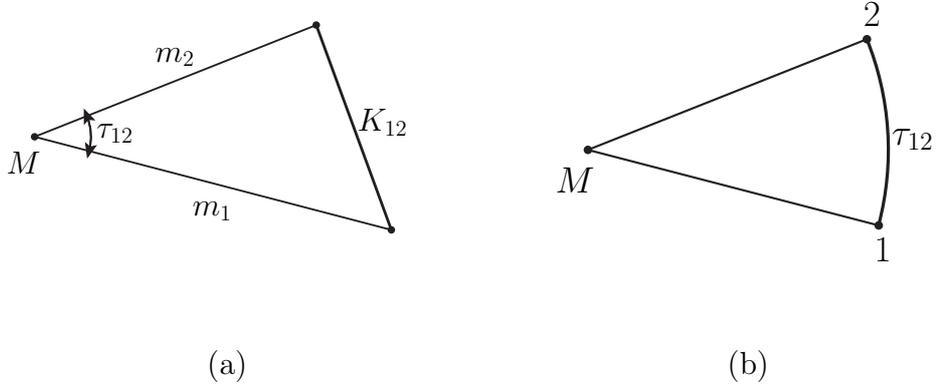
\begin{figure}
\begin{center}
\[
\begin{array}{cc}
{
\begin{picture}(180,150)(-18,-75)
\scalebox{0.7}{
\Vertex(0,0){2}
  \Text(-6,-14)[]{\Large{$M$}}  
  \LongArrowArc(0,0)(30,-14.5,22)
  \LongArrowArcn(0,0)(30,22,-14.5) 
  \Text(43,2)[]{\Large{$\tau_{12}$}}
\Vertex(190,-50){2}
\Vertex(150,60){2}
\Line(0,0)(190,-50)
  \Text(95,-40)[]{\Large{$m_1$}}
\Line(0,0)(150,60)
  \Text(75,43)[]{\Large{$m_2$}}
\SetWidth{1.5}
\Line(190,-50)(150,60)
  \Text(186,8)[]{\Large{$K_{12}$}}
\SetWidth{0.5}}
\end{picture}
} & {
\begin{picture}(190,150)(-35,-70)
\scalebox{0.8}{
\Vertex(0,0){2}
  \Text(-6,-14)[]{\Large{$M$}}
\Vertex(135.5,-35.5){2}
  \Text(138,-47)[]{\Large{1}}
\Vertex(130,52){2}
  \Text(133,64)[]{\Large{2}} 
\Line(0,0)(135.5,-35.5)
\Line(0,0)(130,52)
\SetWidth{1.5}    
\CArc(0,0)(140,-14.5,22)
  \Text(152,5)[]{\Large{$\tau_{12}$}}
\SetWidth{0.5}}
\end{picture}
} \\
{\mbox{(a)}} & {\mbox{(b)}}
\end{array}
\]
\\
\caption{\label{TriangleArc}{\small Two-point case: (a) the basic triangle and
        (b) the arc $\tau_{12}$ \cite{Davydychev2}}}
\end{center}
\end{figure}
In the region between $k_{jl}^2=(m_j-m_l)^2$ and  $k_{jl}^2=(m_j+m_l)^2$, we have $|c_{jl}|<1$, and therefore in this region $c_{jl}$ can be understood as cosines of some angles $\tau_{jl}$,
\begin{eqnarray}
\label{def_costau}
c_{jl} = \cos\tau_{jl} 
= \left\{ \begin{array}{c} \;\; 1, \;\;\; k_{jl}^2=(m_j-m_l)^2 \\
                               -1, \;\;\; k_{jl}^2=(m_j+m_l)^2 
          \end{array} \right. \; .
\end{eqnarray}
The corresponding angles $\tau_{jl}$ are
\begin{eqnarray}
\label{def_tau}
\tau_{jl}= \arccos(c_{jl}) 
=\arccos\left(\frac{m_j^2+m_l^2-k_{jl}^2}{2m_j m_l}\right)
= \left\{ \begin{array}{c}   0, \;\;\; k_{jl}^2=(m_j-m_l)^2 \\
                           \pi, \;\;\; k_{jl}^2=(m_j+m_l)^2
          \end{array} \right. \; .
\end{eqnarray}
Then the expression (\ref{CosinesLaw}) is just the law of cosines represented geometrically by the Figure~\ref{TriangleArc}. If we plot the height of the triangle on the Figure \ref{TriangleArc} (a) and we call it $m_{0}$, we obtain the two right triangles:
\begin{center}
\scalebox{0.7}{
\fcolorbox{white}{white}{
  \begin{picture}(266,116) (223,-111)
    \SetWidth{1.0}
    \SetColor{Black}
    \Line[](224,-108)(304,4)
    \Line[](304,4)(304,-108)
    \Line[](224,-108)(304,-108)
    \Line[](400,4)(400,-108)
    \Line[](400,-108)(480,-108)
    \Line[](400,4)(480,-108)
    \Arc[](294.3,-21.7)(12.735,-164.982,-46.909)
    \Arc[](409.184,-24.289)(10.839,-159.981,-3.759)
    \Arc[clock](235.8,-103)(13.185,80.395,-22.286)
    \Arc[](470.6,-98.76)(12.623,121.524,227.055)
    \Text(222,-66)[lb]{\Large{\Black{$m_{1}$}}}
    \Text(316,-66)[lb]{\Large{\Black{$m_{0}$}}}
    \Text(373,-66)[lb]{\Large{\Black{$m_{0}$}}}
    \Text(454,-66)[lb]{\Large{\Black{$m_{2}$}}}
    \Text(442,-92)[lb]{\Large{\Black{$\beta$}}}
    \Text(410,-50)[lb]{\Large{\Black{$\tau_{02}$}}}
    \Text(282,-50)[lb]{\Large{\Black{$\tau_{01}$}}}
    \Text(252,-90)[lb]{\Large{\Black{$\alpha$}}}
  \end{picture}
}}
\end{center}
so that
\begin{eqnarray}
\dfrac{sin\tau_{12}}{k_{12}} = \dfrac{sin\beta}{m_{1}} = \dfrac{m_{0}}{m_{1}m_{2}}, \label{m0}
\end{eqnarray}
therefore $m_{0}=\dfrac{m_{1}m_{2}sin \tau_{12}}{k_{12}}$. By other side, we have the additional relations 
\begin{eqnarray}
cos \tau_{01}=\dfrac{m_{0}}{m_{1}}  & ; & cos \tau _{02} = \dfrac{m_{0}}{m_{2}} .
\end{eqnarray}
After apply the above substitutions and evaluate the delta function, the integral (\ref{1l2PF-2}) becomes:
\begin{eqnarray}
B_{11}^{(d)} ~=~ \Gamma\left(2 - \dfrac{d}{2} \right) \int_{0}^{1} dx_{1}\dfrac{1}{[x_{1}^{2}m_{1}^{2}+(1-x_{1})^{2}m_{2}^{2}+2x_{1}(1-x_{1})m_{1}m_{2}c_{12}]^{2-\frac{d}{2}}}. \label{1l2PF-3}
\end{eqnarray}
The denominator $\Delta$ must be modified again in a convenient way:
\begin{eqnarray}
&&\Delta = x_{1}^{2}(m_{1}^{2}+m_{2}^{2}-2m_{1}m_{2}c_{12}) + x_{1}(2(m_{1}m_{2}c_{12}-m_{2}^{2})) + m_{2}^{2} \nonumber \\
&& \nonumber \\
&& ~~~= (x_{1}-x_{+})(x_{1}-x_{-})
\end{eqnarray}
where 
\begin{eqnarray}
x_{\pm} = \dfrac{-b^{2} \pm \sqrt{b^{2}-4k_{12}^{2}m_{2}^{2}}}{2k_{12}^{2}},
\end{eqnarray}
with $b ~=~ 2(m_{1}m_{2}c_{12}-m_{2}^{2}) ~=~ m_{1}^{2} - m_{2}^{2} - k_{12}^{2} $. The discriminant $D^{2} ~=~ b^{2}-4k_{12}^{2}m_{2}^{2}$ is just the well known Kallen's function  
\begin{eqnarray*}
D^{2} = \lambda (m_{1}^{2},m_{2}^{2},k_{12}^{2}) = m_{1}^{4}+m_{2}^{4}+k_{12}^{4} -2(m_{1}^{2}m_{2}^{2}+m_{1}^{2}k_{12}^{2}+m_{2}^{2}k_{12}^{2}).
\end{eqnarray*} 
Using that $sin^{2}\tau_{12}~=~1 - cos^{2}\tau_{12}$, one can easily show that
\[m_{1}^{2}m_{2}^{2}sin^{2}\tau_{12}~=~\dfrac{-\lambda (m_{1}^{2},m_{2}^{2},k_{12}^{2})}{4}.\]
Employing the equation (\ref{m0})  we find
\[ \lambda (m_{1}^{2},m_{2}^{2},k_{12}^{2}) = - 4m_{0}^{2}k_{12}^{2}.\]
Thus, we have two complex roots:
\[x_{\pm} = \dfrac{k_{12}^{2}+m_{2}^{2}-m_{1}^{2}}{2k_{12}^{2}} \pm i\dfrac{m_{0}}{k_{12}}.\]
Now, we rewrite the denominator as
\begin{eqnarray}
\Delta = (x_{1}-x_{+})(x_{1}-x_{-}) = (a+ib)(a-ib) = a^{2} + b^{2},
\end{eqnarray}
where 
\begin{eqnarray*}
a = x_{1} - \dfrac{k_{12}^{2}+m_{2}^{2}-m_{1}^{2}}{2k_{12}^{2}}  &~;~& b=\dfrac{m_{0}}{k_{12}}.
\end{eqnarray*}
If we make the substitution $a=btan\theta$, then $dx_{1}=\frac{m_{0}}{k_{12}}sec^{2}\theta d\theta$, $\Delta=\frac{m_{0}^{2}}{k_{12}^{2}}cos^{-2}\theta$ and the integral (\ref{1l2PF-3}) becomes 
\begin{eqnarray}
B_{11}^{(d)} ~=~ \Gamma\left(2 - \dfrac{d}{2} \right) \left(\dfrac{m_{0}}{k_{12}} \right)^{d-3} \int_{\theta_{1}}^{\theta_{2}} d\theta \dfrac{d\theta}{cos^{d-2}\theta}. 
\end{eqnarray}  
To understood the integration limits in the above equation we use the substitution
\[tan\theta = \dfrac{x_{1} - ((k_{12}^{2}+m_{2}^{2}-m_{1}^{2})/2k_{12}^{2})}{m_{0}/k_{12}}.\]
When $x_{1}= 0$ , $\theta = \theta_{1}$ 
\begin{eqnarray}
\therefore ~~ tan \theta _{1} =  -\dfrac{k_{12}^{2}+m_{2}^{2}-m_{1}^{2}}{2m_{0}k_{12}} ~=~ - tan \tau_{02}.
\end{eqnarray}
When $x_{1}=1$ , $\theta = \theta_{2}$
\begin{eqnarray}
\therefore ~~ tan \theta _{2} =  -\dfrac{k_{12}^{2}-m_{2}^{2}+m_{1}^{2}}{2m_{0}k_{12}} ~=~ tan \tau_{01}.
\end{eqnarray}
In this way 
\begin{eqnarray}
B_{11}^{(d)} ~=~ \Gamma\left(2 - \dfrac{d}{2} \right) \left(\dfrac{m_{0}}{k_{12}} \right)^{d-3} \int_{-\tau_{02}}^{\tau_{01}} d\theta \dfrac{d\theta}{cos^{d-2}\theta}. 
\end{eqnarray} 
That can be written finally as:
\begin{eqnarray}
B_{11}^{(d)} ~=~ \Gamma\left(2 - \dfrac{d}{2} \right) \left(\dfrac{m_{0}}{k_{12}} \right)^{d-3}\left\lbrace \Omega_{1}^{(2;d)} + \Omega_{2}^{(2;d)} \right\rbrace,
\end{eqnarray} 
where 
\begin{eqnarray}
\Omega_{i}^{(2;d)} = \int_{0}^{\tau_{0i}}\dfrac{d\theta}{cos^{d-2}\theta}. 
\end{eqnarray}
The integral $\Omega_{i}^{(2;d)}$ is known in terms of hypergeometric functions \cite{Davydychev2}. Expanding, we get:
\begin{eqnarray}
\label{ep-exp}
\int_0^{\tau}
\frac{\mbox{d}\theta}{\cos^{d-2}\theta}
= \sum\limits_{j=0}^{\infty} \frac{(4-d)^j}{j!}
\int_0^{\tau}
\frac{\mbox{d}\theta}{\cos^2\theta}
\ln^j(\cos\theta) 
= \sum\limits_{j=0}^{\infty} \frac{(d-4)^j}{j!}
f_j(\tau),
\end{eqnarray}
where
\begin{eqnarray}
\label{f_j}
f_j(\tau) \equiv (-1)^j
\int_0^{\tau}
\frac{\mbox{d}\theta}{\cos^2\theta}
\ln^j(\cos\theta).
\end{eqnarray}
The three first lowest terms of the expansion are \cite{Davydychev2}
\begin{eqnarray}
f_0(\tau) &=& \tan\tau, \\
\label{f_1}
f_1(\tau) &=& -\tan\tau \ln(\cos\tau)-\tan\tau +\tau, \\
\label{f_2}
f_2(\tau) &=& 
\tan\tau \left[\ln^2(\cos\tau)\!+\!2\ln(\cos\tau)\!+\!2 \right]
-2\tau (1\!-\!\ln 2) - \mbox{Cl}_2\left(\pi\!-\!2\tau\right).
\end{eqnarray}
The Clausen function of order 2\,\, $Cl_2(x)$ has the following integral representation
\begin{equation}
\mbox{Cl}_2(x)=-\int_0^x\log\left(2\sin\frac{t}{2}\right)dt, \label{Clausen}
\end{equation} 
and satisfies the following relations
\begin{equation}
\label{Cl2-L}
\mbox{Cl}_2\left(\theta\right) = - 2L\left(\frac{\pi-\theta}{2}\right)
+ (\pi-\theta)\ln 2 , \hspace{10mm}
L(\theta)= -{\textstyle{1\over2}} 
\mbox{Cl}_2\left(\pi-2\theta\right) + \theta \ln 2 ,   
\end{equation}
with $L(\theta)$ the Lobachevsky function, defined in equation (\ref{Lobachev}) of Appendix \ref{AppIntegralJ}. If we consider dilogarithm of a general complex argument, the Clausen function is related to the imaginary part of the dilogarithm:
\begin{eqnarray}
\mbox{Cl}_{2}(\theta)=Im\left[Li_{2}\left(e^{i\theta} \right) \right]=\dfrac{1}{2}
\left[\mbox{Cl}_{2}(2\theta)+\mbox{Cl}_{2}(2\omega)+\mbox{Cl}_{2}(2\xi) \right]
\end{eqnarray}
with $tan\omega=sin\theta/(1-cos\theta)$ and $\xi = \pi -\theta - \omega$. More detailed information about dilogarithm and related functions can be found in \cite{Lewin}.

\chapter{Evaluation of J$^{(d)}_{111}$($q^{2}=0$)}\label{AppIntegralJ}
\markboth{APPENDIX \ref{AppIntegralJ}}{APPENDIX \ref{AppIntegralJ}} 

In this appendix we solve the integral  $J^{(d)}_{111}(0)$ (with three different masses $m_{1}^{2}=x$, $m_{2}^{2}=y$ and $m_{3}^{2}=z$) using the Tarasov method exposed in Chapter~\ref{cha:TwoLoopCalc}, in conjunction with the method of characteristics used in \cite{Ford-Jack} by C. Ford and I. Jack. 
 
We start from the Tarasov identity (\ref{TarasovRecurrence}) applied to $J^{(d)}_{111}(0)$ for a particular election of parameters $x_{l}$: 
\begin{eqnarray}
0=\int\int d^dk_{1}\,d^dk_{2}\,
\sum_{r=1}^{2}{\partial \over {\partial k_{r}^{\mu}}}\left[{k_{1}^{\mu} - k_{2}^{\mu}\over{(k_{1}^2-x)(k_{2}^2-y)((k_{1}+k_{2})^2-z)}}\right].
\label{IBPoverI}
\end{eqnarray}
From Equation~(\ref{IBPoverI}) it follows that
\begin{eqnarray}
\label{IDiffIBP} (d-3)J^{(d)}_{111} + \kappa(x,y,z) - 2xJ^{(d)}_{211} - (z+x-y)J^{(d)}_{112} = ~~~~~~~~~~~~~~~~~~~~~~\\
\nonumber \\
(d-3)J^{(d)}_{111} - \kappa(x,y,z) - 2yJ^{(d)}_{121} - (y+z-x)J^{(d)}_{112} \nonumber 
\end{eqnarray}
where

$$\kappa(x,y,z)=- A_{2}^{(d)}(z)(A_{1}^{(d)}(x)-A_{1}^{(d)}(y)).$$

Now from Equation~(\ref{IDiffIBP}) and similar equations produced by $(x,y,z)$ cyclic permutations, we obtain the system of equations:
\begin{eqnarray}
\kappa(x,y,z)= xJ_{211}^{(d)} - yJ_{121}^{(d)} + (x-y)J_{112}^{(d)} \nonumber \\
\nonumber \\
\kappa(y,z,x)= yJ_{211}^{(d)} - zJ_{121}^{(d)} + (y-z)J_{112}^{(d)} \nonumber \\
\nonumber \\
\kappa(z,x,y)= zJ_{211}^{(d)} - xJ_{121}^{(d)} + (z-x)J_{112}^{(d)} \label{kappaeqs}
\end{eqnarray}
The $\alpha$-parametric representation $G^{(d)}(0,0,0)$ given by the relation (\ref{represG}) shows that $J^{(d)}_{111}(0)$ is invariant under cyclic permutations of the masses (x, y, z). Therefore, if we perform the sum of the three equations in (\ref{kappaeqs}), we obtain
\begin{eqnarray}
(y-z)J_{211} + (z-x)J_{121} + (x-y)J_{112} = K(x,y,z) \label{KappaRel},
\end{eqnarray}
where we has defined
\begin{eqnarray*}
K(x,y,z) ~=~ \kappa(x,y,z)+ \kappa(y,z,x)+ \kappa(z,x,y).
\end{eqnarray*}
Using the expansion (\ref{IntegralJ}) for $A_{1}^{(d)}$ and his first derivative
\begin{eqnarray}
A_{2}^{(d)}(z) = \dfrac{\partial A_{1}(z)}{\partial z} = i\Gamma(2-{d\over2})(z)^{{d\over2}-2},
\end{eqnarray}
is easy obtain the explicit value of $K(x, y, z)$. The expansion obtained is:
\begin{eqnarray}
 K(x,y,z) = - \Gamma(2-{d\over 2})\Gamma(1-{d\over2})~~~~~~~~~~~~~~~~~~~~~~~~~~~~~~~~~~~~~~~~~~~~~~~~~~~~~~~~~~~~~ \\
 \nonumber \\
 \times \left[(zx)^{d/2 -2}(z-x)+(xy)^{d/2 - 2}(x-y)+(yz)^{d/2 - 2}(y-z)\right]. \nonumber
\end{eqnarray} 
To solve the first-order partial differential equation (\ref{KappaRel}) we use the method of characteristic described in Chapter \ref{cha:Effective Potential}. If a particular parametrization $t$ of the curves is fixed, the equations of the characteristic curve is the system of ordinary differential equations:
\begin{equation}
dt={dx\over{y-z}}={dy\over{z-x}}={dz\over{x-y}}={dJ_{111}\over K}, \label{EqCharact}
\end{equation}
subject to initial conditions $x=X$, $y=Y$ and $z=0$ (at $t=0$), where we will suppose without loss of generality that $X\geq Y$. We then have:
\begin{eqnarray}
J_{111}(x,y,z)=J_{111}(X,Y,0)+\int^t_0dt' \,K(x(t'),y(t'),z(t')). \label{SolutionI}
\end{eqnarray}
Using equation (\ref{KappaRel}) and equation (\ref{EqCharact}), we can rewrite equation (\ref{SolutionI}) as
\begin{eqnarray}
J_{111}(x,y,z)=J_{111}(X,Y,0)-\Gamma'
\biggl[\int^x_Xdx\,(yz)^{\frac{d}{2}-2}+\int^y_Ydy\,(zx)^{\frac{d}{2}-2}+
\int^z_0dz\,(xy)^{\frac{d}{2}-2}\biggr], \label{SolutionI2}
\end{eqnarray}
where
$$\Gamma' = \Gamma(2-{d\over 2})\Gamma(1-{d\over2}).$$
From characteristic equations (\ref{EqCharact}) it follows that {\it for all\/} $t$
\begin{eqnarray}
x^2+y^2+z^2&=d^2=X^2+Y^2 \label{IRel1}\\
x+y+z&=c=X+Y \label{IRel2}
\end{eqnarray}
where $c$ and $d$ are constants. Squaring $c$ we obtain
\begin{eqnarray}
c^{2}-d^{2}=2xy + 2yz + 2xz = 2(x+y+z)z -2z^{2} + 2xy = 2cz - 2z^{2} + 2xy,
\end{eqnarray}
therefore, 
\begin{eqnarray}
xy = z^{2} - cz + \dfrac{1}{2}(c^{2}-d^{2}) = z^{2} - cz + \dfrac{c^{2}}{4} - \left(\dfrac{d^{2}}{2}-\dfrac{c^{2}}{4}\right)~~~~~~~~~~~~~~~~~~~\nonumber \\
\nonumber \\
 = ~~~ \left(\dfrac{c}{2}-z \right)^{2} - \left(\dfrac{d^{2}}{2}-\dfrac{c^{2}}{4}\right)~~~ = ~~~ s^{2} - a^{2},
\end{eqnarray}
where
\begin{eqnarray}
s = \dfrac{c}{2}-z & ; & a=\sqrt{{d^2\over 2}-{c^2\over 4}}={1\over 2}(X-Y)= {1\over2}
(x^2+y^2+z^2-2xy-2yz-2zx)^{1\over2}. \nonumber \\ 
\label{IRel3}
\end{eqnarray}
Similar equations for $yz$ and $zx$ are satisfied. Making the variable change (\ref{IRel3}) (at~$z=0$, $s=c/2$ and at~$z=z~\rightarrow~s=c/2 - z$) hence equation (\ref{SolutionI2}) becomes
\begin{eqnarray}
J_{111}(x,y,z)=J_{111}(X,Y,0)-\Gamma'\Biggl[\biggl(\int^{x-{c\over2}}_a+\int^a_{{c\over
2}-y}+\int^{c\over2}_{{c\over2}-z}\biggr)\ ds\,(s^2-a^2)^{\frac{d}{2}-2}\Biggr]. \label{SolutionI3}
\end{eqnarray}
We must evaluate $J_{111}(X,Y,0)$. By elementary methods it is also tricky to evaluate, hence we employ the method of characteristics once again. $J_{111}(X,Y,0)$ satisfies the PDE 
\begin{eqnarray}
K(X,Y,0) = YJ_{211} -XJ_{121} \label{KappaRel2}.
\end{eqnarray}
Thereby, we most solve the ODE's system 
\begin{equation}
dt={dX\over{Y}}=-{dY\over{X}}={dJ_{111}(X,Y,0)\over K}, \label{EqCharact2}
\end{equation}
subject to initial conditions $Y=0$ and $X=X-Y$ at $t=0$. The general solution is:
\begin{eqnarray}
J_{111}(X,Y,0)=J_{111}(X-Y,0,0)+\int^t_0dt' \,K(X(t'),Y(t'),0), \label{SolI}
\end{eqnarray} 
where 

$$ K(X,Y,0)=-\Gamma'\left\lbrace (XY)^{\frac{d}{2}-2} (X-Y)\right\rbrace.$$

Using the equation (\ref{EqCharact2}), the integral $J_{111}(X,Y,0)$ can be written as:
\begin{eqnarray}
J_{111}(X,Y,0)=J_{111}(X-Y,0,0)+\Gamma'
\biggl[\int^Y_0dY\,(XY)^{\frac{d}{2}-2}+\int^{X}_{X-Y} dX\,(XY)^{\frac{d}{2}-2}\biggr]. \label{SolI2}
\end{eqnarray}
In this case
\begin{eqnarray}
XY=\dfrac{1}{2}(c^{2}-d^{2}) = \dfrac{1}{4}c^{2} - \left(\dfrac{d^{2}}{2} - \dfrac{c^{2}}{4}\right) = s^{2} - a^{2},
\end{eqnarray}
with 
\begin{eqnarray}
s = \dfrac{c}{2} = \dfrac{X+Y}{2}& ; & a=\sqrt{{d^2\over 2}-{c^2\over 4}}={1\over 2}(X-Y). \label{VarChange}
\end{eqnarray}
Making the variables change given by the l.h.s. of (\ref{VarChange}), the integrals in equation (\ref{SolI2}) transforms in 
\begin{eqnarray}
\int^Y_0dY\,(XY)^{\frac{d}{2}-2} + \int^{X}_{X-Y} dX\,(XY)^{\frac{d}{2}-2} ~\longrightarrow ~ 2\int^{\frac{1}{2}(X+Y)}_{\frac{X}{2}} ds\,(s^{2}-\frac{1}{4}(X-Y)^{2})^{\frac{d}{2}-2} \\
= \int^{\frac{1}{2}(X+Y)}_{\frac{1}{2}(X-Y)} ds\,(s^{2}-\frac{1}{4}(X-Y)^{2})^{\frac{d}{2}-2}. \nonumber
\end{eqnarray}
Therefore, we finally obtain
\begin{eqnarray}
J_{111}(X,Y,0)=J_{111}(X-Y,0,0)+\Gamma'\int^{{1\over2}(X+Y)}_{{1\over2}(X-Y)}ds\,
\bigl[s^2-{1\over4}(X-Y)^2\bigr]^{\frac{d}{2}-2}. \label{SolutionI4}
\end{eqnarray}
Substituting equation (\ref{SolutionI4}) in equation (\ref{SolutionI3}) and using equation (\ref{IRel2}) and equation (\ref{IRel3}) we obtain
\begin{eqnarray}
J_{111}(x,y,z)=J_{111}(X-Y,0,0)-\Gamma'\Biggl[\biggl(\int^{x-{c\over2}}_a+\int^a_{c\over
2}-\int^{{c\over2}-y}_{a}-\int^{{c\over2}-z}_{c\over2}\biggr)\ ds\,(s^2-a^2)^{\frac{d}{2}-2}\Biggr] \nonumber\\
\nonumber\\
=~~ J_{111}(2a,0,0)~~+~~\Gamma'\Bigl[F({c\over2}-y)
+F({c\over2}-z)-F(x-{c\over2})\Bigr], ~~~~~~~~~~~~~~~~~~\label{SolutionI5}
\end{eqnarray}
where
\begin{eqnarray}
F(w)=\int^w_ads\,(s^2-a^2)^{\frac{d}{2}-2}.
\end{eqnarray}
Since $J_{111}(2a,0,0)$ can be evaluated by elementary methods, we have reduced the problem to a single integral, $F(w)$. However equation (\ref{SolutionI5}) is only valid in the region $a^2\geq 0$. In the region $a^2\leq 0$, it is possible to derive the following form of the solution:
\begin{eqnarray}
J_{111}(x,y,z)=-J_{111}(2b,0,0)\sin{\pi d\over 2} +\Gamma'\Bigl[G({c\over2}-x)+G({c
\over2}-y)+G({c\over 2}-z)\Bigr] \label{SolutionI6}
\end{eqnarray}
where
\begin{eqnarray}
G(w)=\int^w_0ds\,(s^2+b^2)^{\frac{d}{2}-2}\quad \hbox{and}\quad 
b^2=-a^2.
\end{eqnarray}
Note that for $a^2=0$, which in $x$, $y$, $z$ space is a cone with its apex at the origin, the integral is trivial. By other side, the integral $J_{111}(m,0,0)$ can be computed using the standard Feynman parameters 
\begin{eqnarray}
\dfrac{1}{[k_{2}^{2}-y][(k_{1}+k_{2})^{2}-z]}=\int_{0}^{1}dx\dfrac{1}{\left[x((k_{1}+k_{2})^{2}-z)+(1-x)(k_{2}^{2}-y)\right]^{2}},
\end{eqnarray}
therefore
\begin{eqnarray}
J_{111}(m,0,0)=\dfrac{1}{\pi^{d}}\int d^{d}k_{1}\left[\int_{0}^{1}dx \int d^{d}l \dfrac{1}{[l^{2}-\Delta]^{2}} \right]\dfrac{1}{(k_{1}^{2}-m)},
\end{eqnarray}
where we make the substitutions $l=k_{2}+k_{1}x$ and $\Delta=k_{1}^{2}x(x-1)$. Solving the integral between brackets and using the relations
\begin{eqnarray}
\dfrac{1}{(k_{1}^{2}-m)^{2}(k_{1}^{2})^{2-d/2}} = \int_{0}^{1}dw \dfrac{w^{1-d/2}}{[wk_{1}^{2}+(1-w)(k_{1}^{2}-m)]^{3-d/2}}\times \dfrac{\Gamma (3-d/2)}{\Gamma(2-d/2)\Gamma(1)}, 
\end{eqnarray}
we obtain
\begin{eqnarray}
J_{111}(m,0,0)=\dfrac{1}{\pi ^{d}}\int_{0}^{1}dx \dfrac{i\pi^{d/2}\Gamma (2-d/2)}{\Gamma(2)}\left(\dfrac{1}{x(x-1)} \right)^{2-d/2} ~~~~~~~~~~~~~~~~~~~~ \nonumber \\
\nonumber \\
\times \int_{0}^{1}dw \int d^{d}k_{1} \dfrac{w^{1-d/2}}{[k_{1}^{2}-(1-w)m]^{3-d/2}}\dfrac{\Gamma(3-d/2)}{\Gamma(2-d/2)\Gamma(1)}.
\end{eqnarray}
Recalling that 
\begin{eqnarray}
\int d^{d}k_{1} \dfrac{1}{[k_{1}^{2}-(1-w)m]^{3-d/2}}=\pi^{d/2}\dfrac{(-1)^{3-d/2}i\Gamma(3-d)}{\Gamma(3-d/2)}\left(\dfrac{1}{(1-w)m} \right)^{3-d}
\end{eqnarray}
we find after some algebra:
\begin{eqnarray}
&&J_{111}(m,0,0)= -\Gamma(3-d)\int_{0}^{1}dx x^{d/2-2}(1-x)^{d/2-2}\int_{0}^{1}dw w^{1-d/2}(1-w)^{d-3}\left(\dfrac{1}{m} \right)^{3-d}. \nonumber \\
&& 
\end{eqnarray}
But the Beta function is defined by
\begin{eqnarray}
B(\alpha,\beta) = \int_{0}^{1}dz z^{\alpha - 1}(1-z)^{\beta - 1},
\end{eqnarray}
and satisfies the important relation
\begin{eqnarray}
B(\alpha,\beta)=\dfrac{\Gamma(\alpha)\Gamma(\beta)}{\Gamma(\alpha +\beta)} &~;~& \alpha , \beta > 0, 
\end{eqnarray}
then we can easily find that
\begin{eqnarray}
J_{111}(m,0,0)=-{\Gamma(2-{d\over2})
\Gamma(3-d)\Gamma({d\over2}-1)^2\over{\Gamma({d\over2})}}
\bigl({m}\bigr) ^{d-3}. \label{Jm00}
\end{eqnarray}
Using the equation (\ref{Jm00}) and writing
\begin{eqnarray}
(s^2+b^2)^{\frac{d}{2}-2}=1+(\frac{d}{2}-2)\ln (s^2+b^2)+
{1\over 2}(d/2 - 2)^2\ln^2(s^2+b^2)+...
\end{eqnarray}
we obtain from equation (\ref{SolutionI6}) that
\begin{eqnarray}
 J_{111}(x,y,z)= +{c\over {2(d/2 - 2)^2}} +{1\over{(d/2 - 2)}}\bigl(
{3c\over2}-L_1\bigl)+{1\over2}\Bigl[L_2-6L_1+(y+z-x)
\overline{\ln}y \overline{\ln}z  \nonumber \\
\nonumber \\ 
~+~(z+x-y) \overline {\ln}z\overline{\ln}x +(y+x-z)\overline{\ln}y
\overline{\ln}x +\xi (x,y,z)+c(7+\zeta(2))\Bigr] ~~~~~~~~~\label{SolutionI7}
\end{eqnarray}
where
\begin{eqnarray}
L_m=x\overline{\ln}^{m} x + y\overline{\ln}^{m} y + z\overline{\ln}^{m} z,
\end{eqnarray}
with
\begin{eqnarray}
\overline{\ln}X= ln X + \gamma,  
\end{eqnarray}
and
\begin{eqnarray}
\xi  (x,y,z)=8b\Bigl[ L(\theta_x)+L(\theta_y)+L(\theta_z)-{\pi\over2}\ln2\Bigr]. \label{XiRel}
\end{eqnarray}
$L(t)$ is Lobachevsky's function \cite{Davydychev2}, defined as
\begin{eqnarray}
L(t)=-\int^t_0dx\,\ln\cos x. \label{Lobachev}
\end{eqnarray} 
The angles $\theta_x$, $\theta_y$, $\theta_z$ are given by 
\begin{eqnarray}
\theta_x=\tan^{-1}\bigl({{c\over2}-x\over b}\bigr) \hbox{ etc.}
\end{eqnarray}
Equation (\ref{XiRel}) is valid only in the region $a^2\leq 0$ (ie. inside the
cone). For $a^2>0$, we obtain from equation~(\ref{SolutionI5}) a result identical to equation~(\ref{SolutionI7}) except that now
\begin{eqnarray}
\xi(x,y,z)=8a\left[-M(-\theta_x)+M(\theta_y)+M(\theta_z)\right]
\end{eqnarray}
where
\begin{eqnarray}
M(t)=-\int^t_0dx\,\ln\sinh x
\end{eqnarray}
and $\theta_x$, $\theta_y$, $\theta_z$ are given by
\begin{eqnarray}
\theta_x=\coth^{-1}\bigl({{c\over2}-x\over a}\bigl).
\end{eqnarray}

\end{document}